\documentclass[rivista]{cimento}

\usepackage{amsmath,amssymb}
\usepackage{graphicx}  
\usepackage{epstopdf}
\usepackage{cite}
\usepackage{mathrsfs}
\usepackage{calrsfs}
\usepackage{txfonts}

\newcommand{\ra}{\rightarrow}
\newcommand{\lora}{\longrightarrow}
\newcommand{\vev}[1]{\langle #1 \rangle}
\newcommand{\qhat}{{\hat q}}
\newcommand{\Etrf}{\!{E_\text{\it trf}}}
\newcommand{\bay}{\begin{array}}
\newcommand{\eay}{\end{array}}
\newcommand{\OO}{{\mathcal O}}
\newcommand{\ds}{\displaystyle}
\newcommand{\NN}{{\cal N}}

\providecommand{\mean}[1]{\ensuremath{\left<#1\right>}}
\providecommand{\sqrtsnn}{\sqrt{s_{_{\rm NN}}}}

\def\cO#1{{{\cal{O}}}\left(#1\right)}
\def\pt{p_{T}}
\def\kt{k_{T}}
\def\lqcd{\Lambda_{_{\rm QCD}}}
\def\alphas{\alpha_{_S}}
\def\dd{\rm d}
\def\raa{R_{AA}}
\def\gampi{\gamma$--$\pi^0} 
\def\z{z_{_{\gamma \pi}}}
\def\picut{p_{_{T_\pi}}^{\rm cut}}
\def\gacut{p_{_{T_\gamma}}^{\rm cut}}
\def\ptpi{p_{_{T_\pi}}}
\def\ptgamma{p_{_{T_\gamma}}}
\def\sqrtsnn{\sqrt{s_{NN}}}

\title{Parton Propagation and Fragmentation in QCD Matter 
}
\author{
Alberto~Accardi\from{ins:HU}\ETC,
Fran\c{c}ois~Arleo\from{ins:lapth},
William K. Brooks\from{ins:d}, 
\\ \hspace*{0cm} 
David~D'Enterria\from{ins:c}, 
\atque
Valeria~Muccifora\from{ins:y}\thanks{ valeria.muccifora@lnf.infn.it}}
\instlist{
  \inst{ins:HU} Hampton University, Hampton, VA, 23668, USA, and 
  Jefferson Lab, Newport News, VA 23606, USA 
  \inst{ins:lapth}LAPTH\footnote{Laboratoire d'Annecy-le-Vieux de Physique Th\'eorique, UMR5108}, 
   Universit\'e de Savoie, CNRS, BP 110, 74941 Annecy-le-Vieux cedex, France  
  \inst{ins:d} Departamento de F\'{\i}sica y Centro de Estudios
    Subat\'omicos, Universidad T\'ecnica Federico Santa Mar\'{\i}a, 
    Casilla 110-V, Valpara\'iso, Chile 
  \inst{ins:c} CERN, PH-EP, CH-1211 Geneva 23, Switzerland, and LNS, MIT, Cambridge, MA 02139-4307, USA
  \inst{ins:y} INFN-Laboratori Nazionali di Frascati, via E.Fermi 40, 00044 Frascati (Rome), Italy
}

\PACSes{
\PACSit{24.85.+p}{Quarks, gluons, and QCD in nuclear reactions}
\PACSit{13.87.Fh}{Fragmentation into hadrons}
\PACSit{25.30.-c}{Lepton-induced reactions}
\PACSit{13.85.-t}{Hadron-induced high- and super-high-energy interactions}
\PACSit{25.75.-q}{Relativistic heavy-ion collisions}
\PACSit{12.38.Mh}{Quark-gluon plasma}
}
\begin{document}

\maketitle

\begin{abstract}
We review recent progress in the study of parton propagation,
interaction and fragmentation in both cold and hot strongly interacting matter. 
Experimental highlights on high-energy hadron production in deep inelastic 
lepton-nucleus scattering, proton-nucleus and heavy-ion collisions, as well as 
Drell-Yan processes in hadron-nucleus collisions are presented.  
The existing theoretical frameworks for describing the in-medium
interaction of energetic partons and the space-time evolution of their
fragmentation into hadrons are discussed and confronted to experimental
data. We conclude with a list of theoretical and experimental open issues,
and a brief description of future relevant experiments and facilities.
\end{abstract}

\newpage

\tableofcontents




\section{Introduction}
\label{sec:introduction}

The transition from coloured partons (quarks and gluons) to colourless hadrons -- the so-called 
hadronisation process -- is an exemplary process of the fundamental theory 
of the strong interaction, Quantum Chromo-Dynamics (QCD), which still lacks a quantitative 
understanding from 
first principle calculations. The process by which a highly virtual parton radiates gluons or splits 
into a quark-antiquark pair can be theoretically described by QCD evolution equations such as 
the DGLAP (Dokshitzer-Gribov-Lipatov-Altarelli-Parisi) equations~\cite{Gribov:1972ri,Altarelli:1977zs,Dokshitzer:1977sg}.
However, the final ``bleaching'' of partons into hadrons takes place at a low virtualities 
($Q\approx \Lambda_{\rm QCD}\approx$~0.2~GeV) and so is dominated by nonperturbative 
QCD effects which cannot be theoretically addressed with the existing perturbative techniques. Modeling 
and phenomenology -- e.g. as implemented in the Lund string~\cite{Andersson:1998tv} or 
cluster fragmentation~\cite{Marchesini:1987cf} approaches -- are often used to describe 
hadronisation processes.

One way to study fragmentation and hadronisation is to perturb the environment surrounding the 
hard-scattered parton by introducing a nuclear medium~\cite{Bjorken:1976mk}. The nuclear 
medium provides a sensitive probe of parton evolution through the influence of initial-state (IS) 
and/or final-state (FS) interactions. Such IS and FS may result on modifications of the final hadron
yield distributions compared to ``vacuum'' production and can help us understand for example
the time-scale of the hadronisation process~\cite{Brodsky:1988uf}. 

Nuclear modifications of hadron production have been indeed observed in Deep Inelastic lepton-nucleus 
Scattering nDIS ($\ell^\pm+A$), in hadron-nucleus ($h+A$) and in heavy-ion ($A+A$) collisions, 
compared to ``elementary'' DIS on a proton target or proton-proton collisions. 
In nDIS and $h+A$ collisions, the medium is the nuclear target itself, also called ``cold QCD matter''. 
In $A+A$ reactions, the produced parton must in addition traverse the created hot and dense medium 
(``hot QCD matter''), be it a hadron gas at low temperature, or a Quark-Gluon Plasma (QGP) at high temperatures. 
In all cases, at high enough $p_T$ where hadrons mostly come from parton fragmentation, one typically observes 
two different phenomena: (i) a suppression of hadron multiplicities, 
called hadron or jet quenching, and (ii) a broadening of hadron
transverse momentum spectra, which induces a local enhancement of the hadron
$p_T$ spectrum known as ``Cronin
effect''~\cite{Cronin:1974zm,Antreasyan:1978cw}. Such nuclear effects
are due to elastic and inelastic interactions of the incoming or
outgoing partons and/or of the produced hadrons while traversing the
surrounding medium. 

In nDIS, the target nucleus allows one to test the hadronisation mechanism and 
colour confinement dynamics in a clean environment. Knowledge of partonic in-medium propagation 
gained from nDIS can be used in  Drell-Yan (DY) lepton pair production in $h+A$ collisions 
to factor out FS effects (such as medium-induced gluon radiation) from IS effects (such as nuclear modifications of parton distributions). 
A precise knowledge of parton propagation and hadronisation mechanisms obtained from 
nuclear DIS and DY studies can be very useful for e.g. testing and calibrating theoretical tools used 
to determine the properties of the QGP produced in high-energy heavy-ion collisions, as well as
to reduce the systematic uncertainties in neutrino experiments with nuclear targets.

This review is structured as follows. 
For a non technical overview of motivations, lessons learned, and an
outlook, the remainder of this introduction (where we discuss hadronisation in
elementary collisions, cold and hot QCD matter) can be followed by a
reading of the concluding Section~\ref{sec:Conclusions}. 
The intervening sections detail the experimental and theoretical
status, and discuss open issues and future experimental
possibilities. 
In Section~\ref{sec:variables-observables}, we 
define the relevant observables and kinematic variables for nDIS and
hadronic collisions, 
comparing the phase-spaces for hadron production in these two cases;
and we discuss the space-time development of  hadronisation,
introducing various estimates of the 
hadron formation time.
In Sections~\ref{sec:hadrons-lA}, \ref{sec:hadrons-hA}, and
\ref{sec:hadrons-AA}, we review the most relevant experimental results.
Specifically, in Section~\ref{sec:hadrons-lA} hadron production
in $\nu$-nucleus, $\mu$-nucleus, and $e-$nucleus DIS with emphasis on
recent HERMES and CLAS data are discussed. 
Sections~\ref{sec:hadrons-hA} and \ref{sec:hadrons-AA} are devoted to high-$p_T$ 
hadron production in $h+h$, $h+A$ and $A+A$ collisions, focusing on recent results  
from the Relativistic Heavy Ion Collider (RHIC).
The existing theoretical frameworks for interpreting the experimental
data based on partonic or hadronic degrees of freedom, are discussed respectively 
in Sections~\ref{sec:parton} and \ref{sec:prehadron}.
Specifically, parton propagation and energy loss in both $cold$ and $hot$
QCD matter is addressed in Section~\ref{sec:parton}, where data from DIS, DY, 
$h+A$ and $A+A$ collisions are confronted to different theoretical models.
In Section~\ref{sec:prehadron}, the interaction of  the prehadronic
system and of the formed hadron with the nuclear medium
is discussed. 
Finally, in Section~\ref{sec:future}, we discuss observables sensitive to 
the time scales and different mechanisms involved in parton
propagation and hadronisation, along with experimental
measurements at future facilities that would help to clarify the whole
picture.

\subsection{Parton fragmentation in elementary collisions } \ \\
\label{sec:hadronisation-elementary}

In perturbative Quantum Chromodynamics (pQCD), collinear factorisation
theorems~\cite{Collins:1989gx} allow one to explicitly
separate the short and long distance QCD  
dynamics involved in the mechanism of hadron production from parton
fragmentation.  In a general ${\rm H_1}+{\rm H_2}$
inelastic collision, one writes the inclusive hadronic cross sections 
for production of a hadron $h$ at large momentum transfer or
``hard'' scale $Q^2$, as
\begin{align}\begin{split}
  &\sigma_\text{hard}({{\rm H_1}+{\rm H_2} \,\ra\, h + X}) \\
  &  \qquad = \sum_{f_i,j = \{q,g\}} 
    \left[\!\,\prod_{i=0,N} \phi_{f_i|H_i}(x_i,M^2) \right] 
    \otimes \hat H^{\{ f_i \} \,\ra\,j+X}_\text{hard}
    (\{x_i\},z_j,\mu^2)
    \otimes D_{j\ra h} (z;M_F^2) \ ,
 \label{eq:fact}
\end{split}\end{align}
where $\otimes$ denotes a convolution over the kinematical internal
variables of the process and $N$ is the number of hadrons in the initial state. 
In Eq.~\eqref{eq:fact}, $Q^2$ is the typical 
hard scale of the process and $\hat H_{\rm hard}$ is the
short-distance and  
perturbatively calculable hard coefficient function for the
$\{f_i\}\ra j+X$ partonic process. The long-distance dynamics is
factorised into 
(a) the Parton Distribution Functions (PDF) $\phi_{f_i|H_i}(x_i)$, which
can be interpreted as the probability of finding a parton of flavour $f_i$ and 
momentum fraction $x_i$ inside the projectile ($H_1$) and/or target ($H_2$) hadron, and (b) the
Fragmentation Function (FF) $D_{j \ra h}$, which gives the equivalent
``probability'' that the parton $j$ fragments into the observed hadron $h$
with fractional momentum $z$. 
These functions are non-perturbative and need to be
extracted from experimental data. Typically, PDFs are extracted from 
``global QCD fits'' of inclusive hadron production in lepton-nucleon DIS
[$N=1$ and $D_{j\ra h} (z;Q^2)$~=~1 in Eq.~\eqref{eq:fact}]
At large hadron fractional momenta\footnote{At small $z$, successful QCD resummation 
techniques (e.g. the Modified Leading Logarithmic Approximation, MLLA~\cite{Dokshitzer:1991wu}) 
have been also developed to describe the evolution of a highly-virtual 
time-like partons into final hadrons.} $z=p_{\rm hadron}/p_{\rm parton}\gtrsim 0.1$, 
the FFs obey DGLAP evolution equations and are obtained from 
electron-positron annihilation into hadrons [$N=0$ in Eq.~\eqref{eq:fact}].
The obtained PDFs and FFs are provided by various authors, e.g., 
CTEQ6.6, MRST/MSTW~\cite{Nadolsky:2008zw,Martin:2004ir,Watt:2008hi},
and HKNS, DSS, AKK08~\cite{Hirai:2007cx,deFlorian:2007aj,Albino:2008fy,Albino:2008af,Albino:2008afa}
respectively, to mention the most recent sets.  
Once they are known at a given scale $Q^2_0$ their value at any other
scale can be perturbatively computed by means of the DGLAP evolution
equations \cite{Dokshitzer:1977sg,Gribov:1972ri,Altarelli:1977zs}. The factorisation scales 
$M^2$ and $M_F^2$ entering PDFs and FFs, as well as the renormalisation scale $\mu^2$ 
in the perturbative cross section, should be $\cO{Q^2}$ in order to avoid large logarithmic corrections. 

\begin{figure}
  \centering
  \includegraphics[width=12cm]{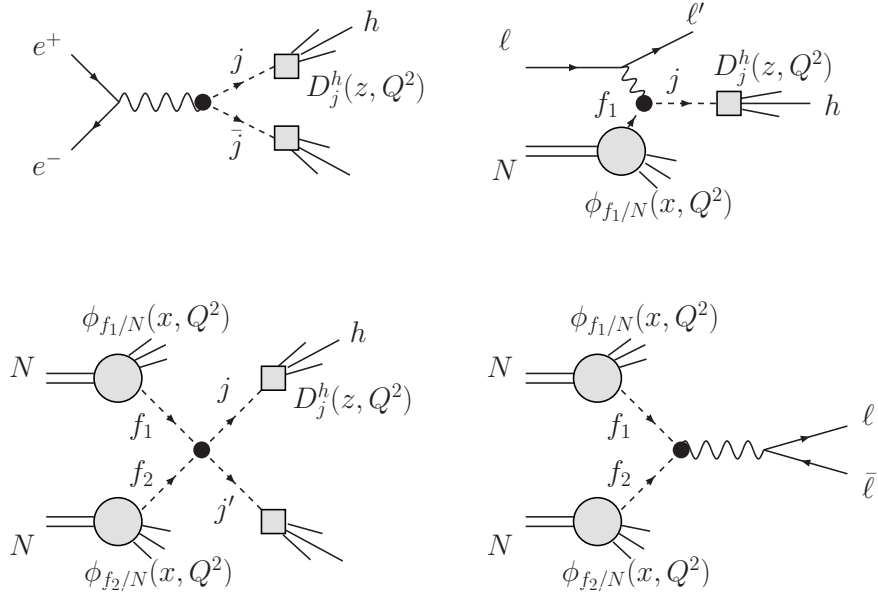}
  \caption{Illustration of universality of PDFs ($\phi_{f/N}$) and FFs ($D_{j \ra h}$) in leading order
    processes. Clockwise from top left: $e^++e^-$ annihilation, Deep
    Inelastic Scattering (DIS), lepton pair (Drell-Yan) emission, and
    hadron production in hadronic collisions. Solid lines indicate
    leptons, dashed lines quarks. The small black disc represents the
    perturbatively calculable hard interaction coefficient $\hat
    H_{\rm hard}$. } 
  \label{fig:universality}
\end{figure}

An important consequence of factorisation theorems is that PDFs and
FFs are universal i.e. process-independent. 
The measured FFs in $e^+ + e^- \ra h+X$ and the 
PDFs in $e^\pm + p \ra h+X$ can then be used to compute observables in
any other process, e.g., hadron spectra in proton-proton collisions [$N=2$ in 
Eq.~\eqref{eq:fact}], or Drell-Yan production of lepton pairs [$N=2$
and $j=\ell \bar\ell$ in Eq.~\eqref{eq:fact}], see
Fig.~\ref{fig:universality}.
When dealing with hadron production with nuclear systems, universality is 
however experimentally observed to breakdown: the details of the 
hadron production cross sections depend on the collision process that yields 
the final particles, as we discuss next.

\subsection{Parton propagation and fragmentation in cold and hot QCD matter} \ \\ 

The basic assumption behind the factorised form of Eq.~(\ref{eq:fact}) is that the characteristic time 
of the parton-parton interaction is much shorter than any long-distance interaction occurring before (among partons 
belonging to the same PDF) or after (during the evolution of the struck partons into their hadronic final-state) 
the hard collision itself. In that case, 
one can treat each nucleus as a collection of free partons, i.e., in the absence of initial-state effects 
the parton density in a nucleus with mass number $A$ is expected to be simply equivalent to that of a 
superposition of $A$ independent nucleons\footnote{In reality, nuclear PDFs are modified compared to 
proton PDFs by initial-state ``(anti)shadowing'' effects (see~\cite{Armesto:2006ph} for a recent review).}:
$\phi_{a/A}(x,Q^2)=A\cdot \phi_{a/N}(x,Q^2)$. In addition,  in the absence of final-state effects the 
parton fragments with universal FFs and, therefore, 
the pQCD factorisation theorem for collisions involving nuclei $A$ predicts that
minimum-bias inclusive hard cross sections scale respectively as
\begin{align}
\begin{split}
  d\sigma_{hard} (l,h+A\rightarrow h + X) 
    & \, = \, A \,\, d\sigma_{hard}(l,h+p\rightarrow h + X)\;, \\
  d\sigma_{hard} (A+A\rightarrow h + X) 
    & \, = \, A^2 \, d\sigma_{hard}(p+p\rightarrow h+ X)\;.
\end{split}
\label{eq:Afactorisation}
\end{align}

\begin{figure}[tb]
 \centerline{
  \includegraphics[height=5.cm]{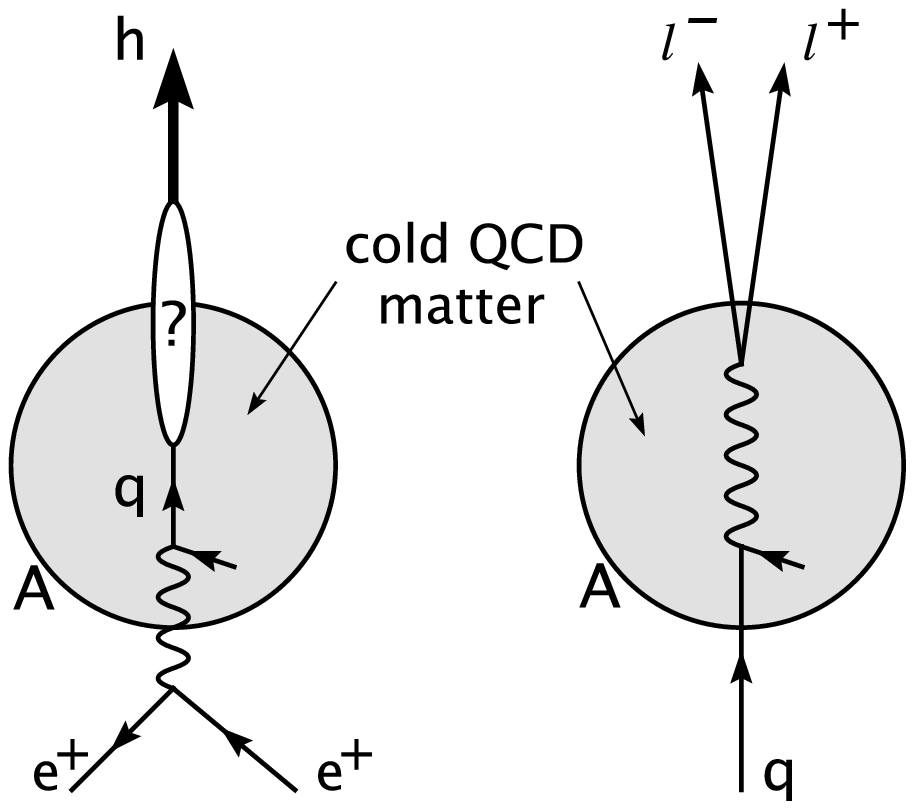}
  \hspace*{.3cm}
  \includegraphics[height=5.cm]{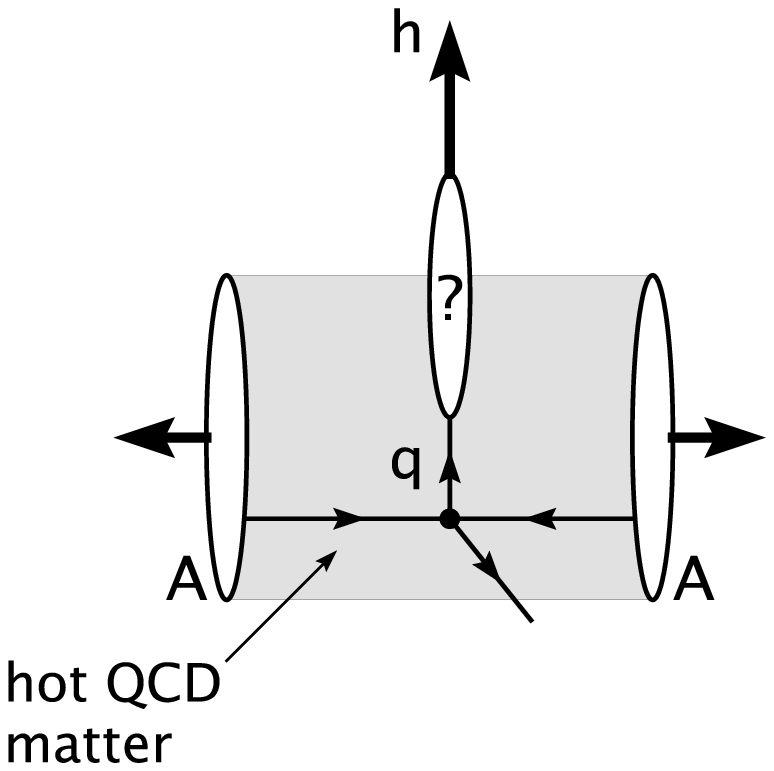}
 }
 \caption[]{Quark propagation inside a target nucleus (``cold QCD matter'')
   in lepton-nucleus ({\it left}) and hadron-nucleus $\ra$ Drell-Yan ({\it centre})
   collisions. {\it Right:} Hard scattered parton traveling through the
   ``hot QCD matter'' produced in a nucleus-nucleus collision. 
 \label{fig:nuke}
 }
\end{figure}

The cleanest environment to test the validity of Eqs.~\eqref{eq:Afactorisation} and
study possible nuclear modifications of hadron production (i.e. ``violations'' of the
expectations given by Eqs.~(\ref{eq:Afactorisation}))
is nuclear Deep Inelastic Scattering (nDIS). In nDIS processes one experimentally controls
many kinematic variables; the nuclear medium (i.e., the nucleus itself) is well known; 
and the particle multiplicity in the final state is low, leading to precise measurements. 
The nucleons act as femtometer-scale detectors of the scattered hadronising quark, allowing one 
to study its space-time evolution into the observed hadrons 
(Fig.~\ref{fig:nuke}, left).  
The relevant observable in semi-inclusive nDIS processes is the ratio of the single hadron multiplicity 
on a target of mass number $A$ normalised to the multiplicity on a deuteron target.
At leading order, this multiplicity ratio corresponds to good
approximation to the ratio of fragmentation functions 
(FF) in cold nuclear matter (the nucleus A) over that in the ``vacuum'' (deuteron). 
Recent HERMES measurements show that this ratio is significantly
below 1 clearly showing a breakdown of universality for fragmentation
functions, see 
Section~\ref{sec:hadrons-lA} and Refs.~\cite{Airapetian:2000ks,Airapetian:2003mi,Airapetian:2007vu,Airapetian:2005yh}.

A complementary means to study parton propagation in cold QCD matter is
by measuring the Drell-Yan (DY) process in hadron-nucleus collisions: 
$h+A \ra \ell^+ \ell^- + X$, where hadronisation does not play a role (Fig.~\ref{fig:nuke}, centre). If the invariant 
mass of the lepton pair is large, the process can be described perturbatively
as a parton-parton scattering producing a virtual photon which
subsequently decays into the lepton pair. Any modification of this process 
will come from initial-state nuclear interactions of the projectile parton inside the target 
(as well as from nuclear modifications of the PDF, which can be isolated by other means).

Tests of pQCD factorisation in hot-dense QCD matter can be carried out studying
high-$p_T$ hadron production in head-on nucleus-nucleus reactions (Fig.~\ref{fig:nuke}, right). 
The suppression of large transverse momentum hadron production in $A+A$  
compared to proton-proton ($p+p$) and hadron-nucleus $h+A$ collisions
at RHIC~\cite{Arsene:2004fa,Back:2004je,Adams:2005dq,Adcox:2004mh}
(see Sect.~\ref{sec:hadrons-AA}), is also indicative of a
breakdown of the universality of the fragmentation process. 
The standard explanation is that the observed suppression is due to
parton energy loss in the strongly interacting matter. This
assumes of course that the quenched light-quarks and gluons are long-lived enough 
to traverse the medium before hadronising, 
which can be expected at large enough $p_T$ because of the 
Lorentz boost of the hadronisation time scales. 
However, dynamical effects may 
alter this argument (see, e.g., Ref.~\cite{Kopeliovich:2003py}), with hadronisation starting at 
the nuclear radius scale or before. In this case, in-medium hadron 
interactions should also be accounted for, 
possibly leading to a different suppression pattern. 
Such mechanisms may be especially important in the case of heavy (charm, 
bottom) quarks which -- being slower than light-quarks or gluons --
can fragment into $D$ or $B$ mesons still inside the
plasma~\cite{Adil:2006ra}. 

In summary, a precise knowledge of parton propagation and hadronisation
mechanisms can be obtained from nDIS and DY data, allowing one to test 
the hadronisation mechanism and colour confinement dynamics. 
In addition, such cold QCD matter data are essential for testing and
calibrating our theoretical tools, and to determine the (thermo)dynamical 
properties of the QGP produced in high-energy nuclear interactions.

\subsection{Hadronisation and colour confinement} \ \\

While not having a direct bearing on the traditional topics of
confinement such as the hadron spectrum,
the hadronisation process nonetheless contains elements that are
central to the heart of colour confinement, as already emphasised 30
years ago by Bjorken~\cite{Bjorken:1976mk}. 
For instance, in the DIS process, a quark is briefly liberated from
being associated with any specific hadron while traveling as a ``free''
particle, and it is the mechanisms involved in hadron formation that
enforces the
colour charge neutrality and confinement into the final state hadron.
The dynamic mechanism leading to colour neutralisation, which is only
implicitly assumed in the traditional treatments of confinement based
on potential models 
\cite{Carlson:1983rw} or lattice QCD~\cite{Hatsuda:1994pi}, 
can be studied quantitatively using the
theoretical and experimental techniques discussed in this review. As
an example, the lifetime of the freely propagating quark may be inferred
experimentally from the nuclear modification of hadron production on
cold nuclei, which act as ``detectors'' of the hadronisation
process. Finally, as already discussed, the behaviour of partons
propagating through the medium created in high energy heavy-ion
collisions can give tomographically insight into the properties of
large-scale deconfined QCD matter (i.e., of the Quark-Gluon Plasma). 
While still at an early stage, the 
understanding of such elements will ultimately provide deeper insights
into the confinement-related properties of QCD.  

\subsection{Hadronisation and neutrino oscillations} \ \\

Neutrino oscillation experiments 
use nuclear targets to enhance the neutrino detection rate. 
Nuclear effects change  the topology and total energy 
of the experimentally measured hadronic final-state and
are known to be one of the largest sources of systematic errors in
current analyses. 
For the lower energy oscillation experiments that use the
quasi-elastic channel, several problems arise among which
the distortion of the knock-out nucleon due to final state
interactions, the contamination from hadron resonances, 
e.g. the $\pi + N \ra \Delta$ process  by which a
final-state pion is absorbed in the nucleus, 
and the unexplained depletion of low-virtuality  
events, much stronger than Pauli blocking can
account for~\cite{:2007ru}. 
Experiments such as MINOS measure the neutrino energy 
adding up the muon and hadronic energies, $E_\nu = E_\mu + E_\text{had}$. 
Experiments such as OPERA \cite{Guler:2000bd} need to
estimate the background to $\tau$-neutrino appearance events due to
charmed mesons production and decay.
It is thus crucial to 
have a good understanding of hadron modifications in the nuclear medium 
and of the space-time evolution of the hadronisation
process~\cite{Adamson:2008zt,Dytman:2008st}.  
However, at the low hadronic invariant mass involved in
these experiments, the theoretical methods discussed in
this review should be supplemented by those described in
Refs.~\cite{Buss:2007ar,Leitner:2006sp,Leitner:2006ww,Leitner:2006pq}.


\section{Kinematics, observables, and hadronisation time estimates}
\label{sec:variables-observables}

In this Section, we cover background material which will be
used throughout the review. We define the kinematic
variables and relevant observables for nDIS and hadronic collisions,
and compare the phase-space for hadron production in both types 
of collisions. Finally, the space-time development of the hadronisation 
process is discussed.

\subsection{Kinematic variables} \ \\
\label{sec:variables}
\label{sec:DISk}
\label{sec:NNk}
\label{sec:DYkinematics}

We discuss the kinematics for hadron production
in nDIS, hadron-hadron ($h+h$), hadron-nucleus ($h+A$) and
nucleus-nucleus ($A+A$) collisions, 
and for the Drell-Yan (DY) process. We will explicitly make
reference to the leading order (LO) processes in perturbative QCD, 
but most of the definitions are of general nature.

Throughout this discussion we use light-cone coordinates
$p^\mu = (p^+,p^-,\vec p_T)$,  where $p^\pm  =  (p^0\pm p^3)/\sqrt 2$
and $\vec p_T = (p^1,p^2)$.  
Our reference frame is such that the $z$ axis is aligned with the
beam, and a particle moving in the positive $z$ direction has large
light-cone plus-momentum. The transverse plane is the plane transverse to the beam and
$p^+ p^- = m^2 + p_T^2 \equiv m_T^2$, where
$m^2 = p^2$ is the invariant mass squared of the particle, and $m_T$
its transverse mass.  

\begin{table}[t]
  \begin{tabular}{rclcll} 
    \hline 
    \it Variable & \multicolumn{2}{l}{\it Definition} & 
      \multicolumn{2}{l}{\it Target rest frame form} &\\\hline
    $\boldsymbol{M^2}$ & = & $P^2$ & & 
      & \parbox[t]{5.cm}{\raggedright Target mass\\ \ }\\
    $\boldsymbol{x_B}$ & = & $\frac{-q^2}{2P\cdot q}$ & &
      & \parbox[t]{5.cm}{\raggedright Bjorken scaling variable\\ \ }\\
    $\boldsymbol{Q^2}$ & = & $-q^2$ & & 
      & \parbox[t]{5.cm}{\raggedright Negative four-momentum 
      squared of \\the exchanged virtual photon}\\
    $\boldsymbol{\nu}$ & = & $\frac{q\cdot P}{\sqrt{P^2}}$ & = 
      & $E_{trf}-E_{trf}^{\prime}$ 
      & \parbox[t]{5.cm}{\raggedright Energy of the virtual  
      photon in the \\target rest frame}\\
    $\boldsymbol{\vary}$ & = & $\frac{q\cdot P}{k\cdot P}$ & = 
      & $\frac{\nu}{E^{trf}_e}$ 
      & \parbox[t]{5.cm}{\raggedright Fractional energy loss 
      of the incident lepton (inelasticity)}\\
    $\boldsymbol{W^2}$ & = & $(P+q)^2$ & = & $M^2 + 2M\nu - Q^2$
      & \parbox[t]{5.cm}{\raggedright  Invariant mass squared of 
      the hadronic final state}\\
    $\boldsymbol{z_h}$   & = & $\frac{p_h\cdot P}{q\cdot P}$ & = & $\frac{E_h}{\nu}$ 
      & \parbox[t]{5.cm}{\raggedright Fraction of the virtual photon 
      energy carried by the  hadron}\\
    $\boldsymbol{p_T}$   & = & $|\vec{p}_{T}|$ & &
      & \parbox[t]{5.cm}{\raggedright Hadron transverse momentum
        (relative to the virtual photon momentum)}\\\hline
  \end{tabular}\\
  \caption{Definitions of the kinematic variables for semi-inclusive
    DIS. The Lorentz invariant definition and its form in the target
    rest frame are provided. Particle 4-momenta are defined in
    Fig.~\ref{fig:NNDISkinematics} and \ref{fig:kine_lep}. 
    All variables are experimentally measurable, hence typeset in
    boldface. Note that $x_B = Q^2 / (2M\nu)$ independently of the
    chosen reference frame.} 
  \label{tab:DISkinvar}
\end{table}

\begin{figure}[t]
  \centering
  \includegraphics[width=5.5cm]{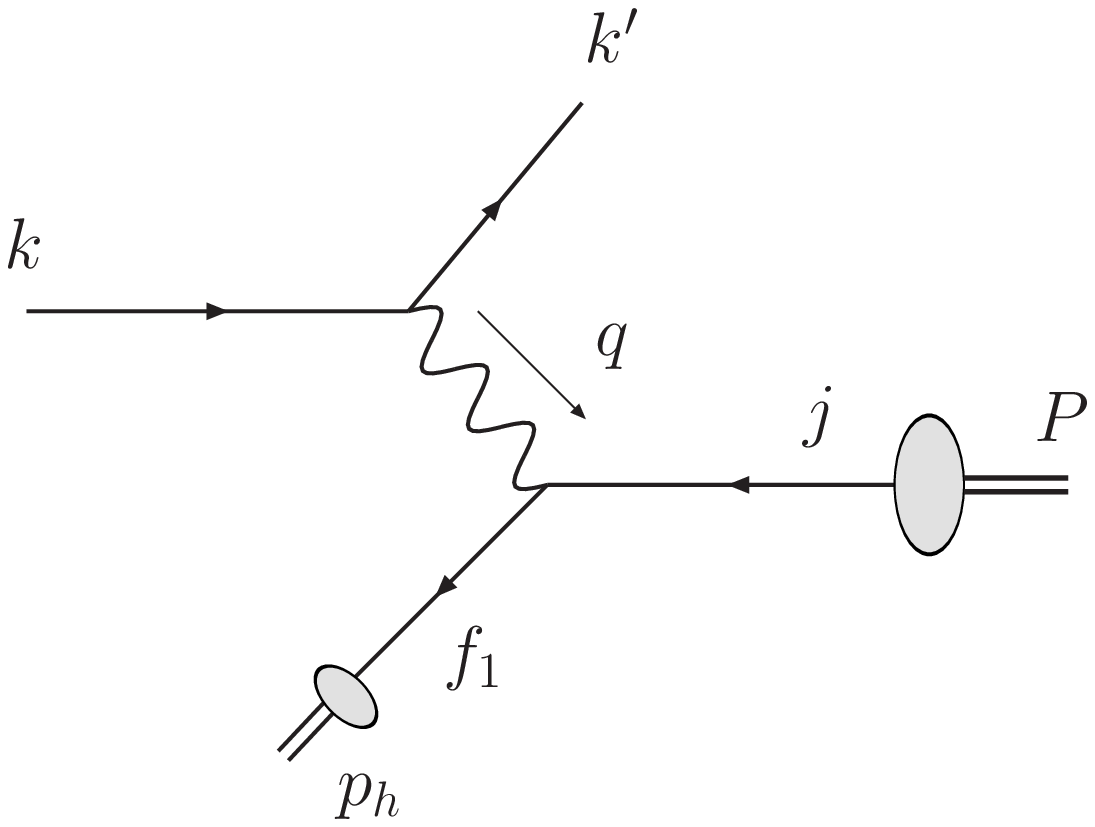}
  \hspace*{0.5cm}
  \includegraphics[width=5.5cm]{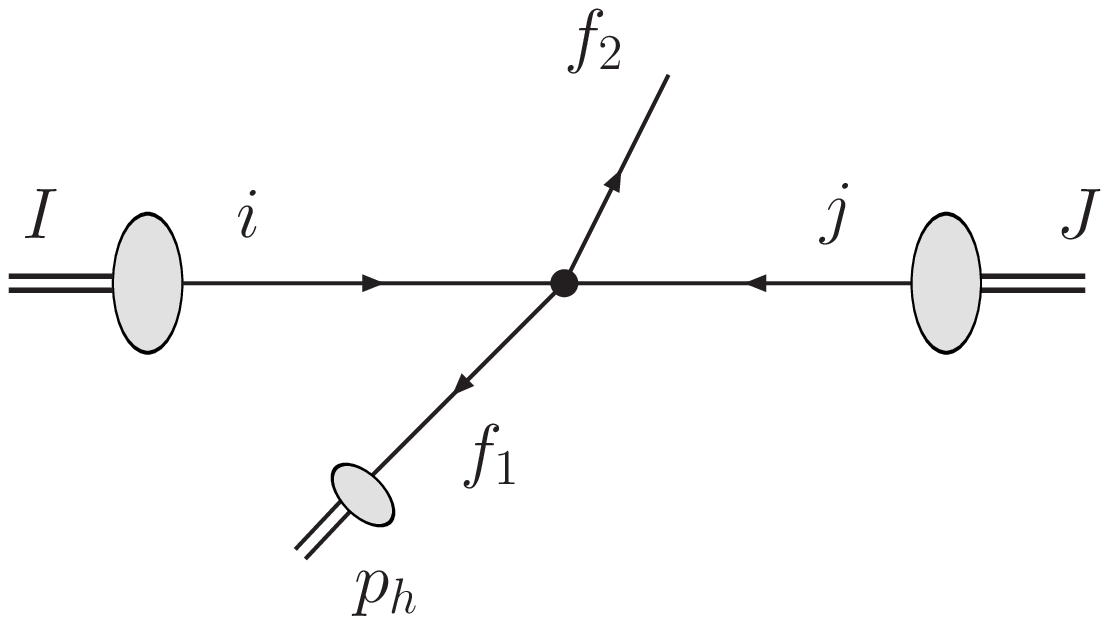}
  \caption{
    LO kinematics for parton production in DIS collisions ({\it left})
    and in hadron-hadron collisions ({\it right}). Double lines indicate
    hadrons or nuclei, thin single lines are partons or leptons. 
    The labels define the particles 4-momenta. }
  \label{fig:NNDISkinematics}
\end{figure}

\subsubsection{{\it Deep inelastic collisions}} 
\begin{figure}[t]
  \centering
  \includegraphics[width=8cm]{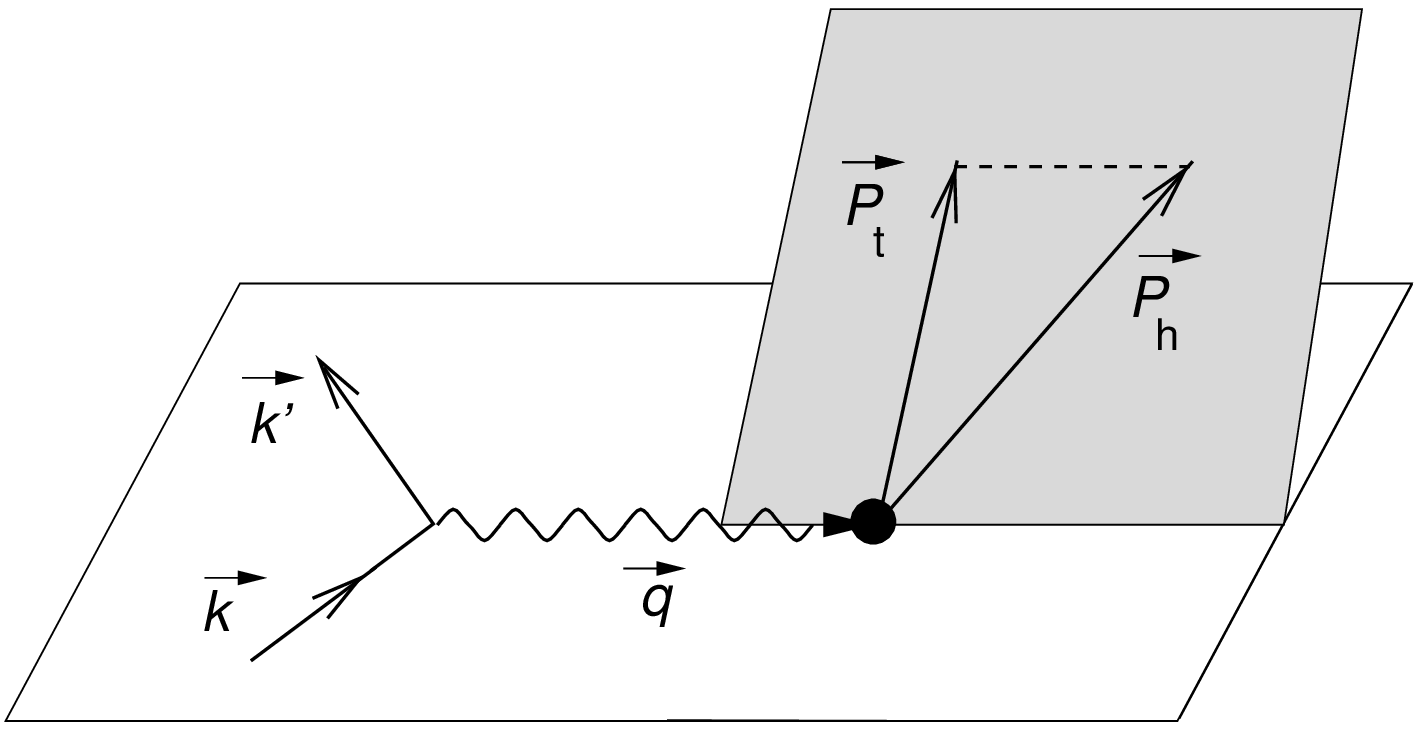}
  \caption{Kinematic planes for hadron production in semi-inclusive
    deep-inelastic scattering and definitions of the relevant lepton
    and hadron variables. 
    The quantities $k$ ($k'$) and $E$ ($E'$) are  the 4-momentum and
    the energy of the incident (scattered) lepton; $p_h$ is the
    4-momentum of the produced hadron, and its transverse component
    relative to the lepton plane is denoted by $\vec p_T$.  
    }
  \label{fig:kine_lep}
\end{figure}

--
Deep inelastic scattering at LO in pQCD proceeds by exchange of a
virtual photon in the $\hat t$-channel 
(Fig.~\ref{fig:NNDISkinematics} left). The DIS Lorentz invariants are
defined in Table~\ref{tab:DISkinvar}. 
Note that the variable $x_B$, $Q^2$ and $\nu$ are not independent but
related through $x_B = Q^2 / (2M\nu)$ in any reference frame. 
Analysis of inclusive DIS is usually carried
out using $x_B$ and $Q^2$, because of the $x_B$-scaling of the total
cross section in the Bjorken limit: $Q^2\to\infty$, $x_B$ fixed. 
Note that in DIS one can experimentally measure all
the listed variables, especially 
$\nu$, $Q^2$ and $z_h$,  because the initial and final state electron is
observable. This is markedly different from the situation in hadronic
collisions, where only final state hadrons can be observed and not 
the partons themselves. 
The hadron transverse momentum in DIS is defined with respect to the
virtual photon direction, see Fig.~\ref{fig:kine_lep}. 
Its analog in hadronic collisions would be the transverse
momentum of a hadron with respect to the beam axis. 

Nuclear DIS experiments have been performed in fixed-target ({\it ft}) conditions
in facilities like Stanford Linear Accelerator Center -- SLAC (E665), Super Proton Synchrotron -- SPS 
(EMC), Deutsches Elektronen Synchrotron -- DESY (HERMES), Jefferson Lab -- JLab (CLAS); 
and are planned in collider mode ({\it cl}) e.g. at the proposed Electron-Ion Collider -- EIC
or Large Hadron-electron Collider -- LHeC.
The colliding nucleon and lepton momenta are
\begin{align}\begin{split}    
  P_{ft} & = \left( \frac{M}{\sqrt{2}}, \frac{M}{\sqrt{2}}, \vec 0_T \right) 
  \ , \quad k_{ft} = \left( \sqrt{2} E,0,\vec 0_T \right) \\
  P_{cl} & = \left( \frac{M^2}{2\sqrt{2}E_N},\sqrt{2} E_N, \vec 0_T \right) 
  \ , \quad k_{cl} = \left( \sqrt{2} E,0,\vec 0_T \right) 
\end{split}\end{align}
where $E$ and $E_N$ are the lepton and nucleon energies measured
in the laboratory frame. To discuss both modes at the same time, it is
convenient to introduce the target rest frame energy of the lepton,
$E_{trf}$:
\begin{align}
  \Etrf = 
    \bigg\{ \bay{ll}
      E & \text{fixed-target} \\[.1cm]
      \frac{2E_NE}{M} \quad\ & \text{collider mode}
    \eay
\end{align}
Then the invariant inelasticity $\vary$ for both modes can be written as $\vary = \nu/\Etrf$.

\subsubsection{{\it Hadron-hadron collisions}} --
Parton production in hadronic collisions at leading order in the
coupling constant $\alpha_s$ proceeds through $2\ra
2$ partonic collisions 
(see Fig.~\ref{fig:NNDISkinematics} right and Table~\ref{tab:NNkvar}).
Several LO processes can contribute to a given $ij\ra f_1f_2$ collision,
represented by a black dot in the cartoon, see
Ref.~\cite{Field:1989uq} for details. The momenta of the two nucleons
colliding in the centre-of-mass-frame (c.m.f.) with energy $\sqrt s /2$ each are
\begin{align}
  I = \left( \sqrt{\frac{\tilde s}{2}},\frac{M^2}{\sqrt {2\tilde s}}
    , \vec 0_T \right) \qquad
  J = \left( \frac{M^2}{\sqrt{2\tilde s}}, \sqrt{\frac{\tilde s}{2}}
    , \vec 0_T \right) 
  \label{eq:IJ}
\end{align}
where $M$ is the nucleon mass and
\begin{align}
  \tilde{s} = s \frac{1+\sqrt{1+M^4/s^2}}{2}  \ .
\end{align}
We will neglect terms of order $\mathcal{O}(M^2/s)$ compared to terms of $\mathcal{O}(1)$,
and will use $\tilde s \approx s$. 
In Eq.~\eqref{eq:IJ}, we explicitly retain the nucleon mass to be able
to perform boosts to the rest frame of either nucleon. 
If we assume the partons to be massless and collinear to their parent
nucleons, their 4-momenta in terms of the parton fractional momenta
$x_i$ read $i = \left( x_1 \sqrt{s}/2, 0, \vec 0_T \right)$ and 
$j = \left( 0, x_2 \sqrt{s}/2, \vec 0_T \right)$.

Particle production (partons or hadrons) is described in terms of the
particle rapidity and transverse momentum. The rapidity of a particle
of 4-momentum $p$ and mass squared $m^2=p^2$ is defined as 
\begin{align}
  y = \frac12 \log \left( \frac{p^+}{p^-} \right) 
    = \log \left( \frac{p^+}{m_T} \right) \ .
\end{align}
Positive rapidity describes a particle moving in the positive $z$
direction, and likewise for negative rapidity. 
In the non-relativistic limit the rapidity coincides with
the particle longitudinal velocity $\beta_L$ in units of the speed of light,
$y \ra \beta_L$. 
Given the rapidity, one can compute
$p^0 = m_T \cosh y $ and $p^3 = m_T \sinh y$. 
Note that under a longitudinal boost of velocity $\beta$, the
rapidity transforms additively: $y' = y-y_\beta$, where 
$y_\beta = 0.5 \log[(1+\beta)/(1-\beta)]$
is the rapidity of the particle rest frame.
As an example, let us boost the target hadron momentum $P$ from the target rest
frame, $P = (M/\sqrt{2},M/\sqrt{2},\vec 0_T)$, to a frame in which
$P' = (\sqrt{s/2},M^2/\sqrt{2s},\vec 0_T)$. This boost is accomplished by 
$\alpha  =  \sqrt{s}/M$. Likewise, boosting a
nucleon from energy $\sqrt{s}/2$ to $\sqrt{s'}/2$ requires $\alpha =
\sqrt{s'/s}$. 

\begin{table}[t]
  \begin{tabular}{rcll} 
    \hline 
    {\it Variable} & & {\it Definition} \\\hline
    $\boldsymbol{s}$ & & &  
      \parbox[t]{8.9cm}{\raggedright Nucleon-nucleon centre-of-mass
        energy squared}\\
    $x_1$ & = & $i^+/I^+$ &  
      \parbox[t]{8.9cm}{\raggedright Initial-state projectile parton 
        fractional momentum}\\
    $x_2$ & = & $j^-/J^-$ &  
      \parbox[t]{8.9cm}{\raggedright Initial-state target parton 
        fractional momentum}\\
    $p_{iT}$ & = & $|\vec{f}_{iT}|$ &
      \parbox[t]{8.9cm}{\raggedright Final state partons transverse 
        momentum (relative to beam)}\\
    $y_i$ & = & $ 0.5 \log(f_i^+/f_i^-)$ &  
      \parbox[t]{8.9cm}{\raggedright Final state partons rapidity}
      \\ 
    $\boldsymbol{y_{cm}}$ & = & $ 0.5 \log\big(\frac{I^++J^+}{I^-+J^-}\big)$ &  
      \parbox[t]{8.9cm}{\raggedright Rapidity of the centre-of-mass}
      \\ \hline
    $z$ & = & $p_h^+/f_1^+$ &  
      \parbox[t]{8.9cm}{\raggedright Hadron fractional momentum
        relative to parent parton $f_1$}\\
    $\boldsymbol{p_{hT}}$ & = & $|\vec{p}_{hT}|$ &  
      \parbox[t]{8.9cm}{\raggedright Hadron transverse momentum
        (relative to beam)}\\ 
    $\boldsymbol{y_h}$ & = & $0.5 \log(p_h^+/p_h^-)$ &  
      \parbox[t]{8.9cm}{\raggedright Hadron rapidity}\\
    $\boldsymbol{\eta}$ & = & $-\log\tan(\theta^*/2)$ &  
      \parbox[t]{8.9cm}{\raggedright Hadron pseudorapidity 
        ($\theta^*$ is the polar angle between the parton and the beam 
        in the centre-of-mass reference frame)}
    \\\hline
  \end{tabular}\\
  \caption{Relevant kinematic variables for semi-inclusive
    parton (top) and hadron (bottom) production. 
    Particle 4-momenta are defined in Fig.~\ref{fig:NNDISkinematics}, right. 
    Boldface variables are experimentally measurable; the others are
    theoretically defined in perturbative QCD.
}
  \label{tab:NNkvar}
\end{table}

Measuring the rapidity of a particle requires measuring two independent
variables, say, its energy and longitudinal momentum. Not in all
experiments this is possible, while just measuring the polar angle $\theta^*$
between the particle trajectory and the beam axis in the centre-of-mass 
frame is easier. This justifies the definition of the
particle pseudorapidity, 
\begin{align}
  \eta = - \log \tan (\theta^*/2) \ ,
\end{align}
such that $|\vec p| = p_T \cosh \eta$ and $p^3 = p_T \sinh \eta$.
For massless particles it coincides with the rapidity: $\eta=y$; for
massive particles, they are approximately equal if $|\vec p| \gg m$ 
(and $\theta$ not too small). Differential particle distributions in $y$ and $\eta$ are related
by
\begin{align}
  \frac{dN}{dydp_T^2} 
    = \frac{dN}{d\eta dp_T^2} \sqrt{1-\frac{m^2}{m^2_T \cosh^2y}} \,=\,\frac{dN}{d\eta dp_T^2} \frac{E}{m_T}\ .
\end{align}
In order to compare collider and fixed-target experiments, and different 
beam energies, it is useful to consider the rapidity in the c.m.f.:
\begin{align}
  y_{c.m.f.} = y - y_{cm} \ .
\end{align}
The backward rapidity region (target hemisphere) 
corresponds to $y_{c.m.f.} < 0$, and the
forward rapidity region (projectile hemisphere) to $y_{c.m.f.}>0$.

At LO, the 4-momenta of the two produced partons can be expressed in
terms of their final state rapidities $y_i$ and transverse momentum
$p_T$ (see Table~\ref{tab:NNkvar} for definitions)
\begin{align}
  f_1 = \left(\frac{p_T}{\sqrt{2}}e^{y_1},\frac{p_T}{\sqrt{2}}e^{-y_1},-\vec{p}_T\right) \qquad
  f_2 = \left(\frac{p_T}{\sqrt{2}}e^{y_2},\frac{p_T}{\sqrt{2}}e^{-y_2},\vec{p}_T\right) \ ,
\end{align}
and their fractional momenta are
\begin{align}
  x_1 = \frac{p_T}{\sqrt{s}} \left(e^{y_1} + e^{y_2} \right) \qquad
  x_2 = \frac{p_T}{\sqrt{s}} \left(e^{-y_1} + e^{-y_2} \right) \ .
\end{align}
Finally, the Mandelstam invariants are defined as follows,
\begin{align}
\begin{split}
  \hat{s} & = (i+j)^2 \\
  \hat{t} & = (i-f_1)^2 = (f_2-j)^2 \\
  \hat{u} & = (i-f_2)^2 = (f_1-j)^2 \ .
\end{split}
\end{align}
and 
$
  \hat s + \hat t + \hat u = 0
$
by momentum conservation.
In terms of rapidities and transverse momentum, the Mandelstam
invariants read
\begin{align}
\begin{split}
  \hat{s} & = x_1 x_2 s \\
  \hat{t} & = -p_T^2 (1+e^{y_2-y_1}) \\
  \hat{u} & = -p_T^2 (1+e^{y_1-y_2}) \ .
\end{split}
\end{align}

Hadronisation in the collinear factorisation framework proceeds
through independent parton fragmentation into a hadron. It is
universal, i.e., independent of the process which produced the
fragmenting hadron, e.g., hadronic or DIS collisions
\cite{Collins:1981uk}. The hadron fractional momentum $z$ is defined
by
\begin{align}
  p_h^+ = z f_1^+ \qquad
  \vec p_{hT} = z \vec f_{1T} \ .
\end{align}
Therefore the on-shell hadron momentum $p_h$ reads
\begin{align}
  p_h = \left(zf_1^+, \frac{m_h^2+z^2f_{1T}^2}{2 z f_1^+}, z\vec f_{1T}\right) \ .      
 \label{eq:zdef}
\end{align}
The parton and hadron rapidities are related by $y_1 = y_h +
\log(m_{hT}/p_{hT})$. 

The partonic variables $p_T$, $y_i$, $x_i$ and
the fractional hadron momentum $z$ are not experimentally measurable,
but are needed in the theoretical computation of the cross section. The
experimentally measurable variables are typed in boldface in
Table~\ref{tab:NNkvar}. 
Note that the hadron transverse momentum $p_T$ in hadron-hadron
collisions is defined with respect to the beam axis, so that at
midrapidity it is the analog of the hadron energy $E_h$ in DIS.

\begin{table}[t]
  \begin{tabular}{rclrll} 
    \hline 
    {\it Variable} & & {\it Definition} && LO in $\alpha_s$ \\\hline
    $\boldsymbol{x_F}$ & = & $2p^*_z/\sqrt{s}$  
      & = &  $x_1 - x_2$ 
      & \parbox[t]{6cm}{\raggedright Feynman $x$}\\
    $\boldsymbol{M}$ & = & $\sqrt{p_{\ell^+}^2 + p_{\ell^-}^2}$  
      & = & $x_1 x_2 s$ 
      & \parbox[t]{6cm}{\raggedright Dilepton invariant mass}\\
    $\boldsymbol{p_T}$ & = & $|\vec p_T|$
      & 
      & & \parbox[t]{6cm}{\raggedright Dilepton transverse momentum}\\\hline
  \end{tabular}\\
  \caption{Kinematic variables for Drell-Yan dilepton
    production. The dilepton momentum is 
    $p = p_{\ell^+} + p_{\ell^-}$, where $p_{\ell^\pm}$ are the
    lepton and anti-lepton momenta. A star indicates momenta measured
    in the centre-of-mass frame. 
    The three DY variables are experimentally measurable, hence typeset in
    boldface. See
    Table~\ref{tab:NNkvar} for the definition of $x_{1,2}$ and $s$. 
  }
  \label{tab:DYkvar}
\end{table}

\subsubsection{{\it Drell-Yan processes}} --
Drell-Yan production of a lepton pair in hadronic collisions occurs,
at zero-th order in $\alpha_s$ {\it via} the quark-antiquark
annihilation channel
\begin{equation*}
q\ +\ \bar{q}\ \to\ \gamma^\star\ \to\ \ell^+\ +\ \ell^-,
\end{equation*}
as shown in the lower-right panel of Fig.~\ref{fig:universality}.
As before, the initial partons carry a momentum fraction $x_1$ and
$x_2$ of the projectile and target hadron, respectively. At this
order, the dilepton is produced with zero transverse momentum, with
invariant mass $M_{l^+l^-}$ and with longitudinal momentum fraction 
$x_F = 2p^*_{z}/\sqrt{s}$, also called Feynman-$x$, 
where $p^*$ is the dilepton momentum in the centre-of-mass frame. 
These variables are summarised in Table~\ref{tab:DYkvar}.
By energy-momentum conservation and assuming $M^2\ll s$, we have
\begin{align}
\begin{split}
  M^2 &= \hat{s} = x_1 x_2 s \\
  x_F &= x_1 - x_2 \ ,
\end{split}
\end{align}
from which it follows immediately that
\begin{equation}\label{eq:x1x2}
  x_{_{1, 2}} = \frac{1}{2} \, \left( \sqrt{x_{_{\rm F}}^2+4 \ M^2/s} \pm x_{_{\rm F}} \right).
\end{equation}
Usually, Drell-Yan production is measured between the charmonium and
bottomonium masses ($4\le M\le 8$~GeV/c$^2$) or above the $\Upsilon$. At
higher orders in $\alpha_s$, new channels open up, such as Compton
scattering, $q\ g\ \to\ \gamma\ q$. The virtual photon acquires a
finite transverse momentum, and it is no longer possible to relate
$x_F$ and $M$ to the momentum fractions of the partons probed in the
projectile and target hadron.

\subsection{Comparison of hadron-hadron and DIS kinematics } \ \\
\label{sec:phasespaces}
\label{sec:dictionary}
\label{sec:NNequiv}
\label{sec:DIS-eqNN}

If we consider parton and hadron production at LO in
hadronic and DIS collisions, it is easy to relate the relevant variables 
in both processes, that allows one to compare their
corresponding phase spaces. The discussion closely follows
Ref.~\cite{Accardi:2007in}, to which we refer for details.
To connect the DIS and hadron-hadron kinematics (Fig.~\ref{fig:NNDISkinematics}) 
we can boost the DIS collision to a 
frame in which the target has energy $\sqrt{s}/2$ per nucleon, and
imagine the lepton to be a parton of a collinear phantom
nucleon of energy $\sqrt{s}/2$ and with 4-momentum 
$P'^\pm = P^\mp$. 
Comparing the left and right parts of
Fig.~\ref{fig:NNDISkinematics} we can identify 
\begin{align}
\begin{split}
  P  & \equiv J , \quad  
  P' \equiv I , \quad  
  k  \equiv i , \quad  
  k' \equiv f_2 .
\end{split}
\end{align}
The virtual photon momentum $q$, the fractional momentum $x_e$ of the
initial-state lepton  and the rapidity  $y_e$ of the final state lepton
are identified as follows 
\begin{align*}
\begin{split}
  q  & = k-k' \equiv i-f_2 , \quad
  x_e = k^+ / P'^+ \equiv x_1, \quad
  y_e \equiv y_2 \ .
\end{split}
\end{align*}
In this way, we can relate the DIS to the hadron-hadron kinematics
discussed in Sect.~\ref{sec:NNk}. 
As an example, it is immediate to see that, in terms of hadron-hadron variables,
$Q^2 = - \hat t$. The full translation ``dictionary'' from DIS to hadron-hadron
variables can be obtained in a straightforward way 
by combining the results of Sect.~\ref{sec:DISk} and the definitions of
Tables~\ref{tab:DISkinvar}--\ref{tab:NNkvar}.

\begin{table}[tb]
  \centering
  \begin{tabular}{cccccc}\hline
                     & SPS  & FNAL & RHIC & RHIC & LHC   \\\hline
    $\sqrt{s}$ [GeV] & 17.5 & 27.4 & 63   & 200  & 5500  \\
    $\Delta y_1$     & 2.4  &  2.0 &  1.2 & 0    & -3.3   \\\hline
  \end{tabular}
  \caption{
    Rapidity shifts
    $\Delta y_1$ of the RHIC-equivalent DIS phase space, tabulated for
    some energies of interest.}
  \label{tab:Deltay1}
\end{table}

First, we can express the DIS invariants in terms of parton rapidities
and transverse momenta. Neglecting target-mass corrections, i.e. up
to terms of $\mathcal{O}(M^2/s)$, we obtain
\begin{align}
\begin{split}
  x_B & = \frac{p_T}{\sqrt{s}} \,\left(e^{-y_2}+e^{-y_1}\right) \\
  Q^2 & = p_T^2 \,\left(1+e^{y_1-y_2}\right) \\
  \nu & = \frac{p_T\sqrt{s}}{2M} e^{y_1} \\
 \vary   & = \frac{1}{1+e^{y_2-y_1}} \\ 
  z_h & = z \ . 
 \label{eq:DIS(NN)}
\end{split}
\end{align}
Note that the first three variables are not independent because 
$Q^2 = 2M x_B \nu$, and that $x_B=x_2$ is interpreted as the struck
parton fractional momentum, as expected in DIS at LO.
Note also that $\nu$ increases with increasing $p_T$ and increasing
$y_1$. This is because a parton of positive and large $y_1$ in the 
c.m. frame travels in the opposite direction as the left-moving
nucleus, considered as the ``target nucleus'' 
(see Fig.~\ref{fig:NNDISkinematics}). Hence in that nucleus rest
frame it is very fast. Conversely, a parton of negative and large $y_1$
travels in the same direction as the target nucleus, which means quite
slow in the target rest frame. 
It is also interesting to note that up to terms of order $\mathcal{O}(M^2/s)$,
the parton and hadron energy in the target rest frame are
$E = \nu$ and $E_h = z_h \nu$ respectively. Finally, we can invert
Eq.~\eqref{eq:DIS(NN)} to obtain the hadron-hadron variables in 
terms of DIS invariants:
\begin{align}\begin{split}
  p_T^2 & = (1-\vary)Q^2 \\
  y_1 & = - \log \left( \frac{Q\sqrt{s}}{2M\Etrf}
    \,\frac{(1-\vary)^{1/2}}{\vary} \right) \\
  y_2 & = y_1 + \log \left( \frac{1-\vary}{\vary} \right) \\
  z & = z_h
  \label{eq:NN(DIS)1}
\end{split}\end{align}
with $\vary = \nu/\Etrf$. 
Note that in DIS, the electron energy $\Etrf$, hence the electron
$x_e$, is fixed by the
experimental conditions; this is different from hadronic collisions where
the parton $j$ has an unconstrained fractional momentum. 
Changing the c.m.f. energy to $\sqrt{s'}$ simply results in a shift of
the parton rapidity,
\begin{align}
  y_1 \xrightarrow[s\ra s']{} y_1 + \Delta y_1
  \label{eq:y1shift}
\end{align}
where
$
  \Delta y_1 = \log(\sqrt{s}/\sqrt{s'})
  \label{eq:Deltay1}
$.
The value of $\Delta y_1$ compared to RHIC top energy $\sqrtsnn = 200$~GeV is listed in Table~\ref{tab:Deltay1} for
the experiments of interest in this paper.

\begin{figure}[tb]
  \centering
  \includegraphics[width=0.49\linewidth]{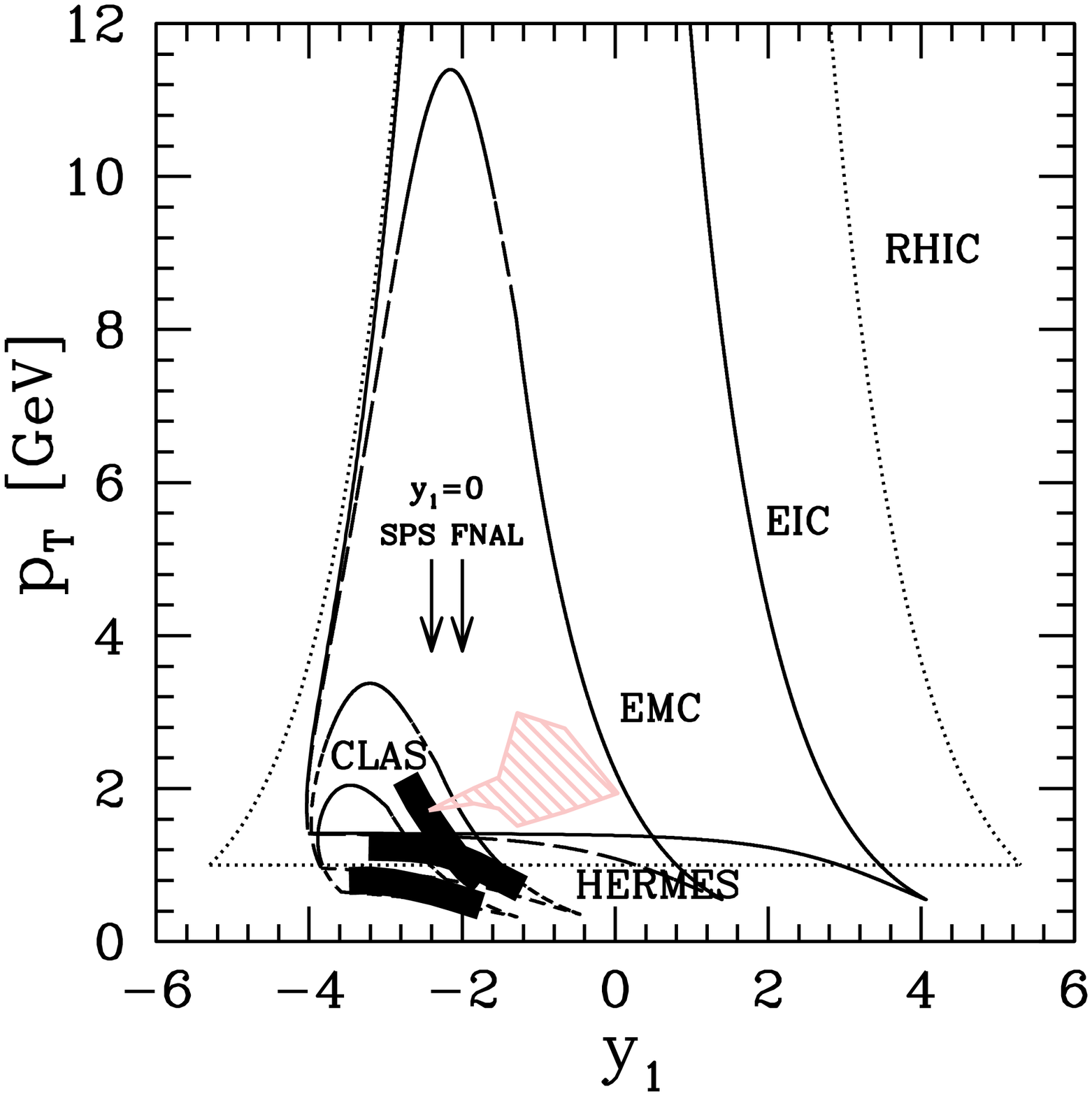}
  \includegraphics[width=0.49\linewidth]{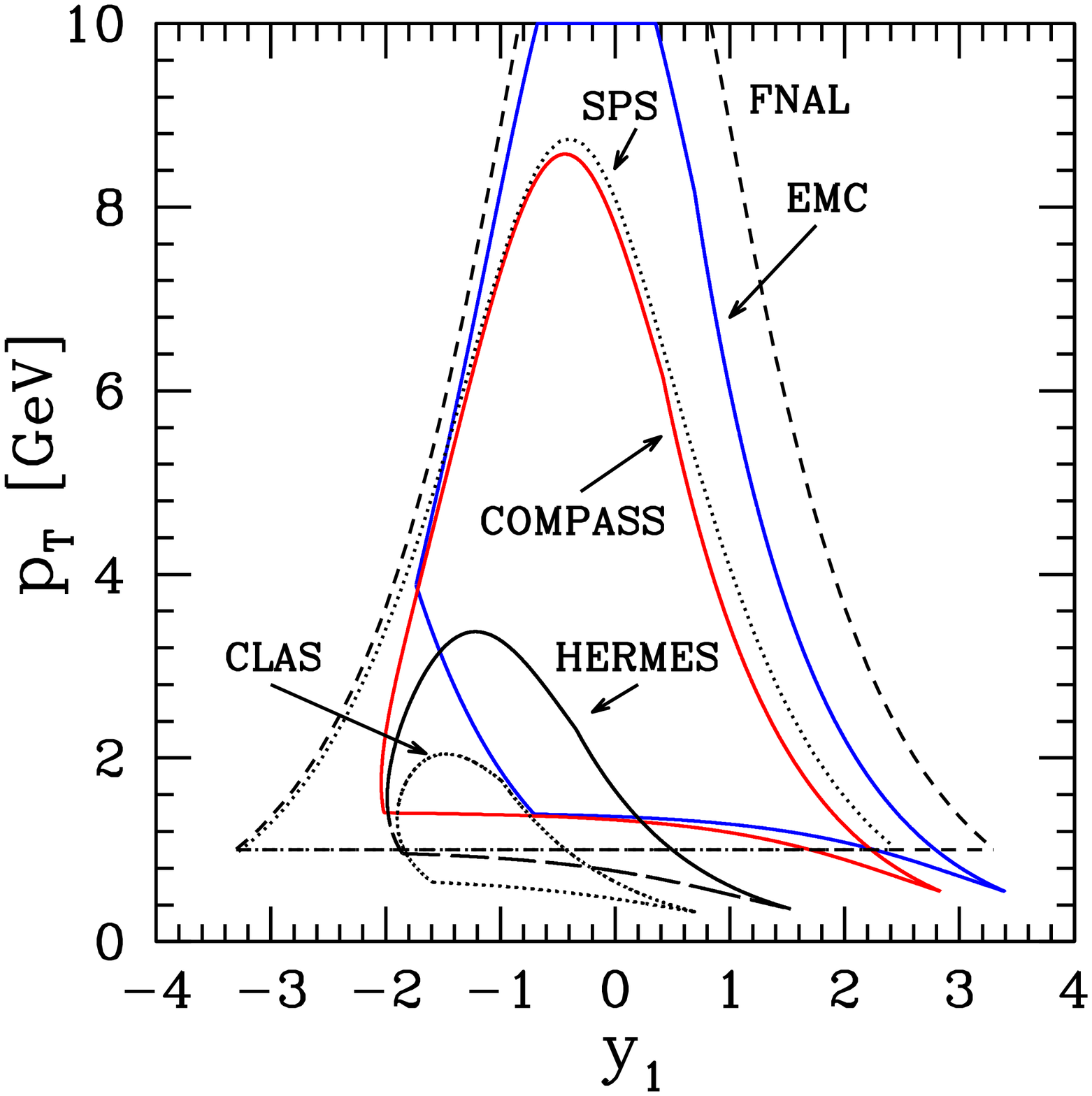}
  \caption{{\it Left:} RHIC-equivalent phase space of nuclear DIS
    experiments at $E_e = 27.6$~GeV (HERMES, solid line), at $E_e = 12$~GeV 
    (HERMES and JLab, dashed line), and at $E_e = 280$~GeV (EMC,
    dot-dashed line). The dotted line shows the borders of the LO pQCD
    phase space in $p,A+A$ at top RHIC energy, $\sqrtsnn = 200$~GeV. 
    The two arrows show the location 
    of the midrapidity region at SPS and FNAL ($p,A+A$) fixed-target
    experiments. The shaded regions show the region of phase-space
    experimentally explored at HERMES~\cite{Airapetian:2007vu,vanderNat:2003au}  
    and EMC~\cite{Ashman:1991cx}.
    {\it Right:} Hadron-hadron-equivalent EMC and COMPASS ($\ell+A$) phase space at
    $\sqrtsnn = 27.4$~GeV, compared to the SPS and FNAL ($p,A+A$) phase spaces. 
  }
  \label{fig:RHICequiv(HERMES)}
\end{figure}

Given a DIS phase space (say a given experiment acceptance
region in the $(\nu,Q^2)$ plane), we define its
hadron-hadron-equivalent phase-space as its image
in the $(p_T,y_1)$ under Eqs.~\eqref{eq:NN(DIS)1}. 
The reason for this definition is that for both hadronic and DIS
collisions we can identify the parton $f_1$ of
Fig.~\ref{fig:NNDISkinematics} with the ``observed'' parton in hadronic and
DIS collisions, i.e., the parton which fragments into the observed 
hadron. Then the variables $p_T$ and $y_1$ fully 
characterise the observed parton. An analogous
definition holds when using $x_B$ instead of $\nu$ as independent
variable. 
As an example, the HERMES DIS phase space in the
$(\nu,Q^2)$ plane is
determined by the values of $W^2_{min}$, $Q^2_{min}$ and $y_{max}$:
\begin{align}\begin{split}
  & \frac{Q^2_{min}+W^2_{min}-M^2}{2M} \leq \nu \leq y_{max}\;\Etrf \\
  & Q^2_{min} \leq Q^2 \leq M^2+2M\nu - W^2_{min} \ .
    \label{eq:DISps} 
\end{split}\end{align}
Additionally, one may impose stronger cuts on $\nu$, e.g., $\nu\geq
\nu_{min}$, as at the EMC experiment, and in some HERMES analysis.
 
Using Eqs.~\eqref{eq:NN(DIS)1} the {\it hadron-hadron equivalent}
DIS phase space in the $(y_1,p_T)$ plane can be determined. 
As an example, in Fig.~\ref{fig:RHICequiv(HERMES)} left, we consider
the RHIC-equivalent phase space of the fixed target $e+A$ experiments
and the planned Electron-Ion Collider (EIC), using $\sqrtsnn$~=~200~GeV.
Note that according to Eq.~\eqref{eq:y1shift}, 
the hadron-hadron-equivalent phase space at other centre-of-mass energies 
can be obtained by a shift $y_1 \ra y_1+\Delta y_1$, see
Table~\ref{tab:Deltay1} and Fig.~\ref{fig:RHICequiv(HERMES)} right
where the Fermilab-equivalent phase space is shown.  
We assume the pQCD formulae used to define the hadron-hadron-equivalent phase
space to be valid for $p>p_0 = \mathcal{O}(1$~GeV/c$)$, see Eq.~\eqref{eq:NNps} below
for details. 
We can see that the HERMES and CLAS experiments, with $\Etrf = 27.6$ and
12~GeV, cover less than one third of the available RHIC $p_T$ range 
at $y_1 \approx -3$, with shrinking $p_T$ coverage at
larger rapidity. In the SPS/FNAL midrapidity region it reaches $p_T$~=~2.5~GeV/c 
at most. Since 
\begin{align}\begin{split}
  y_1 & \leq \log \left( \frac{\sqrt{s}}{2M\Etrf} 
    \frac{p_T}{\vary_{max}}\right),
\end{split}\end{align}
the only way to effectively reach larger values of $y_1$
is to increase the electron beam energy $\Etrf$. Indeed, 
the EMC experiment, with $\Etrf = 100-280$~GeV, covers a larger span
in rapidity and extends to $y_1 \gtrsim 0$ (as would a $\mu+A$ programme
at COMPASS). Moreover, the increased energy allows one in principle to
reach much higher $p_T$ than at HERMES and CLAS. However, only the
$p_T \lesssim 3$~GeV/c region has been explored in actual measurements.  
The proposed Electron-Ion Collider (EIC)
\cite{Deshpande:2005wd,Aidala:2008ic} 
will be able to effectively study the $y_1 > 0$ region and cover most
of the RHIC phase space, but only the $y_1<0$ part of the LHC phase space.

The {\it direct} use of HERMES and CLAS data
to understand nuclear effects in high-$p_T$ hadron production in
heavy-ions collisions is therefore not possible. Instead, one needs to use those
data to understand in detail the hadronisation dynamics and to constrain
the various models in the context of heavy-ion
collisions, by extrapolating them to unmeasured regions of phase space.

The {\it DIS-equivalent hadron-hadron phase space} is defined as the image of
Eqs.~\eqref{eq:NNps} in the $(\nu,Q^2,\vary,z_h)$ space under
Eqs.~\eqref{eq:DIS(NN)}. 
Then, the hadron-hadron phase space at a given
$y_1$ is defined by the kinematic bounds on $2\ra 2$ parton
scatterings~\cite{Eskola:2002kv}:
\begin{align}\begin{split}
  & |y_1| \leq \cosh^{-1} \left(\frac{\sqrt s}{2p_0}\right) \\
  & p_0 \leq p_T \leq \frac{\sqrt s}{2 \cosh(y_1)} \\
  & -\log \left( \frac{\sqrt{s}}{p_T}-e^{-y_1} \right) 
    \leq y_2 \leq 
    \log\left( \frac{\sqrt{s}}{p_T}-e^{y_1} \right) \\
  & \frac{m_{hT}}{\sqrt{s}} e^{y_h} 
    \left(1+\frac{p_{hT}^2}{m_{hT}^2e^{y_h}} \right)   \leq z \leq 1
  \label{eq:NNps}
\end{split}\end{align}
where hard scatterings satisfy $p_T \geq p_0$, with  $p_0 \gtrsim
1$~GeV/c a lower cutoff \cite{Eskola:2002kv,Accardi:2003jh}.

Introduction of next-to-leading order kinematics~\cite{Guzey:2004zp},
would relax somewhat these bounds. At
large rapidity, where the phase space for $2\ra2$ parton processes 
is becoming more and more restricted, 
$2\ra 1$ parton fusion may become the
dominant mechanism because it is sensitive to much
lower fractional momenta $x_i$~\cite{Accardi:2004fi}. Hence, at the
boundary of the hadron-hadron phase space, the presented analysis
becomes unreliable. 

\begin{figure}[tb]
  \centering
  \includegraphics[width=0.49\linewidth]{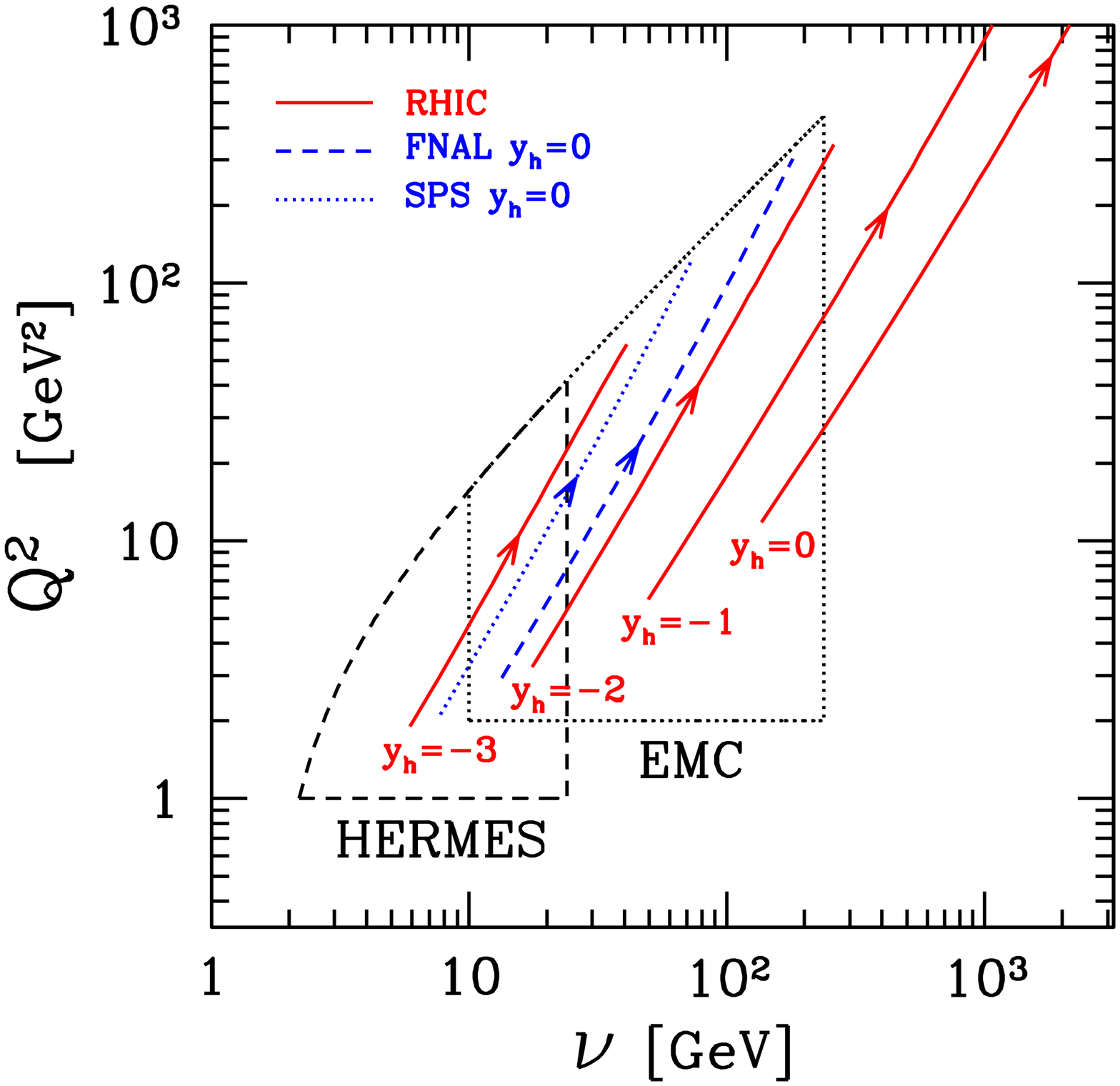}
  \includegraphics[width=0.49\linewidth]{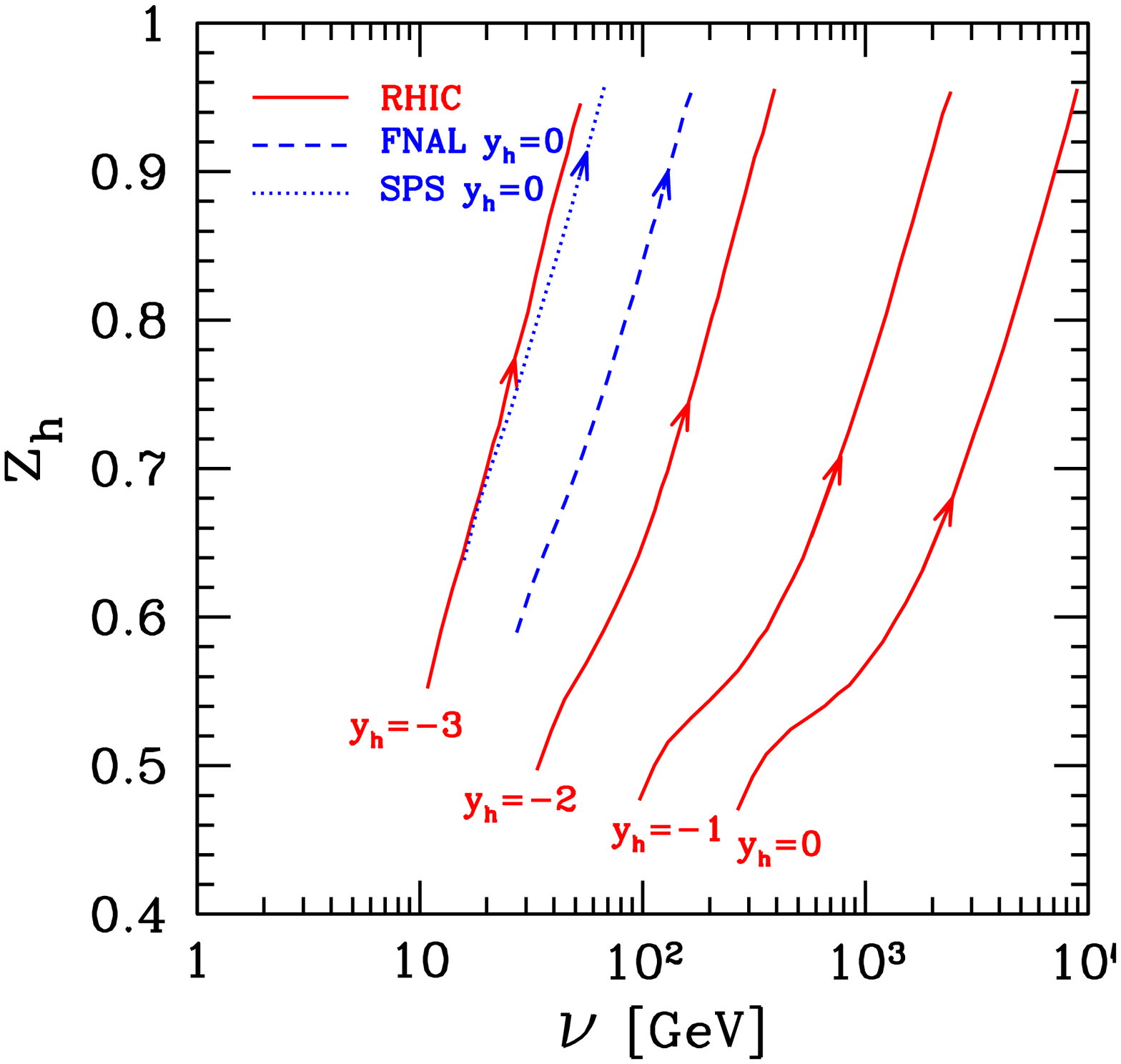}
  \caption{{\it left:} Fixed-$y_1$ hadron-hadron trajectories 
    plotted in the DIS-equivalent $(\nu,Q^2)$ 
    phase space for RHIC at $\sqrtsnn = 200$~GeV and various rapidities,
    for FNAL and SPS at midrapidity. The dashed line encloses the
    HERMES phase space; the dotted line encloses the EMC phase
    space. The arrow indicates the direction of increasing $\vev{p_T}$
    and $\vev{z_h}$. {\it Right:} Trajectories in the $(\nu,z_h)$
    plane. The arrows indicate increasing $p_T$ and $Q^2$.}
  \label{fig:HERMES(RHIC)}
\end{figure}

This phase-space is 4-dimensional and
difficult to directly visualise. A way around this problem is to
define suitable trajectories in hadron-hadron phase space averaged over $y_2$, 
and to project them into the DIS-equivalent ($\nu$,$Q^2$) and
($\nu$,$z_h$) phase spaces. We can define a
$p_{hT}$- and $y_h$-dependent average observable as follows 
\begin{align}
  \vev{\OO}_{p_{hT},y_h} = 
    \frac{ 
      \int dz\,dy_1\,dy_2\, \OO(p_T,y_1,y_2,z) 
      \frac{d\hat\sigma_{AB\ra hX}}{dp_T^2dy_1dy_2dz}
    }{
      \int dz \,dy_1\,dy_2
      \frac{d\hat\sigma_{AB\ra hX}}{dp_T^2dy_1dy_2dz}
    } \ ,
\end{align}
where
\begin{align}
  \frac{d\hat\sigma_{AB\ra hX}}{dp_T^2dy_1dy_2dz} 
    = \sum_{f_1} \frac{1}{z^2} D_{f_1}^h(z) 
    \frac{d\hat\sigma_{AB\ra f_1X}}{dp_T^2dy_1dy_2} \ ,
\end{align}
$d\hat\sigma_{AB\ra f_1X}$ is the LO pQCD differential cross-section
for production of a $f_1$ parton  in a collision of hadrons
$A$ and $B$ (nucleons or nuclei), and $D_{f_1}^h$ is its
fragmentation function into the observed hadron.
Then we can define fixed-$y_h$ trajectories 
$
  \{ (\vev{\nu}_{p_T,\bar y},\vev{Q^2}_{p_T,\bar y})
    ; p_T\geq p_0 \}  
$ 
and 
$
  \{ (\vev{\nu}_{p_T,\bar y},\vev{z_h}_{p_T,\bar y})
    ; p_T\geq p_0 \}  
$
in the DIS-equivalent phase space.

\begin{table}[tb]
  \centering
  \begin{tabular}{ccccccc}\hline
     & SPS 
     & FNAL 
     & \multicolumn{4}{c}{RHIC} \\
     & $\sqrtsnn = 17.5$~GeV
     & $\sqrtsnn = 27.4$~GeV
     & \multicolumn{4}{c}{$\sqrtsnn = 200$~GeV} \\\hline  
 $y_h$          & 0     & 0     & 0     & -1    & -2    & -3   \\
 $p_{hT}$       & 1--8  & 1--12 & 1--90 & 1--60 & 1--25 & 1--9  \\\hline  
  \end{tabular}
  \caption{Range of hadron transverse momentum ($p_{hT}$, in GeV/c) spanned along 
    trajectories at fixed rapidity $y_1$ at RHIC top energy  $\sqrtsnn = 200$~GeV 
    and at fixed-target energies $\sqrtsnn = 17-28$~GeV.}
  \label{tab:pTzhrange}
\end{table}

As an example, in Fig.~\ref{fig:HERMES(RHIC)} 
we considered hadronic collisions at RHIC top energy $\sqrtsnn$~=~200~GeV 
and at fixed-target energies $\sqrtsnn = 17-27$~GeV, and plotted the
fixed-$y_h$ trajectories in the DIS-equivalent phase space.
The range of $p_T$ spanned along each
trajectory is tabulated in Table~\ref{tab:pTzhrange}.
The spanned range in $Q^2$ is limited by the maximum $p_T$ at 
each rapidity, according to Eq.~\eqref{eq:NNps}. 
As expected, the smaller the rapidity $y_h\approx y_1$ 
the smaller the spanned $\nu$. RHIC trajectories with 
$y_h\lesssim -2$ span relatively low values of $\nu \lesssim 60$~GeV and large
values of $z_h \gtrsim 0.5$, where the EMC
and HERMES experiments have shown non negligible cold QCD matter
suppression of hadron production. At higher rapidity, the larger spanned 
values of $\nu$ will make cold QCD matter effects less prominent.
The consequences of these remarks for the interpretation
of hadron production in $h+A$ and $A+A$ collisions will be further
discussed in Section~\ref{sec:coldjetquenching}.

\subsection{Nuclear modification observables } \ \\
\label{sec:observables}

In lepton-nucleus DIS, the experimental results
for hadron production  are usually presented in terms of the {\it hadron 
multiplicity ratio} $R_M^{h}$, which represents the ratio of the number
of hadrons of type $h$ produced per deep-inelastic scattering event on
a nuclear target of mass $A$ to that from a deuterium target (D). 
The multiplicity ratio $R_M^{h}$  depends  on the leptonic variables
$\nu$ and $Q^2$, and on the hadronic variables $z = E_h/\nu$ and $p_T^2$
defined in Section~\ref{sec:variables}. It is defined as the super-ratio
\begin{equation} 
  R_M^{h}(z,\nu,Q^2,p_T^2) = 
    \left(\frac{N_h(z,\nu,Q^2,p_T^2)}{N_e(\nu,Q^2)}\right)_{\ds A}  
      \ \Bigg/ \
    \left(\frac{N_h(z,\nu,Q^2,p_T^2)}{N_e(\nu,Q^2)}\right)_{\ds D} ,
\label{eq:att}
\end{equation}
where  $N_h$ is the yield of semi-inclusive hadrons in a given
($z,\nu,Q^2,p_T^2$)-bin, and $N_e$ the yield of inclusive
deep-inelastic scattering leptons in the same ($\nu$,$Q^2$)-bin. 
Normalising the hadron yield to the DIS yield allows one to cancel, to a
large extent, initial-state nuclear effects such as nuclear modifications of PDFs
and to isolate final-state nuclear modifications of hadron production 
as a deviation of $R_M$ from unity. A suppression of $R_M$ is
experimentally observed to increase with $z$, and to decrease with
$\nu$, and more mildly with $Q^2$ (see discussion in Section \ref{sec:hadrons-lA}). When plotting $R_M$ as a
function of $p_T$, one observes a suppression at small $p_T$ and an
enhancement above  $p_T\approx$~1.5~GeV/c. This behaviour is also known
as ``Cronin effect'' (Sections~\ref{sec:eAdata}
and~\ref{sec:pAdata}). The amount of transverse momentum broadening
(defined with respect to the direction of the virtual photon, see
Fig.~\ref{fig:kine_lep}) is quantified via   
\begin{equation}
 \Delta\langle p_T^2 \rangle^h =  \langle p_T^2  \rangle^h_A -  \langle
 p_T^2 \rangle^h_D \ . 
\label{eq:ptbroadening}
\end{equation}
Here, $\langle p_T^2 \rangle^h_A$ is the average transverse momentum
squared of a hadron of type $h$ produced on a nuclear target $A$
\begin{align}
  \vev{p_T^2}^h 
    = \frac{\sum_{p_T,z,\nu,Q^2} \,p_T^2 \,N_h(z,\nu,Q^2,p_T^2)|_A}
    {\sum_{p_T,z,\nu,Q^2} N_h(z,\nu,Q^2,p_T^2)|_A} \ ,
\end{align}
and $\langle p_T^2 \rangle^h_D$ is the same quantity for a deuterium
target. 

In high-energy proton-nucleus and nucleus-nucleus collisions 
what is usually presented  is the {\it nuclear modification ratio} or ratio of the hadron 
transverse momentum spectrum measured in $A+B$ at a given rapidity $y$ and impact
parameter $b$ (or centrality class) normalised by the nuclear overlap
function $T_{AB}(b)$ -- related to the ``parton luminosity''  at $b$ -- 
over the $p+p$ spectrum:
\begin{equation}
  R_{AB}^h(p_T,y;b) =  
    \frac{1}{T_{AB}(b)} \frac{d^2N_h^{A+B}(b)}{dp_T^2 dy} 
        \, \Bigg/ \,
    \frac{d^2\sigma^h_{p+p}}{dp_T^2 dy} \, . 
  \label{eq:R_AB}
\end{equation}
$T_{AB}(b)$ is computed with a geometrical Glauber eikonal model 
of the nucleus-nucleus collision (see e.g.~\cite{Miller:2007ri}).
In the absence of nuclear effects, one would expect $R_{AB}=1$.
The behaviour of $R_{AB}^h(p_T,y;b)$ as a function of $p_T$ will be discussed in Sections \ref{sec:hadrons-hA} and \ref{sec:hadrons-AA}.

For Drell-Yan processes, one defines an analogous nuclear ratio:
\begin{equation}
  R_\text{DY} =  
    {\frac{1}{A}} \frac{d\sigma^\text{DY}_{h+A}}{dMdx_Fdp_T^2}
    \hspace*{0.2cm} \Bigg/ \,
    {\frac{1}{B}} \frac{d\sigma^\text{DY}_{h+B}}{dMdx_Fdp_T^2} \ .
  \label{DYRatio}
\end{equation}
The dilepton $p_T$-broadening is defined analogously to
Eq.~\eqref{eq:ptbroadening}, with $h\equiv\ell^+\ell^-$. 
Results for $R_\text{DY}$ will be presented in Section \ref{sec:DYdata}.

Medium modifications of hadron production in nDIS and heavy-ion
collisions can also be revealed by means of multi-particle azimuthal
correlations, which are  sensitive to the underlying parton-medium
interaction and to the properties of the medium.
For example, two-hadron correlations measured at RHIC revealed
significant broadening and softening of associated hadrons on the away
side of a triggered high-$p_T$ particle, which is consistent with the
observation of the hadron suppression in the single inclusive
measurements. 
In nDIS processes, the distribution of associated sub-leading hadrons
to the leading hadron challenges various models of
single inclusive hadron suppression.

\subsection{Hadronisation time estimates } \ \\
\label{sec:formationtimes}

Even though hadronisation is a non perturbative process, 
a few features can be extracted from general grounds.
A parton created in a high-energy collision can travel in the vacuum 
as a free particle only for a limited time because of colour
confinement: it has to dress-up in a colour-field of loosely bound
partons, which eventually will evolve into the observed hadron.
The same dressing process can be expected for partons
traveling in QCD matter, yet it will be modified by interactions with the
surrounding medium. In a deconfined nuclear medium such as the QGP,
the dressing process might furthermore be delayed until the medium
cools down and comes closer to the confinement transition. 

\begin{figure}[tb]
  \centering
  \includegraphics[width=10cm]{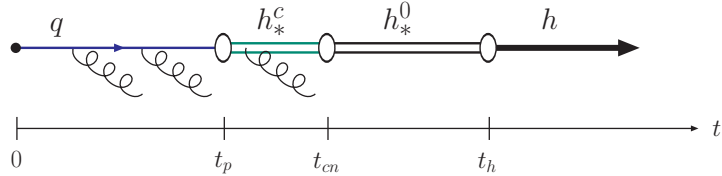}
  \caption{Sketch of the time evolution of the hadronisation process
    with definition of various time scales. A quark $q$ created at
    time 0 in a hard collision turns into a coloured prehadron $h_*^c$,
    which subsequently neutralises its colour, $h_*^0$, and collapses
    on the wave function of the observed hadron $h$. Gluon radiation 
    lasts until colour neutralisation.} 
  \label{fig:hadrosketch}
\end{figure}

The bare parton-medium cross section is dominated by the elastic 
$a+b \ra a+b$ parton-parton scattering and gluon bremsstrahlung.
The gluon-gluon cross-section is of order $\sigma_{gg} = 9/2\, \pi \alpha^2_s/\mu^2 \approx$~2~mb
(for $\alpha_s$~=~0.5 at a $p_T$-cutoff of order $\mu$~=~1~GeV),
but the dressed parton is likely to develop an inelastic cross section 
of the order of the hadronic one, $\mathcal{O}(40$~mb$)$, becoming subject to 
nuclear absorption similarly to the final state hadron. Hence, it can be viewed as a
``prehadron'' (denoted here by $h_*$).
The prehadron may still be for a short time in a coloured state and radiate gluons 
neutralising its colour before its wave function collapses
onto the observed hadron wave function. We can therefore identify three
relevant time scales, see Fig.~\ref{fig:hadrosketch}: 
(1) the ``prehadron production time'' or ``quark lifetime''
$t_{preh}$, at which the dressed quark develops an inelastic cross section,
(2) the ``colour neutralisation time'' $t_{cn}$, at which gluon
bremsstrahlung stops, and (3) the ``hadron formation time'' $t_h$, at
which the final hadron is formed. Typically, model applications further
simplify the process and merge steps (2) and (3) assuming $t_{preh}=t_{cn}$.
Note that the prehadron and the formation times are
introduced as a phenomenological tool, rather than a well defined
quantity, in order to 
distinguish between the stage in which the parton can be described as an
asymptotically free particle and treated in pQCD from the stage in
which colour confinement and non-perturbative interactions kick in and
warrant a treatment in terms of different degrees of freedom. 
Such a phenomenology is well suited to the present status of the
theoretical and experimental investigation, but will need to be
substantiated or replaced by a more fundamental QCD description.
Finally, note that strictly speaking the very question ``is the prehadron
formed within or without the medium?'' is ill-posed: in quantum
mechanics it can happen one way in the amplitude and the other in its
complex conjugate, and the interference between the two may be
non-negligible \cite{Kopeliovich:2008uy}.\\

A simple estimate of the hadron formation time $\vev{t_h}$ can
be obtained by defining it as the time for the 
struck partons to build up its colour field and to develop the
hadronic wave function~\cite{Dokshitzer:1991wu}. 
In the hadron rest frame this time is related
to the hadron radius $R_h$, and in the laboratory frame it is boosted 
by $\gamma = E_h/m_h$:
\begin{align} 
  \vev{t_h} \propto R_h \frac{E_h}{m_h} 
 \label{eq:wangest}
\end{align}
In Table~\ref{table:pertformtime} we show the hadron
formation time estimated with Eq.~\eqref{eq:wangest}
for a typical 7~GeV pion ($R_h\approx$~0.7~fm~\cite{Amendolia:1984nz}) 
at the  kinematics conditions found in HERMES ($z \approx z_h\approx$~0.5, $E_h \approx z_h\nu \approx 7$~GeV) 
and at RHIC mid-rapidity ($z\approx$~0.7, $E_h \approx p_T^h \approx 7$~GeV/c). 
We find $\vev{t_\pi} \approx 35$ fm $\gg R_A$, which points towards a long quark lifetime
with hadron formation outside the medium. However, for the heavier
kaons, $\eta$, and protons (with radii $R_K\approx$~0.6 fm~\cite{Amendolia:1986ui}, 
$R_\eta\approx R_\pi$, $R_p\approx$ 0.9~fm~\cite{Sick:2003gm} resp.) 
we obtain much shorter formations times  $\vev{t_K} \approx 8$~fm, $\vev{t_\eta} \approx 9$~fm
and $\vev{t_p} \approx 6$~fm, which are comparable to the size of the medium. Heavy $D$ and $B$ mesons,
with average radii $R_D$~=~0.57~fm and $R_B$~=~0.5~fm respectively~\cite{Simonov:2007bm}, 
clearly fragment in-medium ($\vev{t_{D,B}}\lesssim$~2~fm).

\begin{figure}[tb]
  \centering
  \includegraphics[height=3cm]{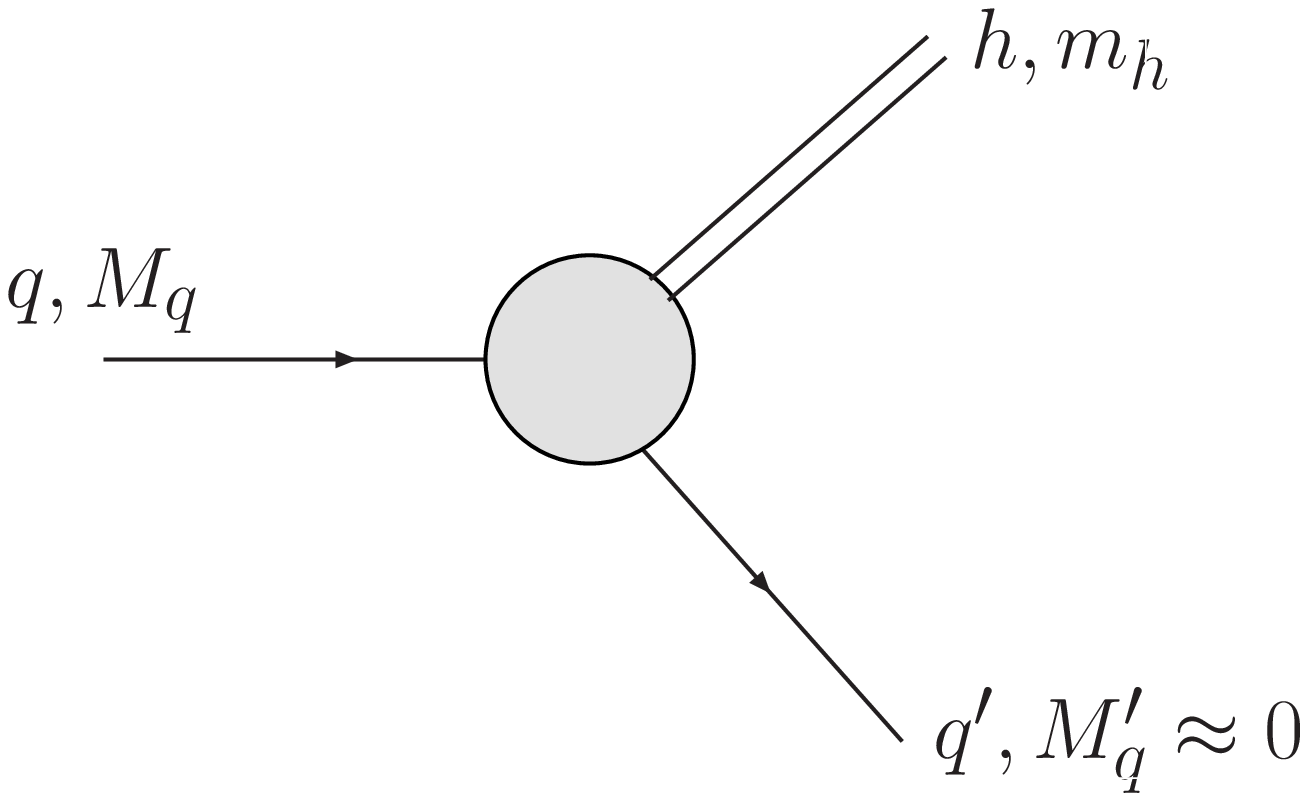}
  \includegraphics[height=3cm]{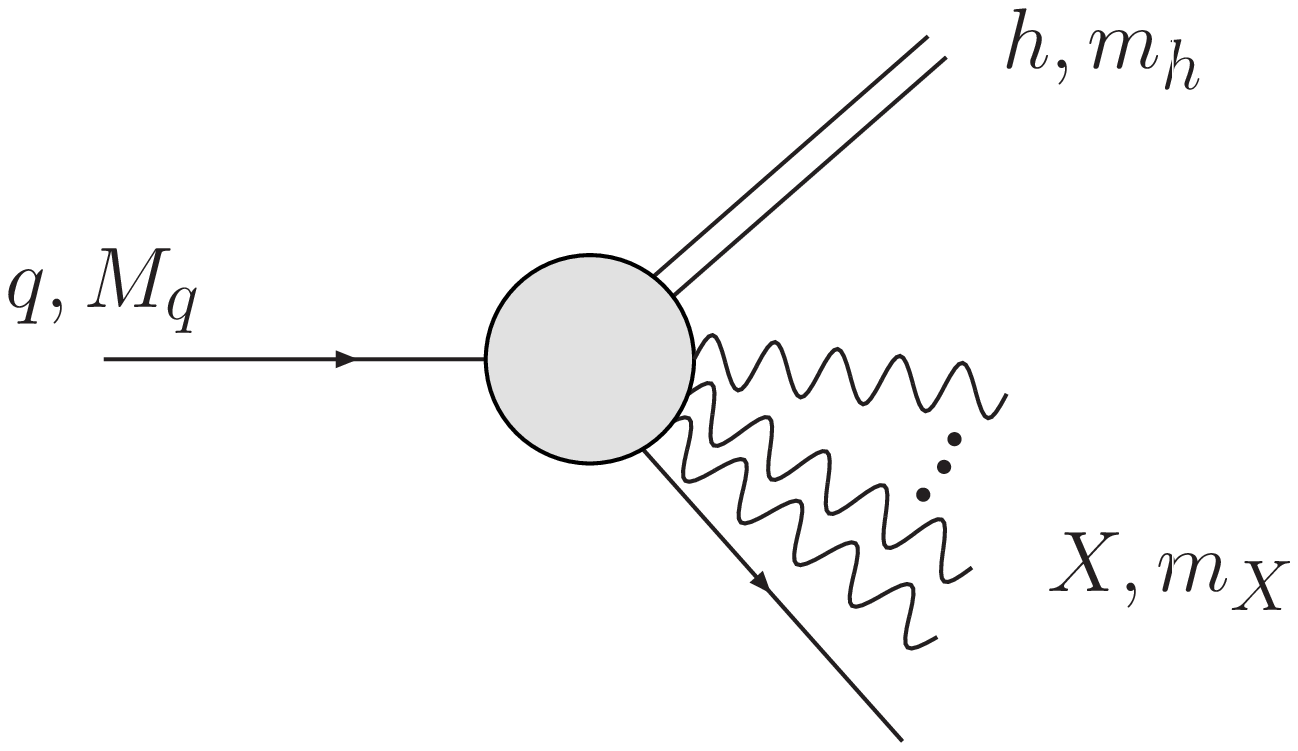}
  \caption{Quark hadronisation by emission of one parton ({\it left}) and
    emission of one parton accompanied by many soft gluons ({\it right}).}
  \label{fig:pertformtime}
\end{figure}

An estimate of the prehadron production time can be obtained by
looking at hadronisation in light-cone coordinates \cite{Adil:2006ra}. 
Consider a relativistic on-shell quark of mass $m_q$ and
plus-momentum $p^+$, hadronising into a hadron
of mass $m_h$ and 4 momentum $p_h^+ = z p^+$.
Minimally prehadron formation, i.e., the formation of a colorless 
partonic object, proceeds by emission of an additional,
(typically light) parton to carry away the initial state colour, see
Fig.~\ref{fig:pertformtime} left. The process in momentum space is
\begin{align}
  \left[ p^+,\frac{m_q^2}{2p^+},\vec 0_T \right] 
    \lora \left[ z p^+,\frac{m_h^2+\vec k^2}{2 z p^+},\vec k \right] 
    + \left[ (1-z) p^+,\frac{\vec k^2}{2 (1-z) p^+}, - \vec k \right]
 \label{eq:pertproc}
\end{align}
where we imposed 4-momentum conservation. In time-ordered
perturbation theory, the light-cone separation $\Delta x^+$ 
between the initial and final state can be estimated by the 
uncertainty principle: 
\begin{align}
  \Delta x^+ \approx 1/\Delta p^- =  
    \frac{2z\;(1-z)\;p^+}{\vec k^2 \;+ \;(1-z)\;m_h^2 -z\;(1-z)\;m_q^2} \ ,
 \label{eq:pertestimate}
\end{align}
where $\Delta p^- = p^-_{q'} + p^-_h - p^-_q$. Since $\Delta x^+ =
(\Delta t_h + \Delta z_h)/\sqrt 2$, and $\Delta z_h=(p_q/E_q)\Delta
t_h$ with $p_q$ $(E_q)$ the momentum (energy) of the quark, the prehadron 
formation time is
\begin{align}
  \vev{t_{preh}} = \frac{\sqrt{2}}{1+p_q/E_q} \;\Delta x^+ \ . 
 \label{eq:thpertest}
\end{align}
This estimate should be used with care since actually prehadron
formation is likely to be accompanied by the emission of many 
soft gluons, see Fig.~\ref{fig:pertformtime}, and the system of
emitted partons has an invariant mass $m_X^2$.  
Taking this into account, we should add an
additional $z m_X^2$ term at the denominator of
Eq.~\eqref{eq:pertestimate}, which would reduce the estimated 
formation time. 
In Table~\ref{table:pertformtime} we show the hadron
formation time from Eq.~\eqref{eq:thpertest} for typical
hadrons at HERMES and RHIC, where we assumed 
$\vec k^2 \approx \Lambda_{QCD}^2$. 
High energy pre-pions are formed outside
the medium ($\vev{t_{pre-\pi}}\approx 25$~fm), while pre-kaons and pre-$\eta$
($\vev{t_{pre-K}}\approx 6-9$~fm) and pre-protons
($\vev{t_{pre-p}}\approx 4$~fm) have formation times comparable to the
medium size, and heavy pre-mesons are produced rapidly inside the medium. 

At least for light mesons, these estimates can be used to
justify the computation of hadron quenching in terms of parton-medium
interactions alone, as done in radiative energy loss models
\cite{Wang:2002ri,Arleo:2002kh,Gyulassy:2000er}, see
Section~\ref{sec:parton}. 

\begin{figure}[tb]
  \parbox{6cm}{
    \includegraphics[width=6cm]{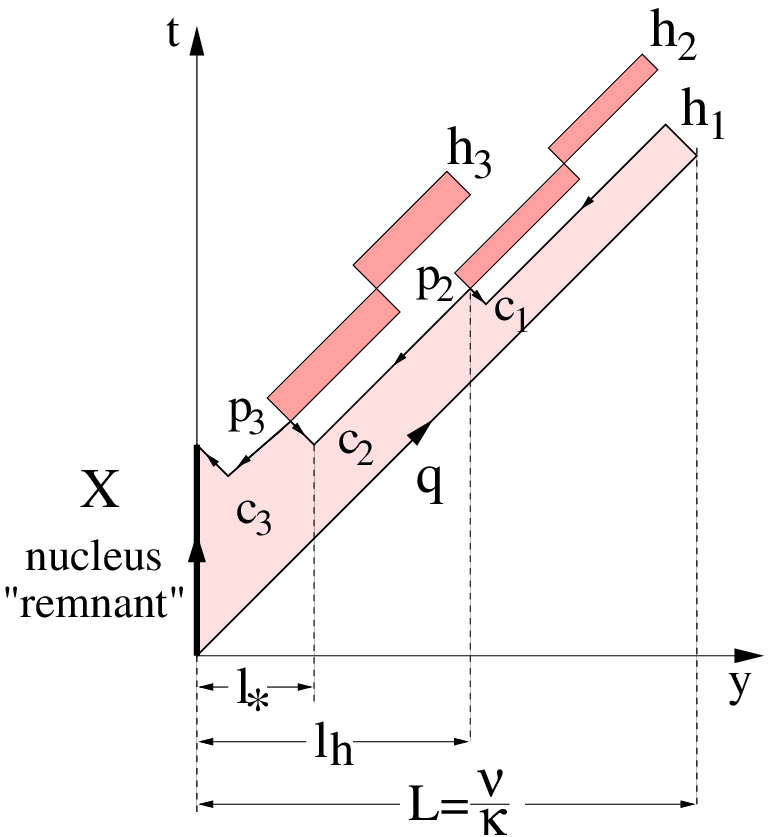}
  }
  \hspace*{0.8cm}
  \parbox{7cm}{
    \includegraphics[width=7cm]{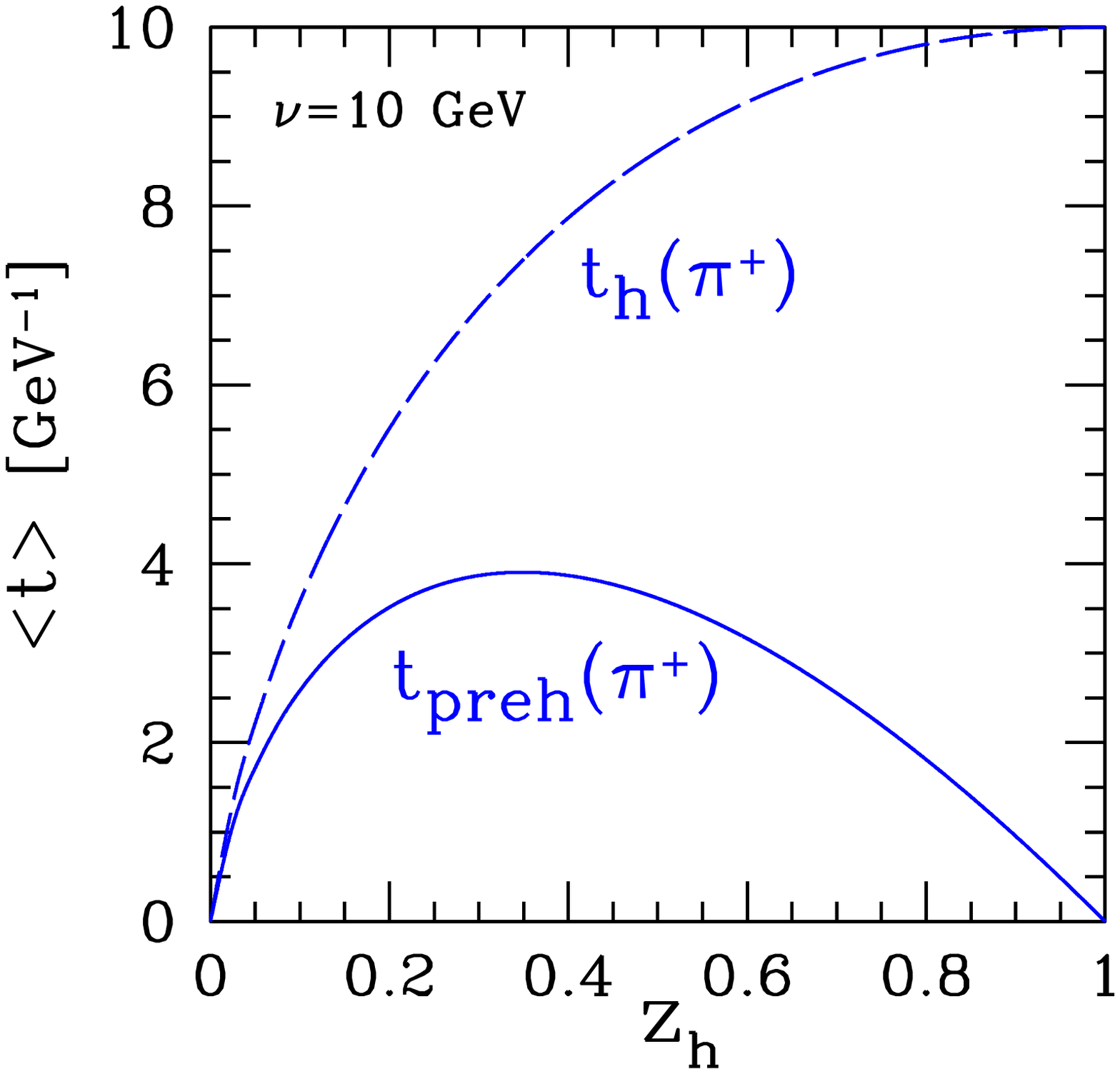} \\[0.8cm]
  }
  \caption{
    {\it Left}: Hadronisation in the Lund string model in the target
    rest frame. {\it Right}: Prehadron production time $\vev{t_{preh}}$,
    and hadron formation times $\vev{t_h}$, for $\nu=10$~GeV quarks
    fragmenting into pions with fractional momentum $z_h$, as computed
    within the Lund string model in Ref.~\cite{Accardi:2005jd}. 
  }
  \label{fig:Lundestimate}
\end{figure}

A successful non-perturbative model of hadronisation is the Lund string
model~\cite{Andersson:1983ia}, see Fig.~\ref{fig:Lundestimate} left.
The confined colour field stretching from the 
struck quark to the rest of the nucleus is modeled as a
string of tension $\kappa_{str} \approx 1$~GeV/fm, and spans the
lightly shaded area in space-time. The prehadron
formation point is identified with the $q\bar q$ pair production point
$C_i$ which breaks the string in smaller pieces~\cite{Bialas:1986cf}.
Hadrons are formed at points $P_i$ when a quark and an antiquark at
the endpoint of a string fragment meet. The subscript $i$ indicates
the so-called rank of the produced (pre)hadron, counted from the right
of the figure. Average prehadron production times can be analytically
computed~\cite{Bialas:1986cf,Accardi:2002tv,Accardi:2005jd} and have the
following general structure:
\begin{equation}
  \vev{t_{preh}} = f(z_h)\; (1-z_h)\; \frac{z_h \nu}{\kappa_{str}} \;\;\;\; , \;\;\; \vev{t_h}  = \vev{t_{preh}} \; + \;\frac{z_h \nu}{\kappa_{str}} \ ,
 \label{eq:lundest}
\end{equation}
where $\nu$ is the struck quark energy, the function $f(z_h)$ 
is a small correction of $\vev{t_{preh}}$, which can be computed
analytically in the standard Lund model~\cite{Accardi:2005jd,Bialas:1986cf}, 
and $\kappa_{str}$~=~(1~GeV/fm)~$R_\pi^2/R_h^2$
with $R_h$ the hadron radius is taken from Ref.~\cite{Accardi:2005jd}.
The factor $z_h \nu$ can be understood as a Lorentz boost factor;
the $(1-z_h)$ factor is due to energy conservation: a high-$z_h$ hadron
carries away an energy $z_h\nu$; the string remainder has a small
energy $\epsilon = (1-z_h)\nu$ and cannot stretch farther than
$L = \epsilon/\kappa_{str}$. Thus the string breaking occurs on a
time-scale proportional to $1-z_h$. 
The resulting pion formation time scaled by a factor $\nu/\kappa_{str}$ is plotted in
Fig.~\ref{fig:Lundestimate} left. A typical pion produced at
HERMES energies (i.e. with fractional energy $z_h\approx$~0.5 from a parent quark with energy $\nu\approx$~14~GeV) 
has $\vev{t_{preh}} \approx 6$ fm $\lesssim R_A$ and 
$\vev{t_h} \approx 13$ fm $\gtrsim R_A$, with
similar values at RHIC at mid-rapidity. Therefore, the final hadron is
typically formed at the periphery or outside the nucleus so that its
interaction with the medium is negligible (see
Table~\ref{table:pertformtime}).   
However, the prehadron is formed inside and can start interacting
with the nucleus.
A detailed space-time analysis of hadronisation in the PYTHIA/JETSET
Monte Carlo implementation of the Lund string model has been performed
in~\cite{Gallmeister:2005ad}, with similar conclusions regarding the
magnitude of the pion prehadron production time.

\begin{table}[tb]
  \centering
  \begin{tabular}{cccccccc} \hline
$\vev{t_h}$ & kinematics & $\pi$& K    &$\eta$ & p     & D      & B      \\\hline
Eq.~\eqref{eq:wangest}  
       & HERMES& 34 fm&  8 fm&  9 fm &  6 fm &  1.9 fm & 0.6 fm \\
Eq.~\eqref{eq:wangest}  
       & RHIC  & 34 fm&  8 fm&  9 fm &  6 fm &  1.9 fm & 0.6 fm \\\hline
Eq.~\eqref{eq:lundest}  
       & HERMES& 11 fm&  9 fm&       & 18 fm &        &        \\
Eq.~\eqref{eq:lundest}  
       & RHIC  &  9 fm&  8 fm&       & 13 fm &        &        \\\hline  
$\vev{t_{preh}}$ \\\hline
Eq.~\eqref{eq:thpertest}
       & HERMES& 28 fm&  9 fm&  7 fm &  3 fm & 0.8 fm & 0.1 fm \\
Eq.~\eqref{eq:thpertest}
       & RHIC  & 18 fm&  7 fm&  6 fm &  3 fm & 0.8 fm & 0.1 fm \\\hline
Eq.~\eqref{eq:lundest}  
       & HERMES&  4 fm&  4 fm&       &  6 fm &        &        \\
Eq.~\eqref{eq:lundest}  
       & RHIC  &  2 fm&  3 fm&       &  1 fm &        &        \\\hline  
\end{tabular}
  \caption{Estimates of typical hadron formation times $\vev{t_h}$ and
    prehadron production times $\vev{t_{preh}}$
    for pions, kaons, $\eta$, protons, and D and B mesons at HERMES
    ($z_h\approx0.5$, $\nu \approx 14$~GeV) and at RHIC at
    mid-rapidity ($p_T^h \approx 7$~GeV/c, $z \approx 0.7$) 
    obtained with Eqs.~\eqref{eq:wangest},~\eqref{eq:thpertest},
    and~\eqref{eq:lundest}.
  }
  \label{table:pertformtime} 
\end{table}

\begin{figure}[t]
  \centering
  \parbox[c]{6.9cm}{\includegraphics[width=7cm]{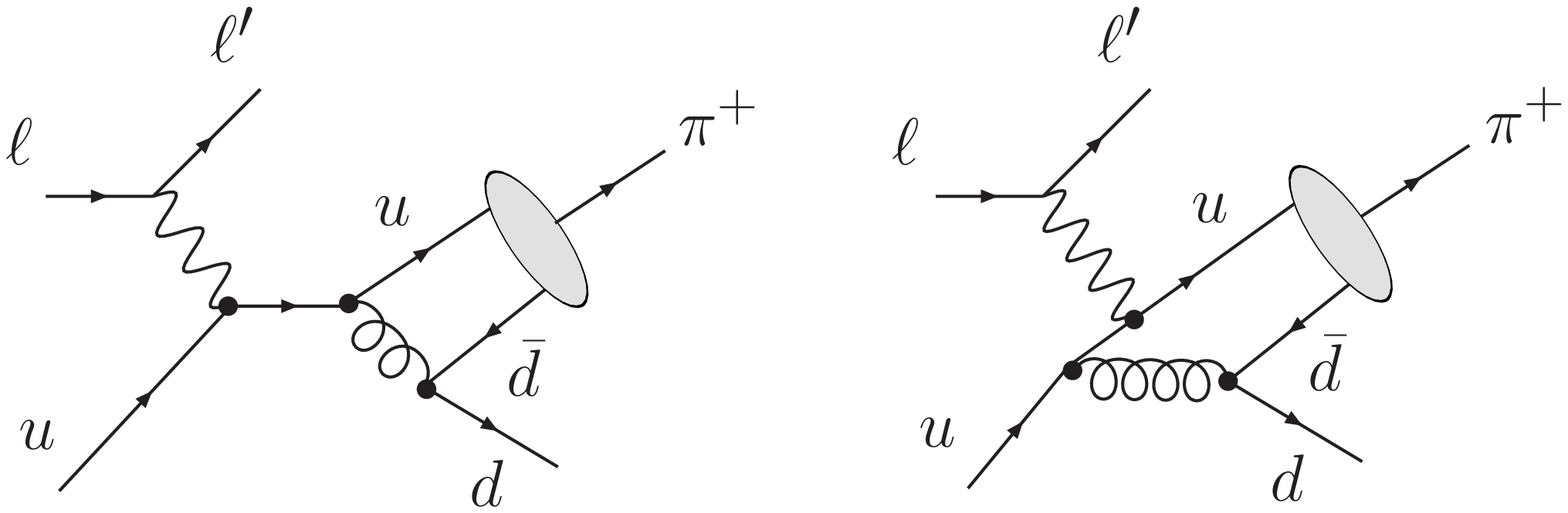}} 
  \parbox[c]{0.5cm}{\vskip0.4cm \text{\Large $\approx$}} 
  \parbox[c]{5.2cm}{\includegraphics[width=5.5cm]{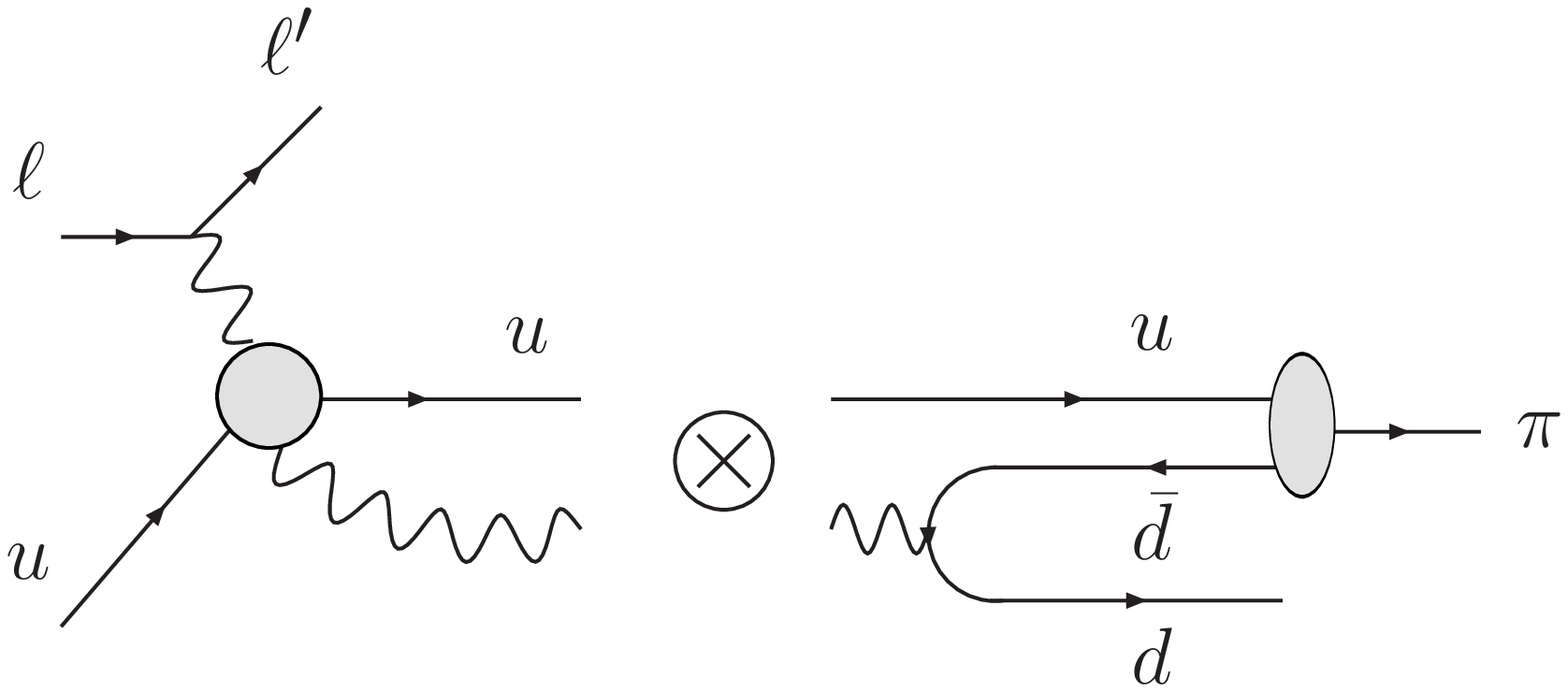}}
  \caption{Diagrams for leading hadron formation in $\ell + A$ collisions: 
    Gauge-invariant set~\cite{Berger:1979xz} ({\it left}), and 
   dipole-model approximation~\cite{Kopeliovich:2003py} ({\it right}).}
  \label{fig:perthadronisation}
  \label{fig:dipmodel} 
\end{figure}

\begin{figure}[t]
  \vspace*{-.2cm}
  \centering
  \includegraphics[width=6cm,origin=c]{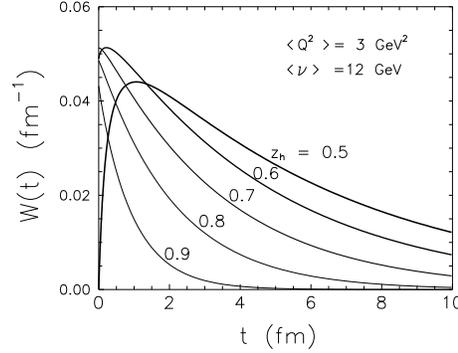}
 \vspace*{-0.3cm}
 \caption[]{
   Probability distribution of the prehadron production time
   $t_{preh}\equiv t$ obtained in the colour-dipole model of~\cite{Kopeliovich:2003py}.
 \label{fig:tpdipolemodel}
 }
\end{figure}

In Ref.~\cite{Berger:1979xz,Berger:1979kz,Kopeliovich:2007yv}
the formation of a leading hadron ($z_h \gtrsim 0.5$) is described in a pQCD
model, see Fig.~\ref{fig:perthadronisation}.
The struck quark radiates a gluon.  The gluon then splits into a
$q\bar q$ pair, and the $\bar q$ recombines with the struck $q$ to
form the leading prehadron, which later on collapses on the hadron
wave function. The cross-section
can be computed from the modulus squared of the sum of the two gauge-invariant 
amplitudes shown in the figure. At $z_h \approx 1$, higher
twist effects spoil this simple mechanism for hadronisation. 
The dipole model of Ref.~\cite{Kopeliovich:2003py} approximates the described 
cross-section as a convolution of a Gunion-Bertsch radiation cross
section~\cite{Gunion:1981qs} with the gluon splitting plus quark recombination process, see
Fig.~\ref{fig:dipmodel} right. The prehadron is identified with the 
$q \bar q$ pair which includes the struck quark, 
and its production time is identified with the time at which
the gluon becomes decoherent from the struck quark. Note
that, strictly speaking, the resolved
quark-gluon system may be in an octet state from the production time 
until gluon splitting occurs. However, in this model, gluon radiation
is neglected during the octet stage.
The model can compute the probability distribution in the prehadron
production time, see Fig.~\ref{fig:tpdipolemodel}, and the average
$\vev{t_{preh}}$ is
\begin{align}
  \vev{t_{preh}} \propto (1-z_h) \frac{z_h \nu}{Q^2} \ .
\label{eq:dipolest}
\end{align}
The interpretation of the $1-z_h$ factor is in terms of energy
conservation: the longer the struck quark propagates, the larger its
energy loss; hence to leave most of its energy to a $z_h \ra 1$
hadron, the quark must be short lived. The scale is set by
$\kappa_{dip} = Q^2$. At HERMES, with $Q^2 \approx 10$~GeV/fm,  
this model obtains for pion production $\vev{t_{preh}} \lesssim 5$ fm
at $z_h>0.5$: pre-pions are formed inside the medium. At
RHIC, where $Q^2 \propto p_T^2$, and $z_h \nu \approx p_T$ for
mid-rapidity hadrons, one obtains the counter-intuitive result that 
$\vev{t_{preh}} \propto 1/p_T$: prehadrons are formed the quicker the higher
their transverse momentum~\cite{Kopeliovich:2003py,Kopeliovich:2006xy},
typically inside the medium. 

In summary, given the model dependences of the theoretical
estimates of the hadron formation time -- and hence of
the length of the partonic phase of the hadronisation process --
it is important to study the kinematic- and flavour-dependences 
of various hadron production processes through a careful analysis of
experimental data and tests of phenomenological models, see
the discussion Section~\ref{sec:future}.




\section{Experimental results 
in lepton-nucleus deep inelastic scattering}
\label{sec:hadrons-lA}

Deep inelastic lepton-nucleus scattering offers a direct way to study the hadronisation process 
which follows the hard scattering.
In contrast to hadronic collisions,
in nuclear DIS no deconvolution of the parton distributions
of the projectile and target particle is needed,
so that the experimental observables can be more directly related to the nuclear 
effects on the quark propagation and fragmentation. 
Moreover, in electron and muon experiments the incoming and outgoing leptons 
are detected and they provide an extra handle to determine the kinematical variables 
of the produced partons/hadrons.

In the past, semi-inclusive leptoproduction of 
undifferentiated hadrons from nuclei has been studied 
with neutrinos at FNAL, CERN and Serpukhov;
with electrons at SLAC~\cite{Osborne:1978ai},
and at CERN and FNAL with high-energy muons by 
EMC~\cite{Arvidson:1984fz} and by
E665~\cite{Adams:1993mu} respectively. 
Recently, HERMES has reported more precise data~\cite{Airapetian:2000ks,Airapetian:2003mi,Airapetian:2007vu} on the 
production of charged hadrons as well as, for the first time, various identified hadrons ($\pi^+$, $\pi^-$, $\pi^0$, $K^+$,  $K^-$, 
$p$ and $\bar{p}$) in deep-inelastic positron scattering off nuclei. Finally, high-statistics data have been collected at JLab~\cite{Brooks:2008wk} with a
5.0~GeV electron beam on targets of carbon, aluminum, iron, tin, and lead at large luminosities ($2\,10^{34}$~cm$^{-2}$s$^{-1}$), 
with detection of several identified hadrons, particularly $\pi^+$, $\pi^0$, $\pi^-$, $K^0$, and $\Lambda$. 
In the next Sections the most significant experimental results  for
DIS of  neutrino, muon, electron and positron beams on nuclei are presented.

\subsection{Hadron production in $\nu$-nucleus DIS } 
\label{sec:nu-Adata}

Neutrinos can interact via the exchange of $Z^0$ (neutral
current), or can turn into charged leptons via $W^\pm$ exchange, 
while at the energies of interest here
electrons scatter primarily through the exchange of photons.
One thus has at hand various methods to extract information on the parton production and propagation 
combining results from experiments encompassing all four (neutral and charged) exchanged bosons. 
In contrast to hadron-induced collisions, which preferentially scatter from the front hemisphere of a nucleus due to 
their strong interaction probability, neutrinos and electrons  directly scatter with partons or nucleons
inside the nuclear target because of their weak interaction probability.
This means that the fragments of the struck parton in neutrino and electron scattering experiments should 
suffer the same final-state interactions.
In general, the laboratory energy $E_{\nu}$ of the incoming (anti)neutrino cannot be determined directly since there is 
substantial energy that goes to undetected neutral particles. Corrections on the estimation of the neutrino 
energy reflect on most of the kinematic quantities and significantly increase the systematic uncertainties for neutrino experiments.

\begin{figure}[tbp]
  \centering
  \includegraphics[width=8.5cm]{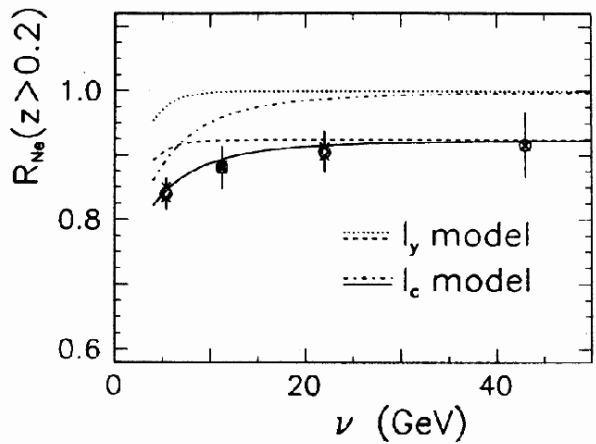}
  \caption{Ratio of the differential  hadron multiplicity distribution
    in (anti)neutrino-$Ne$ over -$He$ collisions as function of the transferred quark energy $\nu$
    for fast ($z_h > 0.2$) charged hadrons~\cite{Burkot:1996sf} compared to 
    model predictions for ``constituent'' ($l_c$) and ``yo-yo'' ($l_y$) lengths (see text for details).} 
  \label{fig:bebc_nu}
\end{figure}

Neutrino-induced hadron production was first studied in the bubble chamber at FNAL~\cite{Berge:1978ie} 
and later in the bubble chamber BEBC at CERN~\cite{Deden:1981pf}. The main objective of these experiments 
was the study of global properties of hadron production in nuclei. Fast hadron production on 
nuclei was found to be attenuated as compared to that in hydrogen target and to depend both on the variables $z$ and $\nu$~\cite{Berge:1978ie}. 

The production of hadrons in charged-current (anti)neutrino interaction was studied with higher accuracy 
by the BEBC WA21/WA59 Collaborations~\cite{Burkot:1996sf}. While in the previous experiments the 
analysis was limited to negative hadrons in order to exclude systematic uncertainties due to 'knock-on' protons, 
charged hadron production was studied in this experiment in terms of $z_h$-distribution normalised to the number 
of events. A small but significant reduction of fast hadron ($z_h > 0.2)$ production was found in a neon target as 
compared to that in a hydrogen target. This is shown in Fig.~\ref{fig:bebc_nu} where the ratio of the normalised 
distribution is presented as function of the transferred energy $\nu = E_{\nu} -E_{\mu}$, where $E_{\mu}$ is the 
laboratory energy of the muon beam. The data indicates a significant (10\% -- 20\%) attenuation ($R_{Ne}<1$) 
of fast charged hadron yields over the whole $\nu-$range with a stronger attenuation at low $\nu$ and high $z_h$.
The experimental results are compared to theoretical predictions of Ref.~\cite{Bialas:1986cf} in which  two 
hadron formation lengths are considered: the constituent ($l_c$) and the ``yo-yo'' ($l_y$) lengths. Specifically, 
the constituent length corresponds to the time after which the first constituent of the hadrons is formed, the yo-yo 
length corresponds to the time after which the quark and antiquark meet to form the color singlet ~\cite{Czyzewski:1990pg}
The $l_y$-model overshoots the value of $R_{Ne}$, while a $l_c$-model give a fair description of the data thus 
pointing out that significant interactions of the hadronising system start as early as at the constituent point.

The role of the formation length for the description of hadroproduction in DIS of neutrino on nuclei has been 
investigated by the NOMAD experiment at CERN~\cite{Astier:2001wd,Veltri:2002jm}. The backward-going 
protons and $\pi^-$ produced in charged current neutrino interaction have been compared with  intra-nuclear 
cascade (INC) models. In these models the production of particles in kinematically forbidden regions
can be seen as the result of multiple scattering and of interactions of secondary hadrons with the other 
nucleons while they propagate through the nucleus. Experimentally it has been observed that the cascade is 
restricted to slow particles only, while the fast ones do not re-interact inside the nucleus. A proposed
 explanation for this effect is that, 
since the formation time is proportional to the hadron energy (via the Lorentz time-dilation factor $\gamma=E/m$), 
the INC process is restricted to slow hadrons which have formation lengths smaller than the nuclear 
radius~\cite{Astier:2001wd}.

\begin{figure}[tbp]
  \centering
  \includegraphics[width=6.5cm]{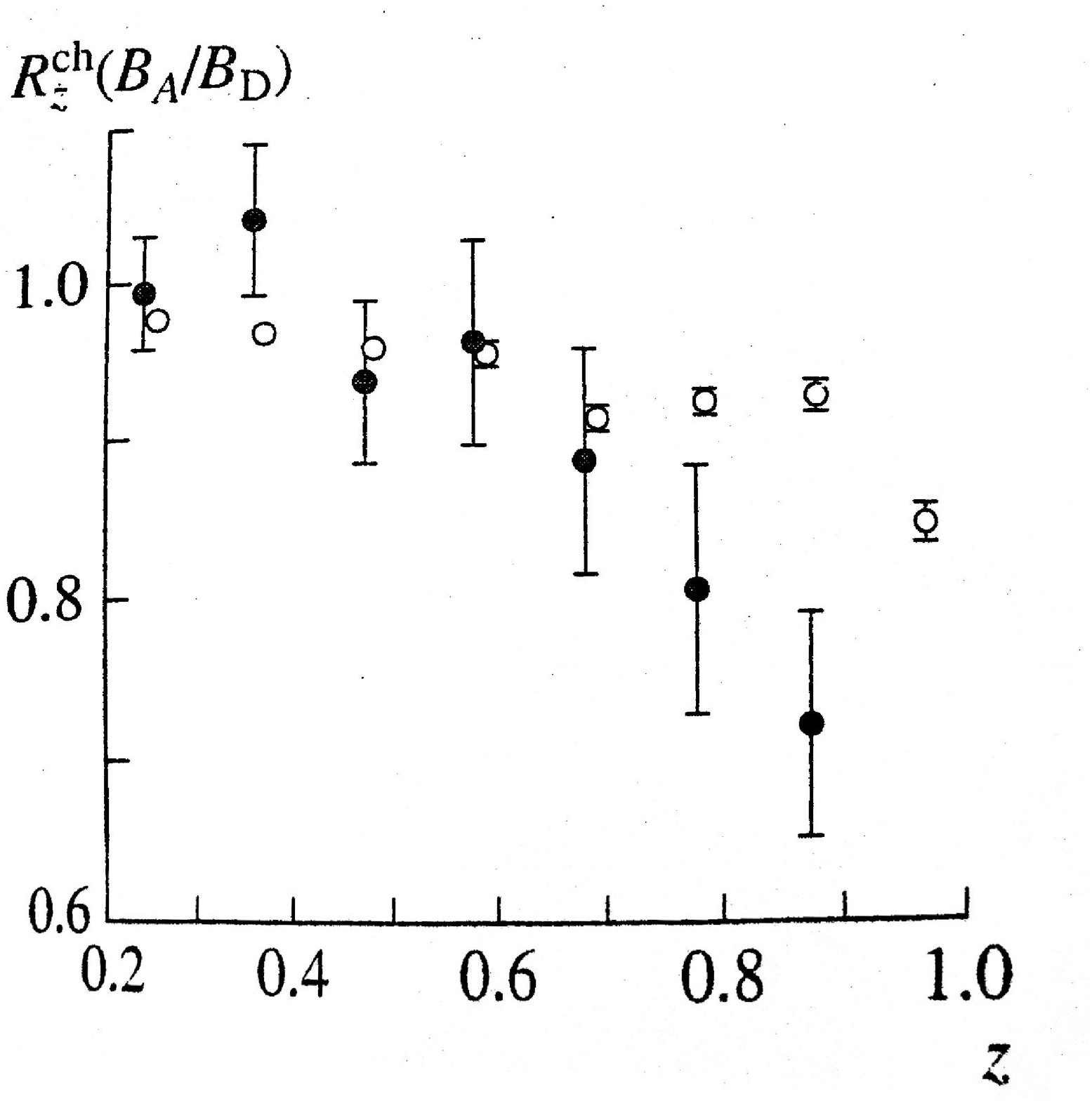}
  \includegraphics[width=6.5cm]{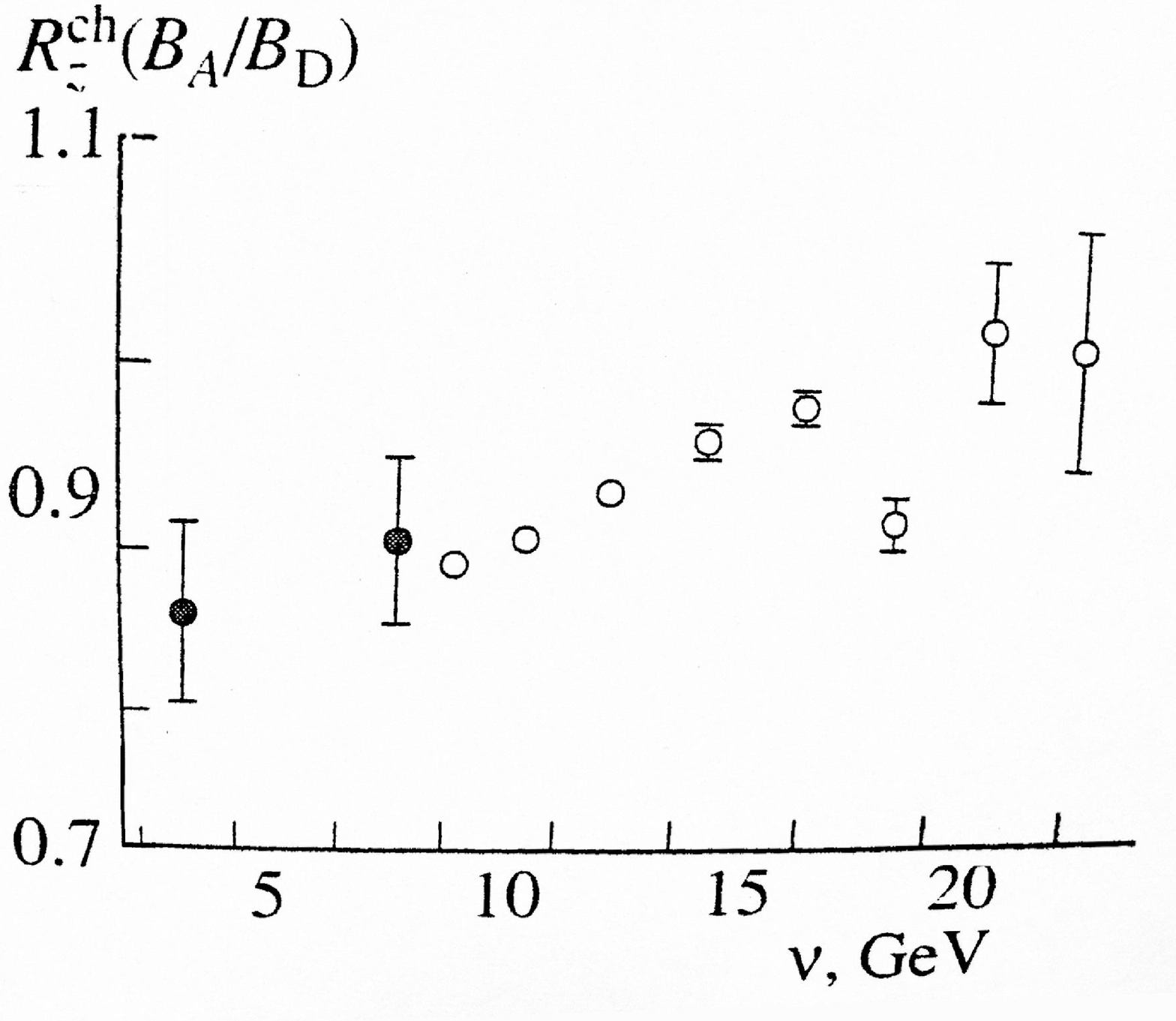}
  \caption{Ratio of the yields of charged hadrons in neutrino DIS on nuclei $A$ over deuteron $D$ targets
    (closed circles)~\cite{Agababyan:2003pe} and in $e^\pm$ DIS on nitrogen and deuteron nuclei
    (open circles)~\cite{Airapetian:2000ks}. {\it Left:} $z_h$-distribution for hadrons produced in processes 
    with transferred energy 2~$<\nu<$~15~GeV. {\it Right:} $\nu$-distribution for hadrons with $z_h>$~0.5. } 
  \label{fig:skat}
\end{figure}

The inclusive spectra of hadrons have been also measured with the aid of the SKAT 
propane-freon bubble chamber irradiated with a beam of 3 to 30~GeV neutrinos from the Serpukhov accelerator~\cite{Agababyan:2003pe}.
The ratio of the yields of charged hadrons in the subsamples  of nuclear interactions $B_A$ and of deuteron interactions
$B_D$  is shown  in Fig.~\ref{fig:skat} as a function of $z_h$ (left) and $\nu$ (right) respectively.
The left part of Fig.~\ref{fig:skat} displays the ratio measured at $\langle A \rangle= 28$  and energies in the range 
2~$< \nu <$~15~GeV, compared with  the multiplicity ratio obtained  from deep inelastic scattering of positron 
on nitrogen nuclei ($A$~=~14) at higher transfer energies of 7~$< \nu <$~24~GeV~\cite{Airapetian:2000ks}.

The suppression of the hadron yield 
appears more pronounced for the most energetic hadrons ($z_h >$~0.6). 
The $\nu$ dependence of the ratio for leading charged hadrons with $z_h>$~0.5 is shown in the right part of 
Fig.~\ref{fig:skat}, which shows the SKAT data at $\langle \nu \rangle$~=~3.3 and 7.7~GeV, along with data 
from $e^\pm$-$^{14}N$ interactions in the region $\nu >$ 8~GeV. 
The reduction of the yields occurs for hadrons produced in processes where the parent quark has the lowest energies.
The neutrino data, within their larger uncertainties, show the same trend as observed with electron beams.

\subsection{Hadron production in $e$-nucleus (SLAC) and $\mu$-nucleus (CERN) DIS} 
\label{sec:eAdata}

In contrast to the neutrino scattering experiments, in  electron- and muon-induced DIS  the incoming 
and scattering leptons are detected and provide a well defined reference system for the measurements 
of the kinematics of the outgoing parton/hadrons.

\begin{figure}[tbp]
  \centering
  \includegraphics[width=7.cm]{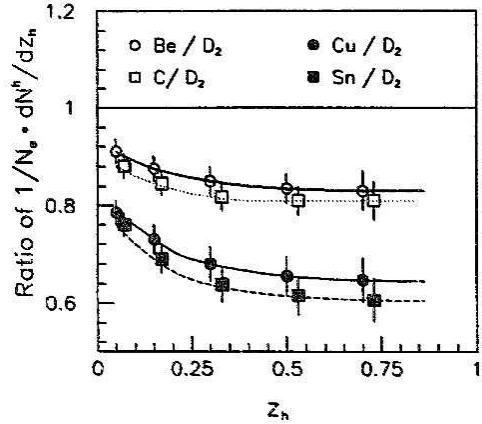}
  \caption{Ratio of differential charged hadron multiplicities as function of $z_h$ measured 
   in $e+A$ interactions at  SLAC~\cite{Osborne:1978ai} and derived in Ref.~\cite{Pavel:1991fj}. The curves 
represent a fit to the data in the functional form $(A/2)^{\alpha(z)}$.} 
  \label{fig:slac}. 
\end{figure}

\begin{figure}[tbp]
  \centering
  \includegraphics[width=12cm]{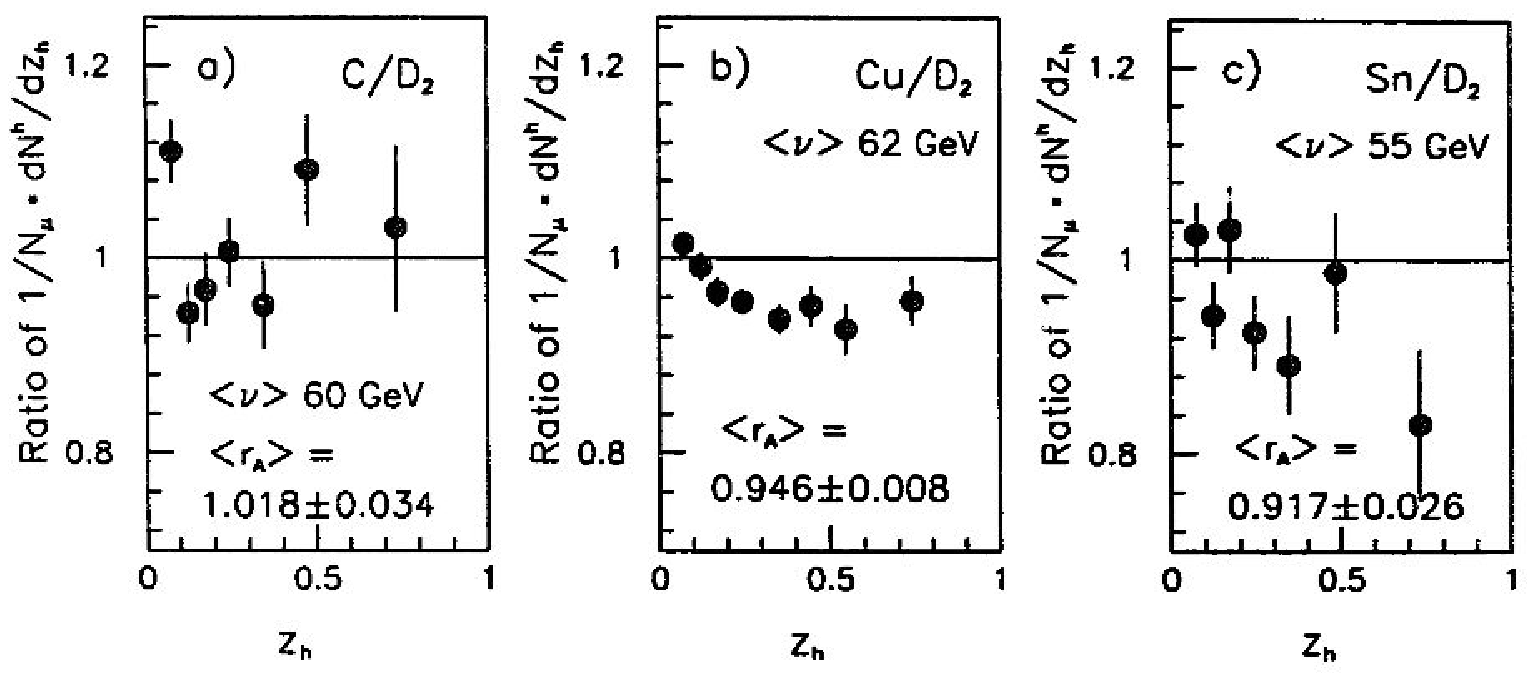}
  \caption{Ratio of the differential  hadron multiplicity distribution
    as function of $z_h$ measured in $\mu$-$A$ interactions by EMC~\cite{Ashman:1991cx}.} 
  \label{fig:emc_z}
  \includegraphics[width=5cm]{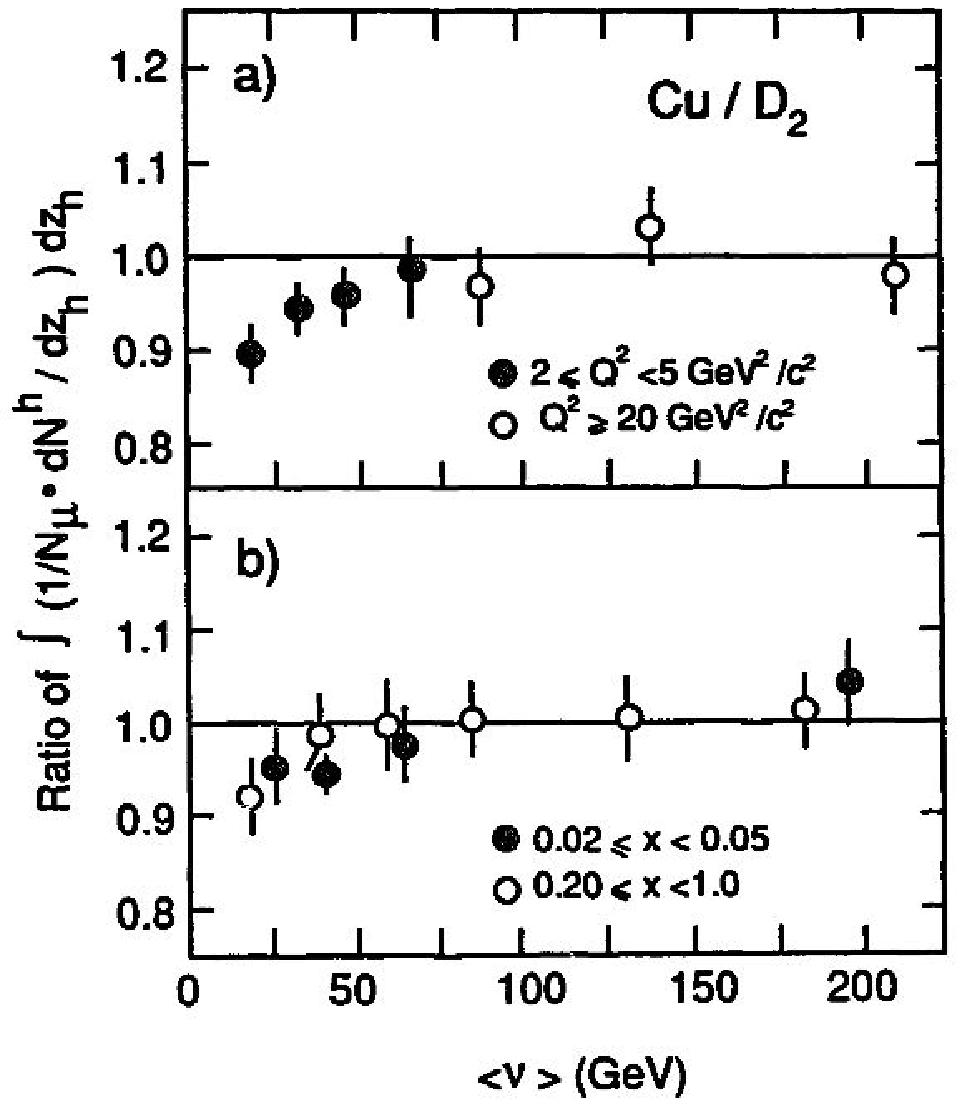}
  \includegraphics[width=8.2cm,height=5cm,angle=359.5]{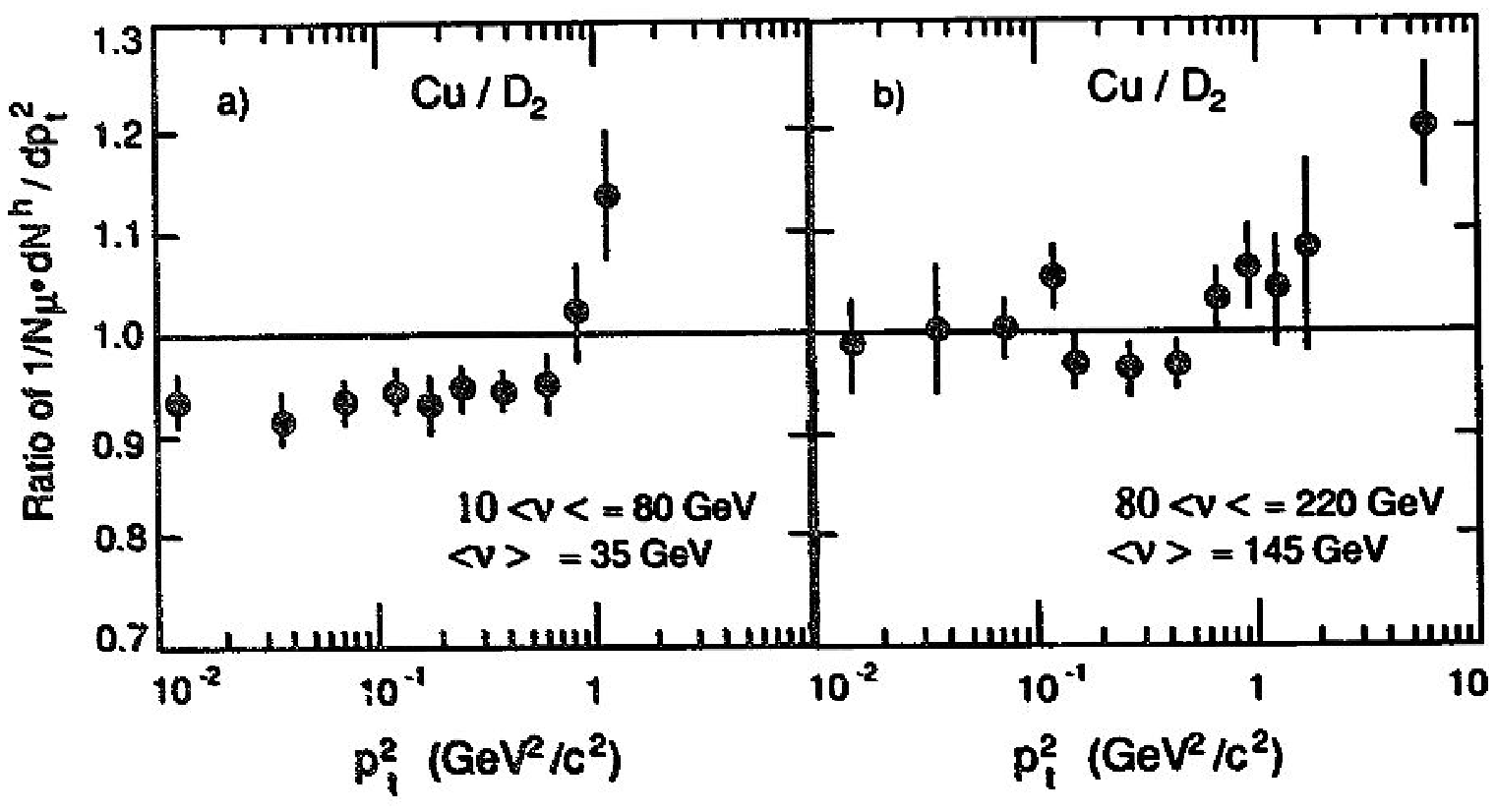}
  \caption{Ratio of the  differential  hadron multiplicity
    distribution as function of $\nu$ ({\it left}) and $p_T^2$ ({\it right}) 
    measured in $\mu$-$A$ collisions at EMC~\cite{Ashman:1991cx}.}  
  \label{fig:emc_nu_pt}
\end{figure}

Electroproduction of hadrons from nuclei was studied for the first time at the SLAC using a 20.5~GeV/c 
electron beam incident on different targets ($H$, $D$, $Be$, $C$, $Cu$, $Sn$) in the late seventies~\cite{Osborne:1978ai}.
A single arm spectrometer was used to measure the scattered electrons and the produced hadrons.
The ratio  between the number of single inclusive hadrons detected per gram per square centimeter 
per incident electron for nucleus  to the analogous number for deuterium was measured.
The results were presented in the original paper~\cite{Osborne:1978ai} as a function of  the 
transverse momentum $p_T$ of the hadron in the photon-nucleon system and 
$z_{c.m.}= p_l/p_{max}$, where $p_l$ is the longitudinal hadron momentum and $p_{max}=\sqrt{s/2}$.
The ratio of differential charged hadron multiplicities as a function
of $z_h$, derived in Ref.~\cite{Pavel:1991fj} from the values published
in the original paper, is shown in Fig.~\ref{fig:slac}.
This experiment  shows an attenuation of the electroproduced hadrons
which clearly increases with the size of the target nucleus. In
addition the nuclear attenuation increases for increasing hadronic
momenta. This measurement performed prior to the discovery of the EMC effect, does not account 
for modifications of the nuclear PDFs and, thus, needs correction because a semi-inclusive cross 
section ratio was measured instead of the multiplicity ratio as defined in Eq.~(\ref{eq:att}). 
An estimate of the correction for this effect based on Ref.~\cite{Airapetian:2000ks} results in a $4 \%$ increase of the original ratio.

After the pioneering measurements with muon beams performed at FNAL~\cite{Hand:1978tx,Adams:1994ri}, 
hadron production in DIS of muons was studied by the EMC-collaboration first in the $\nu$ range 
50-140~GeV~\cite{Arvidson:1984fz} and later on in the  $\nu$ range 20-220~GeV~\cite{Ashman:1991cx}.   
Their large single-arm spectrometer allowed this experiment to carry out a precise determination of the muon momentum 
with a good acceptance for forward produced hadrons. In addition, due to the simultaneous measurements of nuclear 
targets and deuterium, most of the systematics uncertainties cancel in this measurement.

The ratio of the hadron multiplicity distribution as function of $z_h$ is shown in Fig.~\ref{fig:emc_z} for different targets. 
The corresponding mean value of the multiplicity ratios and  of the $\nu$ variable are noted on the figure.
For large nuclei ($Cu$, $Sn$) a small but distinct reduction of the fast hadron production compared to that of deuterium 
is observed, whereas for carbon the ratio is compatible with unity over the whole range in $z_h$.
The left part of Fig.~\ref{fig:emc_nu_pt} shows the hadron multiplicity ratio as a function of $\nu$ in two $Q^2$ bins. 
The same variation with $\nu$ is seen in all intervals, thus no trend in $Q^2$ is observed. The ratios show a gradual 
decrease with decreasing $\nu$ below 60~GeV, whereas they slowly approach unity for higher $\nu$. The depletion 
of the fast hadron multiplicity in muon interaction with heavy targets is only $\sim$10$\%$ even in the low-$\nu$ bin.
The $p_T$ dependence was measured as shown in the right part of Fig.~\ref{fig:emc_nu_pt} for two $\nu$ intervals.
At high $p_T$ the ratio rises above unity in both $\nu$ intervals. The observed trend is consistent with the Cronin 
enhancement reported in hadron-nucleus collisions~\cite{Antreasyan:1978cw,Cronin:1974zm} (see Section~\ref{sec:hadrons-hA}). 
Since in lepton-nucleus, at variance with hadron-nucleus, collisions neither multiple scattering of the incident 
particle nor interaction of its constituents can contribute 
to the Cronin effect, the observed enhancement can be ascribed to rescattering effects in the {\it final-state} only.

In conclusion, the results from the experiments performed at SLAC with electrons and at CERN  with muons  
have shown that the multiplicity ratios mainly depend on the energy of the virtual photon $\nu$, and on
the fraction  $z_h$ of this energy carried out by the final hadron.
These experiments  demonstrate that the nuclear effects on the ratios decrease with increasing $\nu$.
This has been confirmed by the Fermilab experiment E665 ~\cite{Adams:1993mu} in DIS of 490~GeV muons off 
xenon and deuterium targets. No nuclear dependence in the $z_h$-distributions of the forward-produced hadrons 
was found in the $\nu$-range from 50~GeV to 500~GeV. 
Thus we can conclude from the performed measurements that the transfer energies $\nu$
where nuclear effects are the largest are in the range from a few~GeV to few tens of~GeV.
Such an energy dependence is easily understood within parton energy loss models: since the energy lost by quarks 
propagating through QCD media, $\epsilon$, is independent of their energy (i.e. $\nu$ in this context) 
in the high-energy limit, the {\it relative} energy-loss $\epsilon/\nu$ --~which controls the amount of 
hadron suppression in nuclear DIS~-- vanish at high energy making the ratio $R_m^h$ tend to 1.
In the hadron-oriented picture, 
the larger $\nu$ (parton energy) the longer the formation time,
which therefore decreases the amount of nuclear absorption.

\subsection{Hadron production in $e$-nucleus DIS at the HERMES experiment} 
\label{sec:HERMESdata}

The influence of the nuclear medium on lepto-production of  hadrons 
has been recently extensively studied by the  HERMES experiment at DESY in semi-inclusive 
deep-inelastic scattering of 27.6~GeV positrons off deuterium, nitrogen, neon, krypton and xenon targets.
The data were collected in the $\nu$-range 3-23~GeV using high density gas targets internal to the  positron storage ring. 
During these high-density runs HERA operated in a dedicated mode for the HERMES experiment.

\begin{figure}[tbp]  \center
  \centering
  {\includegraphics[width=0.49\linewidth]{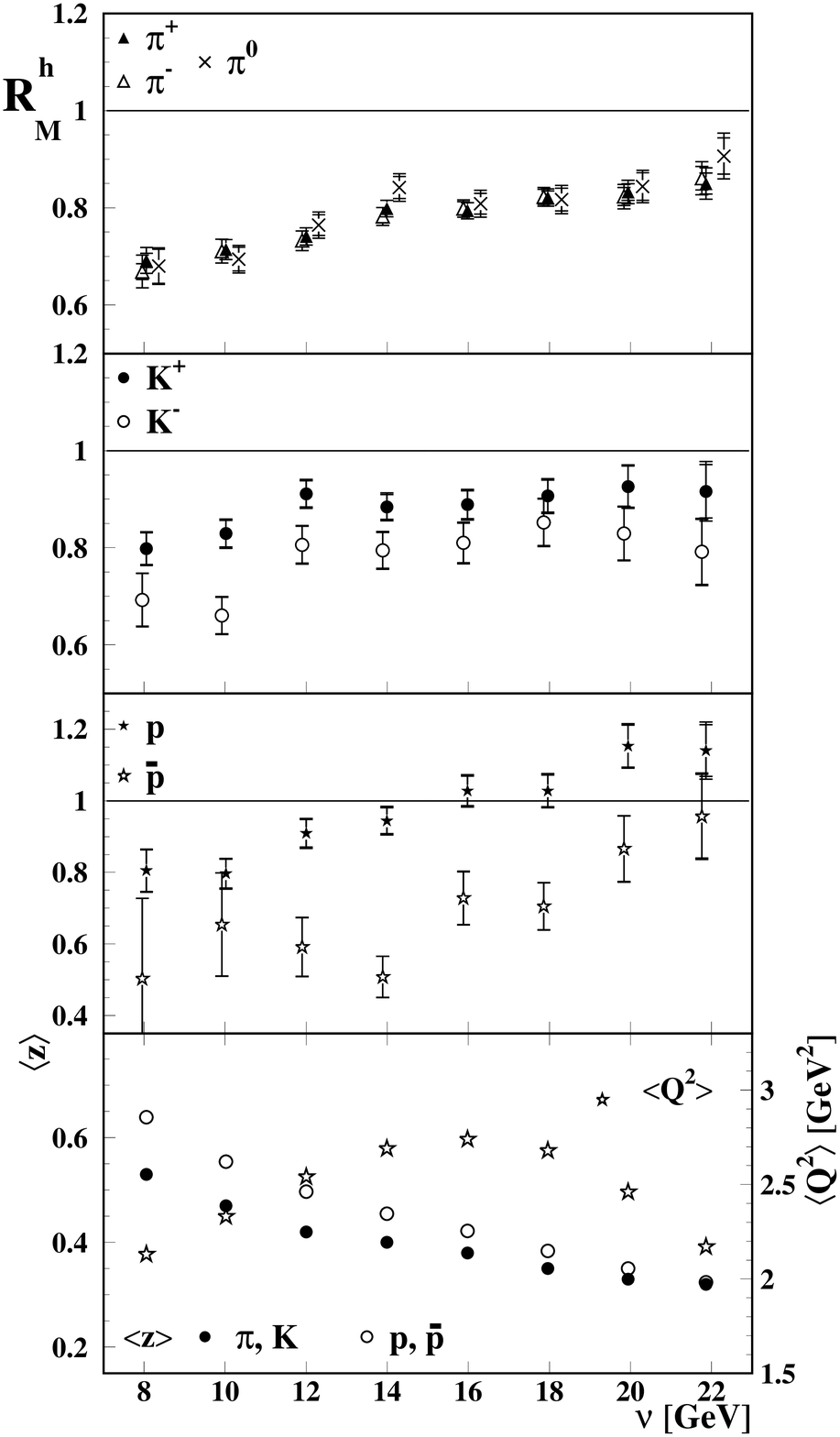}}
  {\includegraphics[width=0.49\linewidth]{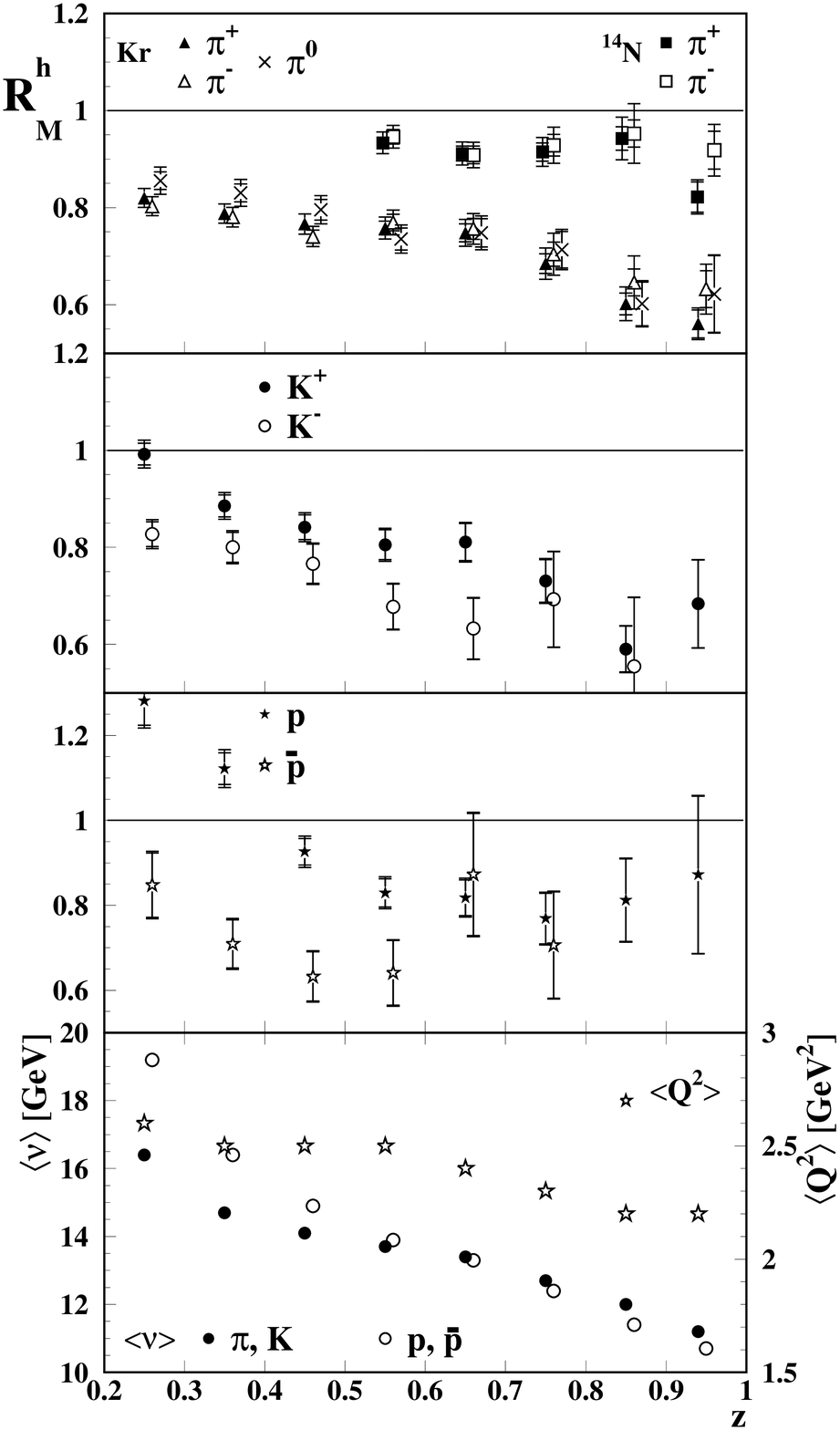}}
  \caption{Multiplicity ratios for identified pions, kaons, protons
    and antiprotons from a $Kr$ target as a function of $\nu$ for  
    $z_h>$~0.2 (left), and as a function of $z_h$ for  $\nu>7$~GeV
    (right). In the upper right panel the multiplicity ratio for
    identified pions  from a $^{14}N$ target are also shown. The
    closed (open) symbols represent the positive (negative) charge
    states, and the crosses represent $\pi^0$ mesons. In the bottom
    panels the average  $z_h$ and $\nu$ values are displayed: pions and
    kaons (protons and antiprotons) are shown as closed (open)
    circles; the average $Q^2$ values are indicated by the open stars
    referring to the right-hand scales. The inner (outer) error bars
    represent the statistical (total) uncertainties. Multiplicity
    ratios for negative kaons and antiprotons at the highest $z_h$-bins
    are not displayed due to their poor statistical significance. } 
  \label{fig:hermes1}
\end{figure}

For the fist time the $\nu$-dependence of 
the multiplicity ratio has been measured 
for  various identified hadrons ($\pi^+$, $\pi^-$, $\pi^0$, $K^+$,  $K^-$, 
$p$ and $\bar{p}$) and neutral pions~\cite{Airapetian:2003mi}, as shown for a krypton target in Fig.~\ref{fig:hermes1}.
The identification of charged pions, kaons, protons and antiprotons 
 was accomplished using the information from the  RICH detector~\cite{Akopov:2000qi}, 
which replaced a threshold  \v Cerenkov counter used in the previously reported measurements 
for the charged hadrons multiplicity on  $^{14}N$~\cite{Airapetian:2000ks}.
The corresponding $z_h$-dependences of  $R_M^h$  
with $\nu>$~7~GeV are shown in the right part of Fig.~\ref{fig:hermes1}. 
In the bottom panels the average values  for $Q^2$ and  $z_h$ or $\nu$ are displayed
for all the presented data. 

\begin{table}[tbp]
\begin{center}
\begin{tabular}{ccc} \hline 
 hadron & $\vev{R_M^h}$ $(z_h>0.2)$&  $\vev{R_M^h}$ $(z_h>0.5)$ \\ \hline
$\pi^{+}$ & $0.775\pm 0.019$ & $0.712 \pm 0.023$ \\
$\pi^{-}$ & $0.770\pm 0.021$ & $0.731 \pm 0.031$ \\
$\pi^{0}$ & $0.807\pm 0.022$ & $0.728 \pm 0.024$ \\
$K^{+}$ & $0.880\pm 0.019$ & $0.766 \pm 0.024$ \\
$K^{-}$ & $0.783\pm 0.021$ & $0.668 \pm 0.036$ \\
$p$ & $0.977\pm 0.027$ & $0.816 \pm 0.029$ \\
$\bar{p}$ & $0.717\pm 0.038$ & $0.705 \pm 0.067$ \\
\hline
\end{tabular}
\nopagebreak
\end{center}
\caption{HERMES multiplicity ratios for various hadron species produced in 
$e+Kr$ over $e$-deuterium collisions integrated over fractional hadron energies 
$z_h>$~0.2 and 0.5 for struck quark energies $\nu >$~7~GeV~\cite{Airapetian:2003mi}.
Total experimental uncertainties are quoted.}
\label{Table1}
\end{table}

The results presented in Fig.~\ref{fig:hermes1} and the average  $R_M^h$  values reported
in Table~\ref{Table1} for  $z_h >$~0.2,  show that the multiplicity ratios for
positive, negative and neutral pions as well as for negative kaons are similar. 
However,   $R_M^{h}$ for positive kaons is significantly larger.
An even larger  difference is observed  between
protons and their antiparticles compared to the meson case.
These differences in  $R_M^{h}$ of positive and negative kaons, as well as those between protons and antiprotons,
 are  still present at  $z_h >$~0.5. This is shown in the last column of Table~\ref{Table1},
where the average $R_M^{h}$   values are  reported for $z_h>$~0.5, i.e. when emphasising leading hadrons.
In addition the  $z_h >$~0.5 range is most suitable to compare  $R_M^{h}$ of mesons and baryons 
as this comparison is performed at the same average $\nu$ as shown in
the bottom right panel of  Fig.~\ref{fig:hermes1}.
The difference observed in mesons and baryons multiplicity ratios resembles the anomalous baryon enhancement 
reported at intermediate $p_T$'s ($p_T\approx$~1.5 -- 8 GeV/c) in proton-nucleus and heavy-ion 
collisions~\cite{Adler:2003kg,Abelev:2006jr} (see Sections~\ref{sec:hadrons-hA} and~\ref{sec:hadrons-AA}).

Recently, new data on neon and xenon and more krypton data have been collected at HERMES~\cite{Airapetian:2007vu}. 
This allows for a multidimensional analysis  of the multiplicity ratio thus reducing the correlation between i.e. the $\nu$ and 
$z_h$ variables shown in the bottom panels of  Fig.~\ref{fig:hermes1}. In addition, the dependence of the $R_M^h$ 
of the other variables like $Q^2$, $p_T^2$ as well the mass number dependence can be studied.

\begin{figure}[tbp]
  \centering
  \includegraphics[width=10cm]{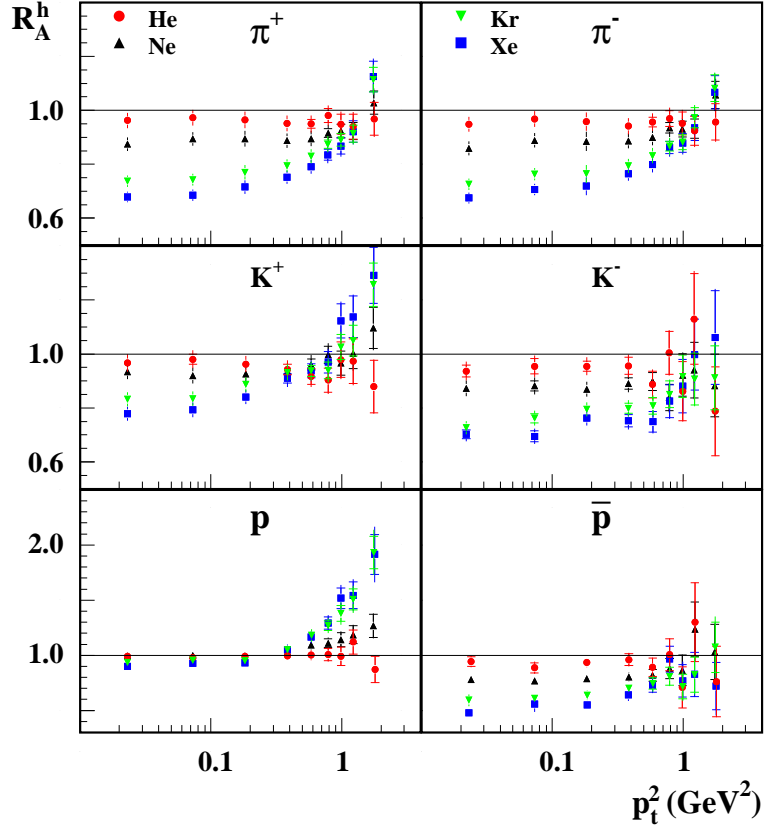}
  \caption{HERMES multiplicity ratio versus $p_T^2$ for identified charged hadrons on krypton, neon and helium.}
  \label{fig:pt_hermes}
\end{figure}

The $p_T$ dependence of the multiplicity ratio for identified charged hadrons in shown in Fig.~\ref{fig:pt_hermes} for different nuclei.
A nuclear enhancement is observed at  $p_T^2 \geq$~0.7~(GeV/c)$^2$
similar to the Cronin effect observed in hadron-nucleus collisions (see Section~\ref{sec:pAdata}).  
The conventional explanation of the Cronin effect in $p+A$ collisions
ascribes this effect to the multiple scattering of projectile partons
within the target nucleus (see Section~\ref{sec:Cronin}). 
HERMES data highlight the role of partonic {\it final-state} multiple
scattering (whereas Drell-Yan data are sensitive to initial-state
parton multiple scattering, see Section~\ref{sec:DYdata}) 
although explanations of the HERMES $p_T^2$ broadening in terms of 
{\it prehadronic} final-state interactions also exist, see Section~\ref{sec:MCstringmodels}. 
The larger Cronin enhancement for protons compared to mesons and antiprotons 
(note also the already discussed difference in the multiplicity ratio of $p$ and $\bar p$)
is also seen in $h+A$ and $A+A$ collisions, where both baryons and
antibaryons are enhanced compared to mesons, see Section~\ref{sec:hadrons-hA}.  

\begin{figure}[tbp]
  \centering
  \includegraphics[width=12.cm]{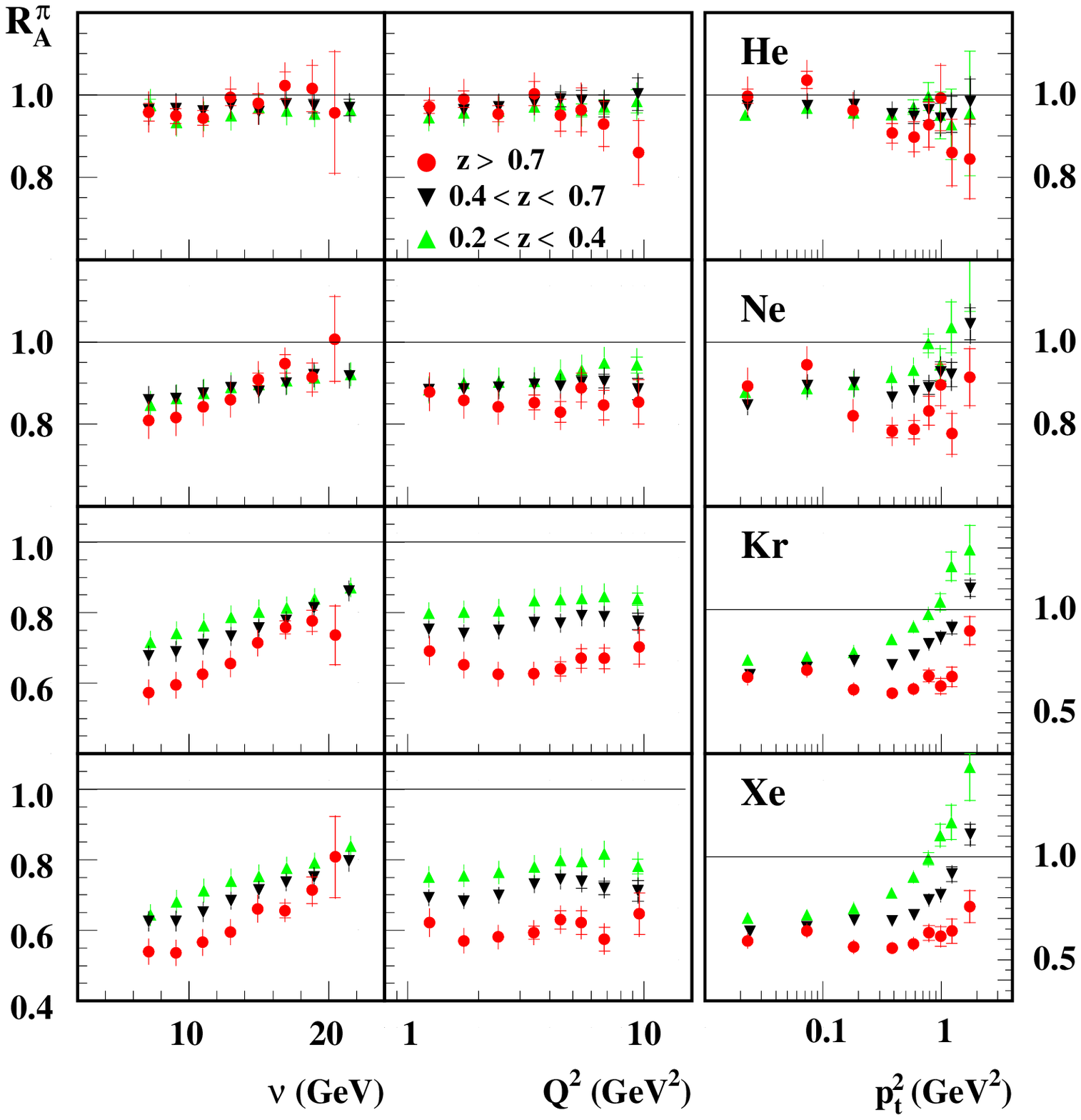}
  \caption{HERMES multiplicity ratio versus $\nu$, $Q^2$ and $p_T^2$
    for charged pions in different $z_h$-bins.} 
  \label{fig:multid_hermes}
\end{figure}

The two-dimensional analysis of the charged pion multiplicity ratio is
presented in  Fig.~\ref{fig:multid_hermes}, where the multiplicity ratio
is shown in three $z_h$-ranges as a function of $\nu$, $Q^2$ and $p_T^2$.
The leftmost panels indicate that the dependence on $\nu$ hardly depends on $z_h$.
The $Q^2$-dependence is similar for the various $z_h$-bins
therefore, the dependence on $z_h$ is not affected when integrating over $Q^2$.
The data in the rightmost plots indicate that the increase of $R_A^h$
for $Kr$ and $Xe$ at large $p_T^2$ is smaller for larger $z_h$. Such a
dependence  was predicted in Ref.~\cite{Kopeliovich:2003py}
and is consistent with the idea that the rise of $R_A^h$ at large $p_T^2$
is of partonic origin.

\begin{figure}[tbp]
  \centering
  \includegraphics[width=11.cm,height=15.5cm,clip]{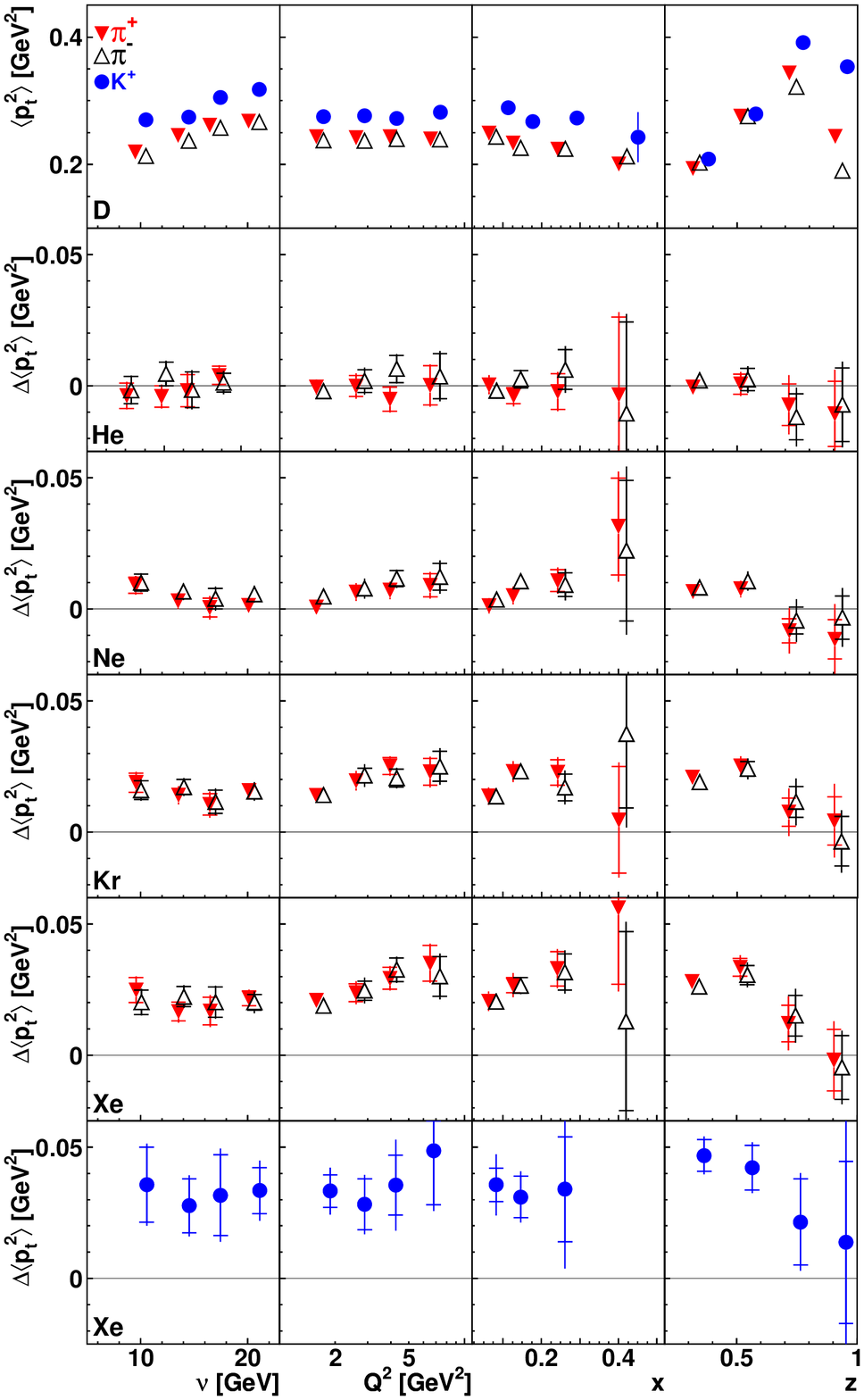}
\caption{From left to right, the $\nu$, $Q^2$, $x$ and $z_h$ dependence of $\langle p_T^2 \rangle $ for deuterium (top row)
and  $p_T$-broadening (remaining rows)  for $\pi^\pm$ produced on  $He$, $Ne$, $Kr$ targets and for $K^+$
    produced on $Xe$ target (bottom row) at HERMES.}
  \label{fig:broad_hermes}
\end{figure}

\begin{figure}[tbp]
  \centering
    \includegraphics[width=5.2cm,height=4.5cm,clip]{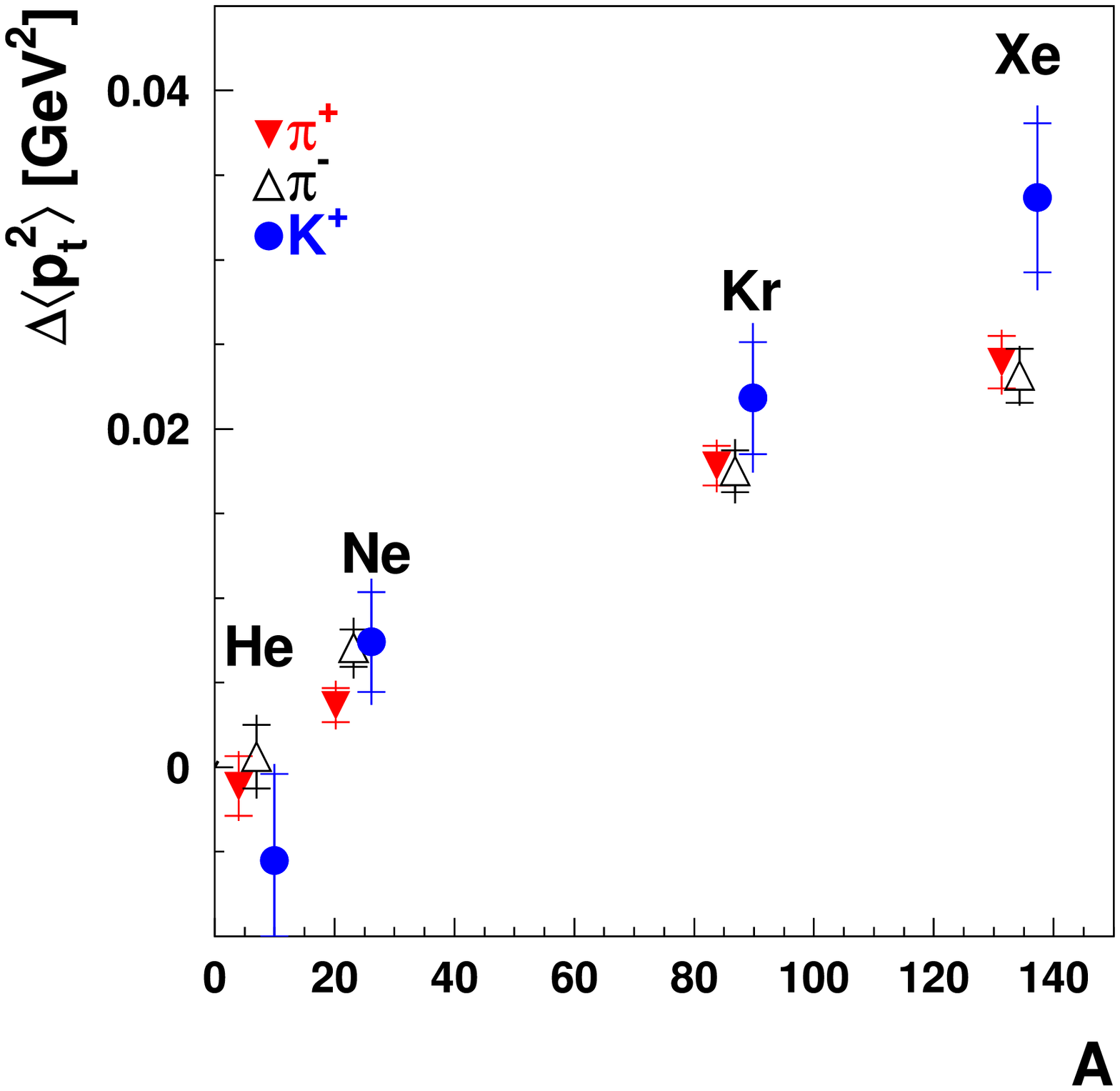}
    \includegraphics[width=8.cm,height=4.5cm,clip]{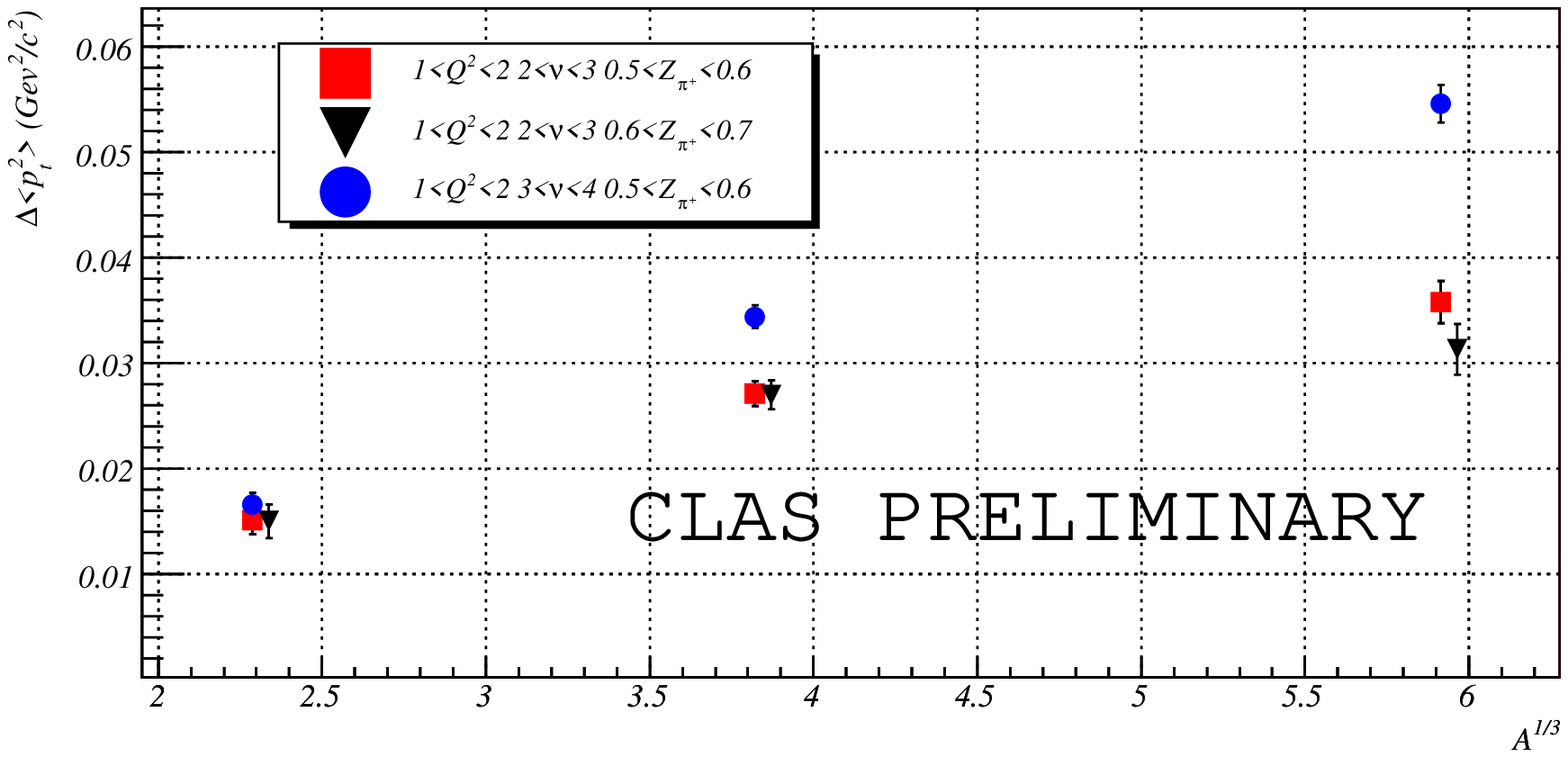}
  \caption{
    Mass number dependence of hadron $p_T$-broadening in nuclear DIS.
    {\it Left:} HERMES data for $\pi^\pm$ and $K^+$ production 
    with $Ne$, $Kr$ and $Xe$ targets. The inner error bars represent the statistical uncertainties, 
the total bars represent the statistical and systematic uncertainties.
    {\it Right:} CLAS preliminary data for $\pi^+$ for $C$, $Fe$ and $Pb$
    targets threefold differential in $Q^2$, $\nu$, and $z_h$ as shown in
    the legend. The errors shown are statistical only.} 
  \label{fig:broad_A_hermes}
  \label{fig:CLAS_data_DeltaPt2}
\end{figure}

An analysis of the hadron $p_T$-broadening has been recently
performed at HERMES~\cite{Airapetian:2009hermes}, which measured the 
observable $\Delta\langle p_T^2 \rangle $= $ \langle p_T^2
\rangle^h_A $- $ \langle p_T^2 \rangle^h_D$, introduced in
Section~\ref{sec:observables}, for different hadron species and nuclear
targets. 
The $p_T$-broadening measurement is expected to provide new insights on  
the space-time evolution of the propagating quark~\cite{Kopeliovich:2003py},
and on the multiparton correlation function inside the nucleus~\cite{Guo:2000eu} 
(see Section~\ref{sec:discussion-ptbroad}). 

The HERMES results for $p_T$-broadening are shown in Fig.~\ref{fig:broad_hermes} 
for $\pi^\pm$ and $K^+$ produced on $He$, $Ne$, $Kr$, and $Xe$ targets.
The panels presented in  Fig.~\ref{fig:broad_hermes} show  $\langle p_T^2 \rangle $ for deuterium (top row)
and the  $p_T$-broadening (remaining rows) as a function of  $\nu$, $Q^2$, $x$ and $z_h$.
The data do not reveal a significant dependence on  $\nu$ in the kinematic range covered.
An increase of the broadening with $Q^2$ is observed, the behavior as function of $x$ is very similar to the $Q^2$ behaviour , due to the strong correlation between $x$ and $Q^2$ in the HERMES kinematics; hence it can not be excluded that the 
$Q^2$ dependence is actually an underlying $x$ dependence or both a $Q^2$ and $x$ dependence.
The $p_T$-broadening is seen to vanish as $z$ approaches unity while the $\langle p_T^2 \rangle $ for deuterium is 0.2 or higher in the highest energy bin.
The results on the mass number ($A$) dependence of the
$p_T$-broadening are shown in the left part of Fig.~\ref{fig:broad_A_hermes}.
 The broadening  is similar for $\pi^\pm$  while is systematically higher for $K^+$; it increases with $A$, however 
the uncertainties of the data do not allow to firmly establish the mass number dependence.
\begin{figure}[tbp]
  \centering
    \includegraphics[width=6.cm,height=5.2cm,clip]{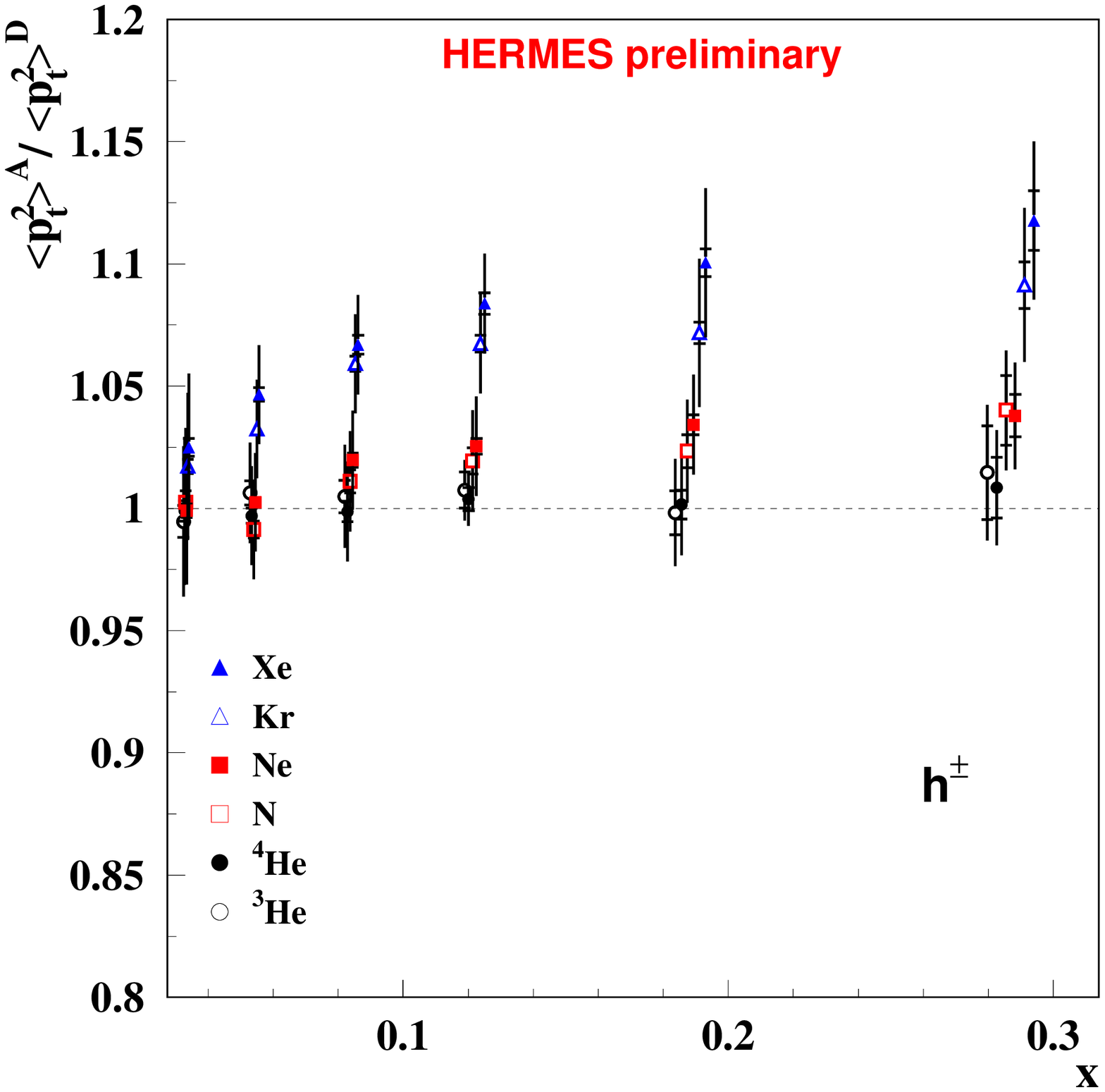}
    \includegraphics[width=6.8cm,height=5.4cm,clip]{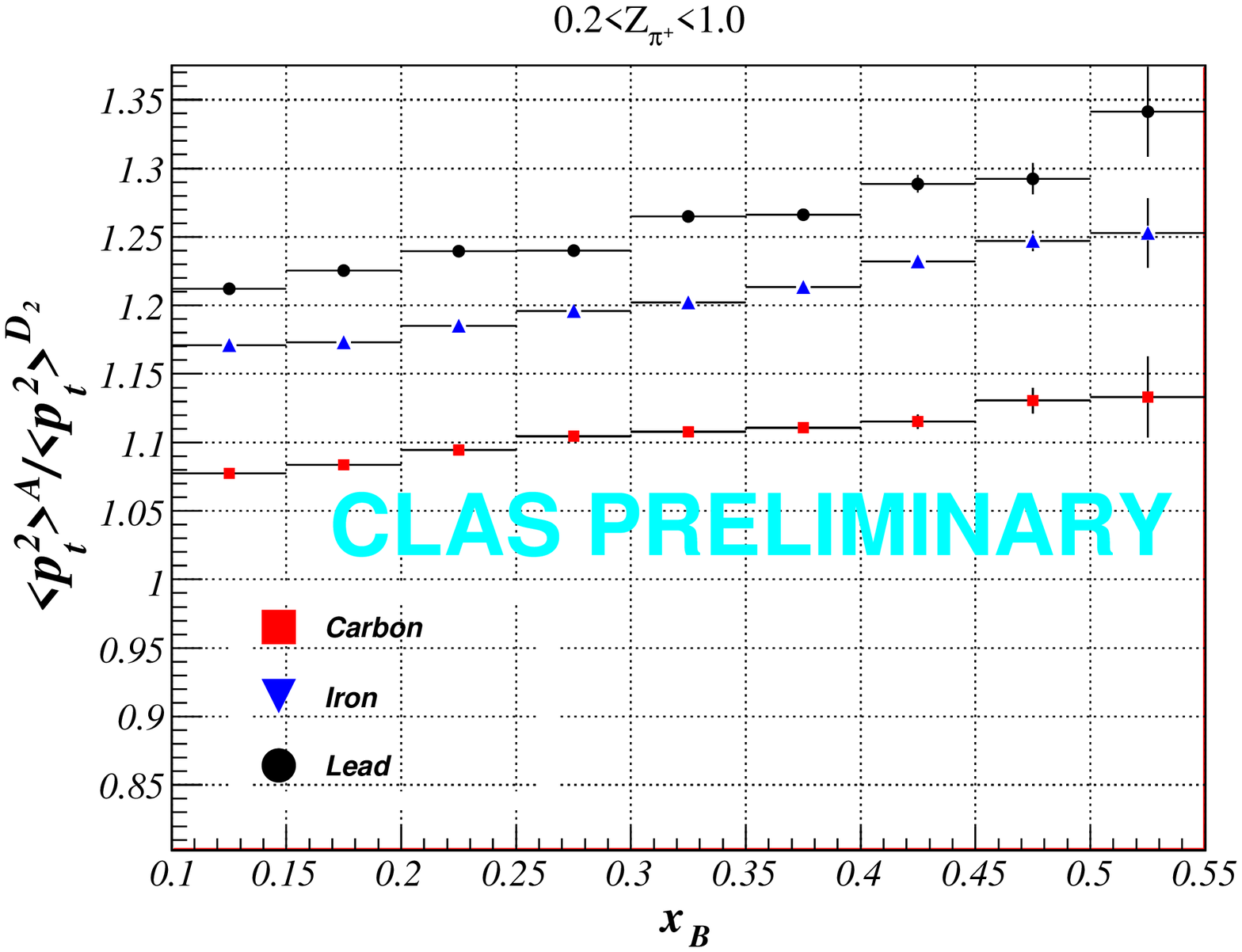}
  \caption{ Preliminary $x_B$-dependence of the $A/D$ average
    $\vev{p_T^2}$ ratio at HERMES ({\it left}) and CLAS ({\it right}). 
For the HERMES data the inner error bars represent the statistical uncertainties, 
the total bars represent the statistical and systematic uncertainties.
For the CLAS data the errors shown are statistical only and the data have not been corrected for acceptance or radiative effects.    
    }
  \label{fig:HERMES-CLAS-xb_broad}
\end{figure}
In addition, the preliminary results for the HERMES ratio  $ \langle p_T^2  \rangle^h_A $ / $ \langle p_T^2 \rangle^h_D$
are displayed in the left part of Fig.~\ref{fig:HERMES-CLAS-xb_broad} as a function of Bjorken $x_B$ for charged hadrons, 
thus showing a non-negligible $x$-dependence of the broadening in the $x$-range explored.
In particular the  $p_T$-broadening appears to increase from low to high $x_B$,
with a tendency to flatten out at $x_B \gtrsim 0.2$. 

As discussed in detail in Section~\ref{sec:prehadron}, the new HERMES data allow one to definitely rule out 
nuclear absorption models based on one-step hadronisation process in which the struck quark propagates in
the nucleus, interacts with the surrounding nucleons with perturbative cross section $\sigma_q$, and fragments 
into a leading hadron in the vacuum. Instead, they support a more complex picture where the space-time evolution
of the fragmentation process -- as encoded e.g. in the Lund string fragmentation model -- is largely modified
by the surrounding matter. However, in spite of the theoretical effort in developing new calculations for 
describing the wide range of data shown in this section, the observed ($\mu$, $z$, $p_T^2$) kinematical 
dependences can be described both in term of interaction of the intermediate prehadronic stage in absorption 
models (see Section~\ref{sec:prehadron}), as well as in terms of parton energy loss calculations as discussed in 
Section~\ref{sec:parton}.
It is therefore important to extend the nDIS kinematical region with new measurements and to analyse other 
observables in order  to disentangle the relative role of perturbative parton energy loss and prehadron absorption.

\subsection{Hadron production in $e$-nucleus DIS at CLAS/JLab} 
\label{sec:CLASdata}

Hadron production has been measured with a 5.0~GeV electron beam with the CLAS detector at Jefferson Lab. 
Approximately 25~fb$^{-1}$ of integrated luminosity was taken among the three primary solid targets, carbon, 
iron, and lead. In each case a 2-cm long liquid deuterium cryo-target was located in the beam simultaneously, 
providing a normalisation for the nuclear ratios.
A small amount of data was also taken on aluminum and isotopically enriched tin. The small spacing of 4-cm between the deuterium and solid targets 
limited the acceptance differences between the two targets for the large CLAS detector. 

Particle identification for hadrons consisted of the standard CLAS instrumentation, which uses tracking 
and time-of-flight (TOF) systems, an electromagnetic shower calorimeter (EC) and a gas \v{C}erenkov counter (CC). 
Identification of positive pions and protons was possible through the full momentum range using a combination of TOF 
and CC with the EC to reject positrons, while a somewhat more limited range of momentum was available for negative 
pions. Neutral pions and $\eta$ mesons were measured in their two-photon decay mode in the EC, while neutral kaons 
were also measured in the $\pi^+\pi^-$ channel, all over the full momentum range, but requiring a background 
subtraction. This very large data set is currently under analysis, and 
preliminary results are only available for positive pions and for neutral kaons at the present. The kinematical range of the data was 
for $\nu=2 - 4$~GeV, $Q^2=1 - 4$~GeV$^2$; the full range of $z_h$ and $p_T^2$ was available for analysis. 
Analysis cuts include $y=\nu/\nu_{max}<0.85$, $W^2>4.0$~GeV$^2$, and target vertex cuts. 
While the CLAS/JLab data have a much more limited range in $\nu$, they offer two orders of magnitude more 
integrated luminosity than the HERMES data. This feature provides access to three-fold dimensional binning 
for at least positive pions, and should provide a first look at hadron formation in some previously unmeasured hadrons, 
such as $\eta$, $\Lambda$, and $K^0$. Because of the limited range in $W^2\approx 4 -10$~GeV$^2$, 
the initial analysis is being focused on the region $z_h=0.4 - 0.7$. 
All the (preliminary) data shown hereafter have not been corrected for acceptance or radiative effects;
however, these corrections are known to be rather small. 

\begin{figure}[tbp]
  \centering
  \includegraphics[width=6.6cm,clip]{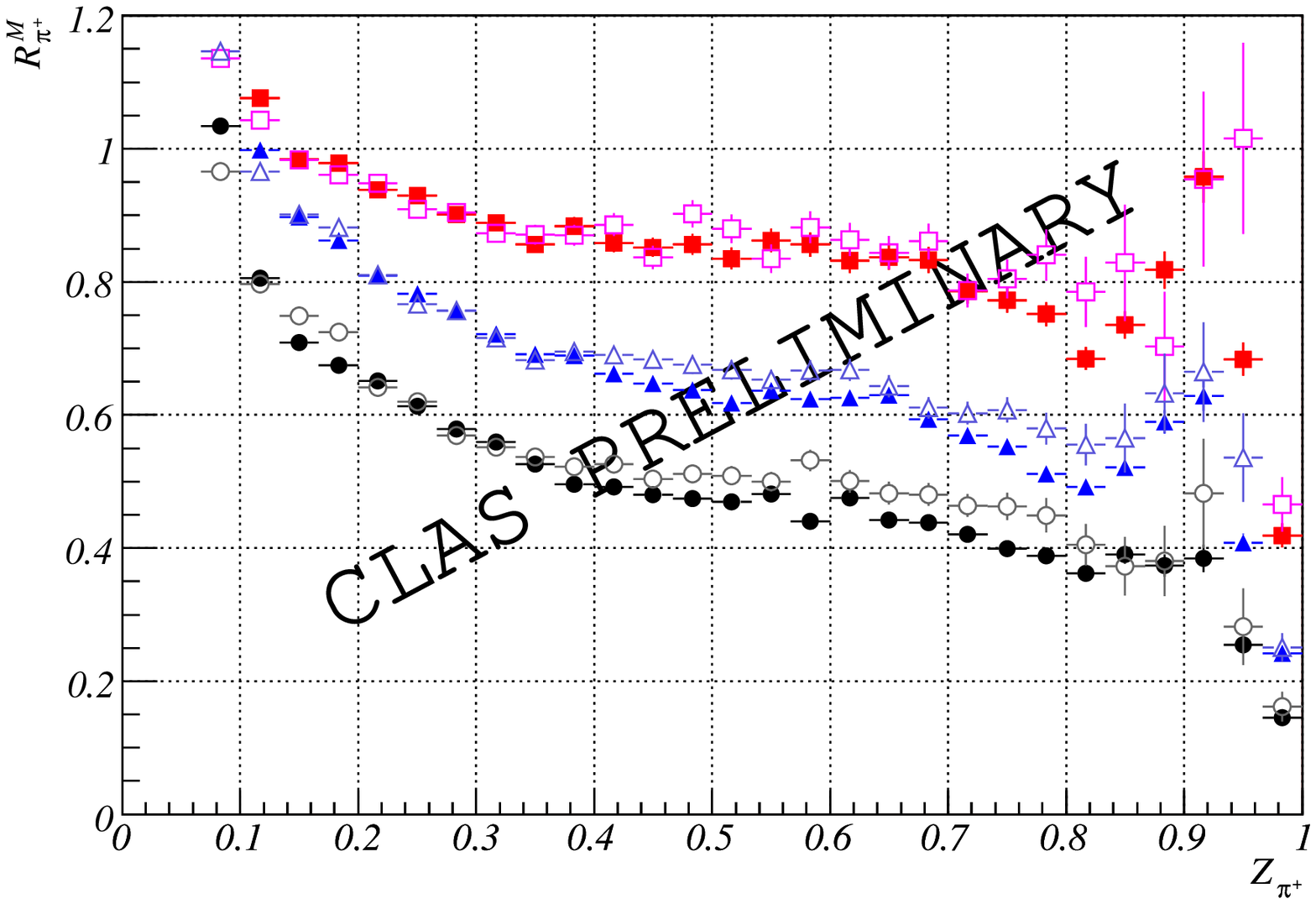}
  \includegraphics[width=6.6cm,clip]{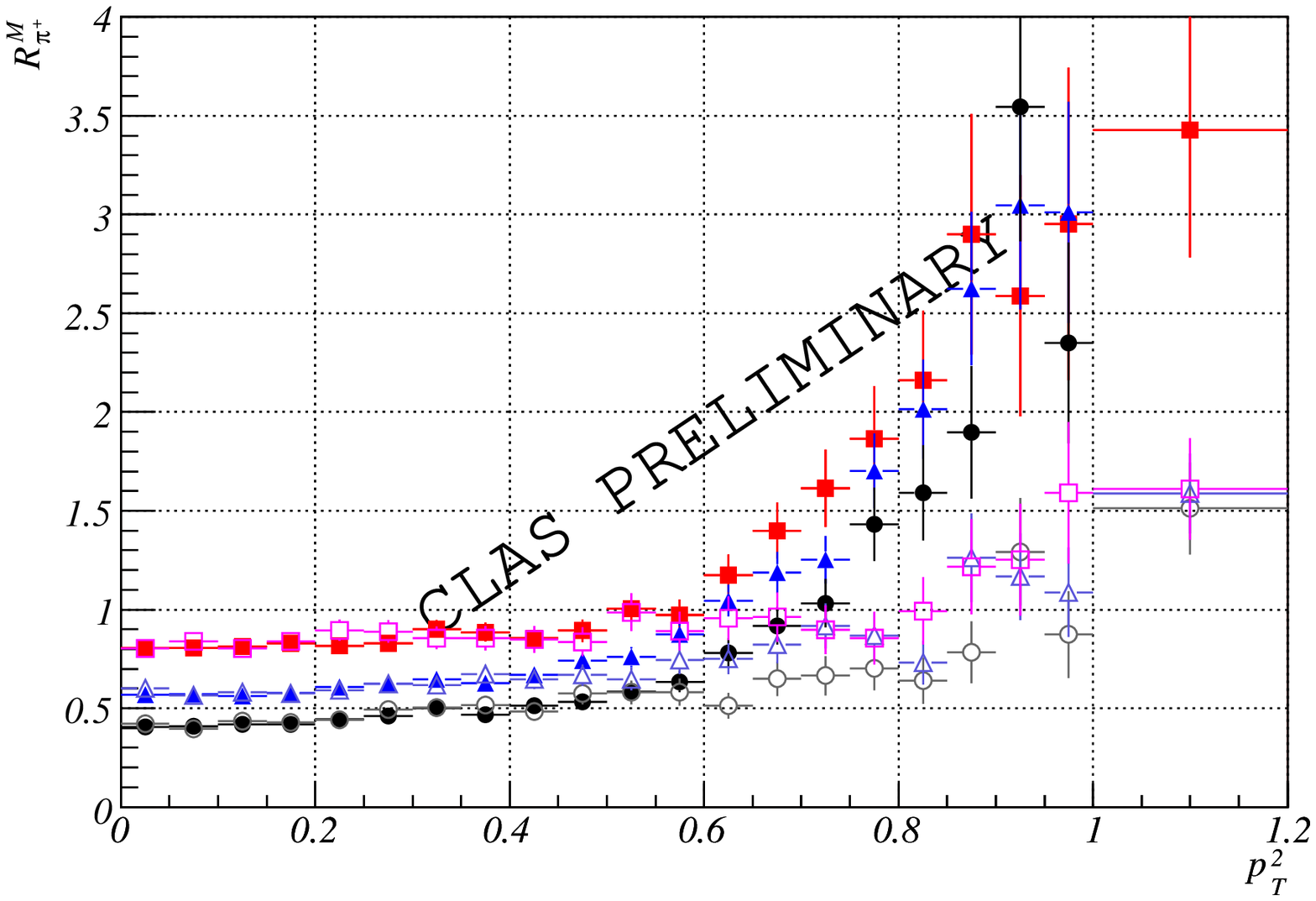}
  \caption{
    {\it Left:} CLAS preliminary data for the $z\equiv z_h$ dependence
    (left) and $p_T^2 \equiv p_{Th}^2$ of the hadronic
    multiplicity ratio for $\pi^+$ for carbon, iron, and lead
    targets. The data with solid (hollow) symbols correspond to the range
    $\nu=2.2 - 3.2$ $(3.2 - 3.7)$~GeV and $Q^2$~=~$1.0-1.3$~GeV$^2$. 
    The errors shown are statistical only. 
    }
  \label{fig:CLAS_data_RvsZ_RvsPt2}
\end{figure}

Figure~\ref{fig:CLAS_data_RvsZ_RvsPt2} shows preliminary data for the CLAS hadronic multiplicity ratio~\cite{HaykThesis:2008}. 
The data are shown in just two plots for compactness, but the statistical accuracy is adequate to divide the data further into 
multidimensional bins in $Q^2$, $\nu$, and $p_T^2$. 
In qualitative terms, these data are remarkably consistent with the HERMES results. The drop in multiplicity ratio with
increasing $z_h$ (and slow rise with increasing $\nu$) as well as the rise with increasing $p_{T}^2$ are 
observed in the JLab data as they are with the HERMES data (Figs.~\ref{fig:hermes1},~\ref{fig:pt_hermes},~\ref{fig:multid_hermes}). The dependence on $Q^2$, 
although visible, is also small in these data, as inferred from the HERMES studies. Thus, it is hoped that the two datasets 
can be inter-compared quantitatively in detailed model studies. 

In Fig.~\ref{fig:CLAS_data_DeltaPt2} right, are shown preliminary CLAS
data for transverse momentum broadening (defined in Eq.~(\ref{eq:ptbroadening})). 
Three sets of data are shown; each data point is binned in a multidimensional bin in $Q^2$, $\nu$, $z_{\pi^+}$ as well as $A^{1/3}$. 
The available statistical sample is adequate to make more than two dozen sets of points with good statistical precision. 
The naive expectation that $\Delta p_T^2$ is linear with the nuclear radius i.e. with $A^{1/3}$, is seen to approximately hold, 
although the data for the heaviest nucleus (lead) appear to undershoot the linear behavior. 
This flattening behaviour suggests the possibility that the partonic-level multiple scattering presumed to cause the broadening 
does not continue uniformly through the largest nucleus. If this picture is correct, then the production length can be estimated from the data in a rather direct 
fashion, using the well-known nuclear densities and sizes. 

In the right part of the Fig.~\ref{fig:HERMES-CLAS-xb_broad} the preliminary $x_B$-dependence of the $p_T$ broadening 
at CLAS is shown for positive pions. Note that the $x_B$ range is different from HERMES.
As can be seen, the $x_B$ dependence is reasonably consistent for these two datasets, which are integrated over all other variables. 
The errors shown are statistical only; the CLAS data shown are not corrected for small effects due to acceptance or radiative processes.

\subsection{Di-hadron correlations in nuclear DIS} \ \\
\label{sec:2partcor_nDIS}

In nuclear DIS, double-hadron leptoproduction on $N$, $Kr$ 
and $Xe$ relative to deuterium 
have been studied~\cite{Airapetian:2005yh} via the ratio
\begin{equation}
  R_{2h}(z_2)=\left( \frac{dN^{z_1>0.5}(z_2)/dz_2} {N^{z_1>0.5}} \right)_A
    \Bigg/ \,  \left( \frac{dN^{z_1>0.5}(z_2)/dz_2}{N^{z_1>0.5}} \right)_D
    \ ,
\label{r2h}
\end{equation}
where $z_1$ and $z_2$ correspond to the leading (largest $z$) and 
sub-leading (second largest $z$) hadrons, respectively.
The quantity $dN^{z_1>0.5}$ is the number of events with at least two
detected hadrons in a bin of width $dz_2$ at  $z_2$ with $z_1>0.5$.  The 
quantity  $N^{z_1>0.5}$ is the number of events with at least one detected 
hadron with $z_1>0.5$.  The label $A(D)$ indicates that the term is calculated 
for a nuclear (deuterium) target.

\begin{figure}[tbp]
  \centering
    \includegraphics[width=4.4cm,height=4.cm,clip]{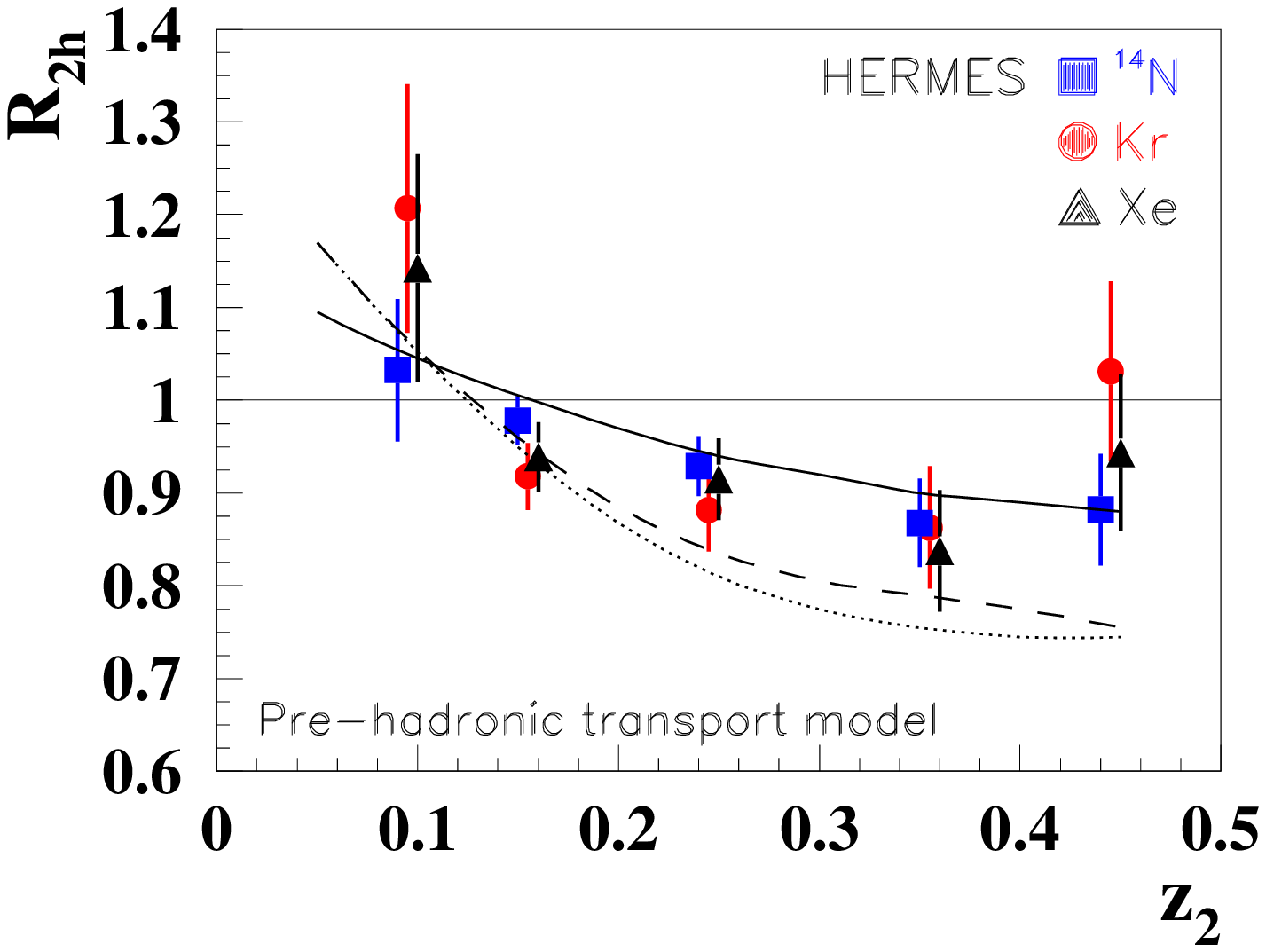}
    \includegraphics[width=4.4cm,height=4.cm,clip]{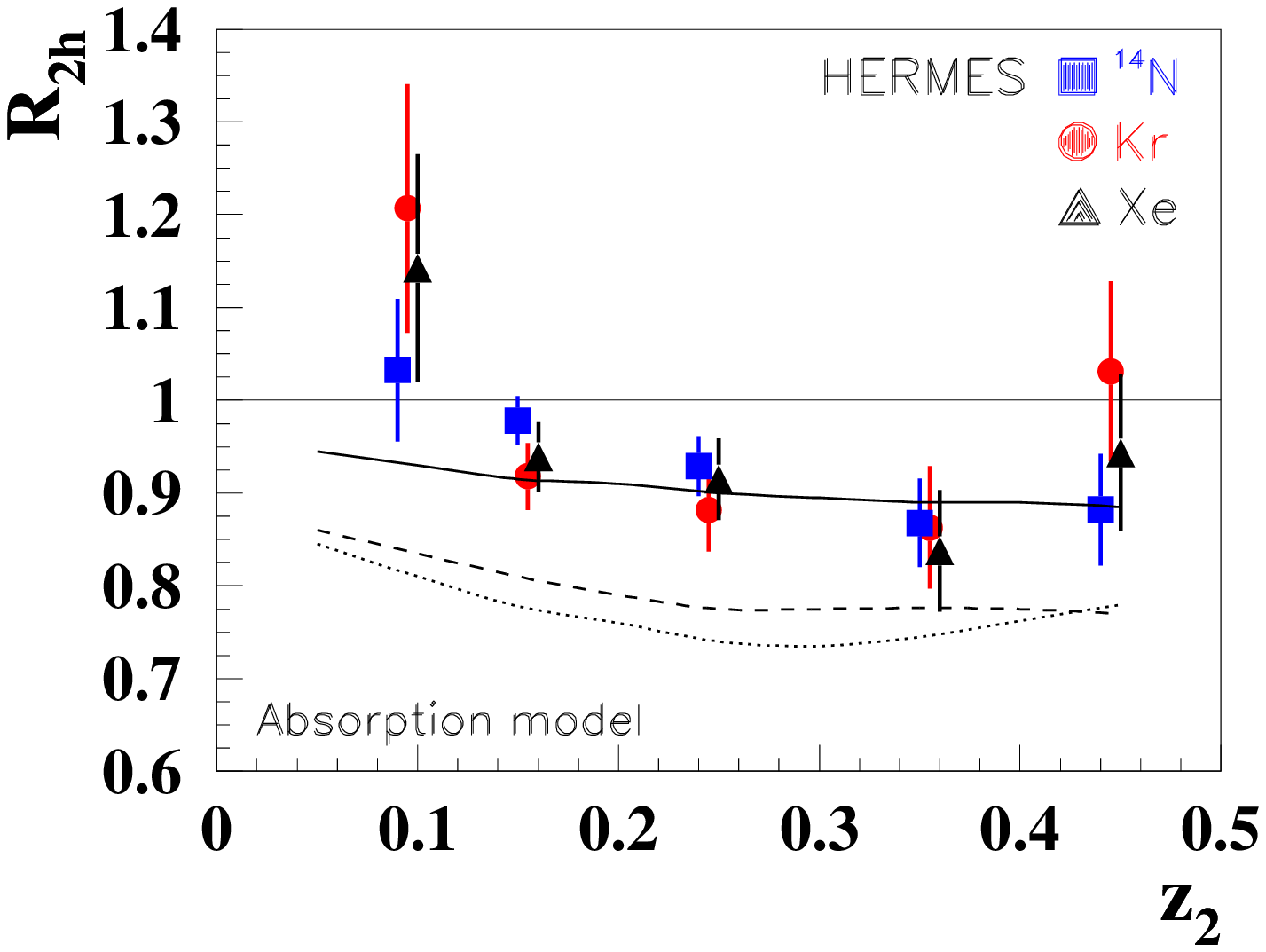}
    \includegraphics[width=4.4cm,height=4.cm,clip]{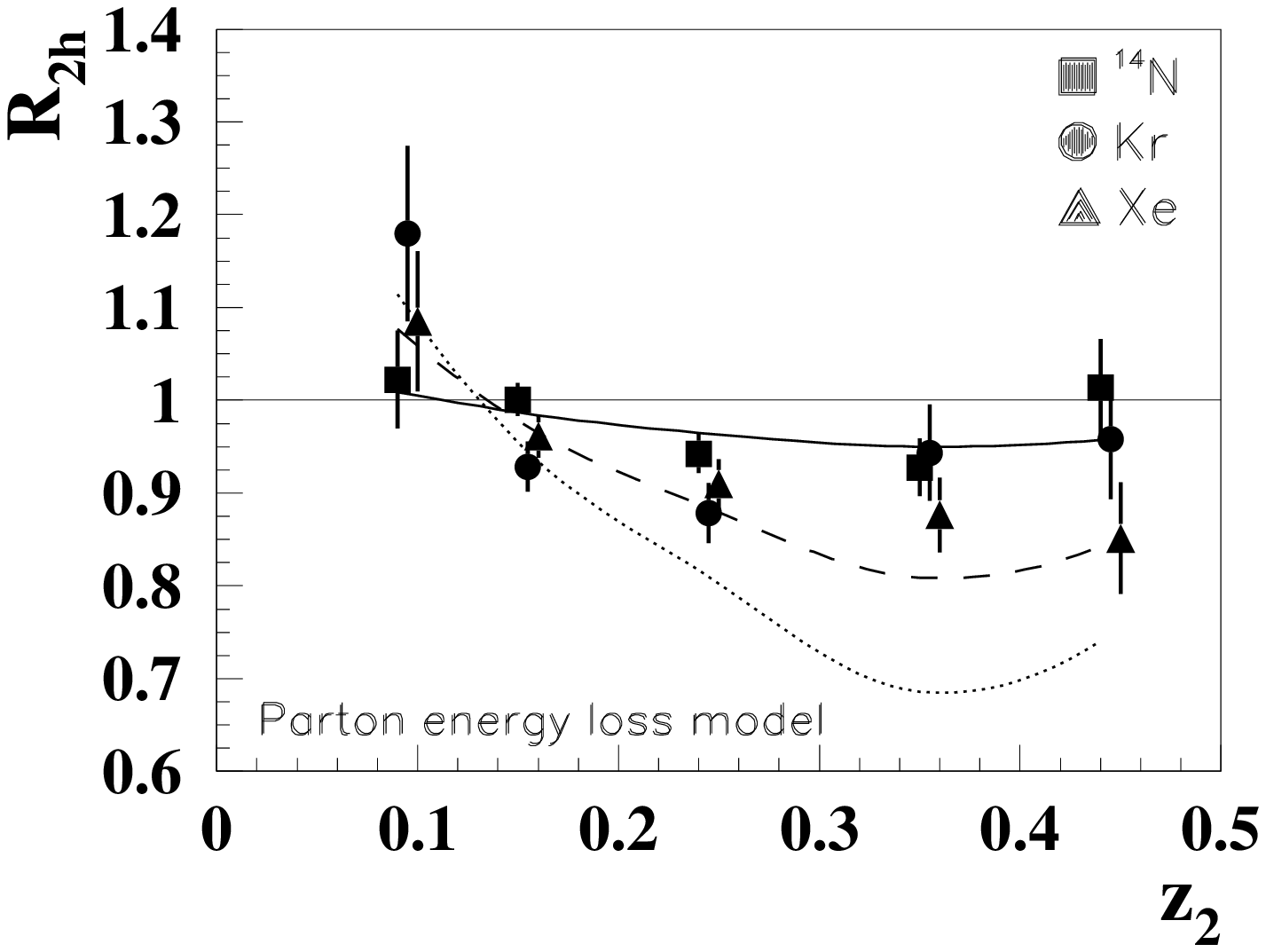}
 \caption{Double-ratio $R_{2h}$ as a function of $z_2$ for
   $^{14}N$ (squares), $Kr$ (circles) and $Xe$ (triangles) with
   $z_1>$0.5, compared to various theoretical predictions
   (solid for  $^{14}N$, dashed for  $Kr$ and dotted for $Xe$):
   BUU prehadronic transport model~\cite{Gallmeister:2004sg} ({\it left}),
   hadron absorption model~\cite{Gallmeister:2004sg} ({\it centre}),
   parton energy loss calculations~\cite{Majumder:2004pt} ({\it right}).} 
  \label{fig:hermes-RHIC-2h}
\end{figure}

If partonic energy loss of the struck quark were the only mechanism
involved, it would be naively expected that the attenuation effect does
not depend strongly on the number
of hadrons involved, and the double-hadron to single-hadron ratio for a
nuclear target should be only slightly dependent on the mass number $A$.
On the contrary, if final hadron absorption is the dominant process, 
the requirement of an additional slower sub-leading hadron that is more 
strongly absorbed would suppress the two-hadron yield from heavier nuclei 
\cite{Kopeliovich:2003py}, so that this ratio should decrease with $A$.

Results from HERMES are presented in Fig.~\ref{fig:hermes-RHIC-2h}, and show 
that the double-hadron ratio $R_{2h}$ is generally below unity with no significant
difference between the $^{14}N$, $Kr$ and $Xe$. The nuclear effect 
is much smaller than for the single-hadron attenuation measured under
the same kinematic conditions. The displayed model computations, based
on prehadron absorption~\cite{Gallmeister:2004sg} or parton energy loss
\cite{Majumder:2004pt,Majumder:2008jy}, can reproduce the general trend
in $z_2$ of the data, but predict a significant $A$-dependence which
is not seen in the data \footnote{Alternative string model approaches to 
two-particle correlations are discussed in~\cite{Akopov:2006av,Akopov:2007cp}.}. 

The small nuclear effect and its relative independence on $A$ may simply
point toward a strong surface bias of the photon-hadron interaction
point. Such a bias is very natural in NLO dihadron production,
in which the virtual photon's hard scattering produces two partons with
energy $\nu_{1,2} \leq \nu$, 
which independently hadronise with fractional momentum 
$\tilde z_{1,2} = z_{1,2} \nu_{1,2}/\nu$.



\section{Experimental results in hadron-nucleus collisions} 
\label{sec:hadrons-hA}

\subsection{Drell-Yan production} \ \\
\label{sec:DYdata}

The production of dileptons with high invariant mass  through the
$q\bar{q}\to l^+l^-$ Drell-Yan (DY) process has been measured
extensively in hadronic collisions, typically focusing on dilepton
invariant masses between the charmonia and bottomonia masses 
($4\lesssim M_{l^+l^-} \lesssim 9$~GeV/c$^2$) or above 
($M_{l^+l^-}\gtrsim 11$~GeV$^2$) -- see
Refs.~\cite{McGaughey:1999mq,Garvey:2001yq,Peng:2008tc} for a review.
NLO pQCD calculations describe well the experimental mass and 
transverse momentum distributions in hadronic collisions \cite{Gavin:1995ch}.
In hadron-nucleus collisions several nuclear effects are expected to
modify the pQCD expectation of a linear dependence 
of the cross section with the mass number $A$ (Eq.~\eqref{eq:Afactorisation}). 
First of all, the modifications of parton densities in nuclei (nPDF)
--~such as shadowing at small parton fractional momentum
$x\lesssim10^{-2}$, or the EMC effect at large $x\gtrsim 0.1$
(see \cite{Armesto:2006ph} for a recent review)~-- affect DY
production in $h + A$  with respect to that in $h + p$ collisions.
On top of nPDF effects, the projectile (anti)quark propagating through the nucleus may
experience multiple scattering and lose some energy before the hard
process takes place. In that sense, DY production data in $h + A$ collisions
is particularly appropriate to study parton propagation in cold
nuclear matter (see theoretical discussion in Section~\ref{sec:enloss_in_DY}).

\begin{figure}[tb]
  \begin{minipage}[c]{0.550\linewidth}
    \begin{center}
      \includegraphics[width=\linewidth]
      {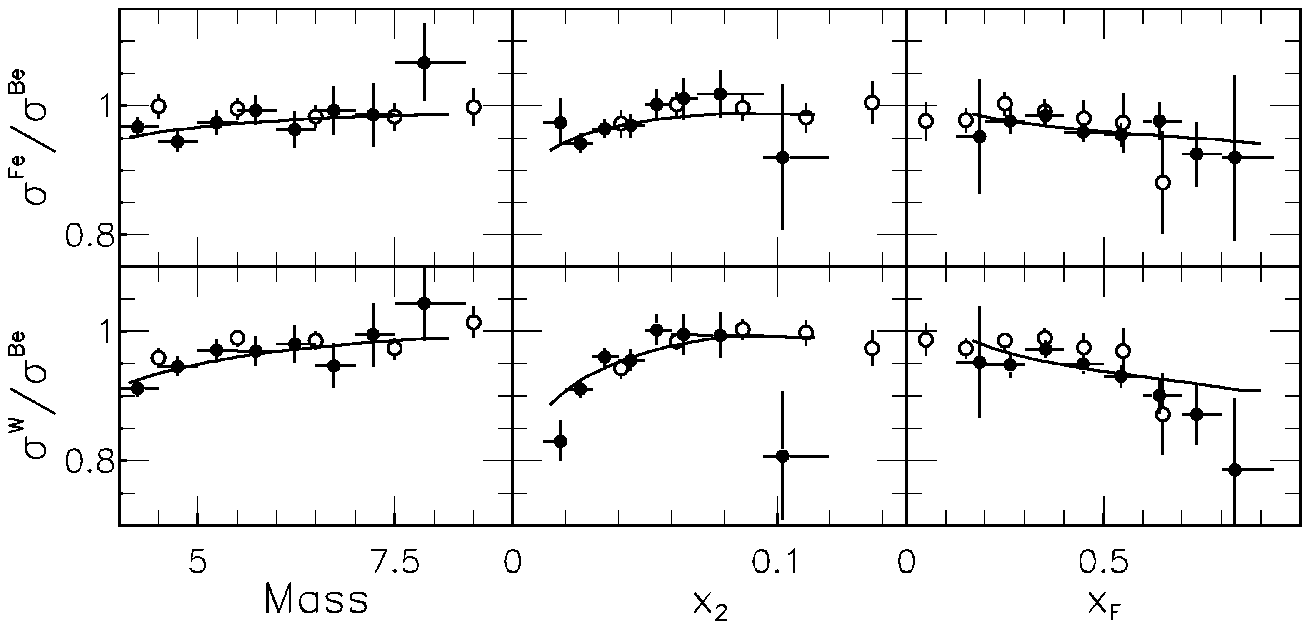}
    \end{center}
\end{minipage}
~\hfill
  \begin{minipage}[c]{0.450\linewidth}
    \begin{center}
       \includegraphics[width=\linewidth]{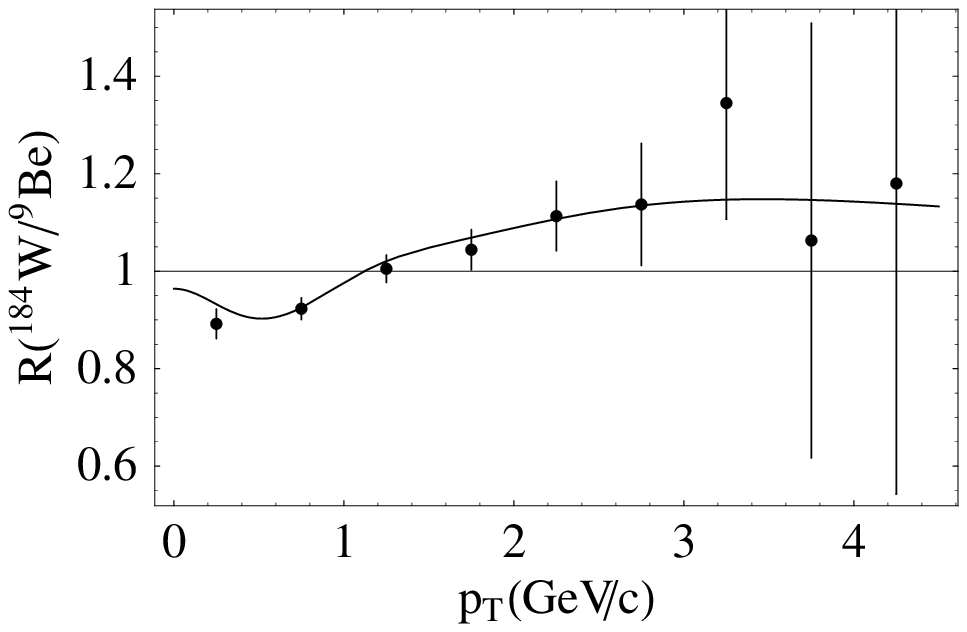}
    \end{center}
  \end{minipage}
    \caption{{\it Left:} 
      Invariant mass, $x_2$ and $x_F$ dilepton distributions in
      $p+Fe$ and $p+W$ scaled by that in $p+Be$ measured by
      E866/NuSea~\cite{Vasilev:1999fa}. Solid curves are computations
      using the EKS98 shadowing parametrisation~\cite{Eskola:1998df}.
      {\it Right:} $p_T$ distribution in $p + ^{184}$W normalised by $p + ^9Be$ 
      measured by E772/E866~\cite{Alde:1990im,Vasilev:1999fa}
      compared to the calculations of~\cite{Johnson:2007kt}. }
    \label{fig:dypa2}
\end{figure}

Drell-Yan lepton pairs have been measured in $p + A$ and $\pi^- + A$
collisions in fixed target experiments at the CERN SPS (NA3~\cite{Badier:1981ci}, 
NA10~\cite{Bordalo:1987cr}, NA38~\cite{Baglin:1991fy}) and at
Fermilab~(E772~\cite{Alde:1990im,Alde:1991sw},
E866/NuSea~\cite{Vasilev:1999fa} and older
experiments~\cite{Ito:1980ev}). The first measurements of 
DY production in nuclei were performed by the NA3 and NA10 collaborations 
using 150~GeV, 200~GeV and 280~GeV pion beams on fixed-target nuclei
($\sqrtsnn=16.5$--$23$~GeV). NA3 reported on the nuclear
dependence of DY production as a function of the longitudinal
momentum-fraction of the incoming parton, $x_1$~\cite{Badier:1981ci},
while NA10 measured the $\pt$-broadening of lepton
pairs~\cite{Bordalo:1987cr}. These data cover a  phase-space range
($0.2 \lesssim x_1 \lesssim 0.9$) ideal to probe energy loss effects,
even though the experimental uncertainties are quite large
at the edge of phase space. Later, the Fermilab E772 experiment
measured similarly DY production on various nuclear
targets ($D$, $C$, $Ca$, $Fe$, $W$) as a function of $x_F$ using the 800~GeV
proton beam (i.e. $\sqrtsnn$~=~38.7~GeV)~\cite{Alde:1990im}. 
More recently, the E866/NuSea collaboration reported on very high-statistics 
data in the same kinematical region of E772 and using $Be$, $Fe$, and $W$ nuclear
targets~\cite{Vasilev:1999fa}. 


\begin{figure}[tb]
  \begin{minipage}[c]{0.530\linewidth}
    \begin{center}
      \includegraphics[width=\linewidth]{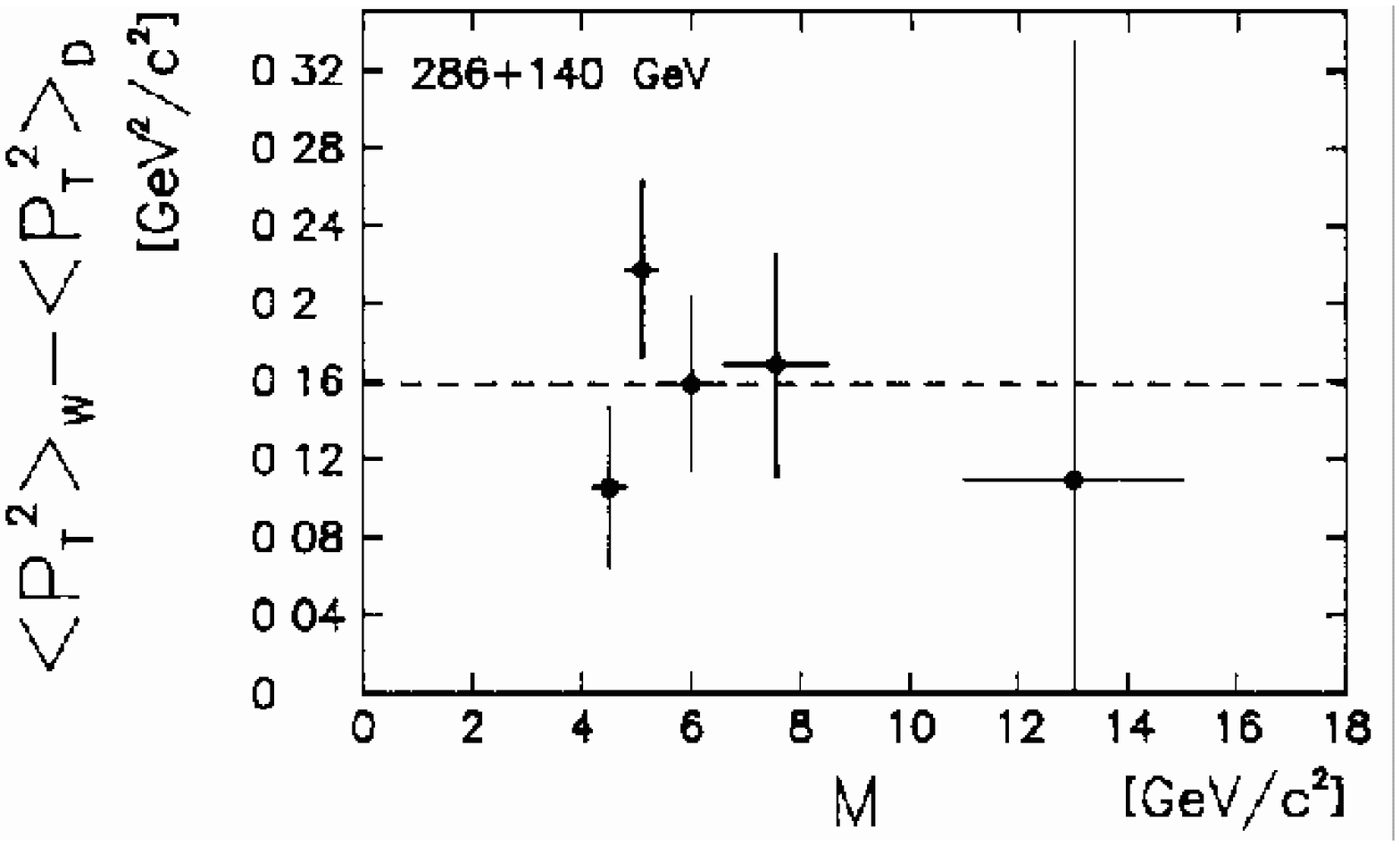}
    \end{center}
  \end{minipage}
~\hfill
  \begin{minipage}[c]{0.400\linewidth}
    \begin{center}
      \includegraphics[width=\linewidth]{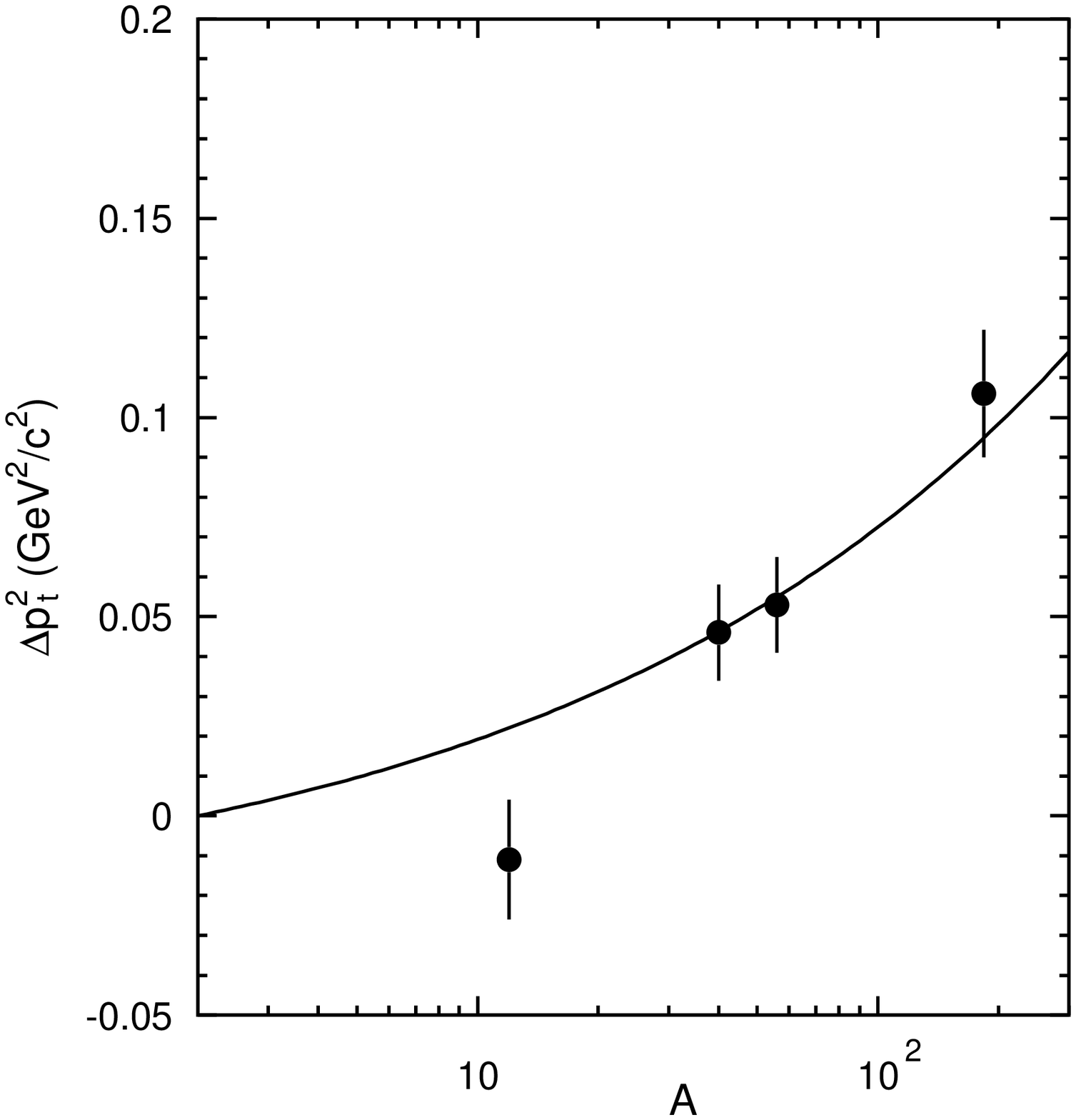}
    \end{center}
    \vskip1.2cm
\end{minipage}
    \caption{Transverse momentum broadening $\langle
      p_{_\perp}^2\rangle_{_{\rm W}}-\langle p_{_\perp}\rangle_{_{\rm D}}$ 
    of Drell-Yan dileptons as a function of the invariant mass (NA10 dimuon 
    data in $\pi+A$ collisions~\cite{Bordalo:1987cr}, {\it left}) 
    and as a function of $A$ (E772 $p+A$ data~\cite{McGaughey:1999mq}, {\it right}). 
    }
    \label{fig:dyptbroad}
\end{figure}


Although the total cross section was found to approximately scale with $A$ and therefore 
shows no nuclear effects within the experimental uncertainties, more
differential measurements such as mean transverse momentum or
longitudinal momentum (or $x_{\rm F}$) distributions actually exhibit
a small, yet significant, nuclear dependence. Moreover, the data show
$p_T$-broadening, also reported in hadron production in semi-inclusive lepton-nucleus DIS 
and hadron-nucleus collisions. Figure~\ref{fig:dypa2} shows the DY distribution
ratios as a function of $M$, $x_2$, $x_F$ and $p_T$ from the E866~\cite{Bordalo:1987cr} 
and E772~\cite{Alde:1990im} experiments. 
At $x_F \gtrsim 0.2$, more than 90\% of the
DY cross-section is due to the scattering of a quark from the hadron
with an {\it antiquark} from the nucleus. Hence the DY
process, unlike DIS, yields direct information on the nuclear
modifications of the antiquark distribution.
The experimental results show no nuclear enhancement of the
$\bar{q}$ distribution at moderate $x_2$, contrary to the predictions
of ``pion cloud'' models which explain the EMC effect in terms of nuclear enhancement of
exchanged mesons~\cite{Norton:2003cb,Piller:1999wx,Geesaman:1995yd,McGaughey:1999mq}. 
Unless a large quark energy loss compensates for the predicted antiquark enhancement, DY
data put strong constraints on nuclear models describing the EMC 
effect~\cite{Bickerstaff:1985ax,Bickerstaff:1985da,Norton:2003cb,Piller:1999wx,Geesaman:1995yd}. 
At variance with the $p_T$-integrated ratios just discussed, the dilepton
$p_T$-broadening is mostly sensitive to the parton rescattering dynamics in the target nucleus. 
Measurements from  NA10 as a function of $M$ and from E772 as a function of $A$ are shown
in  Fig.~\ref{fig:dyptbroad}. While no significant dependence on the DY mass is observed by NA10, 
the clear increase of the $p_T$-broadening with the atomic mass number in the E772 data 
is qualitatively consistent with initial-state parton rescatterings.


\subsection{High-$p_T$ hadron production: ``Cronin effect'' } \ \\
\label{sec:pAdata}

\begin{figure}[tb]
\begin{center}
\includegraphics[width=0.45\linewidth,height=4.9cm]{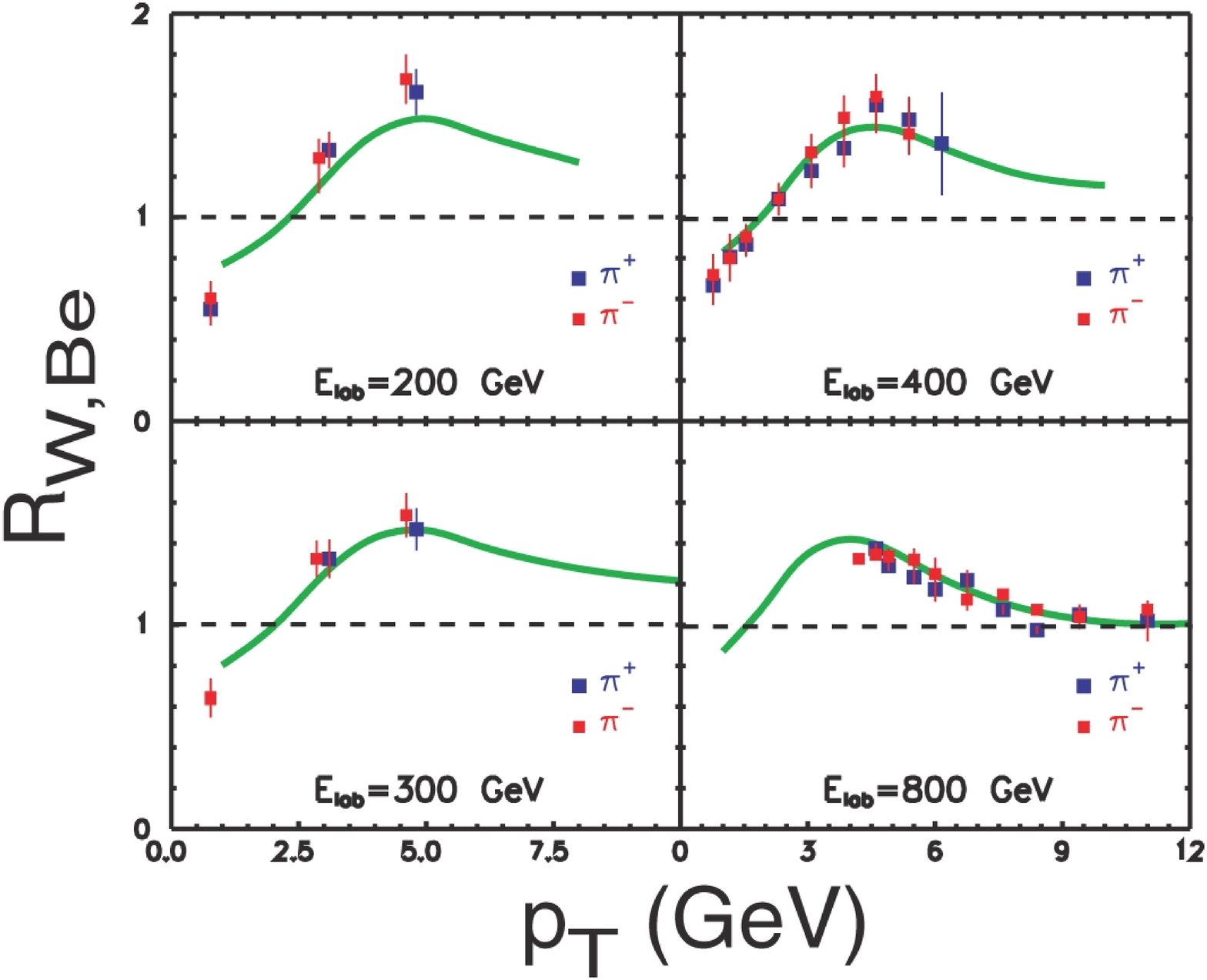}
\includegraphics[width=0.5\linewidth,height=5.cm]{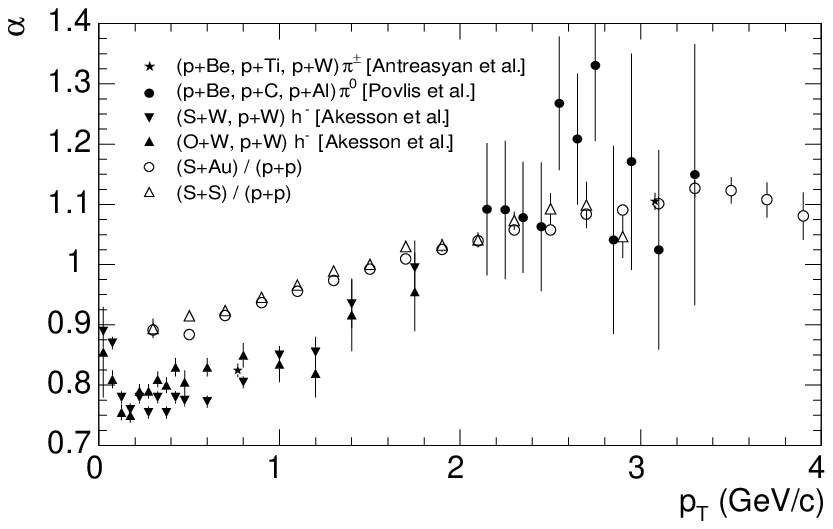}
\end{center}
\vskip-.5cm
\caption{Cronin enhancement in high-$p_T$ hadron production in proton-nucleus collisions 
at FNAL~\cite{Cronin:1974zm,Antreasyan:1978cw,Straub:1992xd}  ({\it left}) compared to the
predictions of~\cite{Kopeliovich:2002yh}, and at FNAL and SPS~\cite{Albrecht:1998yc} ({\it right}).}
\label{fig:cronin}
\end{figure}

High-energy proton-nucleus collisions in fixed-target experiments at 
FNAL~\cite{Cronin:1974zm,Antreasyan:1978cw,Straub:1992xd},
SPS~\cite{Albrecht:1998yc} and at HERA-B~\cite{HERA-B:2008sa} have observed 
an enhancement of the single inclusive hadron production yield relative to 
proton-proton collisions for transverse momenta above $p_T\approx$~1.5~GeV/c
(Fig.~\ref{fig:cronin}). Such a high-$p_T$ enhancement, called ``Cronin effect''~\cite{Cronin:1974zm,Antreasyan:1978cw}, 
is evidenced by  a nuclear modification factor that exceeds unity,
and has also been observed in nDIS, see Figs.~\ref{fig:emc_nu_pt},~\ref{fig:pt_hermes}
and~\ref{fig:CLAS_data_RvsZ_RvsPt2} (right). The fact that $R_{pA}<1$ below $p_T\approx$~1~GeV/c 
is simply because the incoherent binary scaling assumption behind 
Eq.~\eqref{eq:R_AB}, is not valid for soft particle production in hadronic collisions.

\begin{figure}[tb]
\begin{center}
\includegraphics[width=0.48\linewidth,height=5.cm]{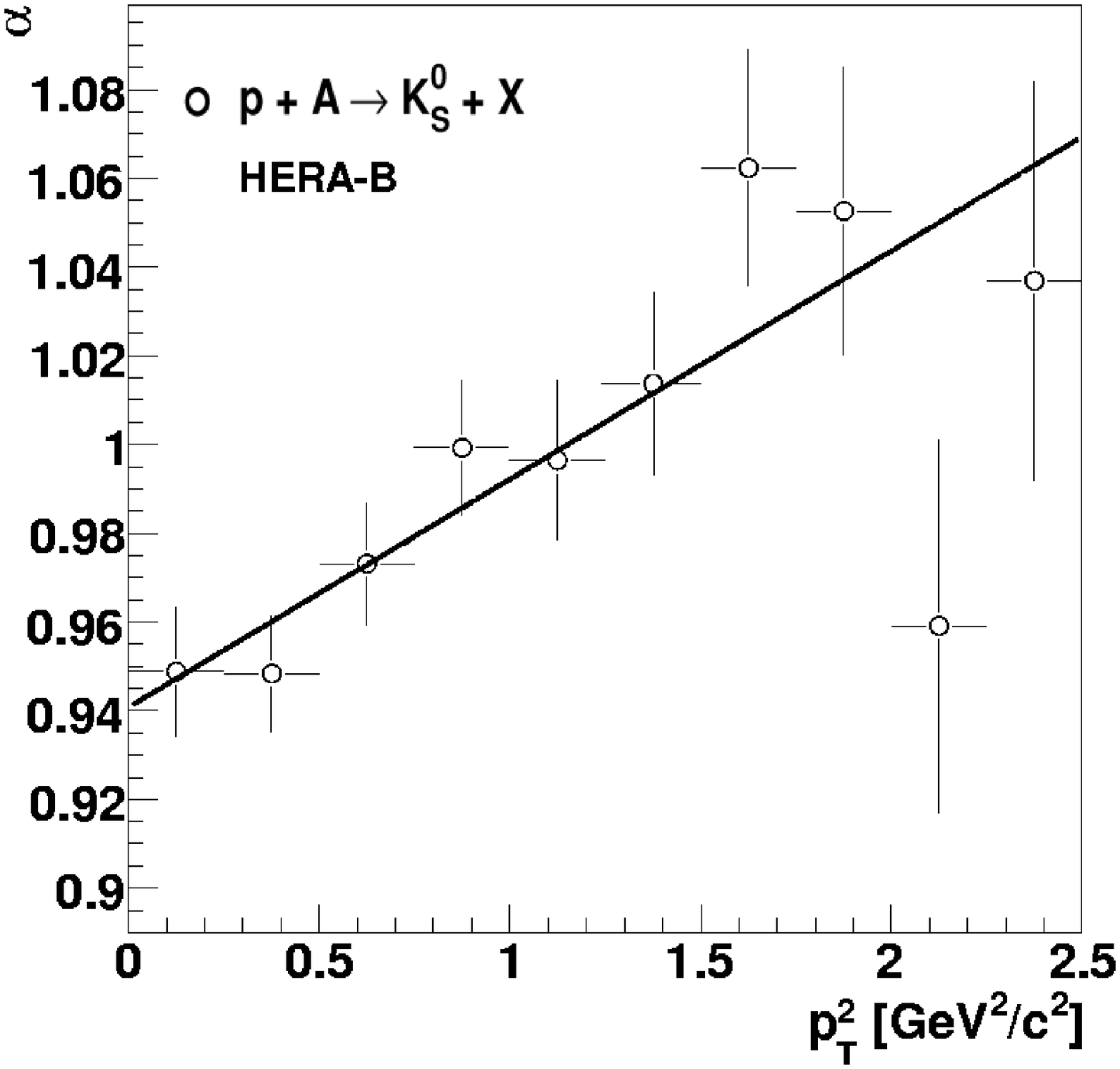}
\includegraphics[width=0.48\linewidth,height=5.cm]{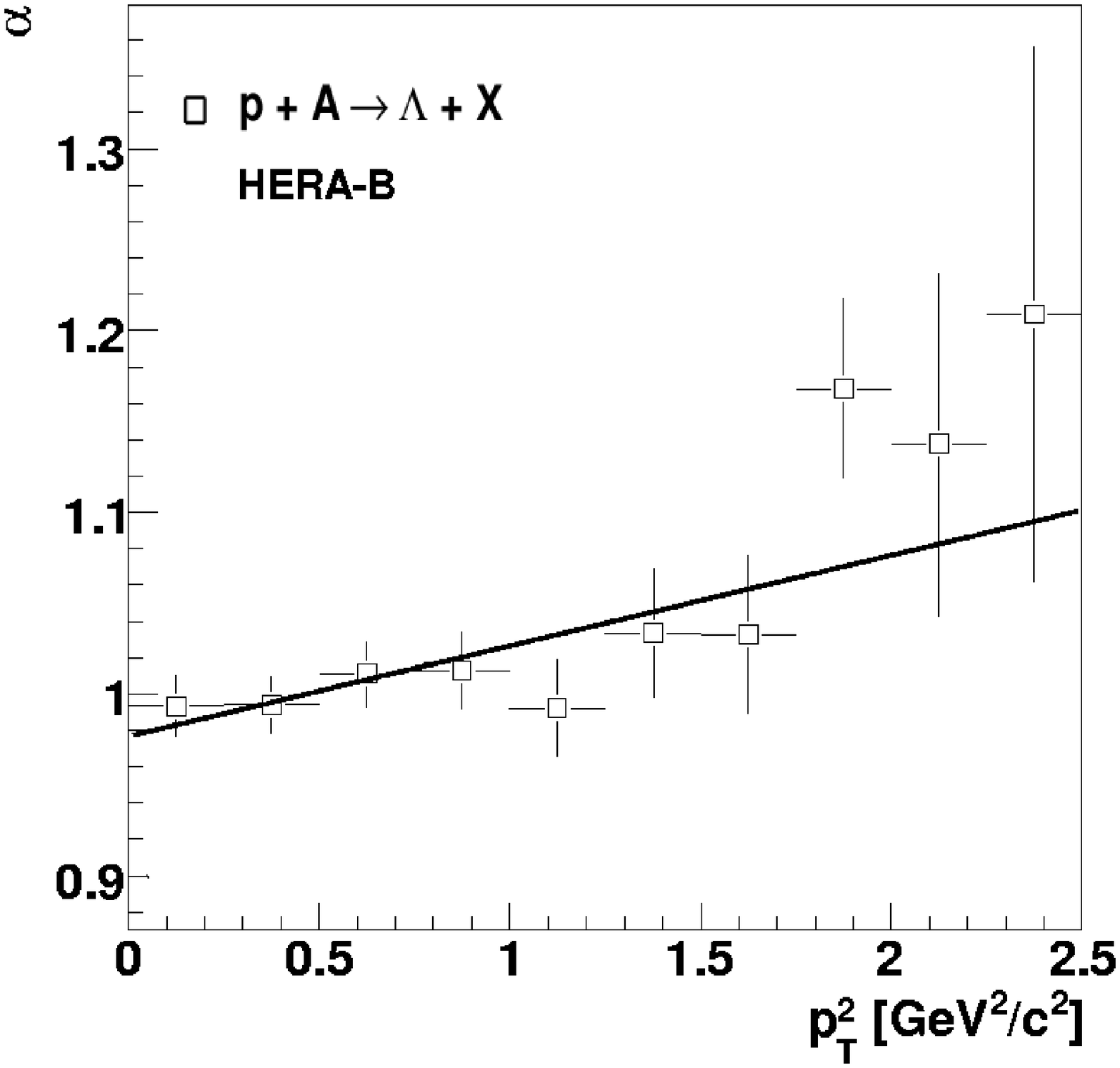}
\end{center}
\caption{Cronin enhancement in high-$p_T$ $K^0_s$ ({\it left}) and $\Lambda$ ({\it right}) 
production in $p$+$A$ collisions measured by HERA-B at $\sqrtsnn$~=~41.6~GeV~\cite{HERA-B:2008sa}.
Note that $\alpha\approx$~1.05
corresponds to $R_{pA} = A^{\alpha-1} \approx$~1.2 for Ti or W (with $A$~=~22, 74).}
\label{fig:cronin_herab}
\end{figure}

The HERA-B collaboration has recently studied the production of
$K^0_s$ mesons and $\Lambda^0$, $\bar\Lambda^0$ baryons
in $p$+C,Ti,W interactions at $\sqrtsnn$~=~41.6~GeV~\cite{HERA-B:2008sa}. The Cronin effect 
is clearly observed for all three species for transverse momenta above $p_T \approx$~1.5~GeV/c (Fig.~\ref{fig:cronin_herab}).
The mass number dependence is parameterised as $\sigma_{pA} = \sigma_{pN} \cdot A^\alpha$ 
where $\sigma_{pN}$ is the proton-nucleon cross section. The values of $\alpha$ are above one, 
in particular for the baryon species ($\Lambda^0$, $\bar{\Lambda}^0$). 

\begin{figure}[tb]
\begin{center}
\begin{tabular}{cc}
   \includegraphics[width=0.48\linewidth,height=4.4cm]{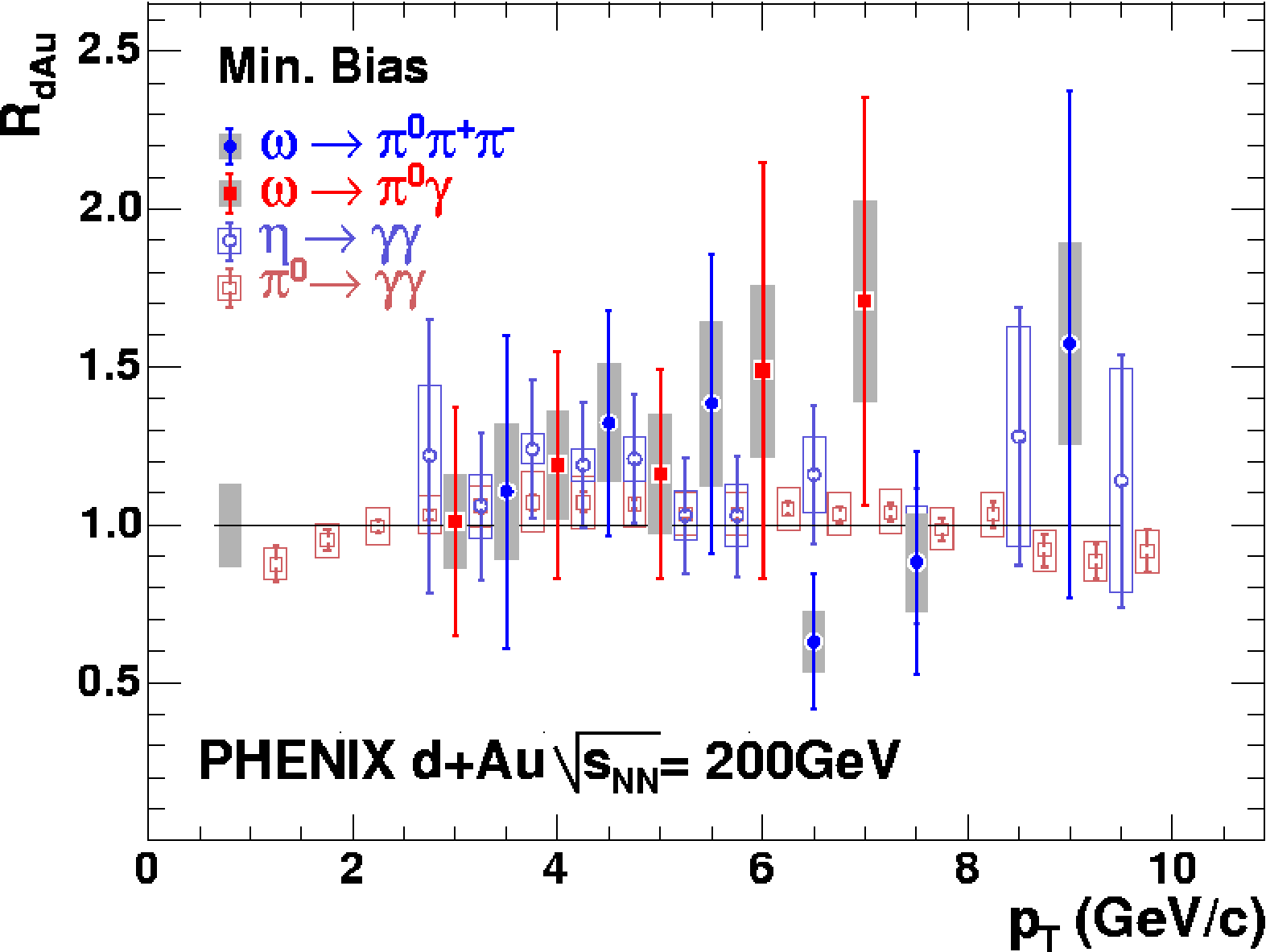} &
   \includegraphics[width=0.48\linewidth]{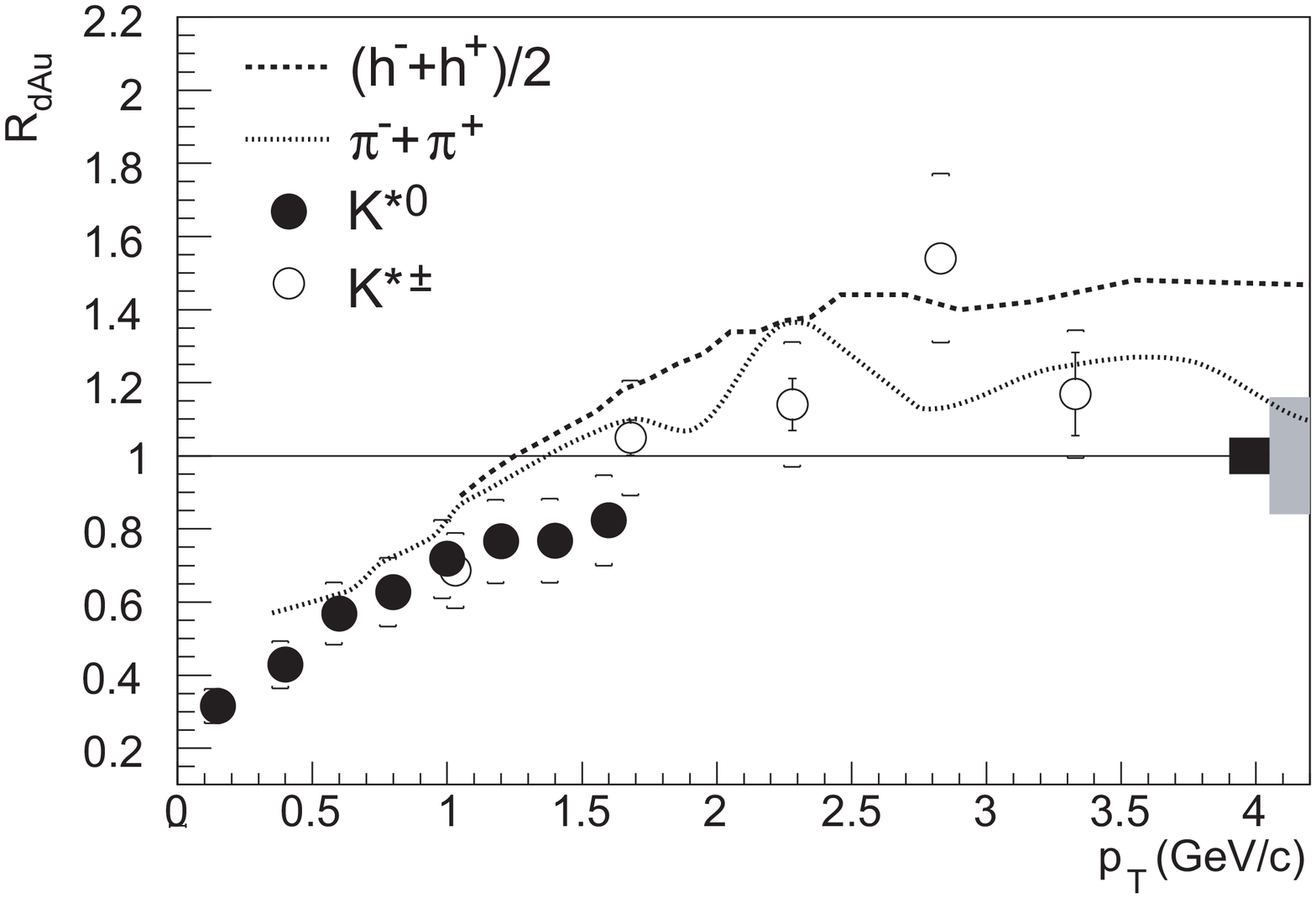} \\
   \includegraphics[width=0.48\linewidth]{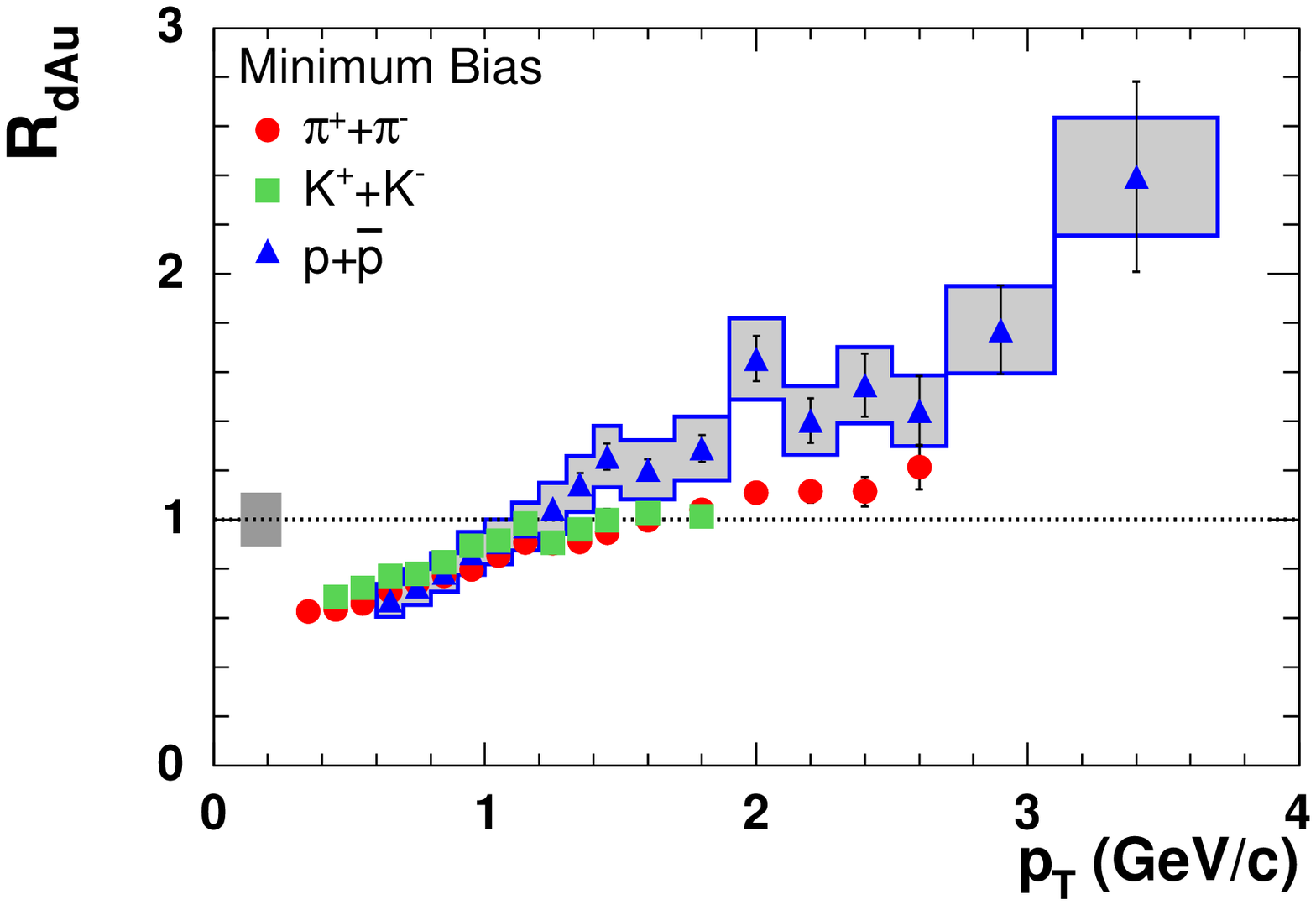} & 
   \includegraphics[width=0.48\linewidth]{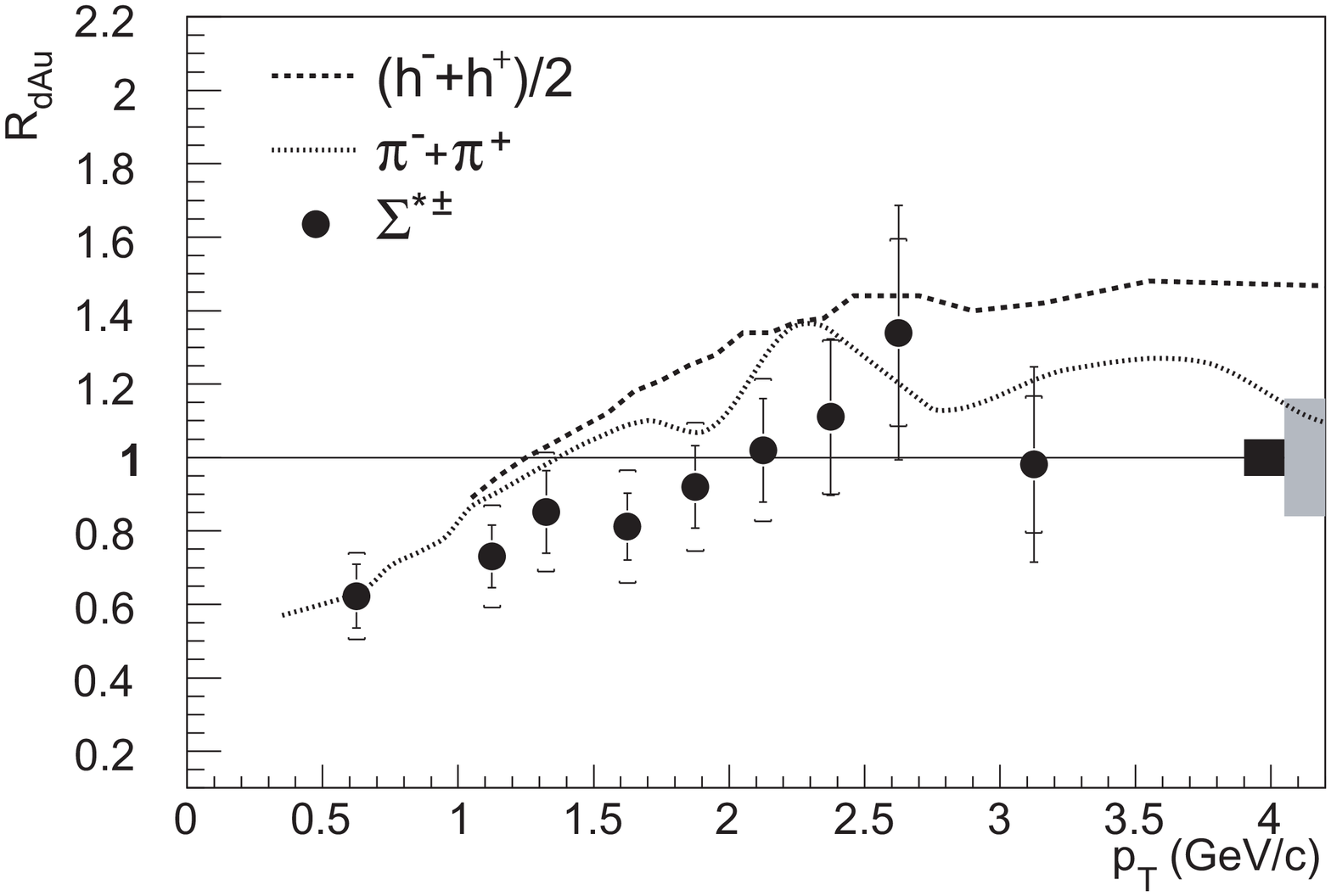} \\
\end{tabular}
\end{center}
\vskip-.3cm
\caption[]{Nuclear modification factor $R_{dAu}(p_T)$ measured in $d+Au$ collisions at $\sqrtsnn$~=~200~GeV at RHIC.
Mesons: $\pi^0$, $\eta$ and $\omega$ (PHENIX~\cite{Adler:2006wg,Adare:2008qa}, {\it top-left}),
and charged pions and kaons (STAR~\cite{Abelev:2008yz}, {\it top-right}).
Hadrons: $\pi^\pm$, $K^\pm$ and $p+\bar p$ (PHENIX~\cite{Adler:2006xd,Adler:2007by}, {\it bottom-left}), 
and $\pi^\pm$, $h^\pm$, and $\Sigma^\pm$ (STAR~\cite{Abelev:2008yz}, {\it bottom-right}).}
\label{fig:RdAu}
\end{figure}

At RHIC, inclusive hadron production in minimum-bias deuteron-nucleus ($d+Au$) collisions 
at $\sqrtsnn$ = 200~GeV also features a small Cronin effect above $p_T$~=~1.5~GeV/c
at mid-rapidity (Fig.~\ref{fig:RdAu}). The enhancement is smaller, or practically absent, 
in the case of mesons ($R_{dAu}\approx$~1) but is visible, of order $R_{dAu}\approx$~1.4, for baryons. 
The enhancement peaks in the range $p_T\approx$~2.5--4~GeV/c, and then starts to decrease and 
disappears beyond 8~GeV/$c$, an observation also apparent in the fixed-target data (Fig.~\ref{fig:cronin}). 
This peak structure is also progressively suppressed as the rapidity of the hadrons is 
increased~\cite{Arsene:2004fa,Adler:2004eh}.

Such a transverse momentum broadening of high-$p_T$ hadrons in hadron-nucleus collisions 
is usually interpreted (see Section~\ref{sec:Cronin}) as due to 
either (i) multiple elastic scatterings of the incoming or outgoing parton inside the
cold nuclear medium~\cite{Accardi:2002ik,Kopeliovich:2002yh}, 
or (ii) recombination at the hadronisation stage of the scattered parton with 
other final state partons created in the collision~\cite{Hwa:2004zd,Hwa:2004yi}. 
The larger Cronin effect for baryons than for mesons is naturally explained by the latter mechanism: 
the combination of three quark momenta boosts up the final baryon spectra more than the two-quark 
coalescence into mesons. The lower Cronin enhancement of all hadrons, in general, at RHIC centre-of-mass 
energies compared to fixed-target results can be explained by the steeper parton spectra at lower energies, 
which makes it easier to get a relatively larger boost for the same amount of $k_T$ ``kick''.
The fast disappearance of the Cronin enhancement with increasing rapidities is well accounted 
for by models based on non-linear QCD evolution of the gluon densities in the 
nuclei~\cite{Kharzeev:2003wz,JalilianMarian:2003mf,Baier:2003hr,Albacete:2003iq}.
The position of the maximum is thus connected to the
rapidity-dependent value of the ``saturation momentum'' $Q_s(y)$ (see
Section~\ref{sec:Cronin}).




\section{Experimental results in nucleus-nucleus collisions} 
\label{sec:hadrons-AA}

\subsection{High-$p_T$ hadron production } \ \\

The production of hadrons at large transverse momentum has been since long proposed as a valuable 
``tomographic'' probe of the hot and dense QCD matter produced in heavy-ion collisions~\cite{Bjorken:1982tu}. 
If the final-state system is dense enough, the hard scattered partons will be attenuated 
while traversing it, resulting in a variety of ``jet quenching''
phenomena, e.g., 
suppression of leading hadron spectra~\cite{Gyulassy:1991xb}, distortion of azimuthal correlations
between back-to-back jets~\cite{Appel:1985dq,Blaizot:1986ma}, modifications of the energy-particle 
flow within the final jets~\cite{Salgado:2003rv,Vitev:2008rz}.
The study of these modifications provides valuable information on the (thermo)dynamical properties 
of the produced system such as the initial gluon density $dN^g/dy$, or the $\qhat$ transport coefficient characterising
the ``scattering power'' of the medium. A detailed recent review of jet quenching results and phenomenology 
can be found in~\cite{d'Enterria:2009am}. Here we highlight the main findings.

The dominant contribution to the energy loss of partons is believed to
be medium-induced gluon radiation as described in the Gyulassy--L\'evai--Vitev (GLV)~\cite{Gyulassy:2000fs,Vitev:2002pf}
and Baier, Dokshitzer, Mueller, Peign\'{e} and Schiff (BDMPS)~\cite{Baier:1996sk,Baier:1998kq,Wiedemann:2000tf} 
(or LPCI~\cite{Zakharov:1997uu}) formalisms (see Section~\ref{sec:hotenloss}). In the GLV approach, 
the initial gluon density $dN^g/dy$ of the expanding plasma (with {\it original} transverse area 
$A_\perp=\pi\,R_A^2\approx$~150~fm$^2$ and thickness $L$) can be estimated from the 
measured energy loss $\Delta E$:
\begin{equation}
\Delta E \propto \alpha_S^3\,C_R\,\frac{1}{A_\perp}\frac{dN^g}{dy}\,L\mbox{ ,}
\label{eq:glv}
\end{equation}
where $C_R$ is the Casimir colour factor of the parton (4/3 for quarks, 3 for gluons).
In the BDMPS framework, the transport coefficient $\qhat$ -- characterising the 
squared average momentum transfer of the hard parton per unit 
path-length: $\qhat \equiv m_D^2/\lambda= \, m_D^2\, \rho \, \sigma$, where 
$m_D$ is the medium Debye mass, $\rho$ its density, and $\sigma$ the parton-matter 
interaction cross section -- can be derived from the average energy loss via:
\begin{equation}
  \langle\Delta E\rangle \propto \alpha_S\,C_R\,\langle\hat{q}\rangle\,L^2.
\label{eq:bdmps}
\end{equation}
For example, for an equilibrated gluon 
plasma at $T=0.4$~GeV with coupling $\alpha_s$~=~0.5 -- i.e. with density 
$\rho_g = 16/\pi^2 \;\zeta (3)\cdot T^3 \approx$ 15~fm$^{-3}$,  
Debye mass $m_D = (4 \pi \alpha_s)^{1/2} T \approx$~1~GeV/c$^2$, and 
LO perturbative cross section $\sigma_{gg}\approx$~1.5~mb, one finds 
$\qhat \simeq 2.2$~GeV$^2$/fm~\cite{Baier:2006fr}. 

Experimentally, the standard method to quantify the medium effects on the yield of a 
large-$p_T$ particle produced at rapidity $y$ in a $A+A$ reaction is given by the 
nuclear modification factor $R_{AA}(p_{T},y;b)$, see Eq.~(\ref{eq:R_AB}),
which measures the deviation of hadron spectra in $A+A$ collisions at
impact parameter $b$ from an incoherent superposition of
spectra in nucleon-nucleon collisions ($R_{AA}$~=~1). 
If the $A+A$ and $p+p$ invariant spectra are both a power-law with exponent $n$, i.e. 
$1/p_{T}\,{d}N/{d}p_{T} \propto p_{T}^{-n}$, the fraction of energy lost $\epsilon_{loss} = \Delta p_{T}/p_{T}$, 
can be (grossly) estimated from $R_{AA}$ via $\epsilon_{loss} \approx 1 - R_{AA}^{1/(n-2)}$~\cite{Adler:2006bw}.

\begin{figure}[tb]
  \centering
  \includegraphics[width=0.53\linewidth,height=5.1cm]{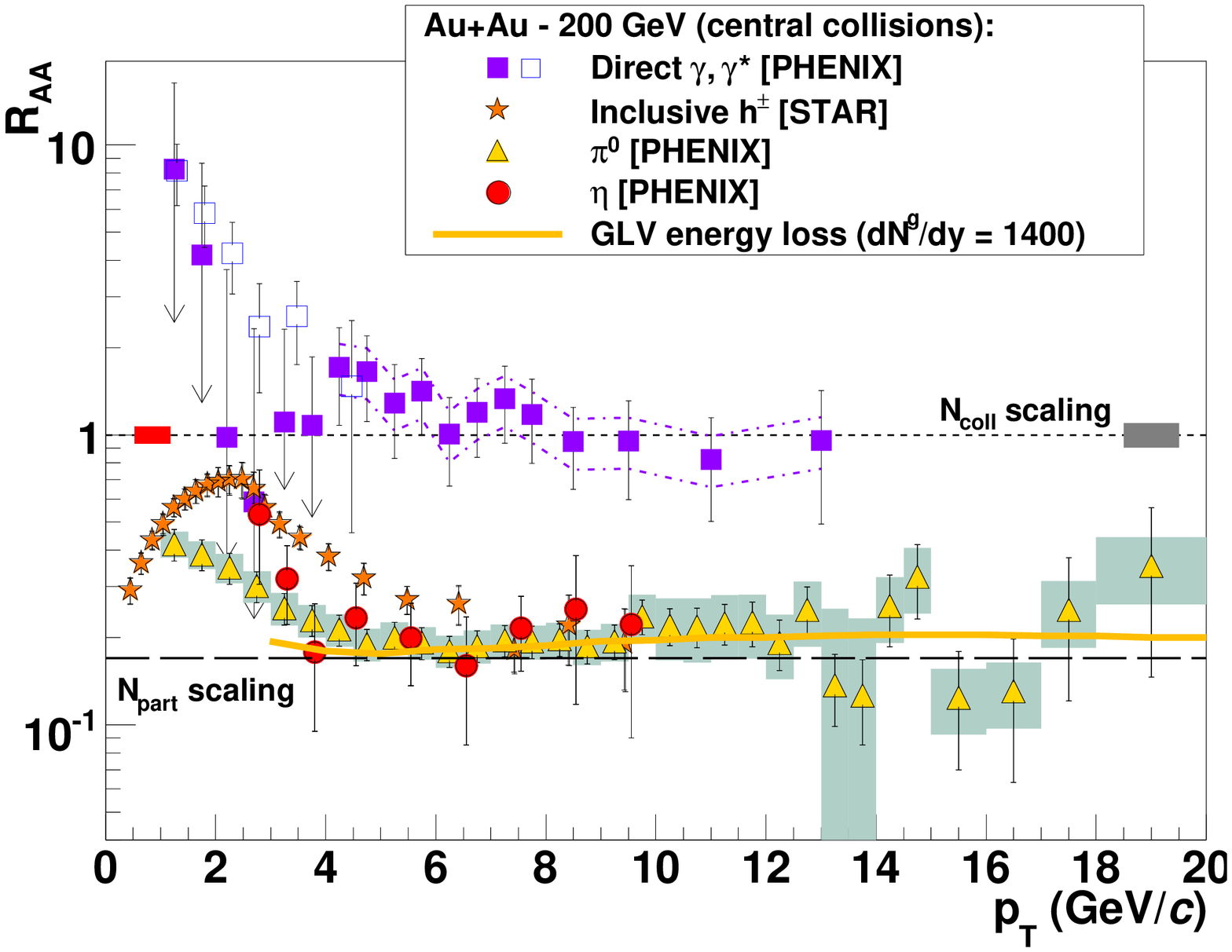}
  \includegraphics[width=0.46\linewidth,height=4.9cm]{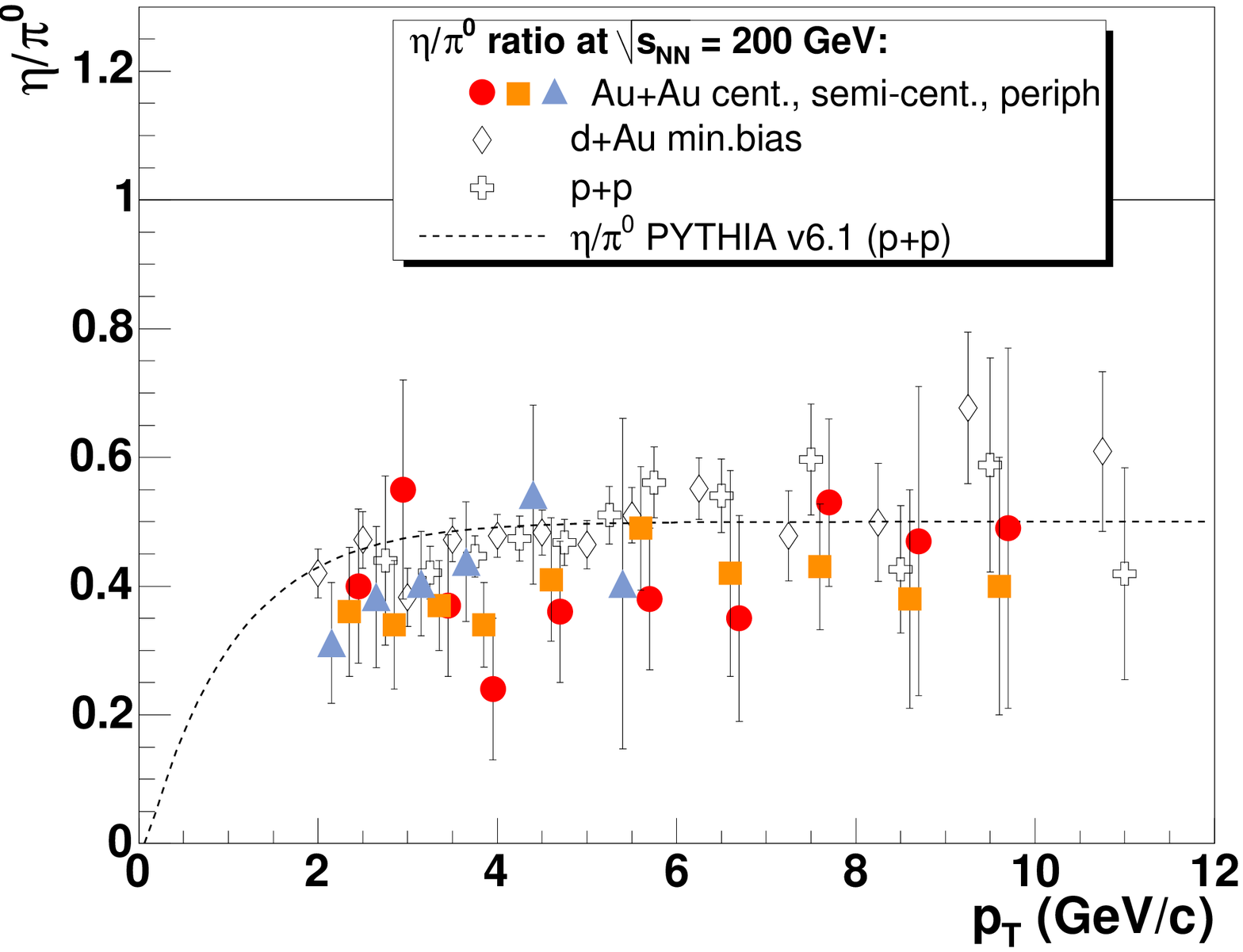}
  \caption{
    {\it Left:} $R_{AA}(p_T)$ in central $Au+Au$ collisions at $\sqrtsnn = 200$~GeV for $\pi^0$~\protect\cite{Adare:2008qa} 
    and $\eta$~\protect\cite{Adler:2006hu} mesons, and for prompt 
    $\gamma,\gamma^\star$~\protect\cite{Adler:2006bv,phenix:2008fq}, and $h^\pm$~\cite{Adams:2003kv} compared to parton 
    energy loss predictions in a medium with gluon densities $dN^g/dy$~=~1400 (yellow curve)~\cite{Vitev:2004bh}.
    {\it Right:} $\eta/\pi^0$ ratios in $p+p$, 
    $d+Au$ and $Au+Au$ collisions measured by PHENIX~\protect\cite{Adler:2006bv} 
    compared to the PYTHIA prediction for $p+p$ (dashed line)~\cite{Sjostrand:2000wi}.  
  }
  \label{fig:RAA_rhic}
\end{figure}

Among the RHIC highlights is the observation (Fig.~\ref{fig:RAA_rhic}, left) of a strong 
suppression of the products of parton fragmentation such as high-$p_T$ neutral mesons 
($\pi^{\circ}$, $\eta$)~\cite{Adcox:2001jp,Adler:2003qi,Adler:2006hu} and charged 
hadrons~\cite{Adler:2002xw,Adams:2003kv,Adcox:2002pe,Adler:2003au}.
The $R_{AA}\approx$~1 perturbative expectation which holds above $p_T\approx$~4~GeV/c
for other hard probes such as ``colour blind'' prompt photons~\cite{Adler:2005ig,Adare:2008fqa} 
(Fig.~\ref{fig:RAA_rhic}, left) or for mesons in $d+Au$ reactions (Fig.~\ref{fig:RdAu}, top), 
is badly broken in central $Au+Au$ collisions where one measures $R_{AA}\approx$~0.2. 
The measured $\pi^0$ power-law spectral exponents of $n\approx$~8 and the $R_{AA}\approx$~0.2 imply an average 
fractional energy loss of the high-$p_T$ hadrons, of $\epsilon_{loss} \approx \Delta p_{T}/p_{T} \approx$~0.2~\cite{Adler:2006bw}.
Such a significant suppression was not observed at SPS where -- even after reevaluating the $p+p$ baseline 
spectrum~\cite{d'Enterria:2004ig,Aggarwal:2001gn} -- the central $Pb+Pb$ meson spectra show an 
$R_{AA}$ around unity (Fig.~\ref{fig:RAA_SPS_RHIC_LHC}).
Yet, this does not exclude the possibility
of energy loss at SPS, since the factor of the $\sim$50\% Cronin enhancement observed in $p+Pb$ at 
$\sqrtsnn\approx$~20~GeV~\cite{Aggarwal:2007gw} likely compensates for
the same amount of final-state
suppression in the hot medium~\cite{d'Enterria:2004ig,Accardi:2005fu} or
in the cold nuclei \cite{Accardi:2007in}. 

As discussed next, most of the empirical properties
of the suppression factor are in quantitative agreement with the
results of the non-Abelian parton energy loss models presented
in more detail in Section~\ref{sec:hotenloss}.

\subsubsection{\it Magnitude of the suppression and medium properties} 
\begin{figure}[tb]
\centering
\includegraphics[width=5.7cm,clip]{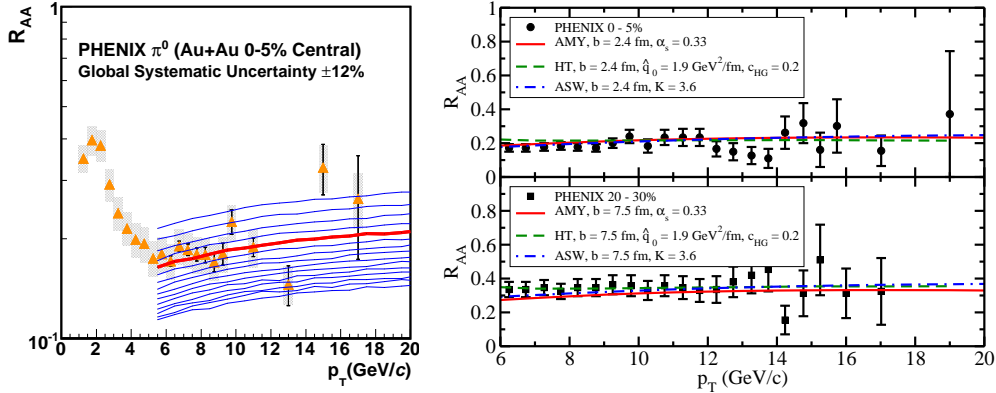}
\includegraphics[width=7.35cm,clip]{figure/RAA_centrality.eps}

\caption{{\it Left:} $R_{AA}(p_T)$ for pions in central $Au+Au$ collisions (triangles)~\cite{Adare:2008qa}
compared to PQM predictions~\cite{Dainese:2004te} for varying values of the $\qhat$ coefficient 
(red curve, best fit for $\mean{\hat{q}}$~=~13~GeV$^2$/fm).
{\it Right:} Central and semi-central pion suppression at PHENIX \cite{Adare:2008qa}
compared to AMY, HT and ASW models within a common space-time evolution
medium profile~\cite{Bass:2008rv}.
}
\label{fig:RAA_vs_PQM}
\end{figure}

--
The $Au+Au$ high-$p_T$ suppression can be well reproduced by parton energy loss calculations
in a very dense system with initial gluon rapidity densities $dN^g/dy\approx$~1400 
(GLV curve in Fig.~\ref{fig:RAA_rhic}, left)~\cite{Vitev:2002pf},
plasma temperatures $T\approx$~0.4~GeV~(AMY model)~\cite{Turbide:2005fk},
and time-averaged transport coefficients $\langle\qhat\rangle\approx$~13~GeV$^2$/fm (PQM model in
Fig.~\ref{fig:RAA_vs_PQM}, left)~\cite{Adare:2008qa,Dainese:2004te}, 
or initial-time transport coefficients $\qhat_0\approx$~10~--~18~GeV$^2$/fm 
(ASW curve in Fig.~\ref{fig:RAA_vs_PQM}, right).

The consistency between the extracted $\qhat$, $dN^g/dy$ and $T$ values in the various models 
has been studied e.g. in~\cite{Majumder:2007iu,d'Enterria:2009am,Bass:2008rv}. 
Whereas the agreement between the fitted thermodynamical variables
$dN^g/dy$ and $T$ is good, the values of the transport parameter $\qhat$ extracted from the data
within the various models  differ by factors of 2~--~3, and are much
larger than 
the LO BDMPS estimate $\qhat\approx$~2~GeV$^2$/fm at $T=0.4$~GeV given before\footnote{It is also interesting 
to note that the BDMPS transport coefficient for
cold nuclear matter $\qhat^\text{cold} \approx 0.6$~GeV/fm$^2$, extracted
via Eq.~\eqref{eq:qhatcold-HERMES} from hadron quenching in nDIS is also
a factor 10 larger than the perturbative estimate $\qhat^\text{cold}\sim
0.05$~GeV/fm$^2$ discussed in Eq.~\eqref{eq:qhatcold2}.}. The discrepancy
with the perturbative estimate is a factor $K=3.6$ in the case of ASW~\cite{Bass:2008rv} and an order of magnitude regarding the value extracted within PQM.
At least part of the uncertainty is due to the relative insensitivity of the $\qhat$ parameter
to the irreducible presence of hadron from (unquenched) partons emitted from the surface of the plasma~\cite{Eskola:2004cr}.
Although relating transport properties to thermodynamical quantities is model- (and medium-) 
dependent, an accord between the fitted $\qhat$ and $dN^g/dy$ values -- via 
$\qhat= \, m_D^2\, \rho \, \sigma$ with $\rho\propto dN^g/dy$ -- can only be 
seemingly achieved for parton-medium cross-sections much larger than the 
$\sigma_{gg}=\mathcal{O}(1$~mb$)$ LO perturbative expectation. Such an observation lends support to the 
strongly-coupled nature of the QGP produced at RHIC~\cite{Gyulassy:2004zy}. It is however 
not satisfactory to obtain parameter values which are typically non-perturbative based on a purely
perturbative framework, see e.g. the discussion in~\cite{Baier:2006fr}.
Additional constraints on $\qhat$ can be placed by requiring also the model reproduction of the
suppressed dihadron azimuthal correlations (see Section~\ref{sec:dihadrons}).

\subsubsection{\it Universality of (light) hadron suppression} -- 
Above $p_T\approx$ 5~GeV/$c$,
$\pi^0$~\cite{Adler:2003qi}, $\eta$~\cite{Adler:2006bv},
and inclusive charged hadrons~\cite{Adams:2003kv,Adler:2003au}
(dominated by $\pi^\pm$~\cite{Adler:2003au}) show all a common factor
of $\sim$5 suppression relative to the $R_{AA}$~=~1 perturbative expectation
which holds for hard probes, such as prompt photons, insensitive to
final-state interactions~\cite{Adler:2005ig} (Fig.~\ref{fig:RAA_rhic}, left).
Such a ``universal'' hadron deficit is consistent with in-medium partonic
energy loss of the parent quark or gluon prior to its fragmentation in the vacuum.
The high-$p_T$ $\pi^0/\eta$ ratio is indeed found to be independent of the collision 
system (or centrality in $Au+Au$ collisions) within uncertainties (Fig.~\ref{fig:RAA_rhic}, right).
Since both hadrons have similar valence quark content but very different masses 
(the $\eta$  is four times heavier than the $\pi^0$), and different 
meson-nucleus cross sections, a natural explanation for the
observation 
is that quenching happens at the parton level: the medium induced energy loss 
experienced by the parent parton is the same independently of the meson which is to be produced, 
and hadronisation occurs in vacuum, according to the same fragmentation functions
extracted from $e^+e^-$ and $p+p$ collisions.  

\subsubsection{\it Centre-of-mass energy dependence} 
\begin{figure}[tb]
\centering
\includegraphics[width=0.60\linewidth]{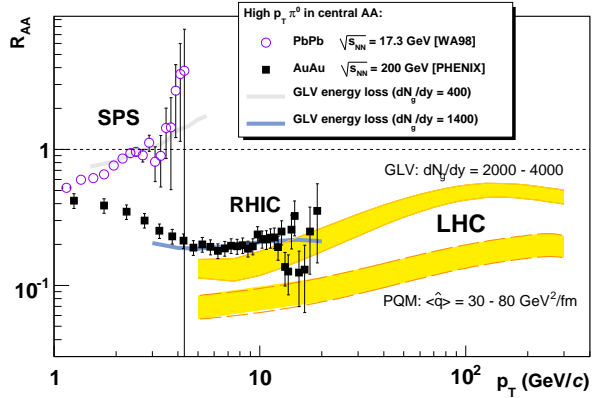}
\caption{Nuclear modification factor, $R_{AA}(p_{\rm T})$, for high-$p_{\rm T}$ $\pi^0$'s at
CERN-SPS~\cite{Aggarwal:2001gn,d'Enterria:2004ig} and RHIC~\cite{Adare:2008qa}
compared to the predictions of the GLV parton energy loss model ($dN^g/dy$ = 400, 1400)~\protect\cite{Vitev:2002pf}. 
The bottom bands show the predicted $R_{AA}$ for high-$p_{\rm T}$ charged hadrons in 
central $Pb+Pb$ collisions at $\sqrtsnn=5.5$~TeV as given by the GLV ($\dd N_g/\dd y$ = 2000--4000) and the PQM 
($\qhat\approx$  30--80~GeV$^2$/fm) models~\protect\cite{Abreu:2007kv}.}
\label{fig:RAA_SPS_RHIC_LHC}
\end{figure}

--
As one increases the collision energy in $A+A$ collisions, the produced medium
reaches higher energy and particle densities, the system stays longer in the QGP phase, and 
correspondingly the traversing partons are more quenched. Figure~\ref{fig:RAA_SPS_RHIC_LHC}
compiles the measured $R_{AA}(p_T)$ for high-$p_T$ $\pi^0$ in central $A+A$ 
collisions at $\sqrtsnn\simeq$17.3 and 200 GeV compared to parton energy loss 
calculations that assume the formation of a QGP with initial gluon
densities 
$dN^g/dy\approx$ 400, 1400~\cite{Vitev:2002pf,Vitev:2004gn} 
or, equivalently, averaged transport coefficients 
$\mean{\hat{q}}\approx$~3.5,~13~GeV$^2$/fm~\cite{Dainese:2004te} respectively.
Table~\ref{tab:dNdy_qhat_sqrts} collects these results as well as those for central
$Au+Au$ reactions at $\sqrtsnn\approx$~62 and 130 GeV~\cite{Buesching:2006ap,Adcox:2001jp}.
For each collision energy the derived values for $dN^g/dy$ are consistent with the final charged 
hadron density $dN_{ch}/d\eta$ measured in the reactions\footnote{The charged particle multiplicity 
itself follows a logarithmic dependence on the c.m. energy~\cite{Adler:2004zn}: 
$dN_{ch}/d\eta \approx 0.75\cdot(N_{part}/2)\cdot \ln(\sqrtsnn\,\mbox{\small [GeV]}/1.5)$
($N_{part}$ is the number of participant nucleons in the collision).}. 
This is expected in an isentropic expansion process, where all the hadrons produced at midrapidity come 
directly from the original gluons released in the collision:
\begin{equation}
\frac{dN^g}{dy}\approx\frac{N_{tot}}{N_{ch}}\,\left|\frac{d\eta}{dy}\right|\,\frac{dN_{ch}}{d\eta}
\approx 1.8\cdot\frac{dN_{ch}}{d\eta}\;. 
\label{eq:dNgdy}
\end{equation}
This relationship is relatively well fulfilled by the data as can be seen by comparing
the fourth and fifth columns of Table~\ref{tab:dNdy_qhat_sqrts}.

\begin{table}[tb]
\caption{Suppression factors measured in central $A+A$ collisions in the range $\sqrtsnn\approx$~20~--~200 GeV,
and initial gluon densities $dN^g/dy$~\protect\cite{Vitev:2002pf,Vitev:2004gn}, and 
transport coefficients $\mean{\hat{q}}$~\protect\cite{Dainese:2004te,d'Enterria:2005cs}
obtained from parton energy loss calculations reproducing the observed high-$p_T$ 
$\pi^0$ suppression at each $\sqrtsnn$. The measured charged particle densities at 
midrapidity, $dN_{ch}^{exp}/d\eta|_{\eta=0}$~\protect\cite{Adler:2004zn}, are also quoted.}
\label{tab:dNdy_qhat_sqrts}
\centering
\begin{tabular}{lccccc}\hline 
\noalign{\smallskip}
\hspace{1mm} & \hspace{1mm}$\sqrtsnn$ (GeV) \hspace{1mm} & \hspace{1mm}$R_{AA}(\pi^0,\,p_T\approx 4\mbox{ GeV/c})$ \hspace{1mm} &\hspace{1mm} $\mean{\hat{q}}$ (GeV$^2$/fm) \hspace{1mm}  &  \hspace{1mm} $dN^g/dy$  \hspace{1mm} &  \hspace{1mm} $dN_{ch}^{exp}/d\eta|_{\eta=0}$  \hspace{1mm} \\
\noalign{\smallskip}\hline\noalign{\smallskip}
SPS  &  17.3  & $\sim$1.0~\cite{Aggarwal:2001gn,d'Enterria:2004ig} & 3.5 & 400  & 312 $\pm$ 21 \\
RHIC &  62.4 & $\sim$0.4~\cite{Buesching:2006ap} & 7. & 800  & 475 $\pm$ 33 \\
RHIC &  130. & $\sim$0.3~\cite{Adcox:2001jp} &  11  & 1000   & 602 $\pm$ 28 \\
RHIC &  200. & $\sim$0.2~\cite{Adler:2003qi} & 13 & 1400 & 687 $\pm$ 37 \\ \noalign{\smallskip}\hline 
\end{tabular}
\end{table}

\subsubsection{\it Transverse momentum dependence} --
At RHIC top energies, the hadron quenching factor remains basically constant from 5~GeV/c up to the highest 
transverse momenta measured so far, $p_T\approx$~20~GeV/c (Fig.~\ref{fig:RAA_rhic}).
As can be seen in Eq.~\eqref{eq:modelff}, the suppression is roughly proportional to
the $z$-slope of the fragmentation functions, $\partial D(z,Q^2)/\partial z$. Consequently, 
even though the relative parton energy loss, $\epsilon/\kt$, becomes smaller at higher $\pt$ 
(leading naively to a smaller suppression and an increase of $\raa$ with $\pt$), the larger steepness 
of the partonic spectrum due to the restricted phase space to produce hight-$\pt$ partons at RHIC 
leads to a significant suppression even at large transverse momenta. 
Indeed, full calculations~\cite{Vitev:2002pf,Dainese:2004te,Jeon:2003gi,Eskola:2004cr}
including the combined effect of (i) energy loss kinematics constraints, 
(ii) steeply falling $p_T$ spectrum of the scattered partons, and (iii) $\mathcal{O}($20\%) $p_T$-dependent 
(anti)shadowing differences between the proton and nuclear parton distribution functions (PDFs), result in an effectively flat $R_{AA}(p_T)$ 
as found in the data. The much larger kinematical range opened at the LHC TeV-energies~\cite{Abreu:2007kv} 
will allow one to test the $p_T$-dependence of parton energy loss, and the associated radiation spectrum,
over a much wider domain than at RHIC.
As can be seen in Fig.~\ref{fig:RAA_SPS_RHIC_LHC} (yellow bands) the
PQM model seemingly predicts a 
slower (and smaller) rise of $R_{AA}(p_T)$ than the GLV model.

\subsubsection{\it Centrality (system-size) dependence} --
The volume of the produced plasma in a heavy-ion collision can be ``dialed'' modifying the 
overlap area between the colliding nuclei either by selecting a given impact-parameter $b$ -- 
i.e. by choosing more central or peripheral reactions -- or by colliding larger or smaller nuclei, 
e.g. $Au$~($A$~=~197) versus $Cu$~($A$~=~63).
\begin{figure}[tb]
\includegraphics[width=1.0\textwidth,height=5.1cm]{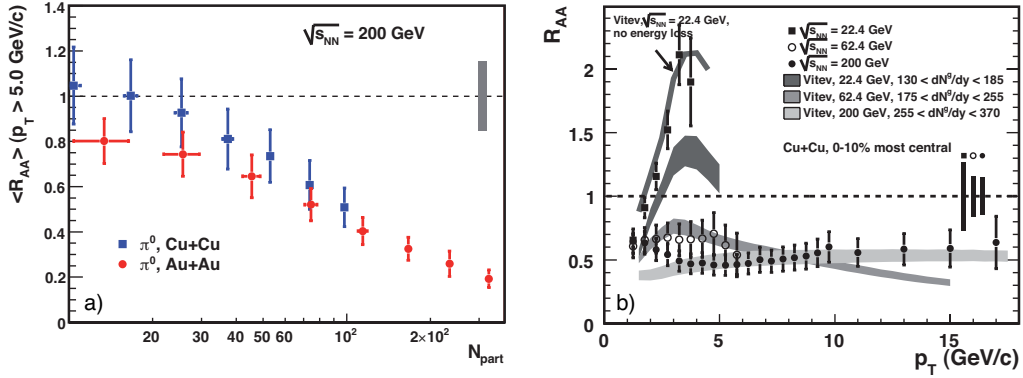}
\caption{{\it Left:} Centrality ($N_{part}$) dependence of the high-$p_T$ $\pi^0$ suppression
in $Cu+Cu$ and $Au+Au$ at 200~GeV~\cite{Reygers:2008pq}. 
{\it Right:} $R_{AA}(p_T)$ for $\pi^0$ in central $Cu+Cu$ collisions at 
$\sqrtsnn=$22.4, 62.4 and 200~GeV compared to GLV calculations with initial gluon 
densities $dN^g/dy\approx$~100~--~370~\cite{Adare:2008cx}.}
\label{fig:RAA_vs_Npart}
\end{figure}
The relative energy loss depends on the effective mass number $A_{\rm eff}$ or, equivalently, 
on the number of participant nucleons in the collision $N_{\rm part}$, as: 
${ \Delta E }/{E} \; \propto \; A_{\rm eff}^{2/3} \; \propto \; N_{\rm part}^{2/3}$~\cite{Dainese:2004te,Vitev:2005he}.  
The measured $R_{AA}(p_T)$ in central $Cu+Cu$ at 22.4, 62.4, and 200\,GeV~\cite{Adare:2008cx} is 
a factor of $(A_{Au}/A_{Cu})^{2/3} \approx$~2 lower than in central $Au+Au$ at the same energies
(Fig.~\ref{fig:RAA_vs_Npart}, right) 
Yet, for a comparable $N_{part}$ value, the suppression in $Au+Au$ and $Cu+Cu$ is very similar
(Fig.~\ref{fig:RAA_vs_Npart}, left). 
Fitting the $N_{part}$ dependence to $R_{AA} = (1 - \kappa\; N_{part}^{\alpha})^{n-2}$ 
yields $\alpha = 0.56 \pm 0.10$~\cite{Adare:2008qa}, consistent also 
with parton energy loss calculations~\cite{Dainese:2004te,Vitev:2005he}.

\subsubsection{\it Path-length dependence} --
The analytical quadratic dependence of the energy loss on the thickness of a {\it static} medium $L$,
Eq.~(\ref{eq:bdmps}), becomes effectively a linear dependence on the {\it initial} value of $L$
when one takes into account the expansion of the plasma, see Eq.~(\ref{eq:glv}). 
Experimentally, one can test the  dependence of parton suppression on the plasma thickness ($L$)
by exploiting the spatial asymmetry of the  system produced in non-central nuclear collisions.
Partons produced ``in plane'' (``out-of-plane'') i.e. along the short (long) direction of the ellipsoid 
matter with eccentricity $\epsilon$ will comparatively traverse a shorter (longer) thickness. 
PHENIX has measured the high-$p_T$ neutral pion suppression as a function 
of the angle with respect to the reaction plane, $R_{AA}(p_T,\phi)$~\cite{Adler:2006bw,Adare:2009iv}.
Each azimuthal angle $\phi$ can be associated with an average medium path-length $L_{\epsilon}$ 
via a Glauber model. 
The energy loss is found to satisfy the expected $\Delta E \propto L$ 
dependence, Eq.~(\ref{eq:glv}), above a ``threshold'' length of $L\approx$~2~fm, interpreted 
in~\cite{Pantuev:2005jt} as due to a geometric ``corona'' effect.

\subsubsection{\it Non-Abelian (colour factor) dependence} 
\begin{figure}[tb]
\centering
\includegraphics[width=0.49\linewidth,height=5.15cm,clip]{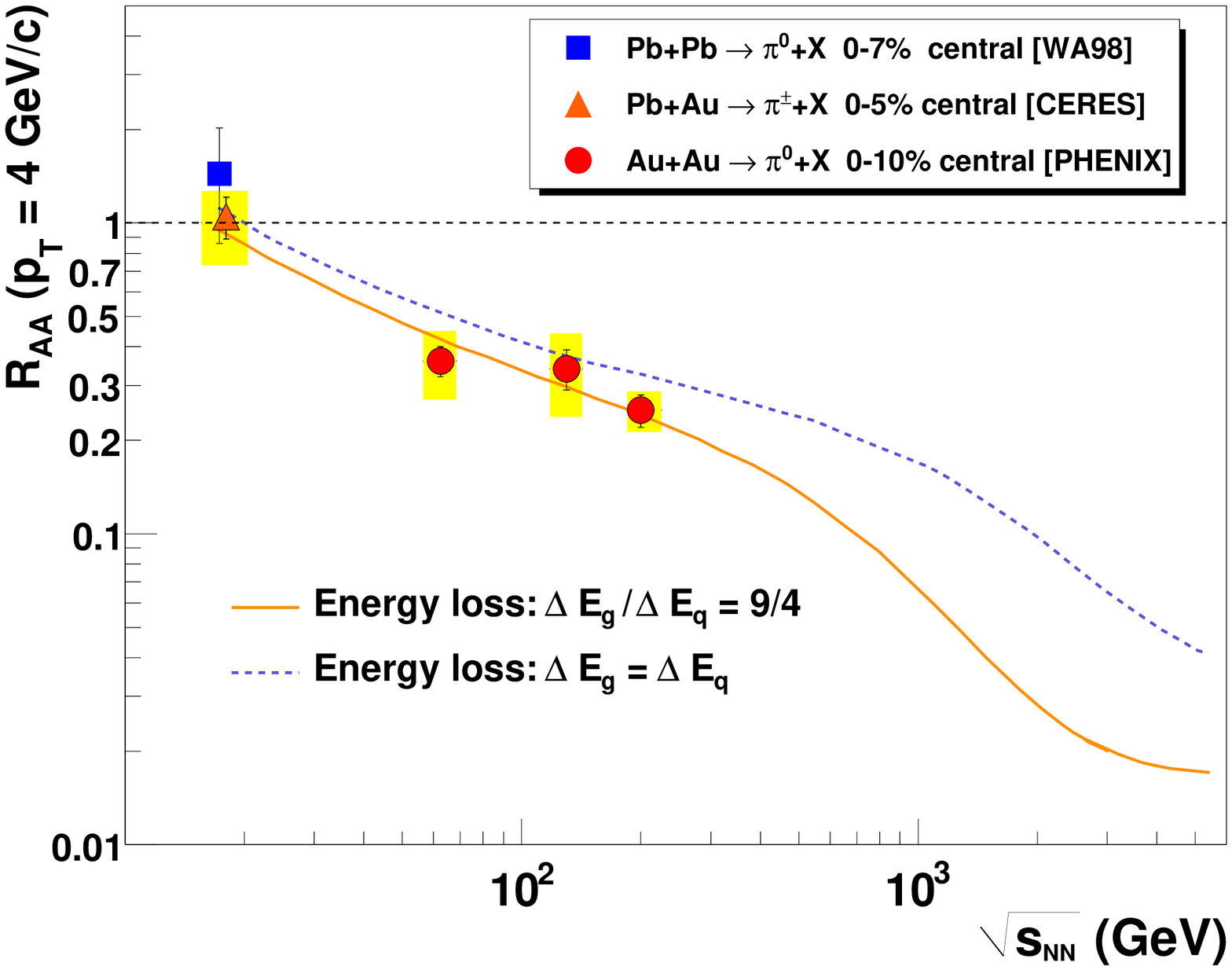}
\includegraphics[width=0.49\linewidth,height=5.cm,clip]{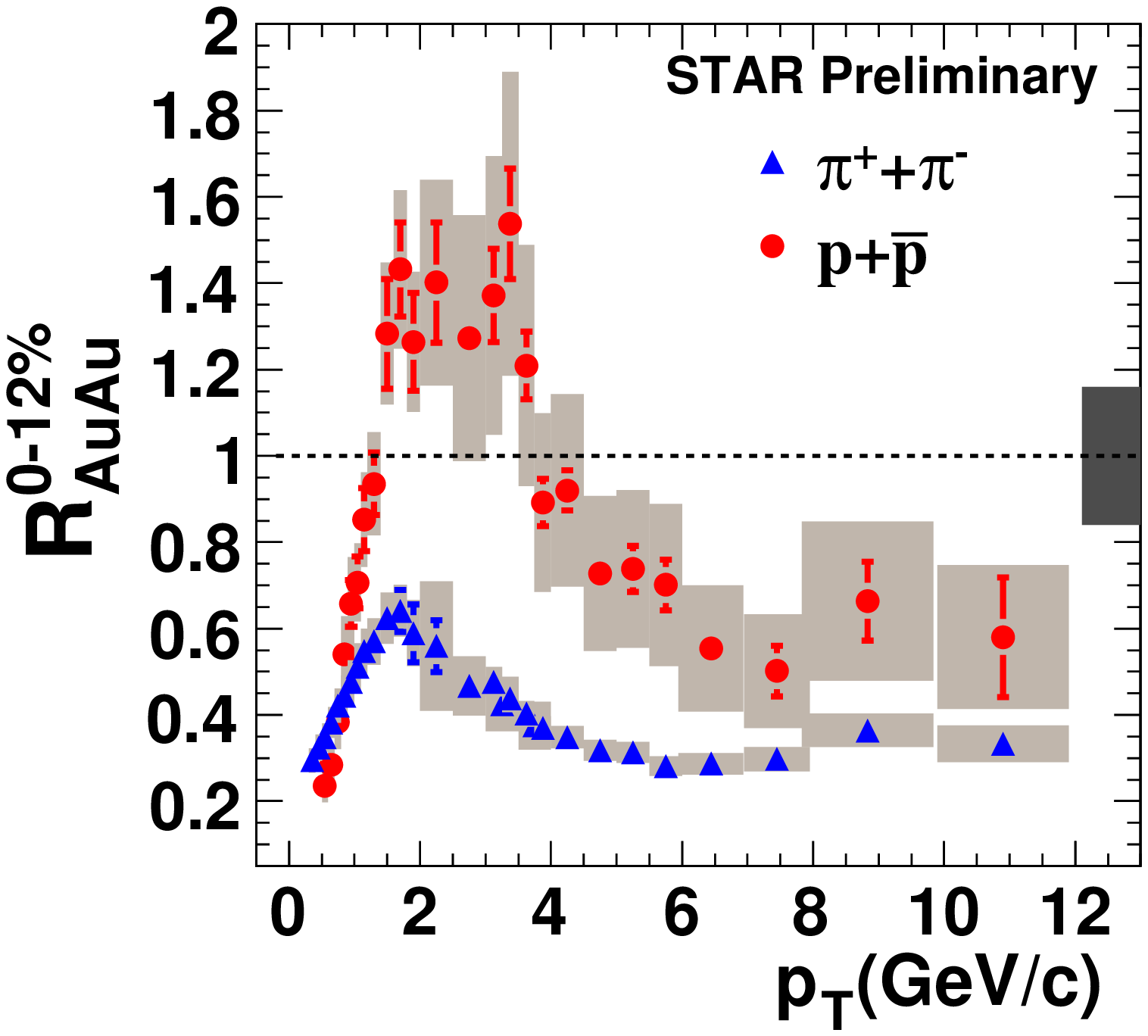} 
\caption{{\it Left:} Excitation function of the nuclear modification factor, $R_{AA}(\sqrtsnn)$, for
$\pi^{\circ}$ in central $AA$ reactions at a fixed $p_T$ = 4~GeV/c~\protect\cite{d'Enterria:2005cs},
compared to non-Abelian (solid) and ``non-QCD'' (dotted) energy loss curves~\protect\cite{Wang:2004tt}.
{\it Right:} $R_{AA}(p_T)$ for pions and (anti)protons in central $Au+Au$ at $\sqrtsnn$~=~200~GeV~\protect\cite{Mohanty:2008tw}. .}
\label{fig:RAA_nonAbelian}
\end{figure}

--
The amount of energy lost by a parton in a medium is proportional to its colour Casimir factor: 
$C_A =$~3 for gluons, $C_F$~=~4/3 for quarks. Asymptotically, the probability 
for a gluon to radiate another gluon is $C_A/C_F$ = 9/4 times larger than for a quark and thus
$g$-jets are expected to be more quenched than $q$-jets in a QGP. One can test such a genuine 
non-Abelian property of QCD energy loss by measuring hadron suppression at a fixed 
$p_T$ for increasing c.m. energy~\cite{d'Enterria:2005cs,Wang:2004tt}.
At large (small) $x$, the PDFs are dominated by valence-quarks (low-$x$ gluons)
and consequently hadroproduction will be dominated by quark (gluon) fragmentation. 
Figure~\ref{fig:RAA_nonAbelian} (left) shows the $R_{AA}$ for 4-GeV/c pions measured 
at SPS and RHIC compared to two parton energy loss curves~\cite{Wang:2004tt}.
The lower (upper) curve shows the expected $R_{AA}$ assuming a normal (arbitrary) behaviour 
with $\Delta E_g/\Delta E_q$~=~9/4 ($\Delta E_g = \Delta E_q$).
The experimental high-$p_T$ $\pi^0$ data supports the expected colour-factor dependence 
of $R_{AA}(\sqrtsnn)$~\cite{d'Enterria:2005cs}.

A second test of the colour charge dependence of hadron suppression has been proposed 
based on the fact that gluons fragment comparatively more into (anti)protons than quarks do.
One would thus naively expect $R_{AA}^{p,\bar{p}}<R_{AA}^{\pi}$.
The STAR results (Fig.~\ref{fig:RAA_nonAbelian}, right) are however seemingly at variance 
with this  expectation: pions appear more suppressed than protons at high-$p_T$~\cite{Mohanty:2008tw}. 
Yet, the use of (anti)protons as a reference for perturbative particle production is questionable.
First, it is worth reminding that the assumption of production from parton fragmentation in vacuum
may well not hold for protons which have estimated formation time-scales a factor $\sim$5 
shorter than for pions (Table~\ref{table:pertformtime}). Second, 
$p,\bar{p}$ are already found to be Cronin-enhanced in $d+Au$ compared to $p+p$ collisions by a factor 
$\sim 50$\% -- 100\% (see Fig.~\ref{fig:RdAu} bottom-left) for $p_T$'s as large as 7~GeV/c~\cite{Adams:2006nd}.
It is likely that there is an extra mechanism of baryon production, based e.g. on in-medium quark 
coalescence~\cite{Hwa:2002tu,Fries:2003kq,Greco:2003xt} (see Section~\ref{sec:Cronin}),
which compensates for the energy loss suffered by the parent partons. 
Finally, another explanation has recently been proposed in Ref.~\cite{Brodsky:2008qp}: protons 
could be produced in a compact colour-singlet configuration (through higher-twist processes) and 
therefore, because of colour transparency, escape more easily the dense medium than pions do.


\subsection{Heavy flavour production } \ \\
\label{sec:heavyflavors}

As seen in the previous Section, most of the empirical properties of the quenching factor 
for light-flavour hadrons -- magnitude, $p_T$-, centrality-, $\sqrtsnn$-
dependences of the suppression -- are in quantitative agreement with
the predictions of non-Abelian parton energy loss models. 
A robust prediction of radiative energy loss models is the hierarchy 
$\Delta E_{\rm  heavy-Q} < \Delta E_{\rm light-q} <  \Delta E_{g}$: 
gluons, which fragment predominantly into light hadrons, 
are expected to lose more energy than quarks because of the larger
color coupling to the radiated gluon;
whereas massive $c,b$ quarks are expected to lose less energy than 
light quarks due to a suppression of small-angle gluon radiation already 
in the vacuum (``dead-cone'' effect)~\cite{Dokshitzer:2001zm,Zhang:2007zz}.
Grossly, the in-medium radiative energy loss is suppressed by a factor 
$\mathcal{O}(m_D/M_Q)$ (where $m_D\sim$~1~GeV/c$^2$ is the medium Debye mass)~\cite{Peigne:2008wu}, 
i.e. it is a factor about 25\% (75\%) less for a charm (bottom) quark 
than for a light-quark.

\begin{figure}[tb]
  \centering
\includegraphics[width=0.44\linewidth,height=4.5cm]{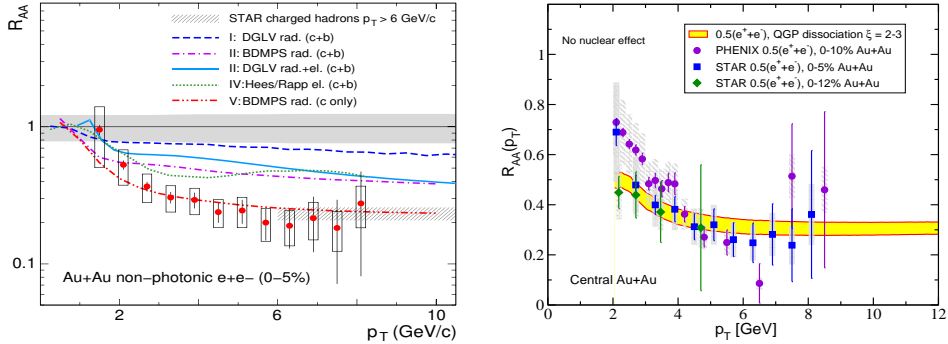}\hspace{0.5cm}
\includegraphics[width=0.44\linewidth,height=4.5cm]{figure/RAA_nonphot_elec_adilvitev.eps}
  \caption{Nuclear modification factor $R_{AA}^e$ of heavy-quark decay electrons in 
    central $Au+Au$ collisions at  $\sqrtsnn = 200$~GeV~\cite{Adler:2005xv,Abelev:2006db} 
    compared to various radiative+elastic energy loss models for $c$ and $b$ quarks ({\it left}) and 
    to a model of heavy-meson ($D$, $B$) dissociation in a plasma ({\it right})~\cite{Adil:2006ra}.
  }
  \label{fig:RAA_heavyQ}
\end{figure}

Yet, RHIC measurements~\cite{Adler:2005xv,Adare:2006nq,Abelev:2006db} 
of high-$p_T$ electrons from the semi-leptonic decays of $D$- and $B$-mesons 
(Fig.~\ref{fig:RAA_heavyQ}) indicate the same suppression factor for light and 
heavy mesons: $R_{AA}(Q)\sim R_{AA}(q,g)\approx$~0.2.
In order to reproduce the high-$p_T$ open charm/bottom suppression,
jet quenching models require either initial gluon rapidity densities
$dN^g/dy\approx$~3000~\cite{Djordjevic:2005db} which are inconsistent 
with the total hadron multiplicities, see Eq.~(\ref{eq:dNgdy}), 
as well as with the $dN^g/dy\approx$~1400 needed to describe the quenched light hadron spectra. 
Various explanations have been proposed to solve this `heavy flavor puzzle'
(see e.g.~\cite{Frawley:2008kk,Vitev:2008jh}).

\begin{itemize}
\item First, precise comparisons between theory and data require a better 
determination of the relative contribution of the $c$ and $b$ quarks to the measured
non-photonic electron yields~\cite{Djordjevic:2005db,Cacciari:2005vx}.
If only $c$ quarks (roughly three times more suppressed than the heavier $b$ quarks) actually contributed 
to the measured high-$p_T$ decay electron spectrum, then one would indeed expect $R_{AA}(c)\approx$~0.2~\cite{Armesto:2005mz}. 
However, indirect measurements from PHENIX~\cite{Adare:2009ic} and STAR~\cite{Mischke:2008qj} have confirmed 
the similar production yields of $e^\pm$ from $D$ and $B$ decays above $p_T\approx$~5~GeV/c 
predicted by next-to-leading-log pQCD~\cite{Cacciari:2005vx,Cacciari:2005rk};

\item The second mechanism 
points to an additional contribution from elastic (i.e. non-radiative) energy loss 
for heavy-quarks~\cite{Mustafa:2003vh,Wicks:2005gt} which was considered negligible
so far~\cite{Gyulassy:1991xb}.
Recent works~\cite{Peshier:2006hi,Peshier:2006az,Peigne:2008nd,Gossiaux:2008jv} 
have shown that a proper evaluation of the QCD running coupling
substantially increases the amount of collisional energy loss 
suffered by the heavy-quarks. $\Delta E_{coll}$ can indeed be a significant contribution 
for heavy quarks (see `rad.+el.' curves in Fig.~\ref{fig:RAA_heavyQ}, left);

\item Two works~\cite{Sorensen:2005sm,MartinezGarcia:2007hf} have argued that in a plasma
the large charm-quark coalescence into $\Lambda_c$ baryons (with a small semileptonic decay branching ratio)
would deplete the number of open-charm mesons, and correspondingly reduce the number of decay electrons, 
compared to $p+p$ collisions;

\item The assumption of vacuum hadronisation (after in-medium radiation) implicit in all parton energy loss
formalisms may well not hold in the case of a heavy quark. The discussed quark-hadronisation time estimates
(see Section~\ref{sec:formationtimes}) are inversely proportional to the mass $m_h$ of the final produced 
hadron: the heavier the hadron, the fastest it is formed. From Eq.~(\ref{eq:pertestimate}), one can see that
in the rest frame of the hot QCD medium of the fragmenting heavy-Q, 
the formation time of $D$- and $B$-mesons~\cite{Adil:2006ra}
is of order\footnote{Note that  in the laboratory system there is an extra Lorentz 
boost factor: $\tau_{_{\ensuremath{\it lab}}} = \gamma_Q\cdot\tau_{_{form}}$.} 
$ \vev{t_h}\approx$~0.4~--~1~fm, respectively. Thus, one may need to account for both
the energy loss of the heavy-quark as well as the possible dissociation of the heavy-quark meson 
inside the QGP. The expected amount of suppression in that case is larger and consistent with the data 
(Fig.~\ref{fig:RAA_heavyQ}, right).
\end{itemize}


\subsection{High-$p_T$ di-hadron correlations } \ \\
\label{sec:dihadrons}

At leading order, the parton-parton $2 \to 2$ scatterings are balanced in $p_T$ i.e. they are back-to-back in azimuthal angle, $\Delta \phi= \pi$, modulo some
smearing due to the partons intrinsic transverse momentum. Such azimuthal correlation is
smeared out if one or both partons suffer rescatterings inside the plasma. 
The dijet-acoplanarity arising from the interactions of a parton in an expanding QGP 
is $ \langle k_{T}^2\rangle_{med} \simeq (m_D^2/\lambda) L\,\ln (L/\tau_0)\propto \qhat L$~\cite{Qiu:2003pm}
and, thus, the final azimuthal correlations between the hadrons issuing from 
the fragmentation of quenched partons will show a dependence on the transport coefficient 
and thickness of the medium: $d^2N_{pair}/d\Delta\phi = f(\qhat,L)$.
Jet-like correlations in heavy-ion collisions have been measured on a statistical basis by selecting 
a high-$p_T$ trigger particle and measuring the azimuthal ($\Delta\phi = \phi - \phi_{trig}$) 
and pseudorapidity ($\Delta\eta = \eta - \eta_{trig}$) distributions of its associated hadrons 
($p_{T}^{assoc}<p_{T}^{trig}$):
$C(\Delta\phi,\Delta\eta) = \frac{1}{N_{trig}}\frac{d^2N_{pair}}{d\Delta\phi d\Delta\eta}$.
In $p+p$ collisions, a dijet signal appears as two distinct back-to-back Gaussian-like peaks at $\Delta\phi\approx$~0, 
$\Delta\eta\approx$ 0 (near-side) and at $\Delta\phi\approx\pi$ (away-side). At variance with such a topology,
early STAR results~\cite{Adler:2002tq} showed a monojet-like topology with a complete 
disappearance of the opposite-side peak for $p_{T}^{assoc}\approx$~2~--~4~GeV/c.

\begin{figure}[tb]
\centering
\includegraphics[width=0.9\linewidth,height=4.cm]{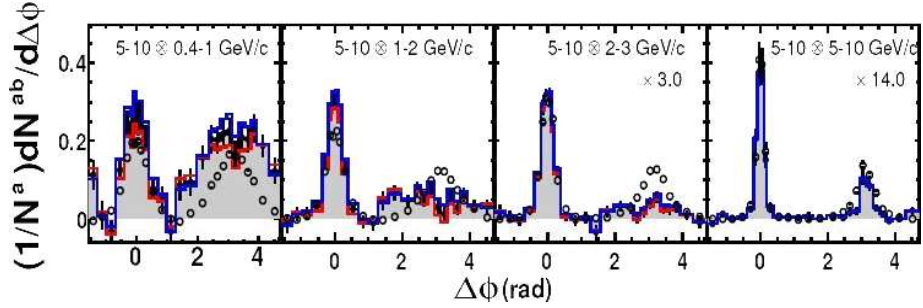}
\caption{Comparison of the azimuthal di-hadron correlation $dN_{pair}/d\Delta\phi d\eta$ for $p+p$ (open symbols) 
and central $Au+Au$ (histograms) at $\sqrtsnn$~=~200~GeV for $p_T^{trig}=$~5--10~GeV/c and increasingly
smaller (right to left) values of $p_T^{assoc}$~\protect\cite{Adare:2008cq}.}
\label{fig:dNdphi}
\end{figure}

\begin{figure}[tb]
\includegraphics[width=0.49\linewidth,height=5.6cm,clip]{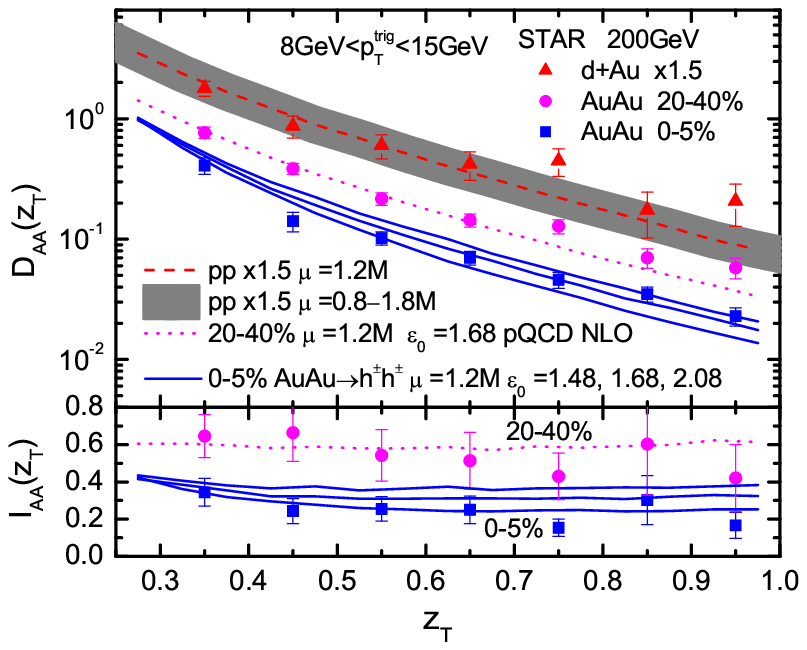}
\includegraphics[width=0.49\linewidth,height=5.8cm,clip]{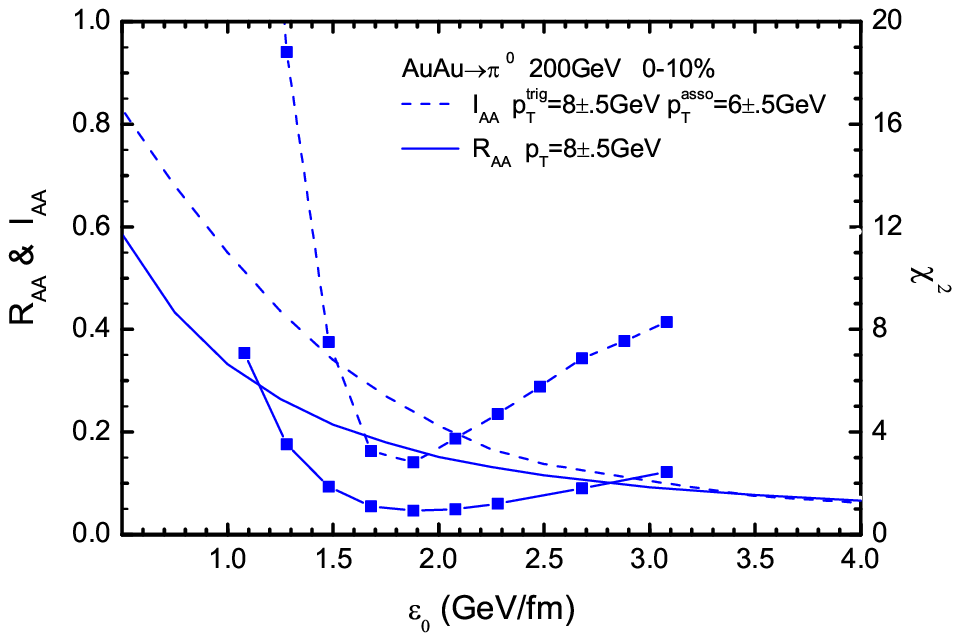}
\caption{{\it Left:} $D^{away}_{AA}(z_T)$ distributions for $d+Au$ and $Au+Au$ 
and $I_{AA}(z_T)$ ratio for central $Au+Au$ at 200~GeV~\cite{Adams:2006yt}, 
compared to HT calculations for varying $\epsilon_0$ energy losses~\cite{Zhang:2007ja}. 
{\it Right:} Data vs. theory $\chi^2$ values for the fitted $\epsilon_0$ parameter~\cite{Zhang:2007ja}.}
\label{fig:IAA_qhat}
\end{figure}

Fig.~\ref{fig:dNdphi} shows the increasingly distorted back-to-back azimuthal correlations 
in high-$p_T$ triggered central $Au+Au$ events as one decreases the $p_T$ of the associated hadrons
(right to left). Whereas, compared to $p+p$, the near-side peak remains unchanged  for all $p_T$'s, 
the away-side peak is only present for the highest partner $p_T$'s but progressively disappears 
for less energetic partners~\cite{Adare:2007vu,Adare:2008cq}.
The correlation strength over an azimuthal range $\Delta\phi$ between a trigger hadron $h_{t}$ and a partner hadron 
$h_{a}$ in the opposite azimuthal direction can be constructed as a function of the momentum fraction 
$z_T=p_{T}^{assoc}/p_{T}^{trig}$ via a ``pseudo-fragmentation function'':
\begin{equation}
D^{away}_{AA} (z_T) =
\int dp_{T}^{trig} \int dp_{T}^{assoc}
\int_{\Delta\phi>130^\circ} d\Delta\phi\; \frac{d^3\sigma_{AA}^{h_t h_a}/d p_{T}^{trig}d p_{T}^{assoc}
d\Delta\phi}{d\sigma_{AA}^{h_t}/d p_{T}^{trig}}\,.
\label{eq:D_AA}
\end{equation}
shown in Fig.~\ref{fig:IAA_qhat} (top-left) compared to 
various values of the $\epsilon_0$ parameter characterising the amount of energy loss~\cite{Zhang:2007ja} 
(see the Higher-Twist model discussion in Section~\ref{sec:Eloss_formalisms}).
Similarly to $R_{AA}(p_T)$, the magnitude of the suppression of back-to-back jet-like two-particle 
correlations can be quantified with the ratio $I_{AA}(z_T) = D_{AA}(z_T)/D_{pp}(z_T)$. 
$I_{AA}^{away}$ is found to decrease with increasing centrality, down to about 0.2~--~0.3 for the most 
central events (Fig.~\ref{fig:IAA_qhat}, bottom-left)~\cite{Adler:2002tq,Adams:2006yt}.

The right plot of Fig.~\ref{fig:IAA_qhat} shows the best $\epsilon_0\approx$~1.9~GeV/fm 
energy loss value that fits the 
measured $R_{AA}$ and $I_{AA}$ factors. Due to the irreducible presence of (unquenched) partons emitted 
from the surface of the plasma, the single-hadron quenching factor $R_{AA}(p_T)$ is in general 
less sensitive to the value of $\epsilon_0$ than the dihadron modification ratio $I_{AA}(z_T)$. 
The combination of $R_{AA}(p_T)$ and $I_{AA}(z_T)$ provides more robust quantitative information on 
the medium properties.

Since energy and momentum are conserved, the ``missing''  fragments of the away-side (quenched) parton 
at intermediate $p_T$'s should be either shifted to lower energy ($p_T\lesssim$~2~GeV/c) and/or 
scattered into a broadened angular distribution. Both, softening and broadening,
are seen in the data when the $p_T$ of the away-side associated hadrons is lowered (see two
leftmost panels of Fig.~\ref{fig:dNdphi}). As a matter of fact, the away-side hemisphere shows a very 
unconventional double-hump angular distribution with a ``dip'' at $\Delta\phi \approx \pi$ and two 
neighbouring local maxima at $\Delta\phi \approx \pi\,\pm$~1.1~--~1.3. 
Such a ``volcano''-like profile has been interpreted as due to the preferential emission of energy 
from the quenched parton at a finite angle with respect to the jet axis. This could happen in a purely radiative 
energy loss scenario due to large-angle radiation~\cite{Polosa:2006hb}, but more intriguing explanations 
have been put forward based on the dissipation of the lost energy into a collective mode 
of the medium in the form of a wake of lower energy gluons~\cite{Ruppert:2005uz}
with Mach-~\cite{Satarov:2005mv,CasalderreySolana:2004qm} 
or \v{C}erenkov-like~\cite{Dremin:2005an,Koch:2005sx} angular emissions 
assuming that the hard parton velocity exceeds the speed of sound or speed of light,
respectively, of the QCD medium. Theoretically, it is unclear if such partonic 
collective wake(s) and cone survive both hadronisation and the final hadronic 
freeze-out~\cite{Betz:2008js}. More involved studies, e.g., accounting
for the plasma expansion and the hadronic phase evolution, are needed
before a final conclusion can be reached.


\subsection{High-$\pt$ photon production } \ \\
\label{sec:photonbrehms}

Photon production in cold and hot QCD media is a particularly
interesting probe because, unlike the gluon,  
photons can escape without final-state interactions,
therefore carrying information about the medium at the location of its
production. Moreover, it avoids the ambiguities related to the
hadronisation process, so that it is in principle a cleaner probe of
the underlying dynamics. However, several mechanisms proposed over 
the last few years showed that prompt photon production could actually 
be somehow affected by the quark-gluon plasma formation in heavy-ion 
collisions, and in principle by cold nuclear matter as well.

\begin{figure}[tb]
  \centering
  \parbox[c]{0.445\linewidth}{
    \includegraphics[width=\linewidth,clip=true]{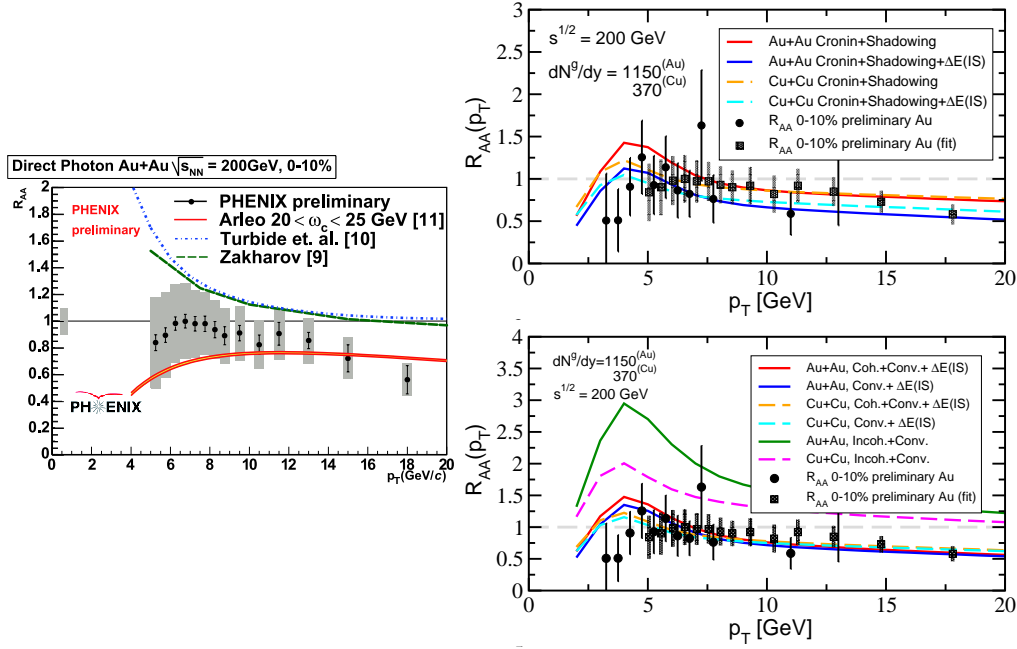}
  }
  \parbox[c]{0.545\linewidth}{
    \includegraphics[width=\linewidth,clip=true]{figure/gammaAA.eps} 
    \includegraphics[width=\linewidth,clip=true]{figure/gammaAAfdbk.eps} 
  }  
  \caption{
    Prompt photon spectrum compared to preliminary PHENIX data.
    {\it Left:} Spectrum  computed in
    Refs.~\cite{Arleo:2007bg,Turbide:2007mi,Zakharov:2004vm}. Plot
    taken from Ref.~\cite{Isobe:2007ku}. 
    {\it Right:} Spectrum computed in the GLV formalism including a
    variety of effects. Plots taken from Ref.~\cite{Vitev:2008vk}.
  }
  \label{fig:photons}
\end{figure}

First, prompt photons are also produced from the collinear fragmentation of quarks 
and gluons produced in the hard process. Such fragmentation
photons are thus sensitive to energy loss of their parent fragmenting parton. The formation time 
needed to produce such a photon--parton system with a small invariant mass exceeds 
by far the typical lifetime of the hot medium: 
the multiple scattering of the hard parton in the medium is followed on a 
much larger time-scale by the parton-to-photon fragmentation process in the vacuum. 
Consequently, one expects that those ``fragmentation'' photons should be as quenched, 
at least qualitatively, as the hadron yield~\cite{Jeon:2002dv,Arleo:2006xb,Arleo:2007bg}.

Second, the multiple scattering incurred by the hard partons traversing the produced
medium induces the emission of soft gluons --~leading to the usual ``jet quenching''~-- as well as 
soft photons~\cite{Zakharov:2004vm,Turbide:2005fk,Arnold:2001ba} (although 
with a probability in principle governed by the much smaller $\alpha_{em}$). Within a 
path-integral picture for the parton energy loss mechanism~\cite{Zakharov:1996fv,Zakharov:1997uu}, 
Zakharov computed this medium-induced photon bremsstrahlung contribution 
at RHIC~\cite{Zakharov:2004bi}. The enhancement of photon production is 
particularly noticeable in the moderate $\ptgamma$ range, say below 20~GeV, 
raising hopes that it could be measured. It has also been proposed that large $\pt$ partons may 
couple to the thermal quarks and gluons in the medium through Compton 
scattering or $q\bar{q}$ annihilation. This so-called jet--photon conversion 
mechanism,~\cite{Fries:2002kt}, 
would enhance the production of photons, leading to a nuclear production ratio, 
$R_{\rm AA}$, larger than one.
Each one of these individual mechanisms, with opposing (quenching and enhancement) 
effects, is unfortunately under poor theoretical control and the medium effects on (prompt) photon 
production still remain rather model-dependent.

Experimentally, the photon 
quenching factor has been measured at RHIC by the PHENIX collaboration, up to large 
$\pt\approx$~20~GeV/c~\cite{Isobe:2007ku}. Despite the still rather large error bars of 
those preliminary data, it seems that the prompt photon $R_{\rm AA}$ is consistent 
with unity in the range $\pt\approx$~4--15~GeV/c, while a suppression is reported, 
$R_{\rm AA}\simeq 0.6$, in the highest $\pt$ bin. Therefore it appears that the photon enhancement predicted either 
due to parton multiple scattering~\cite{Zakharov:2004vm} or jet-photon conversion~\cite{Fries:2002kt} 
is not seen at RHIC. 
The BDMPS energy loss calculation, Ref.~\cite{Arleo:2006xb}, supplemented by the 
proper treatment of protons and neutrons in the $Au$ nuclei (leading to a suppression at 
large $\pt$ due to smaller electric charge of $d$ compared to $u$ valence quarks), is on the contrary 
able to reproduce the shape and magnitude of the PHENIX data.




\section{Parton propagation and energy loss}
\label{sec:parton}

\subsection{In-medium parton propagation and the Cronin effect } \ \\ 
\label{sec:Cronin}

The Cronin effect is the enhancement of single inclusive hadron production at intermediate 
$p_T$'s ($p_T\approx$~1~--~8~GeV/c) observed in hadroproduction in  $e+A$ (Fig.~\ref{fig:pt_hermes}) 
and $h+A$ (Fig.~\ref{fig:cronin}) collisions, as well as in Drell-Yan events (Fig.~\ref{fig:dypa2}, right). 
Theoretically, such an effect has been usually explained in terms of initial- or final-state 
multiple scattering of the parton prior to its fragmentation leading to a broadening of the
transverse momentum of the produced hadrons~\cite{Accardi:2002ik}. More recently,
a modification of the hadronisation mechanism due to the recombination of the scattered 
parton with other partons produced in the collision has been proposed~\cite{Hwa:2004yi,Hwa:2004zd} 
and accounts for the experimental data as well~\cite{Fries:2008hs}. 
We review here both theoretical interpretations of the Cronin enhancement.

\subsubsection{\it Parton multiple scatterings}--
Parton multiple scatterings have been discussed in the pQCD
factorisation formalism, in the colour dipole model and in the Colour Glass
Condensate approach. We briefly review their main features,
similarities and differences.  

\begin{itemize}

\item Glauber-Eikonal models.

The Glauber-Eikonal (GE) approach
\cite{Krzywicki:1979gv,Lev:1983hh,Accardi:2001ih,Accardi:2003jh,Gyulassy:2002yv}
to the Cronin effect treats multiple $2\ra 2$ partonic collisions in
collinearly factorised pQCD. 
The cross-section for the production of a hadron with
transverse momentum $p_T$ and rapidity $y$ in $h+A$ collisions at fixed
impact parameter $b$ is written as~\cite{Accardi:2003jh}
\begin{align}
  \frac{d\sigma_h}{d^2p_T dy d^2b}  
    & = \sum_i \phi_{i/h} e^{\, - \sigma_{\,iN} T_A(b)} \otimes D_{i\ra h} \nonumber \\
    & \otimes \sum_{n=1}^{\infty} \frac{1}{n!} \int d^2b \, d^2k_1 \cdots d^2k_n
    \, \delta\left(\sum _{j=1,n} {\vec k}_j - {\vec p_T}\right)
    \frac{d\sigma_{\,iN}}{d^2k_1} T_A(b) 
    \times \dots \times \frac{d\sigma_{\,iN}}{d^2k_n} T_A(b) 
\label{pAxsec} 
\end{align}
where the crossed-circle symbols denote appropriate integrations and
summations over parton flavours $i$ and $j$, $\phi_{i|h}$ are the PDFs
in the projectile hadron, $d\sigma_{iN}$ is the
cross section for a parton $i$ scattering on a nucleon and $T_A(b)$
the $A$ nucleus thickness function at impact parameter $b$ (see
Ref.~\cite{Accardi:2003jh} for details). 
At moderate $p_T$, the accumulation  of transverse momentum due to
multiple scatterings leads to
an enhancement of transverse momentum spectra, and to a suppression 
of the low-$p_T$ region due to energy-momentum conservation. At
high $p_T$ the binary scaled $p+p$ spectrum is recovered: no
high-$p_T$ suppression is predicted in this approach, except as a
consequence of the nuclear modifications of PDFs.
In early applications, the GE series \eqref{pAxsec}
has been directly evaluated only up to the $n=3$ parton scatterings,
under severe approximations, and 
only for $\sqrt s \leq 40$ GeV
\cite{Krzywicki:1979gv,Lev:1983hh,Straub:1992xd}. 
Instead of evaluating the full GE series, 
other approximated GE models modify the pQCD rates through the
inclusion of a phenomenological 
nuclear broadening of the intrinsic parton momentum 
$k_T$~\cite{Wang:1998ww,Wang:2001cy,Zhang:2001ce,Vitev:2002pf},
ignoring however the unitarity constraints built into the GE
multi-scatterings. 
The GE series is directly computed via a numerical convolution of
elementary parton-nucleon processes, assuming a decoupling of the
transverse and longitudinal kinematics in
Refs.~\cite{Accardi:2003jh,Accardi:2004ut,Accardi:2005fu}, and with 
exact energy-momentum conservation up to 3 scatterings in
Ref.~\cite{Cattaruzza:2004qr}. 
 
The formulation preserves unitarity and is directly constrained to
reproduce the  absolute normalised spectra in $p+p$ collisions, and
quantitatively incorporates kinematic phase-space limitations at large
$p_T$. Thus, the $p_T$-broadening is computed without adjustable
parameters, rather than assumed as an input. 
Midrapidity pion production data at $\sqrtsnn=20-200$ GeV can be well described
(Fig.~\ref{fig:dAu_vs_models} left), but the large hadron suppression reported 
at forward rapidity at RHIC~\cite{Arsene:2007jd,Adams:2006uz}
(Fig.~\ref{fig:dAu_vs_models} right) can be reproduced only within 
gluon saturation approaches~\cite{Kharzeev:2004yx,JalilianMarian:2004xm} or using
extreme nuclear shadowing parametrisations~\cite{Barnafoldi:2008rb}.  

\begin{figure}[tb]
\centering
  \includegraphics[height=5cm]{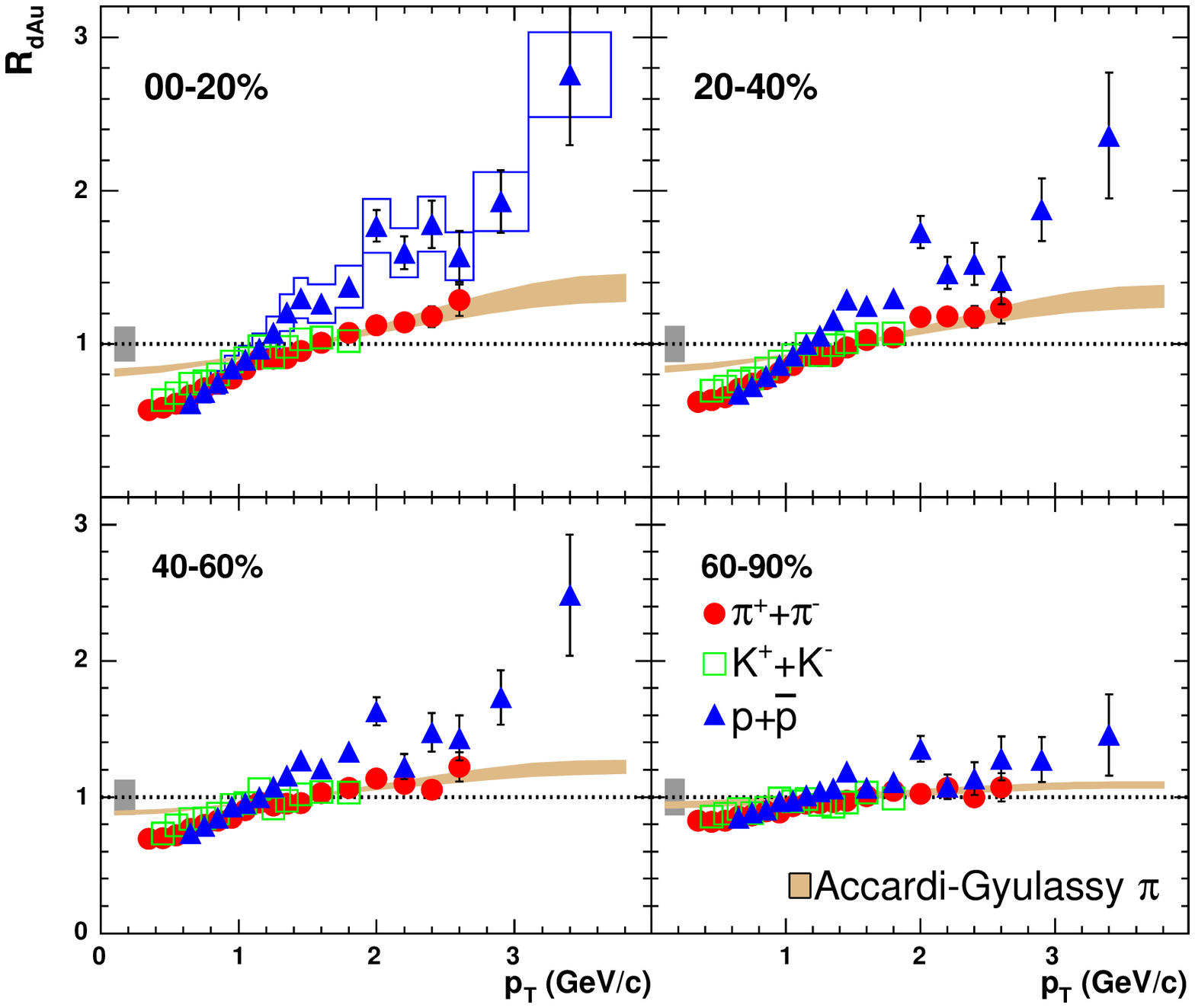}
  \includegraphics[height=4.8cm]{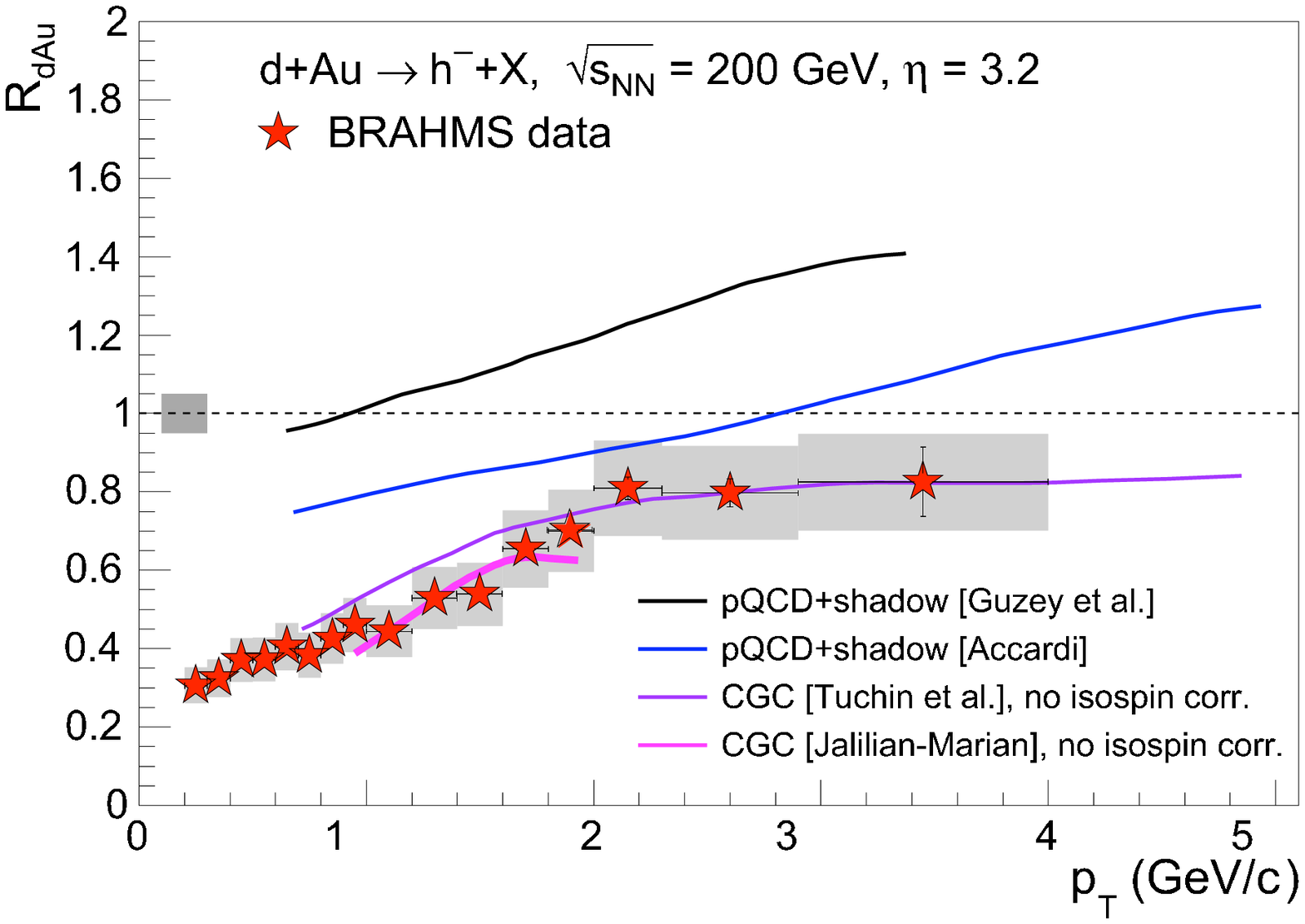}
  \caption{{\it Left}: GE multi-scattering computation~\cite{Accardi:2002ik} 
    of neutral pion
    $R_{dAu}(p_T)$ at midrapidity in different centrality bins compared to
    PHENIX data at $\sqrtsnn = 200$~\cite{Adler:2006xd}. Plot taken from
    Ref.~\cite{Adler:2006xd}.
    {\it Right}: BRAHMS data on negatively charged hadron $R_{dAu}(p_T)$ at
    $\eta = 3.2$ compared to several models. From top to bottom:
    pQCD computation supplemented by nuclear shadowing effects by
    Guzey {\it et al.}~\cite{Guzey:2004zp}; 
    GE model computation by Accardi {\it et al.}~\cite{Accardi:2002ik};
    CGC model computations by Tuchin {\it et al.}~\cite{Kharzeev:2004yx} and by
    Jalilian-Marian~\cite{JalilianMarian:2004xm}. Plot from Ref.~\cite{d'Enterria:2006su}.
  } 
  \label{fig:dAu_vs_models}
\end{figure}

\item The colour dipole model

In Ref.~\cite{Kopeliovich:2002yh,Johnson:2000dm}, the GE series is
formulated in terms of 
the multiple scattering of a colour dipole on the nuclear target. 
The colour dipole cross section is determined phenomenologically by
fits to lepton-proton and proton-proton scattering data
\cite{GolecBiernat:1998js,GolecBiernat:1999qd}, and
hadron production is determined from the overlap of the nuclear broadened
dipole wave function with the hadron light-cone wave function. The
computation of nuclear effects is carried out with no tunable
parameters, and leads to a good description of midrapidity pion data
(Fig.~\ref{fig:cronin} left). This model allows for the inclusion of coherent 
multiple scattering~\cite{Kopeliovich:2002yh}, relevant at RHIC to
LHC energies, and is equivalent to
the GE model in the incoherent scattering limit. 
It can also describe in a unified formalism the $p_T$-broadening of DY
lepton pairs~\cite{Johnson:2006wi,Johnson:2007kt} and that of hadrons produced nDIS~\cite{Kopeliovich:2003py} (see
Section~\ref{sec:perthadrmodel}).  
This approach also allows for a description of the forward rapidity
hadron suppression at RHIC due to energy conservation and Sudakov
suppression at the edge of the phase space, alternative to gluon
saturation~\cite{Kopeliovich:2005ym,Nemchik:2008xy}. Hadron suppression is
predicted at forward rapidity also for lower energy collisions,
$\sqrtsnn =$~63 or 130~GeV, where gluon saturation effects are not
expected to play a major role because of the rather large $x_2$ probed at these energies.

\item Colour Glass Condensate

The Colour Glass Condensate (CGC)
is an effective theory for the gluon field in hadron and nuclei 
at small $x$, see Ref.~\cite{Gelis:2007kn} for a recent review.
At low-$x$ the gluon occupation
number is so large that the 
gluon field can be treated semi-classically and computed as a solution
of the Yang-Mills equation of motion in the presence of random colour sources.
Gluons with momenta lower than a scale $Q_s$ are in the ``saturation''
regime with a density high enough that 
$2\to1$ gluon-gluon fusion processes limit a further
growth; on the contrary at larger momenta the gluon field is in the
DGLAP ``dilute'' regime. 
An intermediate ``geometric scaling window''
extends at momenta $Q_s<p_T<Q_s^2/Q_0$, where quantum effects from the
saturation region further modify the evolution from the perturbative
to the saturation region. 
Observables are computed as an average over the
colour sources density $\rho$ with a weight $W_y[\rho]$, depending on
the gluon rapidity $y = \log(1/x)$.
The quantum evolution of the gluon field with $y$ is
captured by the non-linear JIMWLK evolution
equation~\cite{JalilianMarian:1996xn,JalilianMarian:1997jx,JalilianMarian:1997gr,Iancu:2000hn,Iancu:2001ad}.
Using a Gaussian approximation for the weight $W$, known as
McLerran-Venugopalan model, gluon production in $p+A$ 
collision can be interpreted as multiple $2 \ra 1$ partonic
scatterings~\cite{Iancu:2002aq}. A comparison to the $2\ra 2$ multiple
scatterings included in the GE models is discussed in
Ref.~\cite{Accardi:2004fi}.

\begin{figure}[tb]
\centering
  \includegraphics[width=6cm]{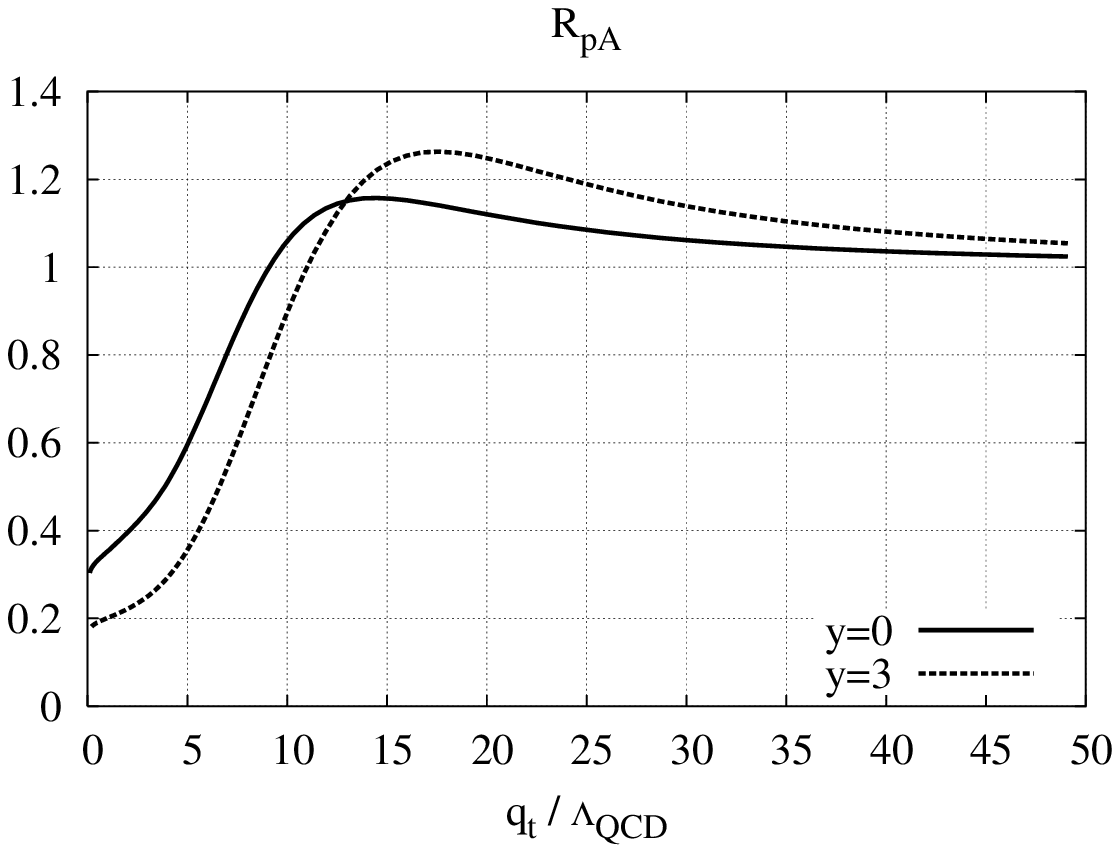}
  \includegraphics[width=6cm]{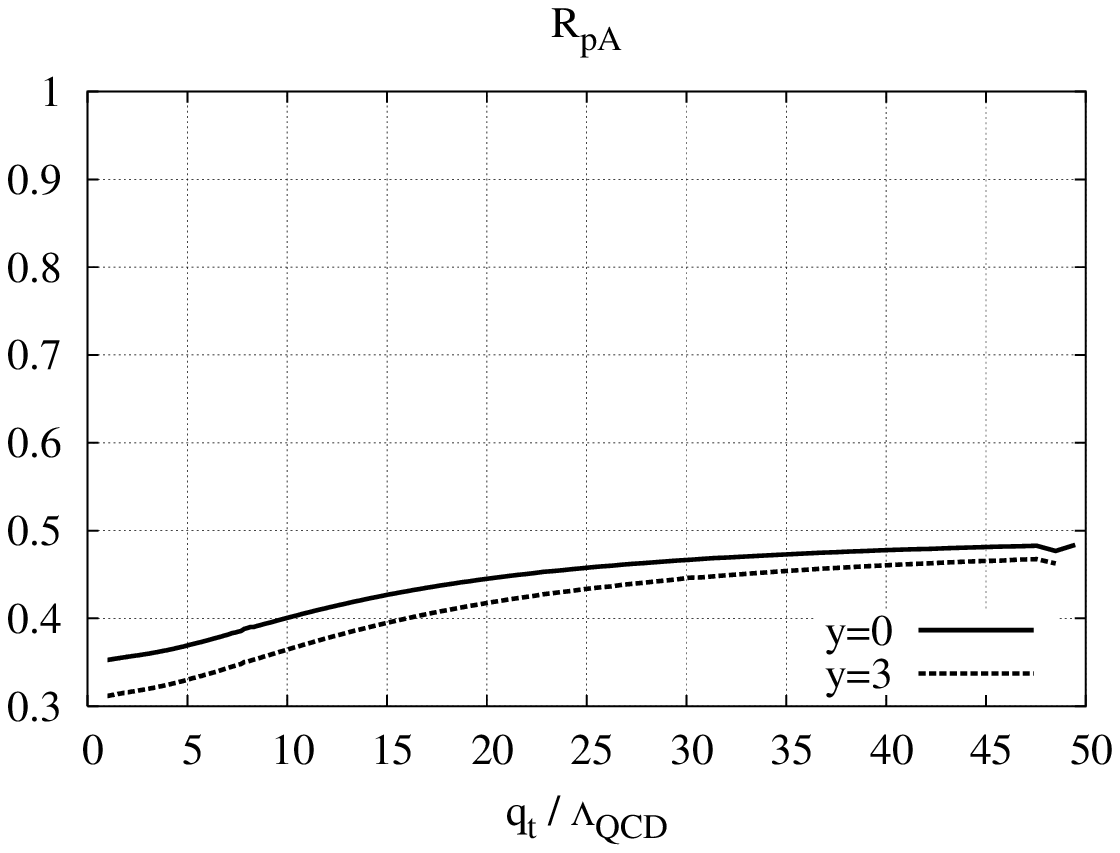}
  \caption{
    Cronin effect in the CGC formalism with naive
    quantum evolution ({\it left}) and with full evolution in the ``deep
    saturation'' limit ({\it right}). Figure taken from
    Ref.~\cite{Blaizot:2004wu}.} 
  \label{fig:dAu_CGC}
\end{figure}

When the correlations in the nuclear
gluon field are local (i.e. in the incoherent scattering
approximation), the gluon 
$p_T$-spectrum shows a peak structure which increases in magnitude and
moves to higher $p_T$ with increasing
rapidity~\cite{Blaizot:2004wu}. This agrees with RHIC data at
midrapidity, but is at variance with forward rapidity measurements.
In the deep saturation regime, the solution of the JIMWLK equation
describes high density partons with
non-local correlations, and suppresses gluon production over the whole
$p_T$ range, irrespective of the rapidity (see Fig.~\ref{fig:dAu_CGC}). 
The resulting picture~\cite{Iancu:2004bx,Baier:2005dz} is that at
the values of $x$ probed at RHIC at midrapidity, the nucleus wavefunction has
not yet reached the saturation regimes -- $Q_s^2$ is
small, at most 2~GeV$^2$ -- hence a description in terms of
pQCD multi-scatterings is valid (and perhaps numerically more
accurate~\cite{Accardi:2004fi}). At larger rapidities, the
saturation scale increases and the nucleus wave function
undergoes a longer quantum evolution, resulting in hadron suppression
on a large $p_T$ interval. Hence, gluon saturation may have indeed
been revealed in the observed forward rapidity hadron suppression at
RHIC.  

To make quantitative contact with phenomenology,  
one approximates the proton as a dilute colour
source, so that gluon production in $p+A$ collisions can be explicitly 
written in a $k_T$-factorised form~\cite{Blaizot:2004wu}. The nucleus
wavefunction is computed in terms of a colour dipole forward
scattering amplitude, which is modeled in order to capture the 
analytic key features of the CGC, and to incorporate the transition to the
semi-classical regime at lower 
rapidity~\cite{Kharzeev:2003wz,Kharzeev:2004yx,JalilianMarian:2004xm,Dumitru:2005gt}. 
Examples from the computations
of Refs.~\cite{Kharzeev:2004yx,JalilianMarian:2004xm}
are shown in Fig.~\ref{fig:dAu_vs_models} (right). 

\end{itemize}

An important limitation of parton multi-scattering models is their
inability to explain the strong flavour dependence of the Cronin effect
and the corresponding ``baryon anomaly'' ($\pi^\pm < K^\pm \ll p,\bar
p$ at intermediate $p_T$) observed at RHIC. Indeed, if the $p_T$-broadening has a partonic 
origin, the only flavour dependence of the Cronin effect can be due to
the different contribution of gluons and quarks to the final-state
hadron, which cannot explain the large difference between protons and
pions~\cite{Accardi:2003jh}.

\subsubsection{\it Final state parton recombination}--
A description of hadronisation as parton recombination, as opposed to
the parton fragmenting into the observed hadron, was proposed long ago to
explain hadron production at large Feynman $x_F$ in $h+A$ collisions~\cite{Das:1977cp}.  
The idea that two or three partons can recombine, or coalesce, into a meson
or baryon has been revived~\cite{Hwa:2002tu,Hwa:2004zd,Hwa:2004yi,Greco:2003xt,Greco:2003mm,Fries:2003vb,Fries:2003kq,Molnar:2003ff}
because of recent experimental findings at RHIC, chiefly, the strong baryon
enhancement (already observed in $h+A$ collisions at Fermilab, see 
Section~\ref{sec:hadrons-hA}), and the scaling of the elliptic flow
of intermediate $p_T$ hadrons with the number of their
constituent quarks~\cite{Adams:2003am,Adare:2006ti}. The main idea is that whenever there is a large
reservoir of partons in the final state, there is a lesser need to
produce additional ones through parton splitting as
assumed in parton fragmentation. Recent reviews of recombination models
can be found in~\cite{Hwa:2008qi,Fries:2008hs}.

\begin{figure}
  \centering
  \includegraphics[height=4.2cm]{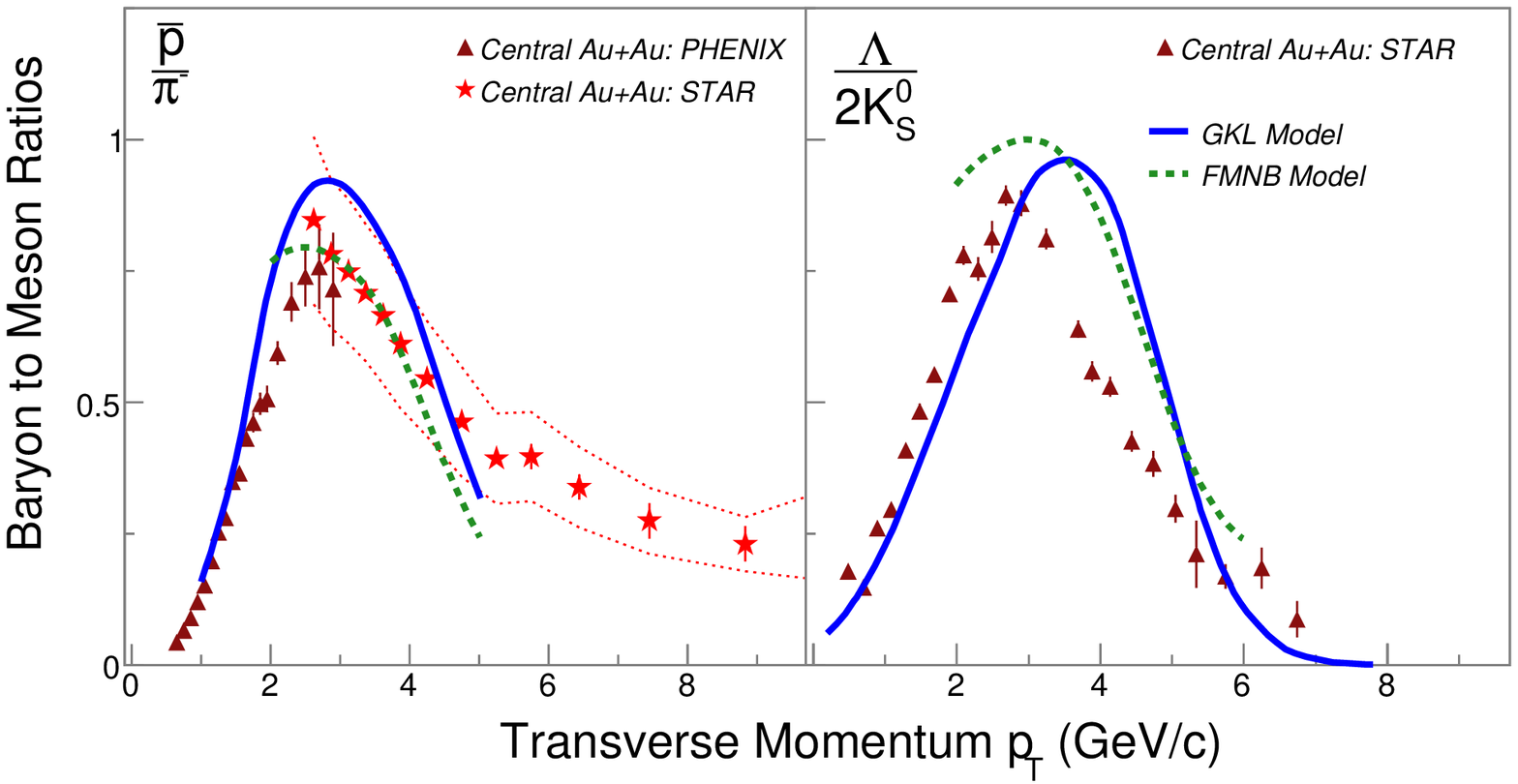}
  \includegraphics[height=4.1cm]{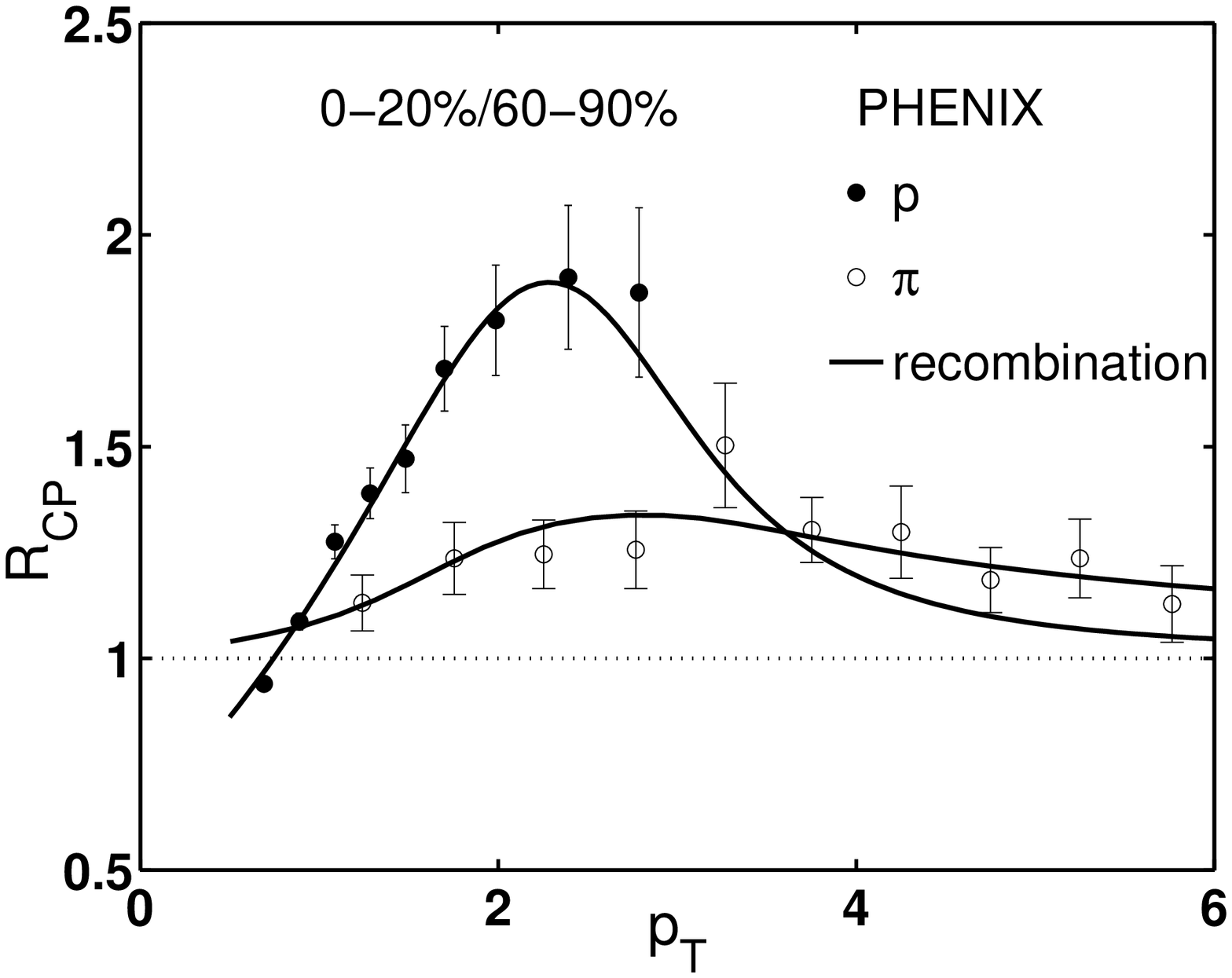}
  \caption{
    {\it Left:} Ratio of baryon yields to meson yields for central $Au+Au$
    collisions at RHIC $\sqrtsnn =$~200~GeV. The recombination model
    computations by Greco, Ko, Levai~\cite{Greco:2003xt,Greco:2003mm}
    and by Fries, M\"uller, Nonaka, Bass
   ~\cite{Fries:2003vb,Fries:2003kq} for $\bar p/\pi^-$ (left) and
    $\Lambda/2K^0_s$ (right) are compared to PHENIX and STAR data. 
    Plot taken from Ref.~\cite{Fries:2008hs}. 
    {\it Right:} Hwa-Yang recombination model computation
   ~\cite{Hwa:2004zd} of the Cronin effect for midrapidity pion and proton in
    $d+Au$ collisions at RHIC $\sqrtsnn =$~200~GeV, compared to
    PHENIX data~\cite{Adler:2006xd,Adler:2006wg}. Plot taken from Ref.~\cite{Hwa:2005ay}. 
  }.
  \label{fig:Cronin_reco}
\end{figure}

In heavy-ion collisions at RHIC, a hot and dense medium made of deconfined 
soft quarks and gluons is created in the first instants after the collision.
The parton reservoir is made of these thermal partons, which are
distributed in momentum according to an exponential
spectrum, $\sim e^{-p/T}$ with $T\approx$~0.2~GeV. At large momenta partons come dominantly
from hard interactions and are distributed according to a power law,
$\sim 1/p_T^n$, with $n\approx7-8$. Parton fragmentation requires 
the production of a hard parton with large momentum $p_T > p_T^h$
to fragment into the hadron. On the contrary, parton recombination is
kinematically favoured 
since 2 or 3 partons of {\it smaller} momenta $p_T < p_h$ can 
coalesce into the observed hadron. At large $p_T$, the fragmentation dynamics 
is supposed to dominate over recombination since 
thermal partons are suppressed, while recombination is expected to be
important at low $p_T$.  
In an intermediate $p_T$ range, the recombination of one hard parton
together with one or two thermal partons  
becomes the dominant channel to produce a meson or a baryon. 
Recombination readily explains the Cronin effect of mesons and baryons
at intermediate $p_T$: the larger the number of soft thermal partons
involved in the recombination process, the larger the momentum  
broadening, leading to a stronger Cronin effect for baryons than for mesons. 
As can be seen in Fig.~\ref{fig:Cronin_reco} left, these models turn
out to reproduce well the data.

In $p+A$ collisions, the parton reservoir is made of 
the soft partons created in sequential nucleon-nucleon
scatterings~\cite{Hwa:2004zd}; in particular, they can recombine among
themselves into
the observed low-$p_T$ hadrons. Fitting the exponential soft parton spectrum to 
the observed yield of low-$p_T$ hadrons in $p+p$ collisions, this
recombination model can explain well the large difference
in the baryon and meson Cronin effect in $d+Au$ collisions at RHIC
(Fig.~\ref{fig:RdAu}). 
Additionally, it can describe
the observed suppression of intermediate $p_T$ hadrons in forward
rapidity bins, and also the 
negative/positive rapidity asymmetry of the Cronin effect at moderate
$\eta$ because its magnitude tracks the (soft) hadron rapidity
distribution, which 
decreases in rapidity at $\eta\gtrsim -2$~\cite{Hwa:2004in}.
The model is quite successful at RHIC
energy in the range $-0.75 < 
\eta < 3.2$. An interesting check would be to measure hadron $p_T$ spectra at
$\eta \lesssim -2$, where the soft hadron yield starts decreasing with
decreasing $\eta$. Therefore, the magnitude of the Cronin effect
should peak at $\eta \approx -2$, and slowly decrease as $\eta$ is
reduced.

\subsubsection{\it Origin of the Cronin effect} --
In the parton recombination model, the Cronin effect  is a
final-state (FS) effect at the hadronisation stage rather than due to
initial state (IS) or final-state parton rescatterings in the
target nucleus.  However, contributions to the Cronin effect from
IS parton rescatterings cannot be entirely discarded. 
In fact, in $e+A$ collisions the multiplicity of soft hadrons in the final
state is much lower than in $p+A$ collisions, making the recombination
mechanism much less effective, even though a large Cronin effect is
observed in the HERMES and EMC data, see Section~\ref{sec:eAdata}.  
(Another potential difficulty of the recombination models in nDIS would be to
account for the anti-proton Cronin effect which is similar to that of
the mesons, unlike in $h+A$ collisions where $p \approx \bar p$.)
The same processes which cause the Cronin effect in $e+A$
collisions are also likely to be at work in $h+A$ collisions in
addition to possible FS recombination effects. Indeed the hard partons
are traversing two media, the cold nuclear target and the soft partons
cloud created in the collisions, and both can contribute to the Cronin
effect. Since these two media are well separated in space-time
because of the formation time $\propto 1/\Lambda_{QCD}$ for the soft
partons, multiple parton scatterings may add to the
$p_T$-broadening from FS recombination. 

The effect of a short quark life time, as indicated by HERMES data in 
nDIS, has not been included so far in recombination model 
computations. At RHIC energy the large DIS-equivalent $\nu =
\mathcal{O}($1000~GeV$)$ (see Section~\ref{sec:phasespaces})
ensures that parton fragmentation starts outside the cold nuclei.
However, at small enough $p_T$, fragmentation may still start inside the 
soft parton cloud, thus reducing the quark path-length in the soft
cloud, therefore reducing the probability of soft parton 
pick-up and the effectiveness of the recombination mechanism. 
At fixed-target energies $\nu = 10 - 100$ GeV hadronisation
may in fact start inside the cold nucleus target, before the parton 
has any time to travel through the soft parton cloud, thus 
preventing altogether the recombination mechanism.

We finally remark that a precise understanding of the mechanisms
underlying the Cronin effect in $h+A$ collisions 
is crucial for measurements of hadron
quenching in $A+A$ collisions at the low SPS energy, $\sqrtsnn =
17.3$~GeV~\cite{Aggarwal:2007gw}.
In $A+A$ collisions, two competing effects modify hadron $p_T$
spectra: (i) the $p_T$ broadening caused by both targets, hence larger 
than in $h+A$ collisions, and (ii) the hadron suppression due to parton
energy loss in the produced medium. The Cronin effect  is observed in $h+A$
collisions to grow 
with decreasing $\sqrtsnn = 200 - 27$~GeV: the steeper transverse
momentum spectra at lower energies translate into a larger relative
enhancement of the yields for the same $k_T$ ``kick''. Hence, at SPS
energy one can expect an even larger enhancement than observed at Fermilab. 
Therefore the $Pb+Pb/p+Pb \approx 0.5$  nuclear modification ratio observed by
the WA98 collaboration~\cite{Aggarwal:2007gw} might be caused by a
large parton energy loss in $Pb+Pb$ that outweighs a large Cronin
effect~\cite{d'Enterria:2004ig,Aggarwal:2007gw}. A numerical
evaluation of jet quenching at fixed-target energies requires a
precise theoretical and experimental control of the underlying Cronin effect
(including baseline $p+p$ $p_T$ spectra at SPS energies).

\subsection{Energy loss in hot QCD matter } \ 
\label{sec:hotenloss}

\subsubsection{\it Formalisms}--
\label{sec:Eloss_formalisms}
The idea of parton energy loss in hot QCD matter was first discussed by Bjorken
in the early eighties~\cite{Bjorken:1982tu}. This process was then revived a decade later when for the
first time Thoma and Gyulassy~\cite{Gyulassy:1991xb} and Gyulassy, Pl\"umer, and Wang ~\cite{Wang:1994fx,Gyulassy:1993hr} computed perturbatively
the radiative energy loss of high-energy quarks in a QGP. Since then many
approaches have been developed to determine the gluon radiation spectrum, $\dd I/\dd\omega$,  of a hard parton
undergoing multiple scattering. For a short discussion and comparison
of the different energy loss formalisms, see
Ref.~\cite{Majumder:2007iu}. Let us briefly recall the main
assumptions made in each of the four main frameworks detailed there:

\begin{itemize}
\item Opacity expansions (BDMPS and GLV)

BDMPS developed the
first perturbative framework to describe the medium-induced gluon
emission process from the soft multiple scattering of hard partons in
both cold~\cite{Baier:1997sk} and hot~\cite{Baier:1997kr} QCD matter. The
calculation assumes that the number of collisions, or opacity, is large:
$n = L/\lambda\gg 1$, where $L$ is the medium length and $\lambda$ the
parton mean free path. The typical momentum exchange in
each scattering is given by the Debye mass of the medium, $\mu$, and
the hard scale of the calculation is $n\times\mu^2 = \qhat L \gg \lqcd^2$,
where the transport coefficient $\qhat\equiv\mu^2/\lambda$ represents the scattering power of
the QCD medium. The Landau-Pomeranchuk-Migdal (LPM) effect --
basically the destructive interference of the gluon radiated on several scattering
centres -- takes place whenever the gluon lifetime $t = \omega/\kt^2$
exceeds its mean free path, that is for gluon energies larger than
$\sim\mu^2\lambda$. Soft gluon emission is assumed,
$\omega\sim\qhat L^2 \ll E$, although corrections $\cO{\omega/E}$ were
proposed in [108]. On the contrary, the formalism by Gyulassy, L\'evai
and Vitev (GLV) first took into account one hard scattering in the
medium~\cite{Gyulassy:1999zd} (i.e. first order in $n$) and from this
a recursive approach has been used to determine the gluon spectrum at
any opacity ~\cite{Gyulassy:2000er,Gyulassy:2000fs}. Such an expansion
has also been rederived by Wiedemann in Ref.~\cite{Wiedemann:2000za}
from the light-cone path-integral (LCPI) approach to energy loss developed by
Zakharov~\cite{Zakharov:1997uu,Zakharov:1996fv} (and equivalent to the
BDMPS framework described above, see~\cite{Baier:1998kq}). 
All these approaches model the medium-modified fragmentation
functions as a convolution of (vacuum) fragmentation functions,
$D_k^h(z, Q^2)$, and an energy-loss probability distribution, 
${\cal P}(\epsilon)$~\cite{Wang:1996yh}: 
\begin{equation}\label{eq:modelff}
  {\tilde D_f^h}(z,Q^2) = \int_0^{(1-z)E_f} d\epsilon \,{\cal P}(\epsilon)\, 
    z^\star\,D_f^h (z^\star, Q^2),
\end{equation}
where 
\begin{align*}
  z^\star = z/(1-\epsilon/E_f)
\end{align*}
is the rescaled
momentum fraction carried away by the hadron $h$ in presence of the
QCD medium, and $E_f$ is the energy of the parent parton in the medium
rest frame. In heavy-ion collisions, for instance, $E_f = k_T$ for partons produced at mid-rapidity, 
while in semi-inclusive DIS to LO in the strong coupling constant the quark energy is given by the photon energy, $E_f = \nu$.
The energy loss probability distributions ${\cal P}(\epsilon)$, also called {\it quenching
weights}, will be discussed in Section~\ref{sec:qw}.

\item Higher-twist formalism (HT)

In the HT approach~\cite{Wang:2001if,Guo:2000nz,Majumder:2004pt},
power corrections proportional to $1/Q^{2}$ (where $Q$ is the typical
radiated gluon virtuality) and enhanced by the medium length $L$ are
included to the leading-twist total cross section in DIS, assuming
that $Q$ is much larger than the energy-scale exchanged $\mu$ between
the hard parton (with energy $E$) and the medium: $E\gg Q \gg
\mu$. 
This approach was first proposed in the context of DIS (see
Section~\ref{sec:nDISHT}) where 
the strength of the higher-twist terms is monitored by a single parameter,
$C$, which can be adjusted to fit the data just like $\qhat$ in BDMPS
or the initial gluon density $dN^g/dy$ in GLV. For hot QCD media, a
phenomenology based on a $z$-shift prescription for the FF,  $z \ra
z/(1-\epsilon/E_f)$ similarly to Eq.~\eqref{eq:modelff}, was developed
afterwards \cite{Wang:2001if}. The energy loss $\epsilon$ can be
related to $C$ and is adjusted to fit the data.

\item Thermal field theory (AMY)

The formalism of Arnold, Moore, and Yaffe (AMY)
addresses the production of thermal photons in a finite-temperature
QCD medium~\cite{Arnold:2002ja,Arnold:2001ms} and was then extended to
describe gluon emission from the scattering of ``hard particles'',
with energy $\cO{T}$, on softer modes, $\cO{gT}$. The photon/gluon emissions that are 
collinearly divergent were resummed to all orders in $\alphas$, leading to a
suppression as compared to the leading-order result because 
of the LPM effect already discussed. This calculation should be
accurate for asymptotically large temperatures, for which all the relevant (hard,
soft, ultra-soft) scales are well separated: $T\gg gT\gg g^2T$ . 

\end{itemize}

An alternative $Q^2$-shift prescription for medium modified fragmentation
functions -- which is quite different from the $z$-rescaling
prescription \eqref{eq:modelff} used in most applications -- has been 
discussed in~\cite{Kopeliovich:2003py,Kopeliovich:2006xy}. Assuming 
that hadronisation occurs outside the medium, as in the above-mentioned
energy loss models, an upper bound on hadron quenching 
can be obtained by considering
\begin{align}
  \tilde D_q^h(z,Q^2) = D_q^h(z,Q^2+\Delta p_T^2) \ ,
\label{eq:Q2shift}
\end{align}
with $\Delta p_T^2$ the parton in-medium $p_T$-broadening
related to $\qhat$ \cite{Baier:1996sk}. Numerical estimates
show that such an effect is much smaller than observed in experimental
data, which is interpreted as showing that mechanisms other
than energy loss are at play in quenching hadron spectra.
A detailed vacuum hadronisation model 
which tries to avoid this ambiguity~\cite{Kopeliovich:2003py}
is briefly discussed at the end of Section~\ref{sec:formationtimes}, and its
consequences for hadron quenching in cold nuclear matter
will be reviewed in Section~\ref{sec:perthadrmodel}.

\subsubsection{\it Quenching weights}--
\label{sec:qw}
The quenching weights appearing in Eq.~\eqref{eq:modelff}
have been computed numerically within the BDMPS approach by Salgado and
Wiedemann (SW) for light partons~\cite{Salgado:2003gb,SWandASWroutines}, 
and later on extended by Armesto, Salgado and Wiedemann (ASW) to include heavy
quarks~\cite{Armesto:2003jh,Armesto:2005iq}. They have also been
studied in the GLV formalism in Ref.~\cite{Gyulassy:2001nm}.

The ASW quenching weights are determined using the Poisson
approximation of independent gluon emission~\cite{Baier:2001yt},  
\begin{equation}\label{eq:quenchingweight}
{\cal P}(\epsilon) = \sum^\infty_{n=0} \, \frac{1}{n!} \,
\left[ \prod^n_{i=1} \, \int \, d\omega_i \, \frac{dI(\omega_i)}{d \omega} 
\right] \delta \left(\epsilon - \sum_{i=1}^n  \omega_i\right)
\, \exp \left[ - \int d\omega \frac{dI}{d\omega} \right].
\end{equation} 
It depends on the characteristic radiated gluon energy $\omega_c = \qhat L /2$ and on the medium size 
parameter $R = \omega_c L$, where $L$ is the parton in-medium path-length. 
The approximation of an asymptotically large medium, considered by BDMPS (supplemented in~\cite{Arleo:2002kh}
with finite energy corrections of order $\cO{\omega/E_f}$)
would correspond to $R\ra\infty$. At finite $R$,
the parton has a probability $p_0$ of not interacting with the medium
and therefore not to suffer any energy loss. Correspondingly, the quenching weight
can be split in a discrete and continuum parts, 
\begin{align}
  {\cal P}(\epsilon) = p_0\,\delta(\epsilon) + P(\epsilon) \ .
\end{align}
The quenching weight is computed for a static and uniform medium. 
In heavy-ion collision, the longitudinal expansion of the medium 
is taken into account by rescaling the transport coefficient according to
an approximate scaling law discussed in~\cite{Salgado:2002cd}:
\begin{equation}\label{eq:dynamicalscaling}
  \hat{q}(L) = \frac{2}{L^2}\int_{\tau_0}^{L}
  d\tau\, \left(\tau - \tau_0\right) \,
  \hat{q}(\tau)
\end{equation}
where $\alpha$ characterises the power-law time-dependence of the medium
number-density, $\rho(\tau)\propto \tau^{-\alpha}$, and $\qhat(\tau) = \qhat(\tau_0) \left(\tau_0/\tau\right)^\alpha$.
The purely longitudinal (or Bjorken) expansion corresponds to
$\alpha = 1$, and is often assumed in phenomenological applications.
When $t_0 \ll L$, Eq.~\eqref{eq:dynamicalscaling} reduces to $\langle
\qhat\rangle\simeq 2\qhat(t_0)\ t_0/L$~\cite{Baier:2002tc}.
In nDIS, the medium is static but non-uniform, and an analogous
scaling law is proposed in Ref.~\cite{Accardi:2007in}. Recently, a  simple 
prescription has been given by Arnold in order to determine $\dd I/\dd\omega$ 
in a finite and expanding medium~\cite{Arnold:2008iy}. Therefore, applying this 
recipe using e.g. hydrodynamical space-time evolution 
will allow eventually for the computation of more realistic quenching weights.
Also, Peshier has proposed a useful way to compute iteratively the quenching 
weights in Ref.~\cite{Peshier:2008bg} avoiding the need to compute inverse Laplace 
transforms (as done e.g. in~\cite{Arleo:2002kh,Salgado:2003gb}).

\subsubsection{\it Phenomenology}--
\label{sec:enlossphen}
Many phenomenological applications of the above formalisms have been carried out in order to describe
or predict the production of large-$\pt$ hadrons or jets in heavy-ion collisions (see Section \ref{sec:hadrons-AA}
and Ref.~\cite{d'Enterria:2009am} for a recent review). All models coincide in characterising the system with 
very large initial gluon densities ($dN^g/dy\approx$~1400), transport coefficients 
$\langle\qhat\rangle\approx$~13~GeV$^2$/fm and/or high temperatures $T \approx$~400 MeV, in order to
reproduce the existing high-$\pt$ suppression. Yet, when trying to compare all model predictions through
a common $\qhat$ coefficient, differences of a factor of 3~--~4 appear~\cite{Dainese:2004te} .
Recently, progress has been made towards a more
realistic implementation of energy-loss scenarios including a full hydrodynamical 
expansion of the produced medium --~constrained by soft/global observables;
first in~\cite{Hirano:2002sc,Hirano:2003hy,Hirano:2003hq} and more recently 
in~\cite{Renk:2006sx,Majumder:2007ae,Qin:2007zz}. These approaches aim at a consistent 
description of soft and hard probes and at a realistic extraction of the medium-parameter 
from fitting the large-$\pt$ suppression data (see e.g. Fig.~\ref{fig:RAA_vs_PQM}).
A recent quantitative comparison of energy loss schemes under identical conditions (i.e. same medium evolution, 
same choice of parton distribution functions and scales, etc.) is presented in \cite{Bass:2008ch}.  

\subsubsection{\it Modified DGLAP evolution}--
\label{sec:mDGLAP}
As an alternative to the discussed energy loss formalisms based on an energy-rescaling of
vacuum fragmentation-functions, attempts to reformulate parton energy loss in pQCD 
as a medium modification of the Dokshitzer-Gribov-Altarelli-Parisi (DGLAP) evolution
of fragmentation functions have been recently
suggested~\cite{Borghini:2005em,Armesto:2007dt,Domdey:2008gp}.   

In the perturbative description of fragmentation processes, the produced parton of time-like 
virtuality $Q$ radiates gluons in order to reduce its virtuality
down to a soft scale $Q_0 = \cO{1 {\rm~GeV}}$ where hadronisation takes place.
The $Q^2$-evolution is governed by the DGLAP equations
\cite{Gribov:1972ri,Altarelli:1977zs,Dokshitzer:1977sg}, which control
the probability that a quark, say, branches into a quark and a gluon in
going from $Q^2+d Q^2$ to $Q^2$. Gluons also can split into a $gg$ or a
$q\bar q$ pair. The offspring partons can in turn split, iteratively.
After each branching the scattered parton gains some $p_T$, which can be
controlled by using $p_T$-dependent evolution equations
\cite{Ceccopieri:2005zz}. 
Physically, the probabilistic picture of DGLAP evolution leads to a parton shower which can be
simulated in a Monte Carlo generator. 

\begin{figure}[tb]
  \centering
  \includegraphics[width=5.cm,height=5.cm]{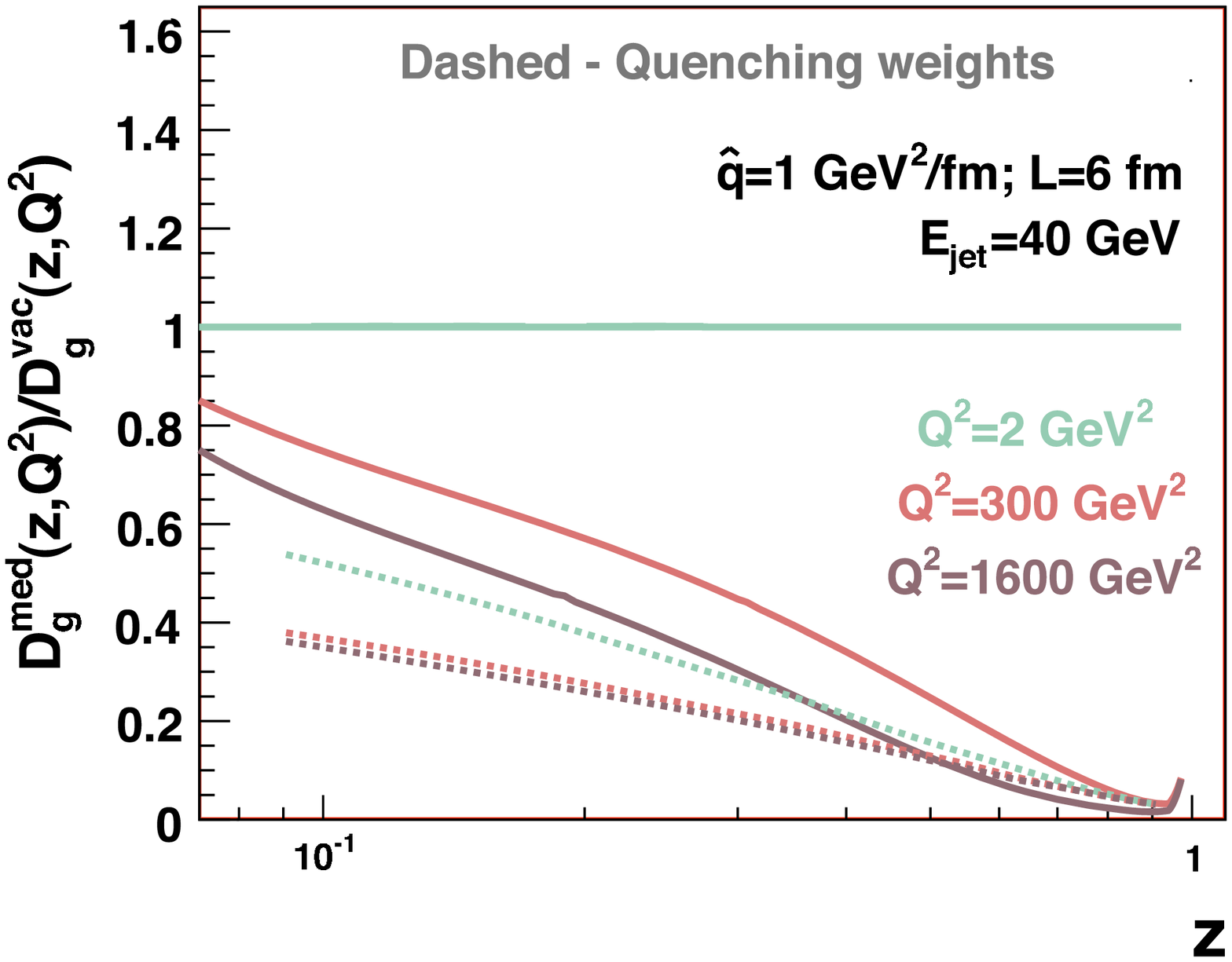}
  \includegraphics[width=8.cm,height=5.cm]{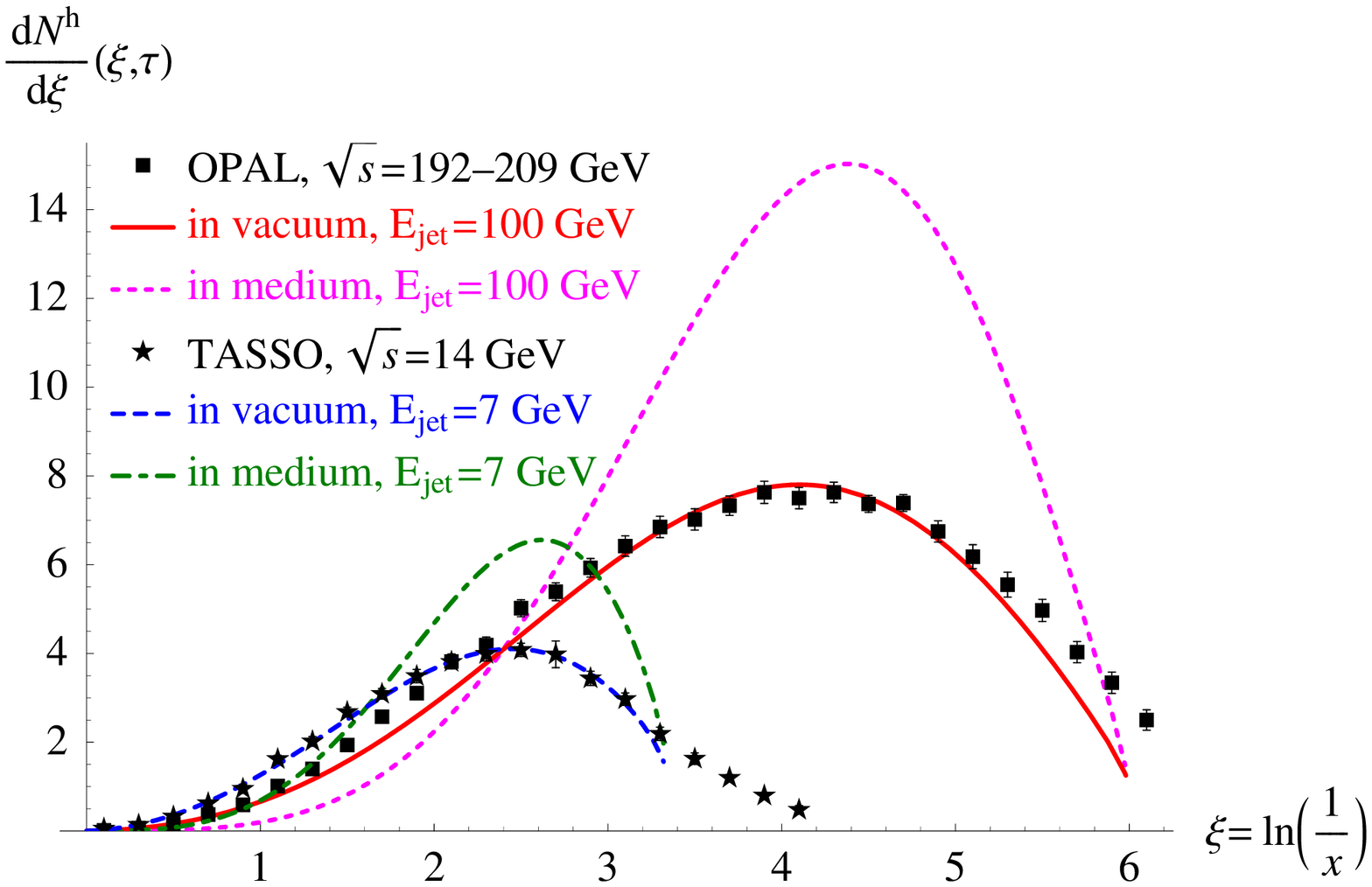} 
  \caption{
    {\it Left:}
    Medium over vacuum ratio of the gluon fragmentation functions 
    plotted for a medium with $\hat{q} = 10$~GeV$^{2}$/fm (green) and
    $\hat{q} = 50$~GeV$^{2}$/fm (red), and for two different medium lengths:
    $L$~=~2~fm (solid) and $6$ fm (dashed).
    Figure taken from Ref.~\cite{Armesto:2007dt}.
    {\it Right:} Single inclusive hadron distribution as a
    function of $\xi = \log \left(E_{\rm jet}/p\right)$. $e^+e^-$ data 
    from TASSO~\cite{Braunschweig:1990yd} and OPAL~\cite{Abbiendi:2002mj}
    compared to vacuum-FFs (solid curves) and to medium-FFs (dashed/dotted curves,
    obtained with $f_{\rm med} = 0.8$ in the LO splitting functions). Figure taken from
    Ref.~\cite{Borghini:2005em}.
  }
  \label{fig:mDGLAP}
\end{figure}

In the presence of a QCD medium, the parton shower may be modified
in basically two ways: (i) the splitting probability is enhanced,
mocking radiative energy loss, and (ii) the partons can
rescatter, leading to a stronger showering process and a
$p_T$-broadening of both the leading particle and the shower.
Various models for the medium-modified splitting enhancement have been
proposed. In~\cite{Armesto:2007dt}, under the assumption that
splittings are independent, the in-medium splitting functions are 
\begin{align}
  P(z) = P^\text{vac}(z) + \Delta P(z) \ ,
\end{align}
where $z$ is the fractional momentum carried away by one of the split
partons, and $P^\text{vac}$ is the vacuum splitting probability. The
additional term $\Delta P$ can be calculated from the medium induced
gluon radiation spectrum, $d I/d\omega$, previously discussed. 
Medium-modified fragmentation functions at a
scale $Q$ are computed by evolving input fragmentation functions at
a scale $Q_0$ according to  the DGLAP equations with medium-modified
splitting functions, see Fig.~\ref{fig:mDGLAP} left. A virtue of this approach is that at large virtualities, $Q\gg\lqcd$, the usual medium-modified fragmentation functions defined by the quenching weights, Eq.~\eqref{eq:modelff} is formally recovered. 
In the approach of Borghini and Wiedemann~\cite{Borghini:2005em}, the medium effect is argued to modify the
splitting functions by enhancing its singular part; for example, the
quark splitting function is
\begin{align}
  P_{qq}(z) = C_F \;\left( \frac{2(1+f_{med})}{(1-z)_+} - (1+z) \right) \ ,
\end{align}
where $f_{med} = 0$ gives back the vacuum $P^\text{vac}$. 
An interesting application is the medium modification of the
single-inclusive energy distribution of hadrons inside a jet, $d
N/d\xi$, where $\xi = \log(E_{\rm jet}/E_h)$. This leads to a
distortion of the usual hump-back plateau, predicted within the
Modified Leading Logarithmic Approximation (MLLA) of QCD and observed
experimentally in $e^+e^-$, DIS and hadronic collisions (for a review
see e.g.~\cite{Khoze:1996dn}). As can be seen in Fig.~\ref{fig:mDGLAP}
(right), the number of highly-energetic particles (small $\xi$) is
suppressed while the soft gluon yield is enhanced at large $\xi$ due
to energy-momentum conservation.  
The inclusion of $2\to2$ elastic rescatterings, which accounts for
elastic energy loss, can be accomplished 
by supplementing the DGLAP evolution equations with a gain and loss
term, which describes partons scattered into and away from 
a given kinematic variable bin, and is also suitable
for a Monte Carlo interpretation~\cite{Domdey:2008gp}.

Several parton showers in the medium have been recently developed~\cite{Zapp:2008gi,Renk:2008pp,Armesto:2008qh,Lokhtin:2008wg,Lokhtin:2005px}. 
As compared to analytic calculations, parton showers have many advantages such as conserving energy-momentum throughout the evolution. 
They allow one to directly compare their multiple-differential hadronic distributions to experimental data, and thus 
give a better access to the microscopic dynamics. They also allow to
study the particle and energy flow inside a jet.
As an example, first results from the JEWEL parton shower~\cite{Zapp:2008gi}, 
indicate that the distribution of 1, 2 and 3 jets events (reconstructed using a given 
granularity parameter, $y_{\rm cut}$) is sensitive to the elastic or inelastic nature of parton rescatterings.

Applications of modified DGLAP equations
have been studied only in the context of $A+A$ collisions (see, e.g.,
\cite{Polosa:2006hb,Sapeta:2007ad}) testable at RHIC and LHC. 
It would be very interesting to also study jet modifications in $e+A$
collisions, which would be accessible at the Electron-Ion Collider
(Section~\ref{sec:EIC}), and the hadron $p_T$-broadening already
under study at HERMES and CLAS (Section~\ref{sec:discussion-ptbroad}),
although the lower virtualities of the latter limit the amount of QCD
evolution (parton radiation) accessible.


\subsection{Energy loss in cold QCD matter } \ \\
\label{sec:coldenloss}

Although the density of scattering centres in cold QCD matter is much
smaller than in a deconfined state of matter such as a quark-gluon
plasma, it is not a priori excluded to observe effects of
parton energy loss in large nuclei~\cite{Baier:1997sk}. As a matter of
fact, several of the above-discussed parton energy loss formalisms
should prove more appropriate to describe the rescattering of a hard
parton off static nucleons rather than off quarks and gluons carrying
thermal momenta, $p = \cO{T}$. Moreover, the nuclear density is well
known and the medium does not expand while the hard parton traverses it,
unlike for the QGP whose energy density drops rapidly with proper time. 
For these reasons, cold QCD matter is an
ideal testing ground to compare different energy loss formalisms and
test the approximations made in their phenomenological applications.\\

One pragmatical approach, consists in comparing the nuclear DIS and $h+A$ hadroproduction
data to NLO pQCD calculations (with nuclear PDFs) and simply encode any observed modifications 
of the final yields for various species in properly fitted medium-modified parton-to-hadron FFs~\cite{Sassot:2009bs}. 
More commonly, however, one tries to get a more physical insight on the mechanisms affecting
hadron production in cold nuclear matter by describing parton multiple scatterings in nuclei 
within the BDMPS framework~\cite{Baier:1997sk}, as done for hot and dense QCD matter. In
such an approach~\cite{Baier:1997kr} one derives a simple relationship between the energy lost by 
the hard parton (per unit-length) and its transverse momentum broadening,~\cite{Baier:1998kq}
\begin{equation}
  \label{eq:broadening}
- \frac{dE}{dz}\ =\ \frac{\alpha_s\ N_c}{4}\  \langle\pt^2\rangle,
\end{equation}
independent of the parton species. The transport coefficient of cold nuclear matter, 
$\hat{q} = \mu^2/\lambda$, governing the amount of energy loss or momentum broadening can be estimated perturbatively.
It is related to the nuclear matter density, $\rho\simeq 0.17$~fm$^{-3}$, 
and the gluon density, $G$, in a nucleon~\cite{Baier:1997sk}:
\begin{equation}
  \label{eq:qhatcold}
  \hat{q}^{\rm cold}\ =\ \frac{4\pi^2\, \alpha_s\ C_R}{N_c^2\ -1}\, 
    \rho\, xG(x,Q^2)   
\end{equation}
where $C_R$ is the colour charge of the parton, $(N_c^2-1)/2N_c$ and
$N_c$ for quarks and gluons respectively. The value $x$ at which $G$
should be evaluated is parametrically $(m_N \ell)^{-1}$, where the
scale $\ell$ is larger than the mean free path of the hard parton,
$\lambda$, and smaller than the medium length, $L$. The virtual scale
$Q^2$ which enters $G$ is of $\cO{\hat{q} L}$. Using $\alpha_s\simeq
1/2$ at such low scales, and $xG(x, Q^2)\simeq 1$, the gluon ($C_R =
C_A = N_c$) transport coefficient is roughly given by 
\begin{align}
 \hat{q}^\text{cold}\simeq 0.05\;\text{\rm GeV}^2/\text{fm} \ .
 \label{eq:qhatcold2}
\end{align}
This value is much smaller than the {\it leading-order} BDMPS estimate
for a hot plasma at RHIC temperatures, $\hat{q}^{\rm hot}\simeq 2.2$~GeV$^2$/fm, 
given in~\cite{Baier:2006fr}. 

\begin{figure}[t]
  \centering
  (a) \includegraphics[width=6cm]{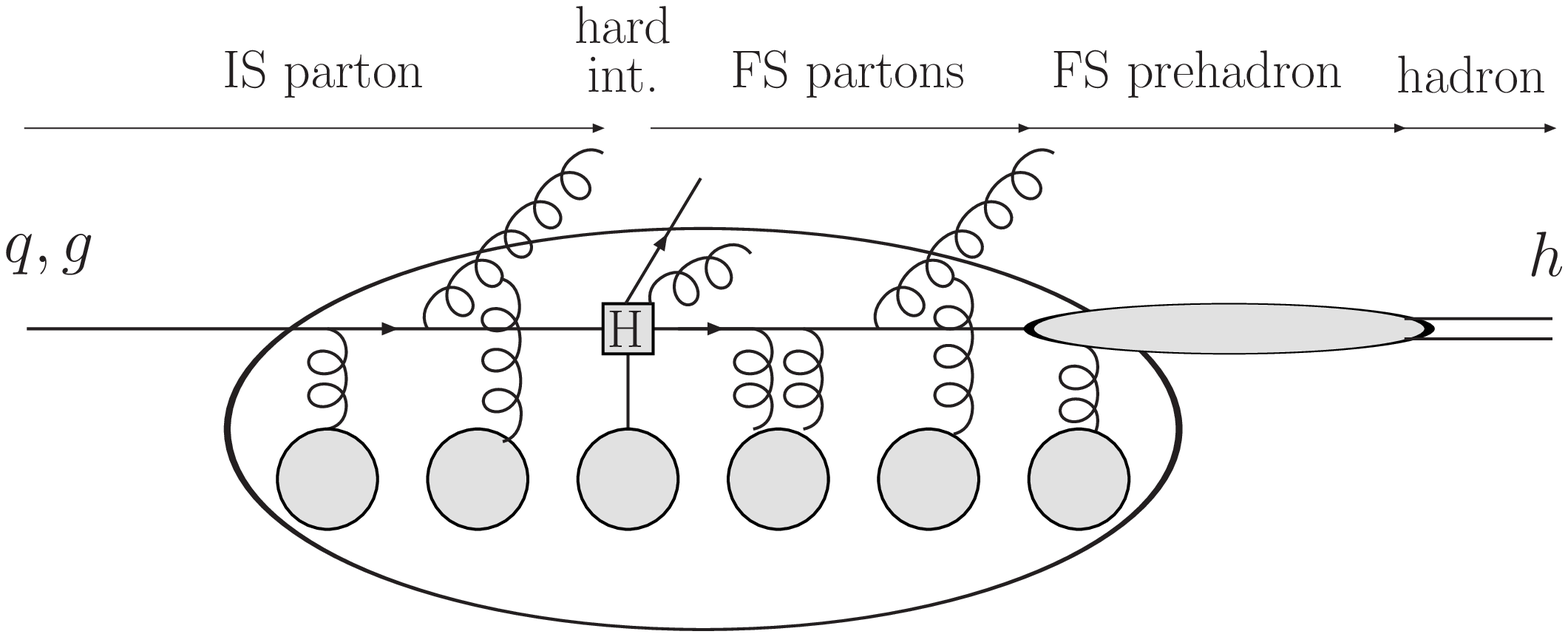}
  (b) \includegraphics[width=6cm]{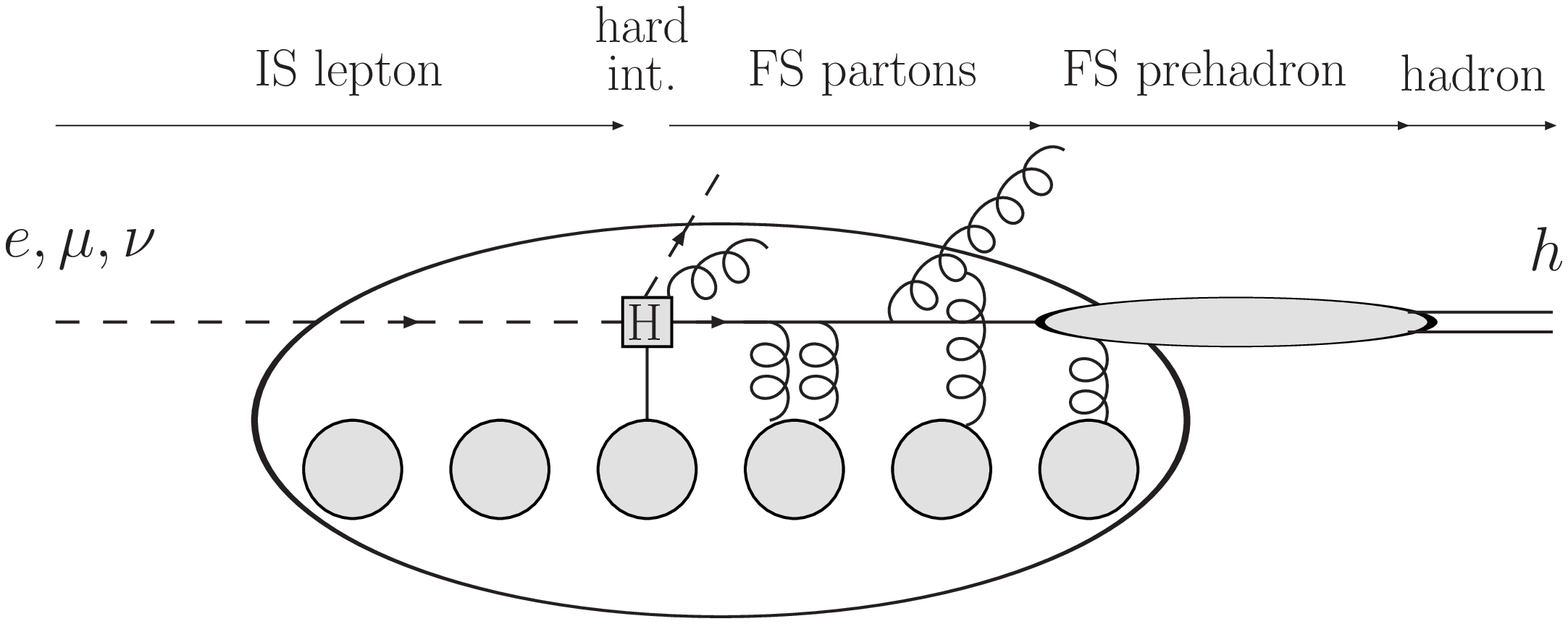}\\[.4cm]
  (c) \includegraphics[width=6cm]{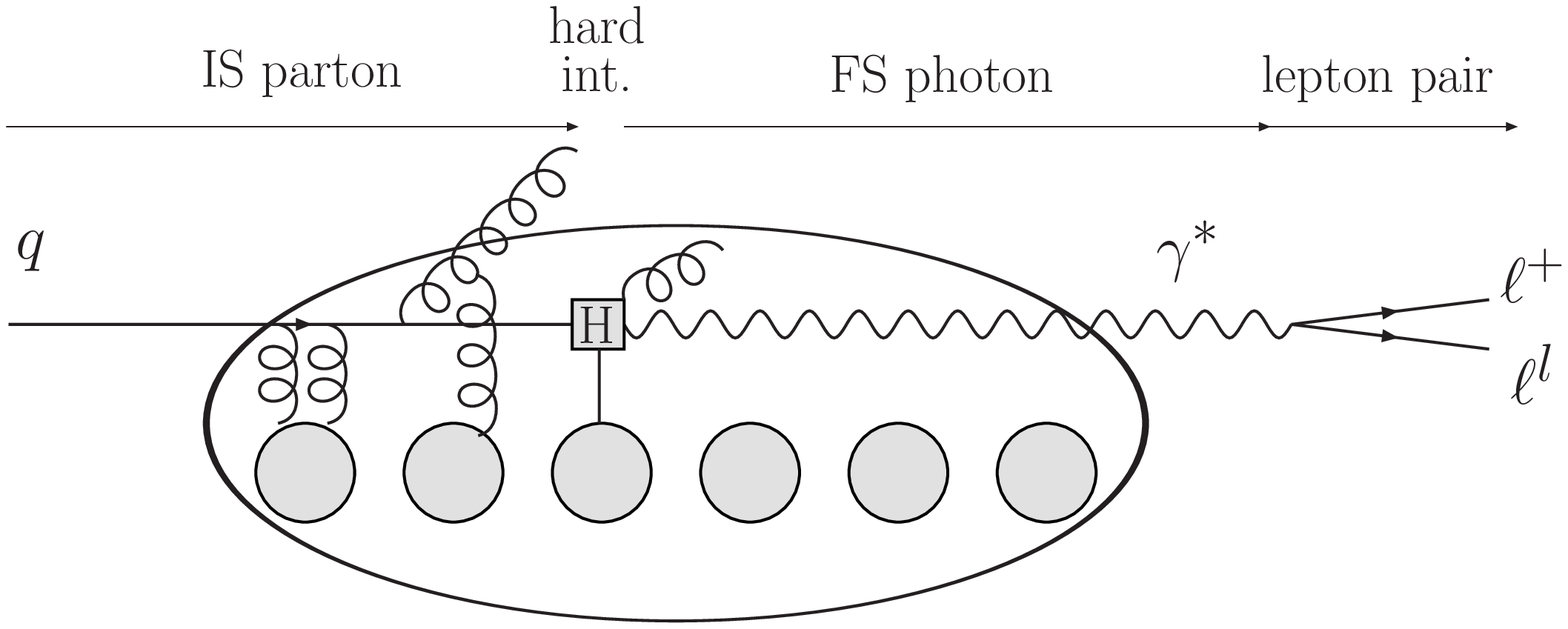}
  (d) \includegraphics[width=6cm]{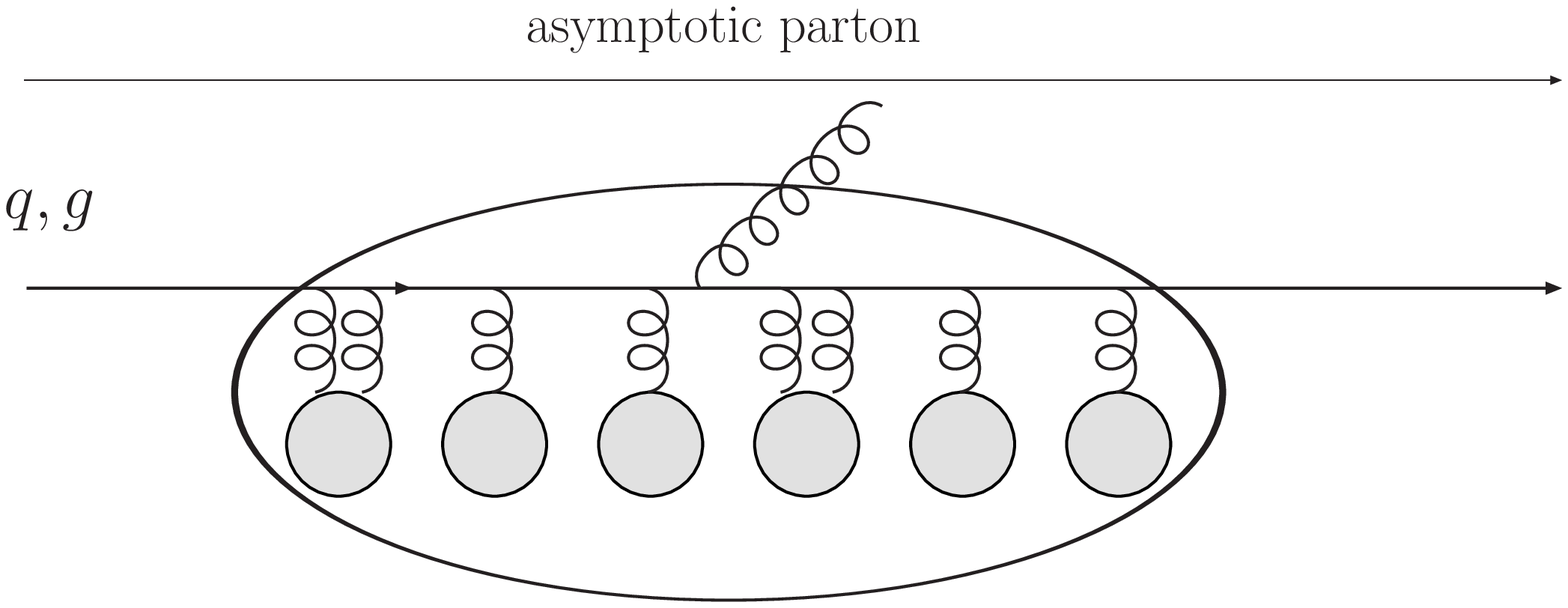}
  \caption{Sketch of parton propagation in cold nuclear matter in the nucleus
    rest frame. 
    (a) Initial- and final- state interactions in $h+A$
    collisions in the nucleus rest frame. (b) Absence of initial state
    interactions in $l+A$ collisions. (c) Absence of final-state
    interactions in DY events. (d) The theoretical
    case of an asymptotic parton penetrating the
    nucleus with no hard interactions. The nucleus is drawn as an
    oblong oval for illustrative purposes. The possibility that
    hadronisation starts inside the target nucleus is considered in this
    cartoon, see Section~\ref{sec:prehadron}.}
  \label{fig:coldenloss}
\end{figure}

When discussing energy loss in cold nuclear matter, however, one should make an
important distinction between initial-state and final-state
energy loss~\cite{Baier:1998kq,Vitev:2007ve}, because the medium-induced gluon
radiation can interfere with the radiation originating at the hard
scattering (see Fig.~\ref{fig:coldenloss}). In $h+A$ and $A+A$ 
collisions, both initial- and final-state interactions occur. If the
rapidity of the parton 
which fragments in the observed hadron is far enough from projectile
rapidity, initial-state and final-state radiation should not interfere,
and may be treated independently. In semi-inclusive DIS, only
final-state interaction may take place because the projectile lepton does
not interact strongly with the nucleus. Similarly, in Drell-Yan
lepton pair production only initial-state interactions occur; an
advantage of the Drell-Yan processes is that conversion rate to a lepton
pair is perturbatively calculable, unlike the non-perturbative fragmentation function
converting a parton into an observed hadron. The theoretical
important case of an asymptotic parton penetrating the nucleus with no
hard interactions is also considered in the figure~\ref{fig:coldenloss} (d).   
Because of the interference with the hard radiation, each of these 
energy-loss problems should be considered and solved separately, in
order to probe the properties of cold nuclear matter such as its transport
coefficient $\hat q$ via measurements of the parton energy loss $-dE/dz$.

Baier {\it et al.} give in Ref.~\cite{Baier:1998kq} the expression for the energy loss of an asymptotic quark:
\begin{equation}
-\frac{d E}{d z} = \frac{\alpha_s\ N_c}{4}\ \hat{q}\ L,
\end{equation}
which is a factor 3 smaller than for a quark produced {\it inside} the medium~\cite{Baier:1997sk}. Indeed, 
a quark coming from $-\infty$ has had the time to construct its gluon field and therefore will only start to 
radiate (medium-induced) gluons only after it has experienced a first single scattering in the medium, after a time
 $t = \cO{\lambda}$. On the contrary, a quark produced in the medium 
immediately radiates soft gluons to get rid of its virtuality; this gluon emission interferes constructively with gluon emission stimulated by the medium.

\begin{figure}[tb]
  \centering
  \includegraphics[width=7cm,clip=true]{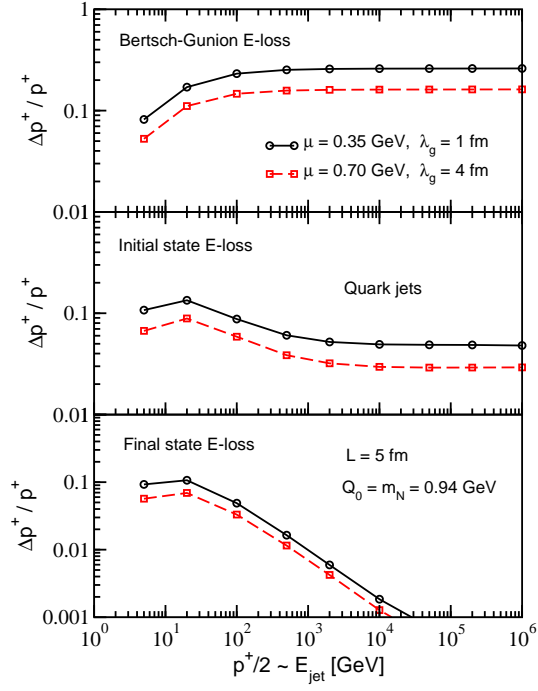}
  \caption{Fractional energy loss $\Delta p^+/p^+$ for massless quark 
    partons versus the parton energy $E_{\rm jet}$, in cold nuclear matter
    of length $L = 5$~fm. Two different sets of typical momentum transfer
    per scattering and gluon mean free path have been used for
    comparison. Plot taken from Ref.~\cite{Vitev:2007ve}.}
  \label{fig:infinenloss}
\end{figure}

In Ref.~\cite{Vitev:2007ve}, initial- and final-state energy loss in
cold QCD matter has been evaluated in the GLV formalism. The
resulting fractional energy loss is shown in Fig.~\ref{fig:infinenloss}
as a function of the parton energy $E_\text{jet}$, for two sets of
average momentum transfer squared $\mu^2$ and parton mean free
path $\lambda$ corresponding to a fixed value of
$\qhat=\mu^2/\lambda=0.12$ GeV$^2$/fm, and compared to the
asymptotic parton case.   
There are large differences in initial- and final-state energy losses, 
which are process-dependent and need to be correctly accounted for 
to experimentally access the properties of the nuclear medium. 
Furthermore, a single parameter such as the transport coefficient
$\hat q = \mu^2/\lambda$ may not adequately  
describe the stopping power of cold nuclear matter, as shown by the
different energy loss of the solid and dashed lines.

Keeping this in mind, we review  in the following the
existing phenomenological applications of energy loss computations 
in cold nuclear matter for semi-inclusive nuclear DIS and DY measurements.
Table~\ref{tab:dedxcomp} and Fig.~\ref{fig:dedxcomp}
summarise various estimates of $-dE/dz$ from the existing data 
(see also the short reviews in Refs.~\cite{Garvey:2002sn,ArleoYR}).

\subsubsection{\it Nuclear DIS in the BDMPS formalism }--
\label{sec:enlossBDMS}
In the  BDMPS approach of Ref.~\cite{Arleo:2003jz,Arleo:2002kh}, extended in
\cite{Accardi:2005mm,Accardi:2007in} to include a realistic treatment 
of the nuclear geometry, the reduced quark energy at the time of
hadronisation is translated into a shift of $z_h$ in the vacuum
fragmentation function $D\to \tilde D$~\cite{Wang:1996yh} via
quenching weights. At leading order (LO) in
perturbative QCD, the hadron multiplicity is then 
computed as follows:
\begin{equation}
\begin{split} 
  \frac{1}{N_A^{DIS}}\frac{dN_A^h(z_h)}{dz_h} & = \;
    \frac{1}{\sigma_{\ell\,A}}
    \int dQ^2\,d\nu\, \sum_f e_f^2 \;q_{f/A}\left(x,Q^2\right) 
    \frac{d\sigma_{lq}}{dQ^2 d\nu} \tilde D_{f/A}^h\left(z_h,Q^2\right) \ ,
 \label{eq:DISxsec}
\end{split}
\end{equation}
where $\tilde D$ is the medium-modified fragmentation function,
Eq.~(\ref{eq:modelff}), computed with $E_f = \nu$.

\begin{figure}[tb]
  \centering
  \includegraphics[width=6cm]{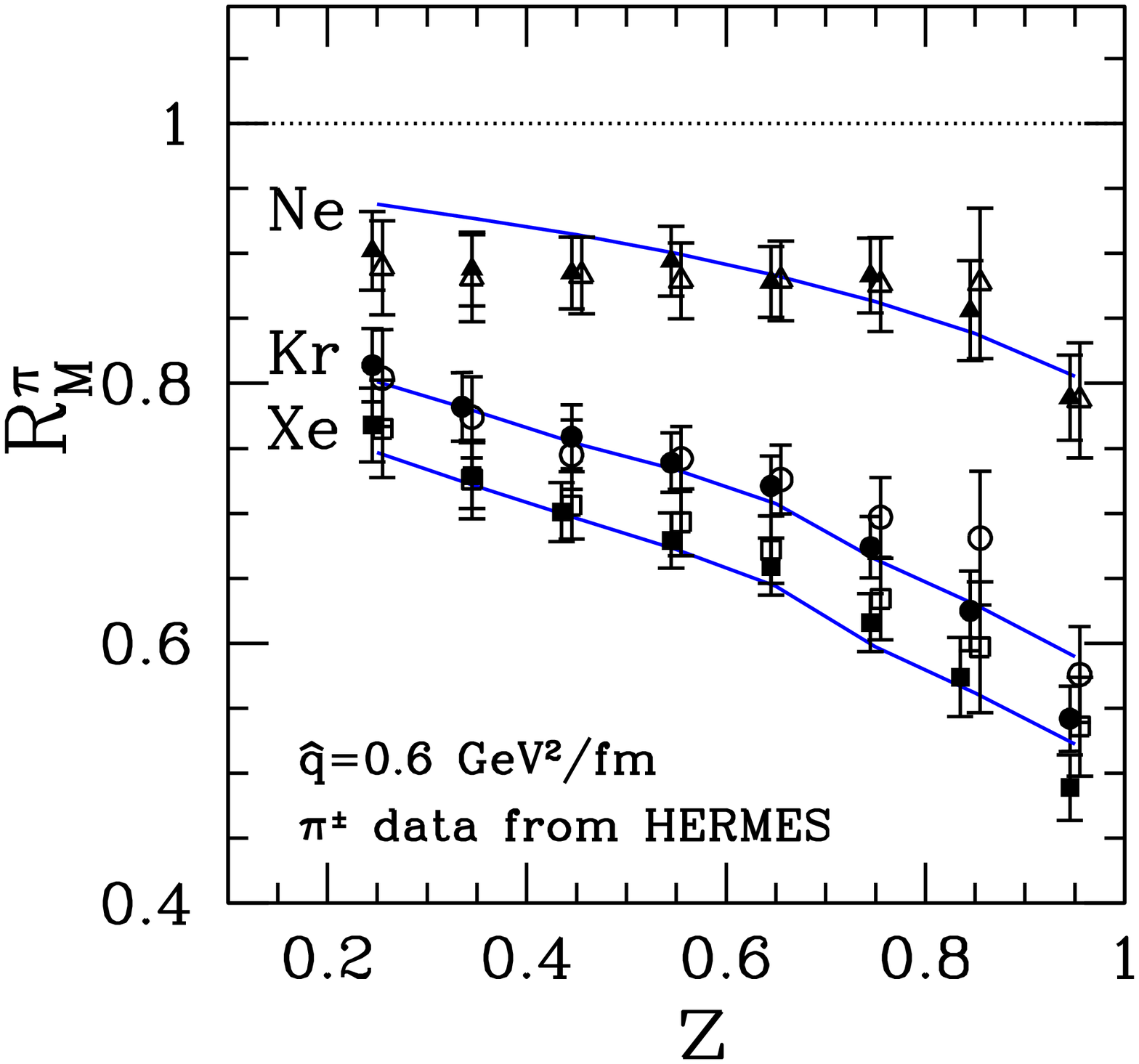} \hspace*{0.5cm} 
  \includegraphics[width=6cm]{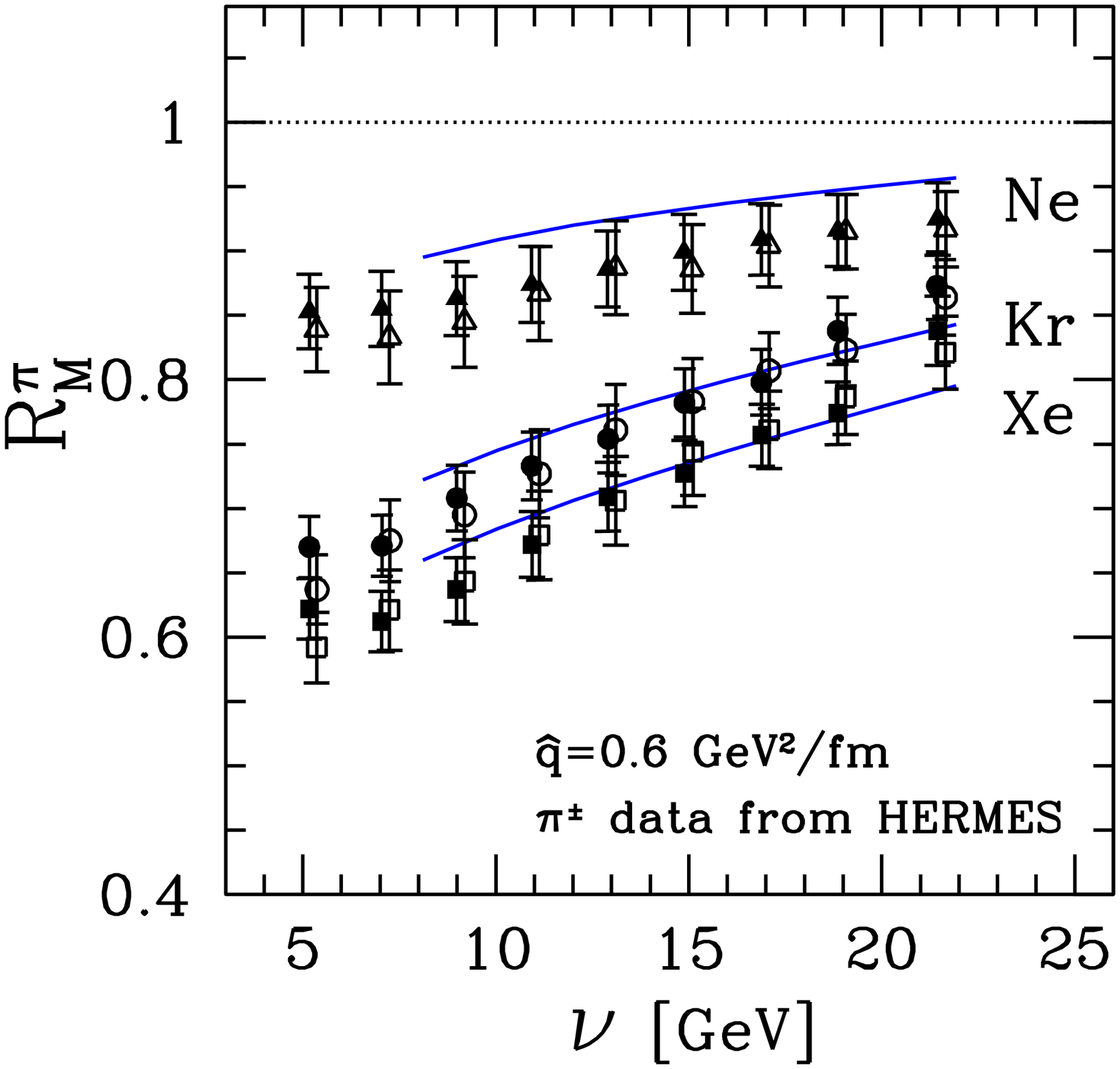} 
  \vskip-.3cm
  \caption{HERMES $\pi^+$ (full symbols)
    and $\pi^-$ (open symbols) data~\cite{Airapetian:2007vu} 
    compared to BDMPS energy loss calculations using quenching
    weights following \cite{Accardi:2005mm,Accardi:2007in}.  
  }
  \label{fig:AAmodel}
\end{figure}

The LO computation is known to underestimate the experimentally
measured average $\vev{\nu}_z$ and $\vev{Q^2}_z$ in each $z_h$ bin 
\cite{Accardi:2002tv}. The problem, is likely to be solved at NLO. 
An effective way of circumventing it at LO is to approximate 
Eq.~\eqref{eq:DISxsec} by
\begin{align} 
  \frac{1}{N_A^{DIS}}\frac{dN_A^h(z_h,Q^2)}{dz_h} & \approx \;
    \frac{1}{\sigma_{\ell\,A}} \sum_f e_f^2\; q_f
    \left(\vev{x}_z,\vev{Q^2}_z\right) 
    \frac{d\sigma_{lq}}
    {d\vev{Q^2} d\vev{\nu}}
    \tilde D_{f/A}^h\left(z_h,\vev{Q^2}_z \right) \ .
 \label{eq:DISxsecapprox}
\end{align}
The value of the average variables $\vev{x}_z$ and $\vev{Q^2}_z$ 
in each $z_h$-bin is taken from the
measured values. This procedure is used in all computations based on
LO cross-sections
\cite{Arleo:2003jz,Accardi:2005mm,Wang:2002ri,Kopeliovich:2003py,Accardi:2005jd}. 
The only parameter of the computation, namely the transport
coefficient at the centre of the nucleus, is found to be
\begin{align}
  \qhat = 0.6 \text{\ GeV}^2/\text{fm} 
\label{eq:qhatcold-HERMES}
\end{align}
in order to reproduce the latest pion quenching data on the $Kr$
target from HERMES~\cite{Airapetian:2007vu}, see Fig.~\ref{fig:AAmodel}. 
When comparing the theoretical $\nu$-distribution with data, mind
that $\vev{z_h}_\nu\approx 0.3$, where the model tends to slightly 
overestimate the data. Note that the extracted cold matter $\qhat$ value is a factor 10
larger than the perturbative estimate $\hat q \approx 0.05$~GeV$^2$/fm 
from Baier {\it et al.}~\cite{Baier:1997sk}. 

\begin{figure}[tb]
  \centering
  \includegraphics[width=6cm]{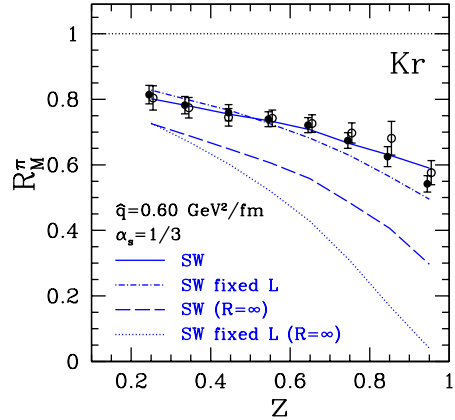} 
  \vskip-.3cm
  \caption{Realistic and
    approximate treatments of the nuclear geometry
    with quenching weights in the model of
    Ref.~\cite{Accardi:2005mm,Accardi:2007in}.  
    The upper two lines
    are computed with $R = \omega_c L$, the lower two lines with
    $R\ra\infty$. Solid and dashed lines: variable quark
    path-length. Dotted and dot-dashed lines: fixed $L = (4/3) R_A$.
    Experimental data are for quenching of $\pi^+$ (full circles)
    and $\pi^-$ (open circles) on $Kr$ from HERMES
   ~\cite{Airapetian:2007vu}.} 
  \label{fig:AAgeom}
\end{figure}

An appropriate treatment of the medium geometry is important to extract
the quenching parameter from the data~\cite{Accardi:2005mm,Accardi:2007in}.
In Fig.~\ref{fig:AAgeom}, we compare experimental data for pion production 
on $Kr$ to realistic and approximate geometries, using SW quenching weights.
The crudest approximation is to use an average quark path-length
$L\approx (3/4) R_A$, asymptotic quenching weights ($R\ra\infty$) and
a constant nuclear density, corresponding to a constant transport
coefficient $\hat q$. This approximation is commonly
considered, but cannot reproduce the $z$-dependence of the data (dotted line)
even if  $\hat q \approx 0.15-0.20$~GeV$^2$/fm is adjusted for the
curve to touch the data.

We can improve these approximations in several 
steps: using asymptotic quenching weights with a variable medium
length (dashed line), non-asymptotic quenching weights using a finite
$R = \omega_c L$, either with fixed average $L$ (dot-dashed line), or with
variable $L$ depending on the position of the $\gamma^*$-quark
interaction point (solid line). The largest effect is given by the use of
non-asymptotic quenching weights, mainly because of a non-zero
probability of no energy loss, see Section~\ref{sec:qw}. 
Modelling the full geometry instead of assuming a mean length gives a
smaller but still important effect. In particular, when geometry is properly 
taken into account there seems to be no need to invoke a finite quark
lifetime to explain the large-$z$ data, as proposed in
Ref.~\cite{Arleo:2003jz}. 

Energy conservation, namely the constraint
$\epsilon \leq E_q$, is not always fulfilled in the quenching weights
because of the approximations involved in their determination. It 
can be imposed from the outside by cutting the quenching weight at
$\epsilon = E_q$ and reweighting it to conserve 
probability~\cite{Dainese:2004te}. Alternatively, one can cut the
single gluon radiation spectrum at $\omega = E$, and consider
$\mathcal{O}(\omega_c/E_q)$ corrections~\cite{Arleo:2002kh}.  These
corrections tend to reduce the quenching. Their effect increases with
$\qhat$ and $A$, and decreases with $z_h$. Their magnitude is about
5-10\% for $Kr$ targets, and typically comparable to the experimental error
bars in the figure. Thus they are subleading compared to the effect of
correctly implementing the nuclear geometry.

The flavour dependence of hadron quenching in the BDMPS formalism has
been discussed in~\cite{Arleo:2003jz,Arleo:2005wq,Arleo:2003yf}. 
We can approximate the vacuum FF at large $z_h$ by $D_q^h(z_h) \propto
(1-z_h)^{\beta_q^h}$, and read the value of the exponents from the
global fits of Refs.~\cite{Kniehl:2000fe} (Table 2 there). 
Taking into account that
$\pi^\pm$ production is dominated by $u,d$ quarks, and $K^\pm$ 
production is dominated by $u,s$ quarks, at $Q^2=2$ GeV$^2$ we have 
\begin{align}
  & D_{u,d}^{\pi^\pm}(z_h) \propto (1-z_h)^1 \\
  & D_{u,s}^{K^\pm}(z_h) \propto (1-z_h)^{0.9} \ .
\end{align}
So at HERMES ($Q^2\approx 2.5$ GeV$^2$) we may expect a slightly
stronger suppression for large-$z_h$ pions than for kaons (this is
opposite to what is asserted in Ref.~\cite{Kopeliovich:2003py}, which
assumes $\beta^\pi \approx 0.5$ and  $\beta^K \approx 0.8$ on the
basis of Regge phenomenology). 
Using the flavour separated global fit of
Kretzer~\cite{Kretzer:2000yf} one has $\beta^{K^+} < \beta^{K^-}$,
which translates into a stronger quenching for $K^-$ than for $K^+$,
compatible with data. The described flavor dependence of hadron
attenuation in the BDMPS formalism is in agreement with HERMES data.

\begin{figure}[tb]
  \centering
  \parbox[c]{5.5cm}{
    \includegraphics[width=4.6cm]{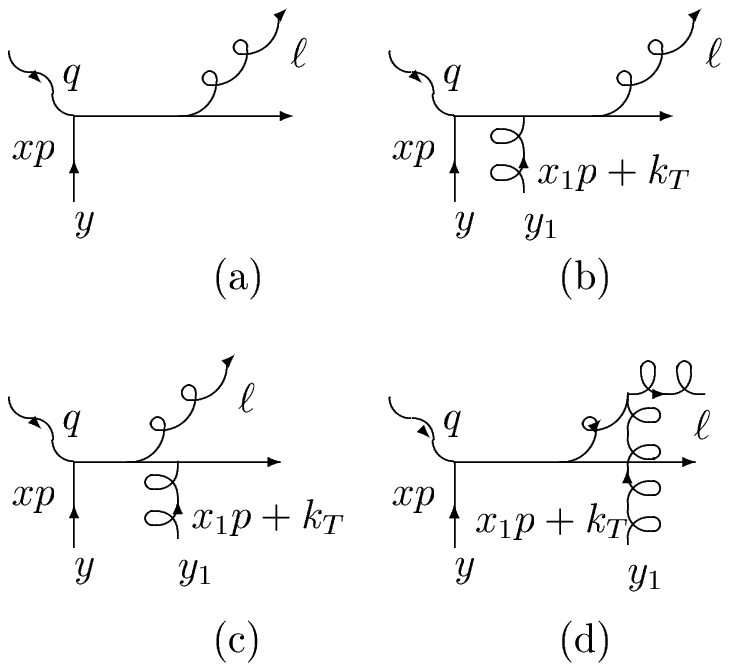} }
  \parbox[c]{6cm}{
    \includegraphics[width=6cm]{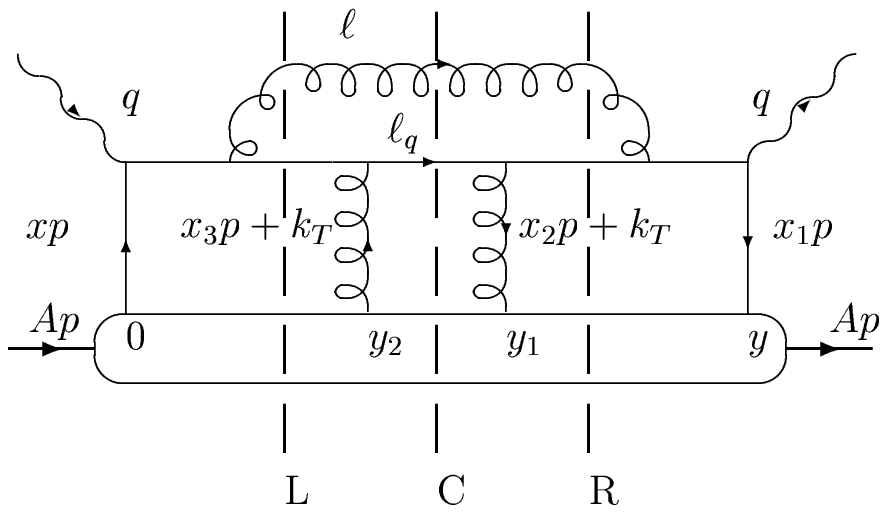} }
  \caption{{\it Left}: Gluon radiation from a single scattering (a) and from
    double scattering (b-d). {\it Right}: a sample diagram for quark-gluon
    rescattering processes with 3 possible cuts. Figures taken from
   ~\cite{Wang:2001if}.}
  \label{fig:htdiag}
\end{figure}

\subsubsection{\it Nuclear DIS in the higher-twist formalism }--
\label{sec:nDISHT}
In Refs.~\cite{Wang:2002ri,Wang:2001if,Guo:2000nz} the medium
modifications of the fragmentation functions are computed from leading
twist and twist-4 contributions to the leading order 
DIS cross-section, including diagrams with one elastic quark-nucleus 
scattering and one radiated gluon, see Fig.~\ref{fig:htdiag}. 
Both the struck quark and the
radiated gluon are allowed to fragment according to vacuum FF. The
obtained modified FF, $\tilde D$, can be modeled to a good accuracy by
shifting $z_h$ in the leading-twist fragmentation function
\begin{align}
  \tilde D (z_h) \, \lora \, \frac{1}{1-\Delta z_h} 
    D\left(\frac{z_h}{1-\Delta z_h} \right)  
 \label{eq:zshift}
\end{align}
where $\Delta z_h = \Delta E_q / \nu$ is the quark's fractional energy
loss, and $\Delta E_q \approx 0.6 \vev{z_g} \nu$ with the
average fractional energy  $\vev{z_g}$ carried away by the radiated
gluon computed diagrammatically~\cite{Wang:2002ri,Wang:2001if,Guo:2000nz} 
\begin{align} 
 \vev{z_g} & \approx \alpha_s^2(Q^2)\;\tilde C(Q^2) \;m_N \;R_A^2\;
    \frac{1}{\nu} \; f_g(1-z_h) \ ,
\end{align}
where $m_N$ is the nucleon mass and $R_A$ the nuclear radius, and
$f_g$ is a function of $(1-z_h)$ because of energy conservation
\cite{Accardi:2006ea}. The average gluon energy 
depends on one parameter, $\tilde C(Q^2)$, which represents 
the strength of parton-parton correlations in the nucleus. We can note
the dependence on the square of the medium size, typical of the LPM
effect in QCD. Inclusion of
quark-quark double scatterings beside quark-gluon double scatterings
leads, except for pions, to a different quenching of positively and
negatively charged hadrons~\cite{Zhang:2007mw}:
\begin{align*}
  & R_M^{\pi^+} \simeq R_M^{\pi^-} \simeq R_M^{\pi^0} \\
  & R_M^{K^-} < R_M^{K^+} \, ; \quad
    R_M^{\bar p} < R_M^{p} \, ; \quad
    R_M^{h^-} < R_M^{h^+}  \
\end{align*}
These features agree qualitatively with HERMES data
\cite{Airapetian:2003mi,Airapetian:2007vu,vanderNat:2003au}, and with
the BDMPS model previously 
discussed. A generalisation of the higher-twist formalism to include
heavy-quark energy loss has been discussed in
\cite{Zhang:2003wk,Zhang:2004qm}. 

\begin{figure}[tb]
  \centering
  \includegraphics[height=5cm]{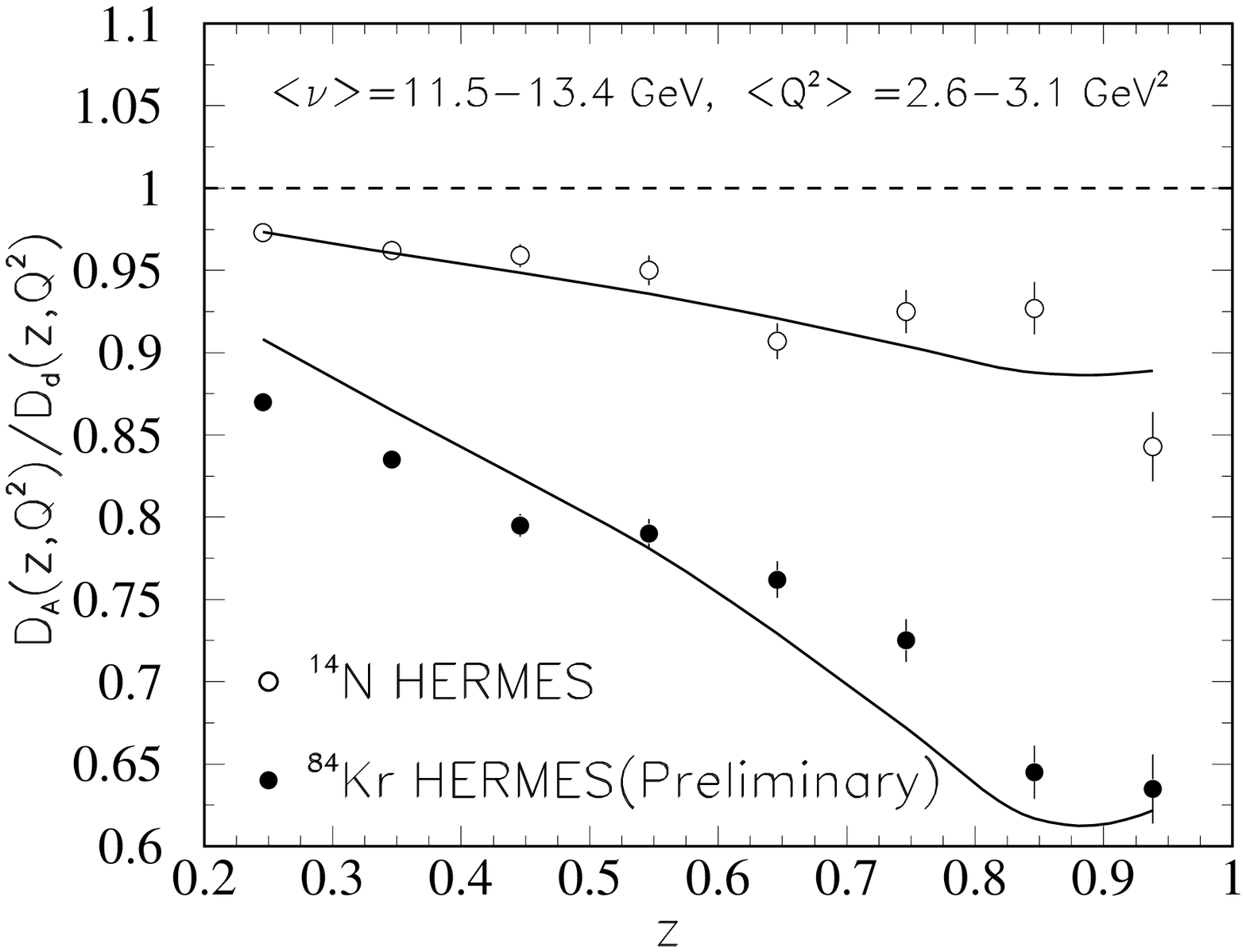}
  \includegraphics[height=5cm]{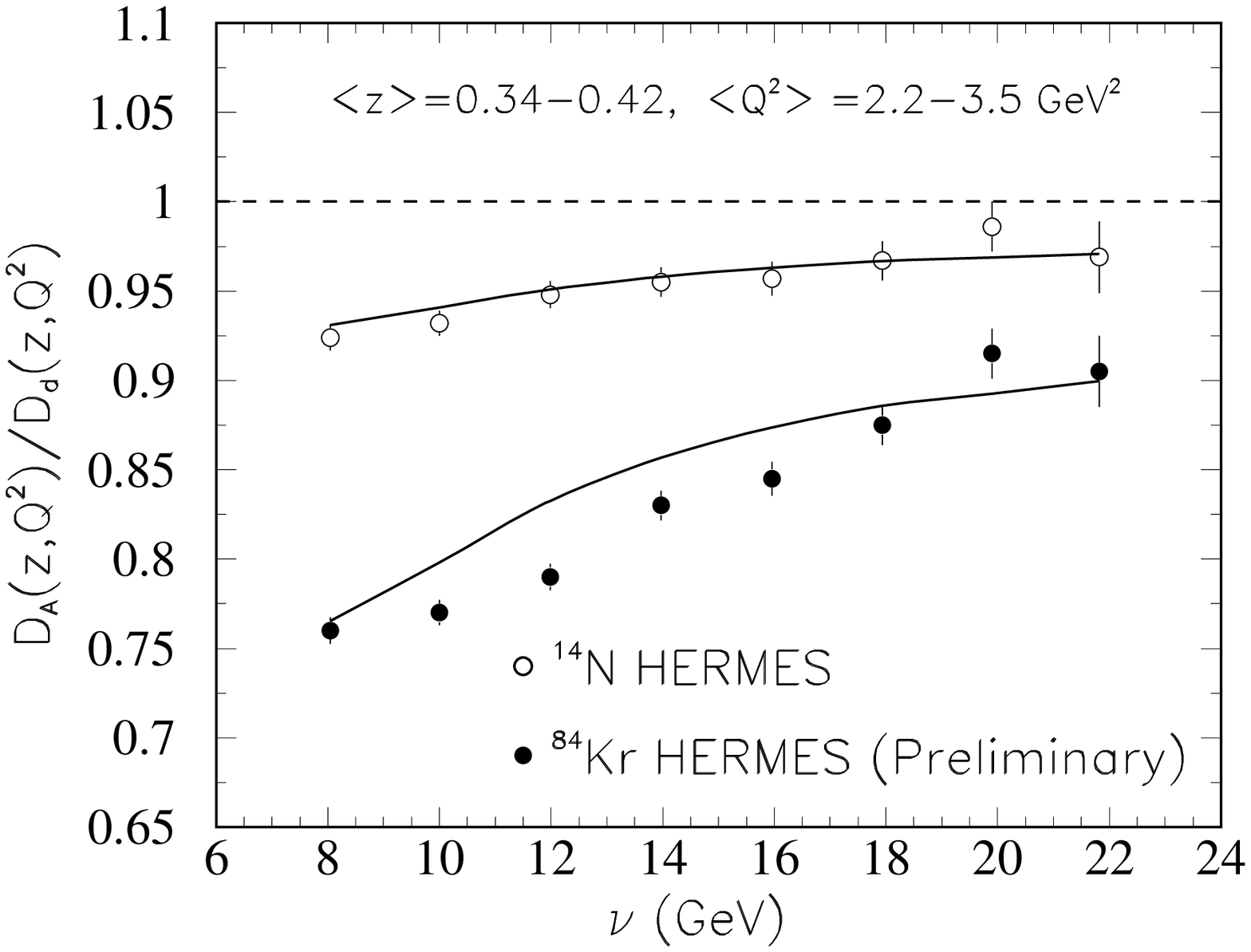}
  \caption{Multiplicity ratio computed in the high-twist formalism for
    parton radiative energy loss in nDIS. Figures taken from
   ~\cite{Wang:2002ri}.} 
  \label{fig:RMwang}
\end{figure}

To apply the model to HERMES data, the parameter $\tilde C$ is fitted
to the overall suppression of unidentified charged hadron on a nitrogen target:
$\tilde C(Q^2) = 0.0060$~GeV$^2$, with $\alpha_s(Q^2) = 0.33$ at $Q^2=3$
GeV$^2$, which corresponds to an energy loss $dE/dz = - 0.5$
GeV/fm for a $Au$ nucleus~\cite{Wang:2002ri}, or equivalently to a
transport coefficient $\qhat = 0.12$~GeV$^2$/fm~\cite{ArleoYR}. The
multiplicity ratio for other targets 
and parton species can then be computed without further adjusting the
parameter, see Fig.~\ref{fig:RMwang}. See also
Ref.~\cite{Majumder:2008jy}. One observes an overall
agreement with data, but with a tendency to overestimate
the slope of $R_M(z_h)$ for heavy nuclei. As discussed above in
connection with the BDMPS approach, this can be caused by a too
schematic treatment of the medium geometry. 
Another important remark, is that the
computations of Refs.~\cite{Wang:2002ri,Majumder:2008jy} include only
twist-4 diagrams, i.e., up to one parton rescattering in the medium. This
might not be sufficient for large targets such as $Kr$. A computation up
to twist-6 (two rescatterings) has been carried out in
\cite{Guo:2006kz}, however without comparison to experimental data.
(An all-twist resummation for radiative processes is unfortunately very
hard to achieve, but the first preliminary steps have been taken in
Refs~\cite{Majumder:2007hx,Majumder:2007ne}.)
Higher-twist effects on the DGLAP evolution of the fragmentation
functions have been studied in Ref.~\cite{Majumder:2009zu}, and reduce
the slope of $R_M$ at large $z_h$.   
We should finally note that coherent multiple parton scatterings may
lead to an additional suppression of quark production
\cite{Guo:2007ve}, which would also reduce the slope of $R_M$.

\subsubsection{\it Drell-Yan processes }--
\label{sec:enloss_in_DY}
Since the lepton pair does not interact strongly with the nucleus, the Drell-Yan process offers a clean probe 
of the initial multiple scattering of the projectile parton (a quark to leading order in $\alphas$) in the 
target nucleus before the hard process, $q\bar{q}\to\ell^+\ell^-$, takes place. 
The effects of energy loss on the Drell-Yan cross section can be estimated from its production cross section 
at leading order,
\begin{align}
  \frac{d\sigma^{DY}_{p+A}}{dx_1dx_2} 
    = \frac{4\pi \alpha ^2_{em}}{9 s} \frac {x_1 x_2} {x_1 +x_2} 
      \sum_{i} e^2_i \left[ q_{i/p}(x_1) \bar q_{i/A}(x_2) + 
      \bar q_{i/p} (x_1) q_{i/A}(x_2) \right] \ ,
  \label{eq:DYxsec}
\end{align}
where the fractional momenta $x_{1,2}$ are related to the observable
Feynman $x_F$ and dilepton mass $M$ as in Eq.~\eqref{eq:x1x2} (see Section~\ref{sec:DYkinematics}). 
At large $x_1>0.5$ and for $x_2 = 0.1-0.3$, where $\bar q_{i/D} \approx
\bar q_{i/p}$, the ratio of the DY cross section on a nucleus $A$ and on
deuterium $D$ can be approximated by
\begin{align}
  \frac {d\sigma^{DY}_{p+A}}{dx_1dx_2} 
    \bigg/ \frac {d\sigma^{DY}_{p+D}}{dx_1dx_2}
    \sim \frac{q^{u/p}(x^A_1)}{q_{u/p}(x_1)}
    \sim \frac{(1-x^A_1)^\eta}{(1-x_1)^\eta} \ ,
\label{eq:DY2}
\end{align}
where the large-$x$ exponent is $\eta\simeq 3$ from quark counting rules~\cite{Garvey:2002sn}.
Assuming that the projectile quark experiences an initial-state energy
loss per unit length, $\alpha = -dE/dz$, we can set $x^A_1 = x_1 + \alpha
\vev{L}_A/E$, where $E$ is the quark's energy, for the sake of a qualitative discussion. Hence, one obtains
\begin{align}
  \frac{d\sigma^{DY}_{p+A}}{dx_1dx_2} 
    \bigg/ \frac {d\sigma^{DY}_{p+D}}{dx_1dx_2}
    \sim 1 - \frac {3 \alpha \langle L \rangle_A}{E_p (1-x^p_1)},
\label{eq:DY3}
\end{align}
where $E_p$ is the incident proton's energy in the nucleus rest frame. 
This equation shows that the energy loss increasingly suppresses the DY
cross-section the larger the nucleus, and the larger $x_1$, or
equivalently the larger $x_F = x_1-x_2$ when $x_2\ll1$. A related observable is the dilepton $p_T$
broadening, which is related to the $p_T$-broadening of the incoming
quark caused by the parton initial state multiple scatterings (see
Eq.~\eqref{eq:broadening}).

\begin{table}[tbp]
\begin{center}
\begin{tabular}{ccccc} \hline
Reference & $-dE/dz$ (GeV/fm) & Observable \\
\hline
\cite{Guo:2000nz} & $\sim1.2$ & Nuclear modification of $e+A$ fragmentation functions \\
\cite{Wang:2002ri} & $\sim0.5$ & Nuclear modification of $e+A$ fragmentation functions \\
\cite{Baier:1997sk} & $\sim0.4$ & $p_T$ broadening of $p+A$ D-Y yield\\
\cite{Arleo:2002ph} & $0.20 \pm 0.15$ & Nuclear dependence of 150~GeV $\pi + A$ D-Y cross sections \\
\cite{Vasilev:1999fa} & $< 0.44$ & Nuclear dependence of 800~GeV $p+A$ D-Y cross sections \\
\cite{Johnson:2000ph} & $1.12 \pm 0.15 \pm 0.21$ & Nuclear dependence of 800~GeV  $p+A$ D-Y cross sections \\
\cite{Johnson:2001xfa} & $0.95 \pm 0.21 \pm 0.21$ & Nuclear dependence of 800~GeV  $p+A$ D-Y cross sections \\
\cite{Baier:1998yf} & $\sim2.8$ & $p_T$ broadening of $p+A$ jets \\
\hline
\end{tabular}
\end{center}
\caption{List of published parton energy loss values in cold nuclear matter 
extracted from various $e+A$ and $h+A$ observables (adapted from Ref.~\cite{Garvey:2002sn}).}
\label{tab:dedxcomp}
\end{table}

The wealth of experimental data, reviewed in Section~\ref{sec:DYdata},
makes likely a precise estimate of the cold nuclear matter transport
coefficient $\hat{q}^{\rm cold}$, or more generally the amount of
energy lost by fast quarks in heavy nuclei via expressions like
Eq.~\eqref{eq:DY3}. 
Note, however, that 
the nuclear modifications of the parton distribution functions
(shadowing, anti-shadowing and the EMC effect depending on the typical
values of
$x$)~\cite{Armesto:2006ph,Norton:2003cb,Piller:1999wx,Geesaman:1995yd} 
play a role in the nuclear dependence of DY production, and would need to be better constrained.

In Ref.~\cite{Johnson:2001xf}, the E772 and E866/NuSea data have been analyzed as a 
function of the momentum fraction $x_1$ as well as a function of the mass of the Drell-Yan 
pair, $M_{_{\ell^+ \ell^-}}$. The DY process is calculated using a dipole approach in the rest 
frame of the target nuclei, which is seen as the bremsstrahlung of a massive photon from 
the fast going incident quark, $q\to q\gamma^*$. This models allows the authors to compute 
equally the nuclear modifications of parton density in nuclei when the momentum fraction of the 
target parton, $x_2$, is small, $x_2\ll 1$ (shadowing). Performing a global fit of the E772 
and E866/NuSea measurements, a huge energy loss has been extracted:
\begin{equation}
  \label{eq:cold_johnson}
  - \frac{dE}{dz} = 2.7 \pm 0.4 \pm 0.5 \ {\rm~GeV}/{\rm fm},
\end{equation}
that is much larger than the above BDMPS-based expectation 
$-d E/d z \ll 1$~GeV/fm, but in fact this value includes ``vacuum
energy loss''~\cite{Johnson:2001xf} which should be removed to obtain
the medium-induced energy loss. 
It was later shown, however, that the large uncertainty of the sea quark nPDF in the E772 and E866/NuSea data makes 
it difficult a model-independent extraction of energy loss in cold nuclear matter from those measurements~\cite{Arleo:2002ph}. 
Rather, it was proposed to use the lower energy pion-beam data from the NA3 collaboration 
which is sensitive to {\it valence quark} densities (while anti-quarks from the {\it sea} are probed in $p$+A collisions)
for which nuclear modifications are known to be small. From the  statistical analysis of these data, 
a much smaller quark energy loss in nuclei was obtained~\cite{Arleo:2002ph}
\begin{equation}
  \label{eq:cold_arleo}
  - \frac{dE}{dz} = 0.20 \pm 0.15\ {\rm~GeV}/{\rm fm},
\end{equation}
equivalent to $\qhat = 2 (dE/dz)/L \approx 0.14 \pm 0.11$~GeV$^2$/fm~\cite{ArleoYR}, and in good
agreement with the BDMPS estimate.

\begin{figure}[tb]
  \centering
  \includegraphics[width=12.cm,height=6cm,clip=true]{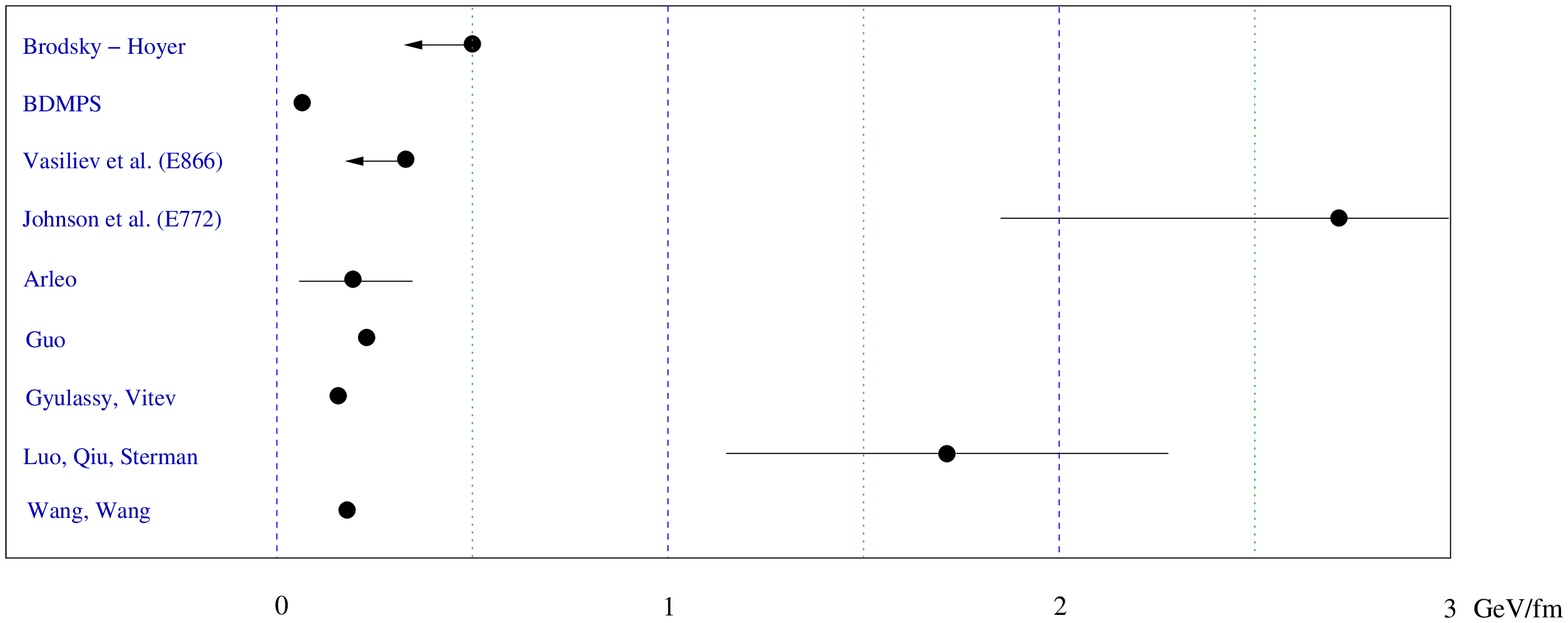}
  \caption{Compilation of the different estimates for the magnitude of an 
incoming quark mean energy loss per unit length, 
$\left(-dE/dz\right)_{\mathrm{in}}$, in a $L = 5$~fm 
nucleus. Taken from Ref.~\cite{ArleoYR}.}
  \label{fig:dedxcomp}
\end{figure}

Various estimates on cold nuclear energy loss extracted phenomenologically from a variety of observables, 
are summarised in Fig.~\ref{fig:dedxcomp} and Table~\ref{tab:dedxcomp}. Most of them point out to a 
rather small energy loss, $d E/d z\lesssim 0.5$~GeV/fm. In order to clarify this issue, Garvey and Peng 
proposed~\cite{Garvey:2002sn} to measure the nuclear dependence of DY production in fixed-target $p+A$
collisions at low beam energies, $E_{lab}$~=~50, 120~GeV (respectively $\sqrtsnn$~=~9.8, 15~GeV), 
where the effects of energy loss prove the strongest.

\subsubsection{\it Hadron-nucleus and nucleus-nucleus collisions } \ \\
\label{sec:coldjetquenching}
The discussion of jet quenching in $A+A$ is typically
focused on parton radiation in the hot medium created in the collision,
and neglects energy losses in the nuclear target. Similarly, in
$h+A$ collisions, where no hot and spatially extended medium is
created, energy loss effects are totally disregarded. However, as
discussed in Section~\ref{sec:coldenloss}, the cold QCD matter in the
target nucleus can induce sizeable IS or FS parton energy loss in DY
processes and $e+A$ collisions. Here we discuss cold matter energy loss
effects on hadron production in $h+A$ and $A+A$, and review the
phenomenological studies performed in
Refs.~\cite{Vitev:2006bi,Accardi:2007in}. 

\begin{figure}[tb]
  \centering
  \includegraphics[width=6cm,clip=true]{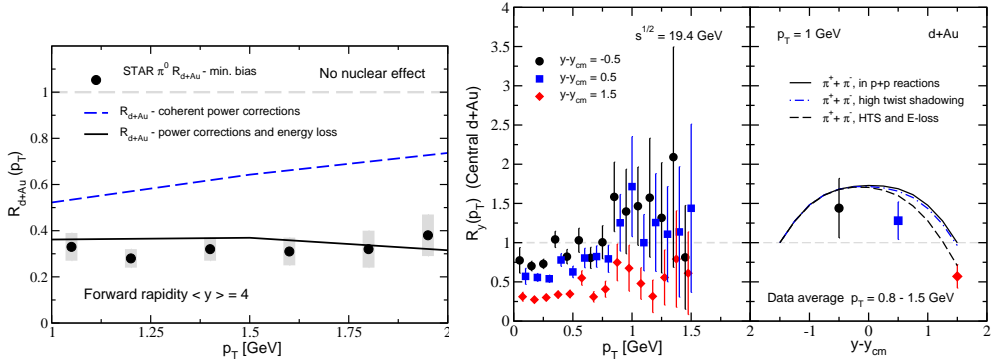}
  \includegraphics[width=7cm]{figure/d+Au.eps}
  \caption{Effects of IS energy loss and shadowing (in the ``high-twist'' formalism) 
  for hadron production in $d+Au$ collisions:
    $R_{dAu}(p_T)$ for forward pions measured in STAR at $\sqrtsnn = 200$~GeV ({\it left}) 
    and pion NA35 data at various rapidities at $\sqrtsnn = 19.4$~GeV ({\it right}). 
    Plots taken from Ref.~\cite{Vitev:2006bi}.} 
  \label{fig:ISenloss}
\end{figure}

The initial state parton suffers multiple scatterings and
medium-induced gluon radiation. In a simple phenomenological model
\cite{Vitev:2006bi}, the resulting energy loss may be
accounted for by a shift of the incoming parton fractional momentum,
$x_1 \ra x_1(1-\epsilon)$, with $\epsilon = \kappa A^{1/3}$ the
fractional IS energy loss. The effect of such energy loss is felt
in a kinematic region where the flux of incoming partons varies rapidly
with $x_1$, typically at large rapidity. In~\cite{Vitev:2006bi}, the suppression factor
$R_{dAu}(p_T)$ for charged hadrons was computed including coherent parton
multiple scatterings (higher twist shadowing) and energy loss. A value 
$\epsilon = 0.0175$ was fitted to STAR data on $R_{dAu}$ for charged
hadrons at $\eta = 4.1$ and $\sqrtsnn = 200$~GeV, see
Fig.~\ref{fig:ISenloss}. Applying the same formalism to CERN NA35 data
at $\sqrtsnn = 19.4$~GeV, one obtains a reasonable description of the
rapidity dependence of $R_{dAu}$, and sees that
IS state energy loss becomes
relevant only at forward rapidity $y-y_{cm} \gtrsim 0$. According to
the rapidity shifts listed in Table~\ref{tab:Deltay1}, 
we may expect a similar conclusion to hold for 
$y-y_{cm} \gtrsim 2 (5)$ at RHIC (LHC).

\begin{figure}[tb]
  \centering
  \includegraphics[width=5.5cm]{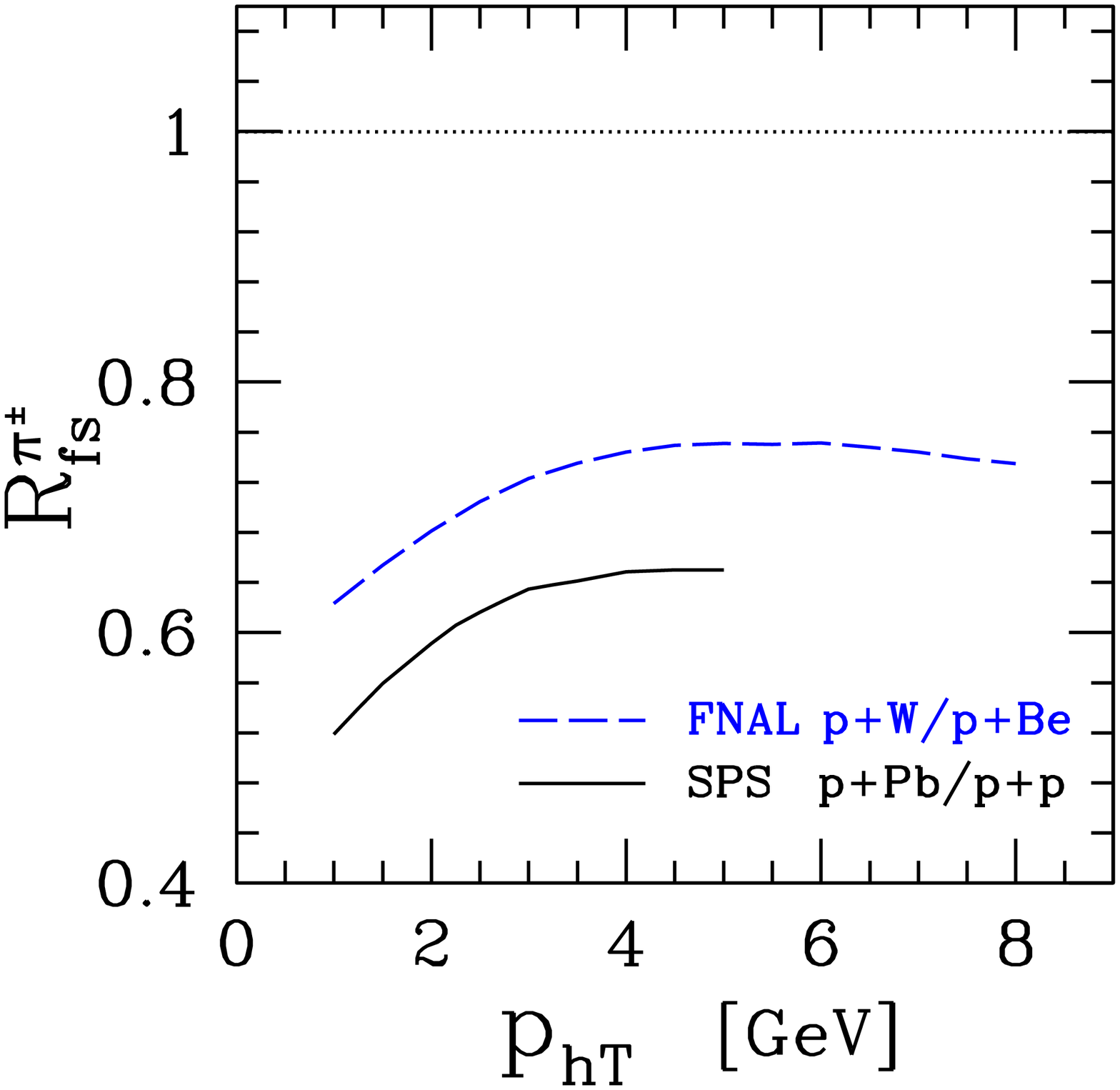}
  \includegraphics[width=5.5cm]{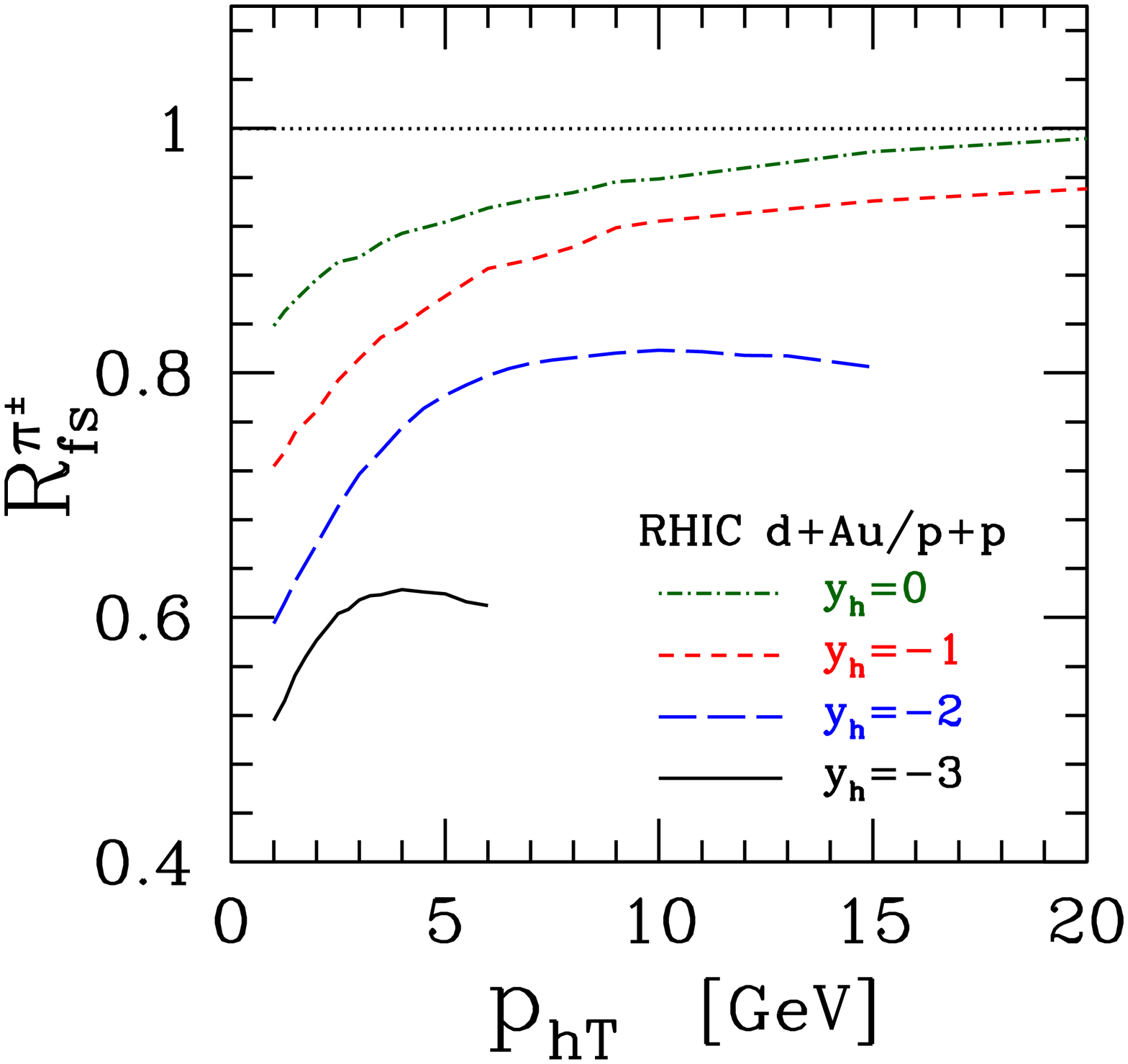}
  \caption{Energy loss model estimates of final-state hadron quenching
    in cold nuclear matter for midrapidity pions produced in $p+A$
    collisions at SPS and FNAL fixed-target energies, and several negative rapidities at
    RHIC.
  }
  \label{fig:coldquenchtheory}
\end{figure}

As discussed in Ref.~\cite{Accardi:2007in} and Section~\ref{sec:phasespaces}, 
a parton scattered at negative rapidity, $y-y_{cm}<0$,  
in a $h+A$ collision travels in the same direction as the
target nucleus: seen in the nucleus rest frame, it appears to move
slowly and corresponds to a low value of $\nu$ in the language of
$\ell+A$ collisions. Therefore, 
based on the observed suppression of hadron production in
lepton-nucleus DIS
\cite{Airapetian:2003mi,Airapetian:2000ks,Airapetian:2003mi,Ashman:1991cx,Osborne:1978ai}   
at low $\nu$ (see the kinematic analogy between DIS and $h+h$
collisions at LO in Section~\ref{sec:phasespaces}), we can expect
non-negligible hadron suppression due to FS interactions in cold
nuclear matter also in $h+A$ and $A+A$ collisions.
Using Eq.~\eqref{eq:NN(DIS)1}, it is possible to plot the experimental
nDIS data on $R_M^h$ from HERMES and EMC in terms of the $h+h$ kinematic
variables $p_T$, $y_1$ and $z$, which provides a rough estimate of
final-state hadron attenuation in $h+A$ collisions~\cite{Accardi:2007in}.
The obtained quenching is not small, and increases
with decreasing rapidity $y_1$ as expected from the kinematic analysis
of Section~\ref{sec:phasespaces}. However, the $p_T$ range covered
by HERMES and EMC is quite limited compared to the $p_T$ for which
hadron production in $h+A$ and $A+A$ can be measured.
Moreover the value of 
$z\approx z_h$ is not experimentally accessible in hadronic
collisions and not easily correlated to the measured variables
\cite{Accardi:2007in}. Furthermore, the $A$-dependence of hadron
quenching in nDIS is non trivial, and very different from a naive 
$A^\alpha$ power law as often assumed, see Section.~\ref{sec:Adep}. For
these reasons, a theoretical estimate is needed for $h+A$ and $A+A$
collisions. This can be obtained in the energy loss model described in
Section~\ref{sec:enlossBDMS}~\cite{Accardi:2007in}, and is displayed in
Fig.~\ref{fig:coldquenchtheory}, which quantifies final-state 
``cold'' hadron quenching by the ratio
\begin{align}
  R_{fs}^h(p_T,\bar y) 
    = \frac{1}{A}\cdot \frac{d\sigma_{pA\ra hX}}{dp_T^2dy_1dy_2}
      \left[ \frac{1}{B}\cdot \frac{d\sigma_{pB\ra hX}}{dp_T^2dy_1dy_2} \right]^{-1} \ .
\end{align}
The plots show a substantial final-state hadron quenching already
for midrapidity hadrons at SPS and FNAL energy. At RHIC it is sizeable
for $y_h\lesssim-2$, where it may play a role in understanding the
evolution of the Cronin effect at backward rapidity,
is still present at $y_h = 0$, where it is of order 5\% at $p_T\gtrsim
10$~GeV/c consistent with nuclear PDF modifications, and quickly
disappears at forward rapidity. In $p+Pb$ at the LHC, we may 
expect negligible final-state effects at $y_h \gtrsim -3$ because of
the rapidity shift $\Delta y$ in Table~\ref{tab:Deltay1}. The rapidity 
regions in rapidity (measured in the centre-of-mass 
frame) where we can expect IS and FS hadron quenching in $h+A$
collisions are summarised in Table~\ref{tab:ISvsFS}.

In $A+A$ collisions the scattered partons
can traverse two distinct QCD media: first the two colliding nuclei, then the hot 
medium formed after the collision. The two media are well separated in time since
the time-scale for QGP formation, $\mathcal{O}(1$~fm$)$ is (much) larger than the 
crossing time of the colliding nuclei: 
$\tau_{cross}=2R_A/\gamma\approx$~1.5, 0.15, 5$\,$10$^{-3}$~fm at SPS, RHIC and LHC respectively. 
The degree to which the parton interacts with the 2 cold nuclei
depends on rapidity: at mid-rapidity to the same degree; at large rapidity
more strongly with the comoving nucleus and more weakly with the
opposite moving nucleus.
A rough estimate of the energy loss in the cold nuclei 
can be obtained by multiplying the $R^h_{fs}$ values in
Fig.~\ref{fig:coldquenchtheory} at $\eta$ and $-\eta$.
At RHIC midrapidity, cold quenching is less than 10\% at $|y|\lesssim 1$, much
smaller than the observed factor 4--5 hadron quenching observed in
central $Au+Au$ collisions.
At LHC, due to the even larger longitudinal boost, cold nuclear 
matter effects will be smaller than 10\% in the $|y|\lesssim 4$
rapidity range.
At SPS energy, cold quenching at midrapidity is of order 50\% and may
in fact be comparable to hot quenching: they both need to be taken
into account in any QCD tomographic analysis to characterise the
properties of the produced matter. 

\begin{table}[t]
  \centering
  \begin{tabular}{cccccc}\hline
                     & SPS/FNAL & RHIC & RHIC & LHC   \\\hline
    $\sqrtsnn$ [GeV] & 17-38    & 63   & 200  & 5500  \\
    IS               & $\eta\gtrsim 0$  & $\eta\gtrsim 1$ 
                     & $\eta\gtrsim 2$  & $\eta\gtrsim 5$ \\
    FS               & $\eta\lesssim 2$ & $\eta\lesssim 1$ 
                     & $\eta\lesssim 0$ & $\eta\lesssim -3$ \\\hline
  \end{tabular}
  \caption{Regions of rapidity where IS and FS effects in cold nuclear
    matter may play a role in quenching hadrons in $p+A$ 
    collisions at various energies.In $A+A$ collisions the effects
    should be symmetrised in rapidity.} 
  \label{tab:ISvsFS}
\end{table}

\subsection{Jet quenching from AdS/CFT duality } \ \\
\label{sec:AdS/CFT}

The energy loss computations discussed so far are based on specific
models of the medium rooted in perturbative QCD.  
On the other hand, indications come from RHIC data that
the medium produced in heavy-ion collisions is strongly coupled, so that
non-perturbative effects can become important~\cite{Shuryak:2007zz}. However, non perturbative 
methods like lattice QCD have an intrinsic difficulty in computing dynamical properties, 
which are connected to the Minkowski
geometry of space-time, and not easily simulated in Euclidean
space-time as required by most lattice QCD methods. 

A recent theoretical development, the Anti-de
Sitter / Conformal Field Theory (AdS/CFT) duality
\cite{Maldacena:1997re,Witten:1998qj,Witten:1998zw,Gubser:1998bc}, has
given access to analytical calculations of the dynamical properties of a plasma
in $\NN = 4$ supersymmetric SU($N_c$) Yang-Mills (SYM) theory --
rather than QCD -- in the limit of strong coupling and large number of
colours $N_c$, by relating them to a weakly-coupled string theory
living in a 10 dimensional space. The AdS/CFT duality has been later
extended to a large class of supersymmetric 
theories, and is also known under the name of gauge/string duality.
Even though QCD and supersymmetric Yang-Mills theories are quite
different in terms of running coupling and matter
contents, recent experimental and theoretical considerations suggest
that the many-body physics of the QGP near the phase transition, where
experimental findings at RHIC suggest that the coupling is strong,
is similar in the two theories~\cite{Natsuume:2007qq}.

More in detail, the computation of an observable in the quantum
gauge theory at strong coupling can be rephrased as a classical 
computation of a related observable in a higher-dimensional
gravity theory at weak coupling, where it is perturbatively calculable:
\begin{align*}
  \text{
    $\NN$ = 4 SYM 
    $\leftrightarrow$ 
    type IIB string on AdS$_5\times S^5$
  } 
\end{align*}
Here, $\NN = 4$ means that the theory contains 4 supercharges, AdS$_5$
is a 5-dimensional space with constant and negative curvature, $S^5$
is a 5-dimensional sphere. For finite temperature gauge theories,
relevant to the description of a QGP, the
gravitational equivalent contains a special kind of black hole:
\begin{align*}
  \text{
    $\NN$ = 4 SYM at finite $T$ 
    $\leftrightarrow$ 
    type IIB string on (Schwarzchild-AdS$_5$ black hole)
    $\times$ $S^5$ 
  } 
\end{align*}
The details of the computation of dynamical quantities such as the drag force
coefficient of a moving heavy-quark in the plasma, the quenching parameter
$\hat q$, and the jet transverse momentum broadening in the AdS/CFT
duality framework are reviewed and summarised
in~\cite{Natsuume:2007qq,CasalderreySolana:2007zz,Edelstein:2008cp}.
Although still highly speculative, the connection of these
theoretically computable strong-coupling quantities with the physics
of real-life QGP is opening new ways of understanding the deconfined
phase of QCD near the critical temperature at a fundamental
level.



\section{Hadron formation, propagation and interaction }
\label{sec:prehadron}

In hadron absorption models, hadronisation is typically 
assumed to happen in two stages as shown in Fig.~\ref{fig:2-step}. One 
effectively identifies the prehadron production time and the colour neutralisation
time: (i) the struck quark neutralises its colour and forms a so-called ``prehadron'', 
which then (ii) collapses into the asymptotic hadron $h$ (see Fig.~\ref{fig:hadrosketch} 
and Section~\ref{sec:formationtimes}). 
When the (pre)hadron is formed
inside the nucleus, it can reinteract with the surrounding
nucleons. The space-time evolution of the hadronising system 
can be computed using the non-perturbative Lund string model 
or in a pQCD-based approach.
The various realisations of the two-stage model differ 
by the assumptions concerning the prehadron production
time $t_{preh}$, the hadron formation time $t_h$, the (pre)hadron 
interaction with QCD matter, and by the treatment or neglect of the
subleading effects due to the quark propagation stage. 
Most models address absorption in cold nuclear
matter, especially in $\ell+A$ collisions. We will mostly specialise
the discussion to this case, and briefly comment about applications to
$h+A$ and $A+A$ collisions. 

\subsection{Early string-based absorption models} \ \\
\label{sec:earlystrings}

\begin{figure}
  \centering
  \includegraphics[width=6.5cm]{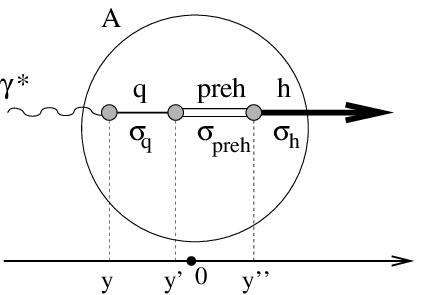}
  \qquad \includegraphics[width=5.5cm]{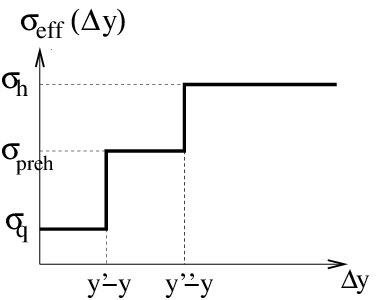}
  \caption{{\it Left:} Sketch of the time evolution of the hadronising system in
    the two-step hadronisation model: a quark $q$ is struck at point $y$ which
    evolves into a prehadron and hadron at point $y'$ and $y''$ respectively
   (in the Lund string model, $y''=y'+z_h\nu/\kappa$). 
    {\it Right:} Sketch of the effective cross section \eqref{eq:sigmaeff}.}
  \label{fig:2-step}
\end{figure}

The first nuclear absorption models 
\cite{Nikolaev:1979an,Bialas:1980at,Bialas:1983kn} assumed a
one-step hadronisation process, in which the struck quark propagates in
the nucleus and interacts with the surrounding nucleons
with cross section $\sigma_q$. After a time $\vev{t_h} \propto E_h
\propto \nu$ the hadron is formed and interacts with cross section
$\sigma_h$. However, the EMC collaboration showed in \cite{Ashman:1991cx}  
that for any choice of the parameters $\sigma_q$ and $\vev{t_h}$ only
a poor description of charged hadron data in $\mu+A$ collisions could be achieved. 

\begin{figure}
  \centering
  \includegraphics[width=6cm, height=5cm]{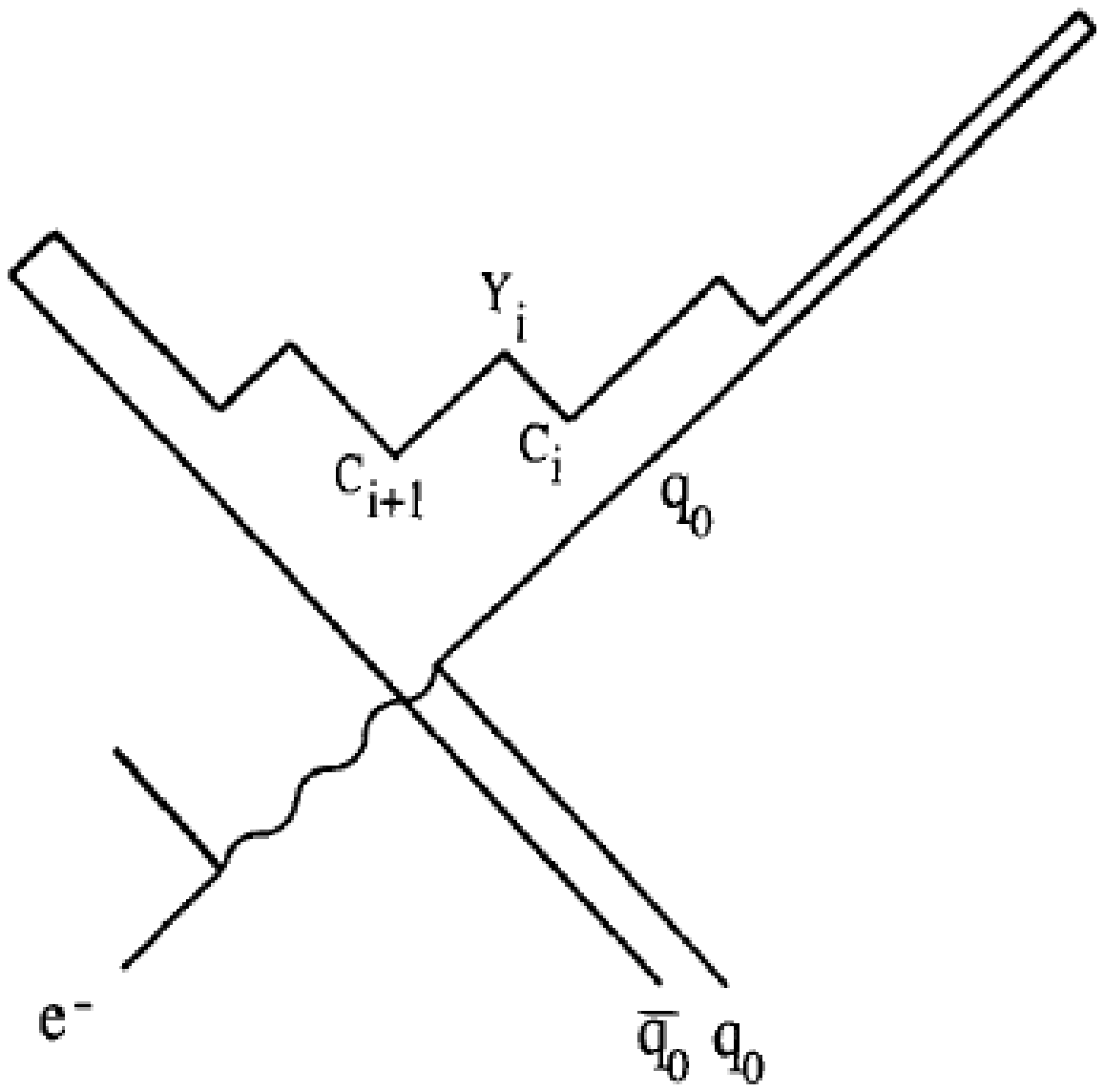}
  \includegraphics[width=6cm, height=5cm]{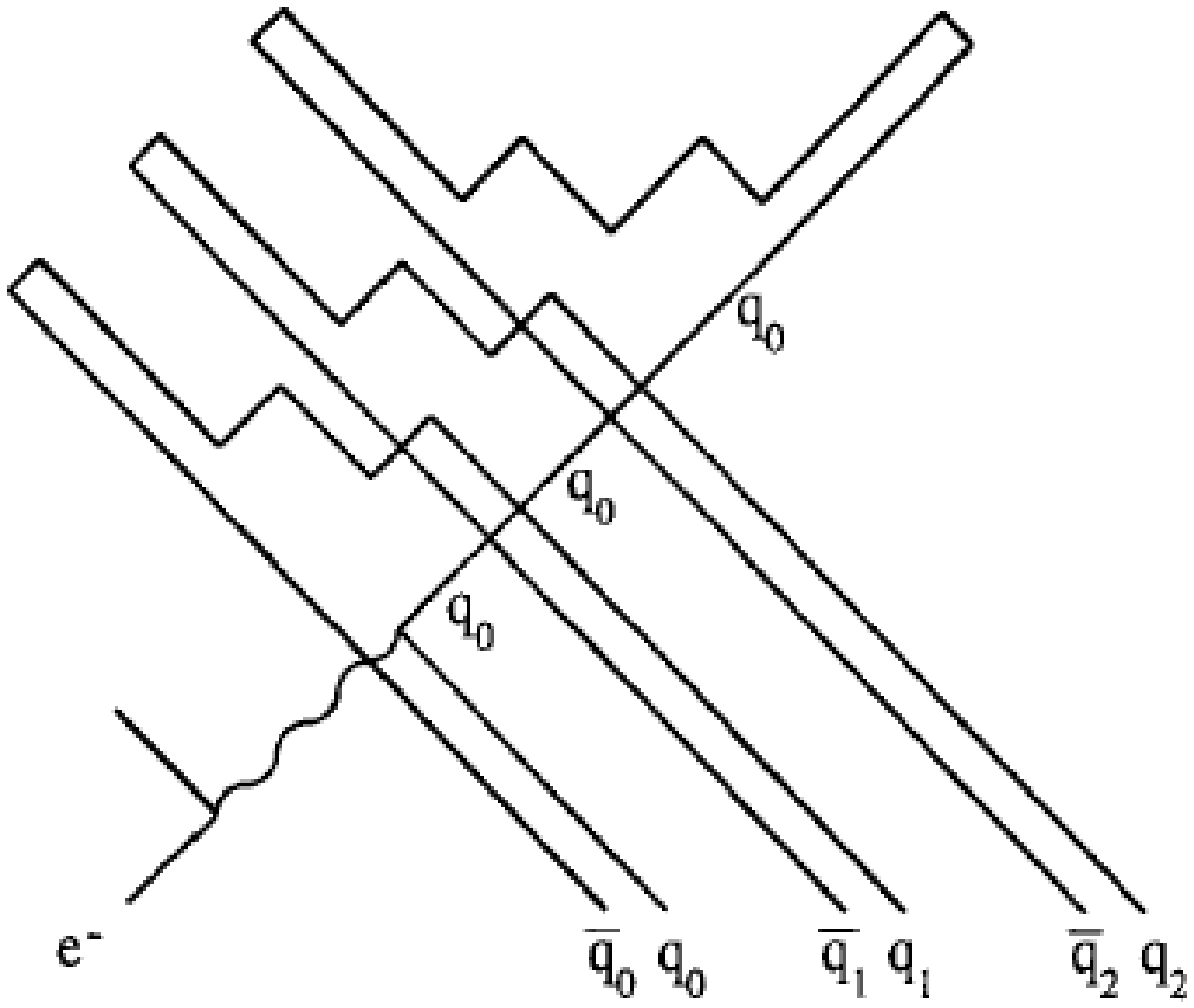}
  \caption{
   {\it Left:} Sketch of the independent Lund fragmentation
    model. $C_Y$ and $Y_i$ are the constituent (prehadron) formation
    point and ``yo-yo'' (hadron) formation point, respectively.
    {\it Right:} Sketch of the string-flip model in cold QCD matter.
    The $\bar q_i$ are diquark configurations initially connected to
    the quark $q_i$ in one of the nucleons. Colour exchange
    interactions rearrange the string end-points in several 
    short stringlets and a leading one with reduced energy.
    Figures taken from \cite{Gyulassy:1990dk}.
    }
  \label{fig:Lund-ind-sf}
\end{figure}

The idea of an intermediate ``prehadronic'' stage  
stage between the quark and the hadron was proposed in
Ref.~\cite{Bialas:1986cf} in the context of the Lund string model. 
Hadronisation was imagined to proceed first through the breaking of
the colour string stretched between the struck quark and the nucleus,
then through the evolution of the string pieces, whose end-point
quarks take some time to come together and finally recombine to form
the final hadron (Fig.~\ref{fig:Lund-ind-sf}), see
Section~\ref{sec:formationtimes} for more details. 
Typically, it is assumed that the
string breaking process does not depend on the nature of the target.
The nuclear dependence comes from the interaction of the hadronising
system with the surrounding nuclear medium. 
In principle, the hadronising system is allowed to
interact at all stages with cross-sections $\sigma_q$, $\sigma_{preh}$,
$\sigma_h$ (Fig.~\ref{fig:2-step}, left). 
In the case of $\ell+A$ collisions, the hadron survival probability 
$S_A$, i.e., the probability that the produced (pre)hadron does not interact with
the nucleus, can be approximated as \cite{Bialas:1986cf}
\begin{align}
\begin{split}
  S_A (z_h) & = \int db^2 \int_{-\infty}^\infty dy\, \rho_A(\vec b,y) \\
    & \times \int_y^\infty dy'\, D(z_h;y'-y) 
    \left[ 1-\int_y^\infty d\tilde y\, \sigma_{\text{\it eff}}(\tilde y,y')
    \rho_A(\vec b,\tilde y) \right]^{A-1} \ ,
\end{split}
\label{eq:hadsurvprob}
\end{align}
where $\rho_A$ is the nuclear density normalised to 1, and
$D(z_h;y'-y)$ is the probability for a string breaking (i.e. prehadron
formation) at a distance $l_c \equiv y'-y$, called
``constituent length'',  from the photon-quark
interaction point. In the Lund string model, the hadron is then formed
at a distance $l_y \equiv y'+z_h\nu/\kappa - y$, also called
``yo-yo'' length. The two-step dynamics is contained
in the effective cross section 
\begin{align}
  \sigma_{\text{\it eff}}\left(\tilde y,y'\right) 
    = \sigma_q \,\theta\left(y'-\tilde y\right) 
    + \sigma_{preh} \,\theta\left(\tilde y-y'\right)\,\theta\left(y'+\frac{z_h\nu}{\kappa}-\tilde y\right)
    + \sigma_h \,\theta\left(\tilde y - y' - \frac{z_h\nu}{\kappa}\right) \ ,
\label{eq:sigmaeff}
\end{align}
depicted in Fig.~\ref{fig:2-step} right. 
Note that the {\it ansatz} \eqref{eq:hadsurvprob}-\eqref{eq:sigmaeff} 
neglects elastic scatterings of the
system, which induce an energy loss and feed down to lower
$z_h$, so that its validity is confined to $z_h \gtrsim 0.4$. 
This limitation can be overcome in Monte Carlo implementations, see
Ref.~\cite{Gyulassy:1990dk} and Section~\ref{sec:MCstringmodels}.

\begin{figure}
  \centering
  \includegraphics[width=8cm,clip=true]{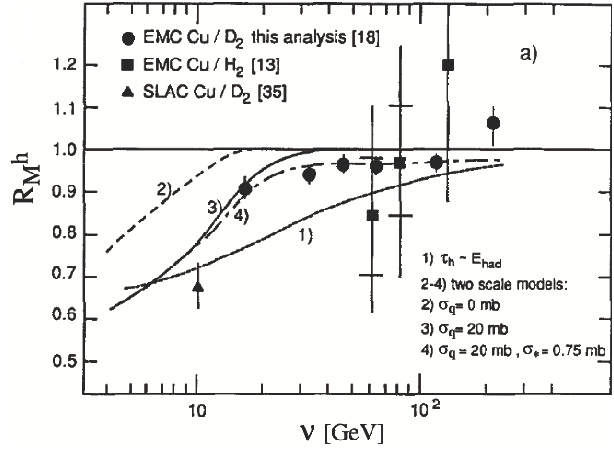}
  \caption{Comparison of EMC $R_M^h$ data to the one- and two-step models 
    with varying quark-nucleus and prehadron-nucleus cross sections. 
    Adapted from Ref.~\cite{Ashman:1991cx}.}
  \label{fig:EMCfits}
\end{figure}

The early applications  of the independent string fragmentation picture, 
\cite{Bialas:1986cf,Bialas:1988mn,Czyzewski:1989ur} focused on
the relevance of the constituent vs. yo-yo lengths (quark-to-prehadron
vs. prehadron-to-hadron formation times) in describing  hadron production data in 
$p+A$ and $\ell+A$ collisions assuming $\sigma_{preh}=\sigma_h$ 
or $\sigma_q=\sigma_{preh}$. 
The key point is that at large
$z_h$, $l_y = l_h \propto z_h \nu$, while $l_c \propto (1-z_h) \nu$.
EMC data on leptoproduction of hadrons 
confirmed that it is in fact the constituent length $l_c$ which controls 
hadron attenuation on nuclear targets, with a negligible contribution from the struck quark
interactions: data could be well fitted with 
$\sigma_q \lesssim 0.75$~mb and $\sigma_{preh}=\sigma_h \approx 20$~mb
\cite{Ashman:1991cx}, see Figure~\ref{fig:EMCfits}.
In other words, 
the hadronising system starts interacting inelastically with
QCD matter already at the prehadronic stage, well before the final
hadron is formed.

The effect of the nuclear medium on the string breaking process has
been considered in Refs.~\cite{Kopeliovich:1990sh,Gyulassy:1990dk}.
The struck quark $q_0$ (see Fig.~\ref{fig:Lund-ind-sf} right) 
can interact with other nucleons with a cross-section of the order of the 
hadronic one because of the non-zero transverse size of the string
attached to it \cite{Kopeliovich:1990sh}. When it does so, there is a
probability of 
colour exchange with the nucleon, so that the $q_0$ reconnects with the
diquark $\bar q_1$, and the quark $q_1$ with the diquark $\bar q_0$
originally attached to $q_0$, and so on. This process was called a
``string-flip''~\cite{Gyulassy:1990dk}, and leads to the
creation of a number of different string configurations:
\begin{align}
  \gamma^*A \ra (\bar q_0 q_1) + (\bar q_1 q_2) + \ldots 
    + (\bar q_{n-1} q_n) + (\bar q_{n} q_0) + (A-n-1) 
  \nonumber
\end{align}
Assuming a color-exchange cross-section $\sigma_{ce}$, the mean free path is
$\lambda_{ce}=(\sigma_{ce}\,\rho_A)^{-1}$, with $\rho_A\approx 0.17$ fm$^{-3}$.
The independent string fragmentation model is obtained in the limit
$\lambda_{ce} \gg R_A$. In the opposite limit, $\lambda_{ce} \ll R_A$,
the struck quark $q_0$ exits the medium reconnected to $\bar q_n$
(leading string) close to the nucleus surface. Therefore all the
hadron components are produced outside the nucleus and do not suffer
nuclear absorption.  
The intermediate 
stringlets have typically a low mass $M\approx \sqrt{2\kappa m_N
  \lambda_{ce}} \approx$~2~GeV/c$^2$, decay in a few low-energy hadrons, and
do not contribute at large $z_h$.  
Suppression of large-$z_h$ hadron production is then due to the energy
lost by the leading string because of colour reconnections, and is
independent of the details of the hadron or prehadron formation time.
Defining the probability for a leading hadron of energy $E_h = z_h \nu$
to fragment out of a string with average energy-loss $\epsilon=\lambda_{ce}\kappa$, 
and assuming it to be independent of the energy of the string, one obtains a medium 
modified fragmentation function 
\begin{align}
 \tilde D_A (z_h) = D(z_h') =
    D\left(\frac{z_h}{1-\epsilon/\nu}\right) \ .
\label{eq:stringenloss}
\end{align}
For fragmentation functions decreasing with $z_h$, this leads to
hadron attenuation: $\tilde D_A < D$. Eq.~\eqref{eq:stringenloss} is a
close analogue of the $z_h$-shifted medium modified 
fragmentation function, see e.g. Eq.~\eqref{eq:modelff},
used in the gluon bremsstrahlung energy-loss
models considered in Section~\ref{sec:parton}.
The hadron attenuation caused by colour exchange can be naively
interpreted as quark absorption in the nucleus
\begin{align}
  \tilde D_A(z_h) = D(z_h) \exp\left( - \sigma_q \int_y^{t_{preh}}
    dy'\rho_A(\vec b,y') \right) \ .
\end{align}
However, the effective quark cross section $\sigma_q$ is process-dependent. 
For lepto-production of light flavours it is rather small,
in agreement with the phenomenological EMC fit
\cite{Kopeliovich:1990sh}. 

Applications of the string-flip model to EMC data have been considered
in Monte Carlo \cite{Gyulassy:1990dk} and analytic
\cite{Czyzewski:1992iw} implementations 
in \cite{Gyulassy:1990dk}, a rather large cross-section $\sigma_{ce} =
30$~mb was considered, assuming a string to have a typical
hadronic transverse size, so that $\lambda_{ce} \gg R_A$. Therefore,
the leading string is produced at the surface of the nucleus, and
hadron attenuation is purely due to the energy loss induced by
colour exchanges, irrespective of the prehadron production time. 
EMC hadron attenuation could be well reproduced both as a
function of $\nu$ and of $z_h$. It leads to a slightly larger hadron
suppression at $\nu \lesssim 20$ GeV and $z_h \gtrsim 0.5$ compared to 
the independent fragmentation model using the constituent length,
otherwise giving comparable results. In Ref.~\cite{Czyzewski:1992iw},
the colour exchange cross section was identified with the constituent-quark cross
section in the additive quark model: $\sigma_{ce} \approx \sigma_q =
0.5 \sigma_{\pi N} \approx 10$~mb. The smaller $\sigma_{ce}$ leads
to an intermediate model between the independent string fragmentation
and pure string-flip model. However, computations turn out to
underestimate the suppression in the EMC kinematics. 

Finally, we should mention that extensions of string models to
address double hadron attenuation have also been discussed
\cite{Czyzewski:1990pg,Bialas:1989uq}.
 
\subsection{Modern string-based absorption models} \ \\
\label{sec:modernstrings}

Modern string-based absorption models build upon the early models
discussed in the previous Section, exploring different possibilities
for the effective cross section $\sigma_\text{\it eff}$ and/or for the  
hadron survival probability \eqref{eq:hadsurvprob}. Their computations
have also been compared to the more recent HERMES data on $e^-+A$ collisions
at $E_e = 12-27$ GeV~\cite{Airapetian:2000ks,Airapetian:2003mi,vanderNat:2003au,Airapetian:2007vu}. 

\begin{figure}[tb]
  \vspace*{-.2cm}
  \begin{center}
  \includegraphics[height=6.5cm,origin=c]{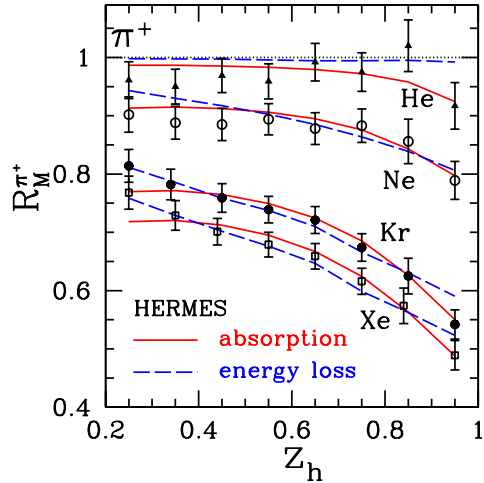}
  \end{center}
 \vspace*{-0.3cm}
 \caption[]{
   Lund string model based absorption model 
   \cite{Accardi:2005mm,Accardi:2002tv} (solid) and energy-loss model 
   \cite{Accardi:2005mm,Accardi:2007in} (dashed) compared to HERMES data
   on $\pi^+$ production \cite{Airapetian:2007vu}. The prehadron cross
   section is $\sigma_{preh}=0.8\sigma_h$ and the transport coefficient is 
   $\hat q = 0.6$ GeV$^2$/fm. Experimental statistical and
   systematic errors have been added in quadrature.
 \label{fig:absmodels}
 }
\end{figure}

The AGMP model~\cite{Accardi:2002tv,Accardi:2005jd} is based
on the Lund string model estimate of the formation times discussed in
Section~\ref{sec:formationtimes}, and neglects the interactions of the struck quark
($\sigma_q=0$~mb) in agreement with fits to EMC data~\cite{Ashman:1991cx}. 
In the HERMES kinematics one typically finds a production time
$\vev{t_{preh}} \approx 4$ fm $< R_A$ and a hadron formation time 
$\vev{t_h} \approx 6-10$ fm $\gtrsim R_A$: the hadron is
typically formed at the periphery or outside the nucleus so that its
interaction with the medium is negligible.
On the contrary, after its formation, the prehadron is allowed to 
interact with the surrounding nucleons with a cross section
$\sigma_{preh}(\nu) = 0.80 \, \sigma_h(\nu)$ proportional to the
experimental hadron-nucleon cross section $\sigma_h$ (taking e.g.
 $\pi^+$ production data on a $Kr$ target \cite{Airapetian:2007vu}).
The probability $S^A_{f,h}(z,\nu)$ that the (pre)hadron -- produced
from the fragmentation of a quark scattered at point $(b,y)$ -- 
does not interact, can be computed using
transport differential equations \cite{Accardi:2005jd}:
\begin{align} 
  S_{f,h}^A(z,\nu) = & \int db^2\,dy\,\rho_A(b,y)  \nonumber \\
  & \times 
    \int\limits_y^{\infty}dx'\int\limits_{y}^{x'}dx\, 
    \frac{e^{-\frac{x-y}{\left< l_{preh} \right>}}}
    {\left< l_{preh}\right>}\;e^{-\sigma_{preh}\int\limits_x^{x'}dsA\rho_A(b,s)}\,
    \frac{e^{-\frac{x'-x}{\left< \Delta l \right>}}}
    {\left< \Delta l\right>} \;
    e^{-\sigma_h\int\limits_{x'}^{\infty}dsA\rho_A(b,s)}
\end{align}
where $\Delta l = l_h-l_{preh}$, and $\rho_A$ is the nuclear density.
The hadron multiplicity is computed, at leading order in pQCD, as
\begin{align} 
  \frac{1}{N_A^{DIS}}\frac{dN_A^h(z)}{dz} =\; &  
       \frac{1}{\sigma_{\ell\,A}} \hspace{-0.2cm}
       \int\limits_{\mbox{\footnotesize exp. cuts}}
       \hspace{-0.4cm}
       dx\,d\nu\,
   \sum_f e_f^2 \; q_f(x,Q^2) 
      \frac{d\sigma_{\ell\,f}}{dx d\nu} S_{f,h}^A(z,\nu) D_f^h(z,Q^2) \ .
\end{align}
Here $\sigma_{\ell\,f}$ and $\sigma_{\ell\,A}$ are the lepton-quark and
lepton-nucleus cross sections, $q_f$ is the
$f$-quark distribution function, and $D_f^h$ its fragmentation
function. The model is applicable for $0.4 \lesssim z \lesssim 0.9$,
where it describes EMC \cite{Accardi:2002tv} and 
HERMES \cite{Accardi:2002tv,Accardi:2005jd} experimental data on a wide
range of hadron flavours and targets, see Fig.~\ref{fig:absmodels}. 
As it stands, the pure-absorption AGMP model does not predict a
dependence of the attenuation ratio $R_M$ on $Q^2$, because neither
the prehadron production time $t_{preh}$ derived in the Lund model, nor 
its assumed cross section $\sigma_{preh}$ depend on it. This
assumption does not
contradict the slightly rising $R_M(Q^2)$ measured at HERMES
\cite{Airapetian:2007vu}, see Section~\ref{sec:eAdata}.

A different variation of the two-scale model has been proposed in 
Refs.~\cite{Fialkowski:2007pt,Fialkowski:2007yf}. The authors argue that in
3+1 dimensions, the yo-yo time is ill-defined because the end quarks
of a string snippet never really meet together,
and propose instead that the hadron is formed
after a proper time for quark recombination, $\tau_r$, following string 
breaking at a time $t_{str}$ taken from the PYTHIA generator.
Then, the hadron formation time is simply set to $t_h =
(t_{preh} + \tau_r )v_{str}\gamma_{str}$, and boosted according to
the string velocity $v_{str}$ and Lorentz factor $\gamma_{str}$.
The system does not interact in the nucleus until hadron formation, at
which time it starts interacting with the full hadron-nucleus cross
section analogously to the AGMP approach. 
The model does a fair job in describing pion and kaon attenuation on $N$
and $Kr$ \cite{Airapetian:2000ks,Airapetian:2003mi} at $z_h 
\gtrsim 0.4$, with a flavour-dependent recombination proper time
$\tau_r(\pi) = 0.8$ fm and $\tau_r(K^-) = 0.4$ fm. It is 
interesting to note that the fitted values are such that
$\tau_r(h) \approx 1/m_h$.
However, this model overestimates the recently measured 
suppression on $Xe$ targets \cite{Airapetian:2007vu}, and no comparison
to EMC data is provided.

Extensions of the two-scale model based on
Eqs.~\eqref{eq:hadsurvprob}-\eqref{eq:sigmaeff} have been considered
in Refs.~\cite{Czyzewski:1990pg,Bialas:1989uq}, and more recently
compared to double hadron attenuation at HERMES \cite{Airapetian:2005yh}
in Refs.~\cite{Akopov:2006av,Akopov:2004ap,Akopov:2007cp}.

\subsection{The colour dipole model} \ \\
\label{sec:perthadrmodel}

In Ref.~\cite{Kopeliovich:2003py} the formation of a leading hadron ($z_h
\gtrsim 0.5$) is described in a pQCD-inspired approach based on
Refs.~\cite{Berger:1979xz,Berger:1979kz,Kopeliovich:2007yv}, see also 
Section~\ref{sec:formationtimes}:
a hard gluon radiated at the interaction point 
splits into a $q\bar q$ pair, and the $\bar q$
recombines with the struck $q$ to form the leading
prehadron, which evolves into the observed hadron.
The time development of hadronisation is included in this
picture by observing that the radiated gluon can be physically
distinguished from the struck quark only after a coherence time $t_c = 
2E\alpha(1-\alpha)/k_T^2$ , where $k_T$ and $\alpha = E_g/\nu$ are the
transverse momentum and fractional energy of the emitted gluon.
If the leading hadron is produced at large $z_h$ and contains the
struck quark, none of the radiated gluons can be emitted with $\alpha >
1-z_h$ by energy conservation. The time-dependence of the
energy radiated into the emitted gluons is computed as
\begin{align}
  \Delta E(t,z_h,Q^2) = \int_{\Lambda_{QCD}^2}^{Q^2} \hspace*{-.2cm} dk_T^2 
    \int_{1-z_h}^1 \hspace*{-.4cm}dz\, 
    \alpha\nu \frac{dN_g}{dk_T^2dz} \Theta\left(t-t_c\right) \ ,
 \label{eq:avenKop}
\end{align}
where $dN_g/dk_T^2dz = \alpha_s(k_T^2)/(3\pi)\, 1/(z k_T^2)$  
is the Gunion-Bertsch spectrum of radiated gluons
\cite{Gunion:1981qs}. The upper limit is 
imposed by the fact that gluons with $k_T>Q$ should be considered part
of the struck quark \cite{Dokshitzer:1991wu}. Next, the gluon (with
momentum $k_T$) splits into a $q\bar q$ pair. In the large-$N_c$ approximation, 
the antiquark and the struck quark form a colourless dipole, which
is identified as a prehadron. 
The prehadron production time is
identified with the coherence time of the gluon (rather than with the
$q\bar q$ splitting time) and
hadron formation is computed by the overlap of the
$q\bar q$ dipole with the hadron light-cone wave function $\Psi_h$.
Assuming that the $q$ and the $\bar q$ in the pair share the same
amount of gluon energy and transverse momentum,
one can compute the probability distribution $W(t,z_h,Q^2,\nu)$
that the prehadron is formed at a time $t$ after the $\gamma^*q$
interaction:
\begin{align}\begin{split}
  W(t,z_h,Q^2,\nu) = & 
    N \int_0^1 \frac{d\alpha}{\alpha} 
    \int_{\Lambda_{QCD}^2}^{Q^2} \frac{dk_T^2}{k_T^2}     
    \delta\left[z_h-\left(1-\frac{\alpha}{2} \right) \frac{E_q(t)}{\nu} \right]
    \frac{\exp(-t/t_c)}{t_c} \\
  & \times 
    \left|\Psi_h\left(\frac{\alpha}{2-\alpha}, \frac 34 k_T \right)\right|^2 
    \exp\left(-\tilde N_g(z_h,t,Q^2,\nu)\right) \ ,
\label{eq:Wkopel}
\end{split}\end{align}
where, $E_q(t)=\nu-\Delta E(t)$, $\Psi_h$ is the hadron light-cone
hadron wave function, $\tilde N_g$ is the number of gluons
radiated within a time $t$, and $N$ is a normalisation factor.
Numerical results are presented in Fig.~\ref{fig:tpdipolemodel}.    
Integrating $W$ over $t$ one obtains the vacuum fragmentation function for
the leading hadron which compares favourably with global fit FF from
Refs.\cite{Kniehl:2000fe,Kretzer:2000yf} in the range $z_h=0.5-0.9$
and $Q^2=2-10$ GeV$^2$.

\begin{figure}[tb]
  \vspace*{-.2cm}
  \begin{center}
  \includegraphics[width=6cm,origin=c]{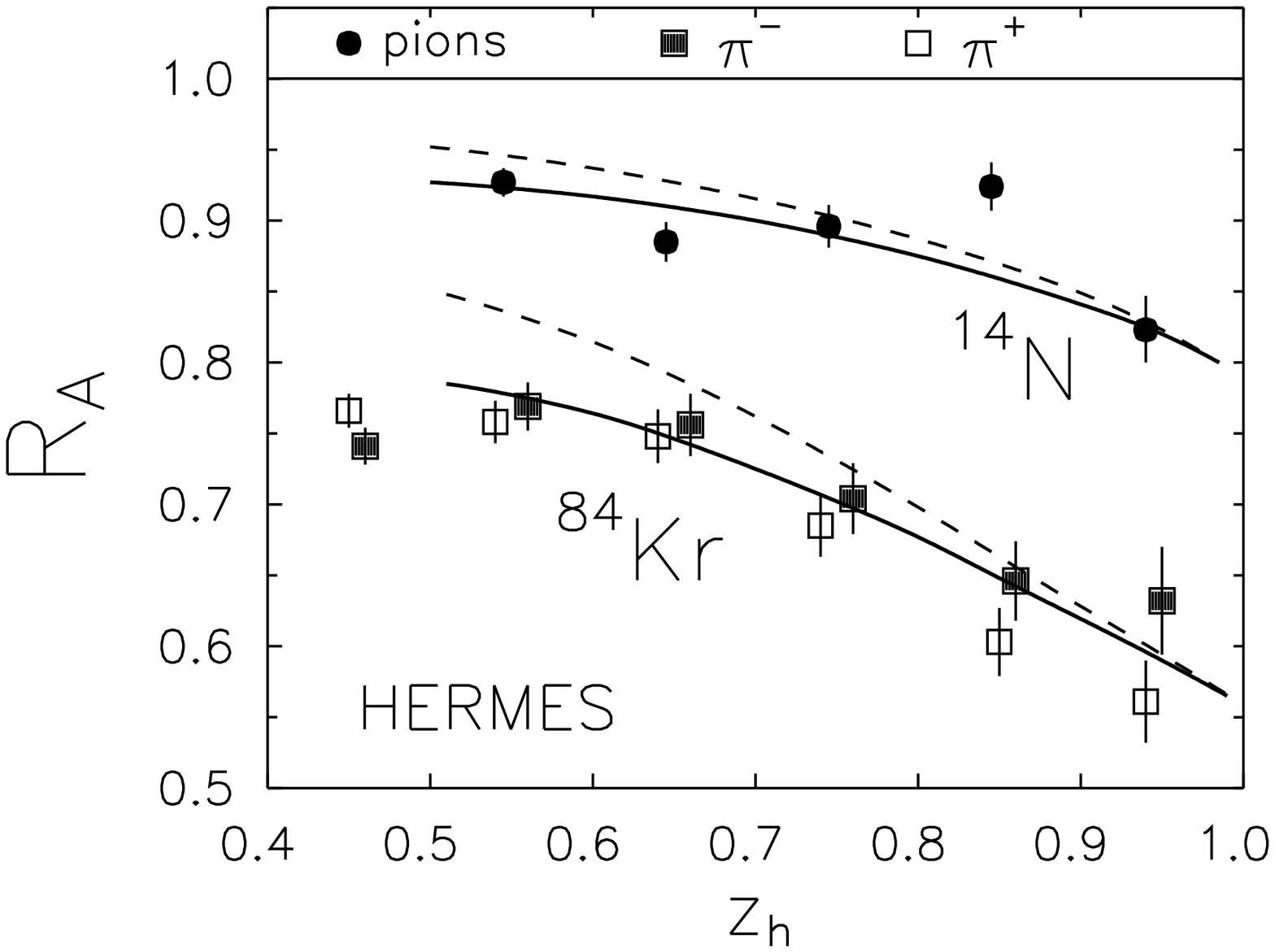}
  \includegraphics[width=6cm,origin=c]{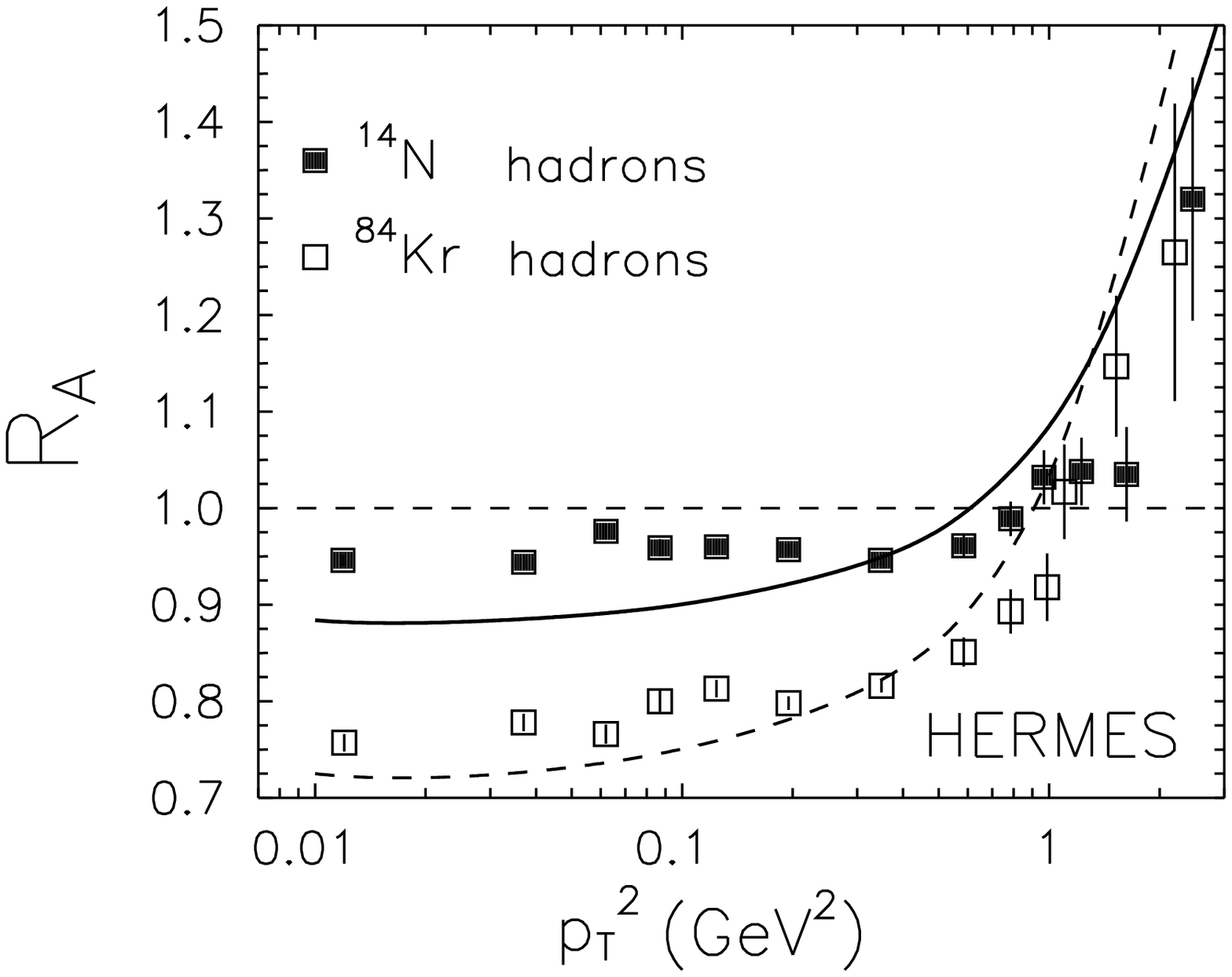}
  \end{center}
 \vspace*{-0.3cm}
 \caption[]{
   Colour dipole model~\cite{Kopeliovich:2003py} (dashed: absorption only, 
   solid: absorption and induced energy loss) compared to
   HERMES hadron multiplicity ratios on $N$ and $Kr$ targets~\cite{Airapetian:2003mi}. 
   {\it Left:} $z_h$ distribution. {\it Right:} $p_T$ distributions. 
 \label{fig:RMdipolemodel}
 }
\end{figure}

The $q\bar q$ dipole, which is assumed to be formed with a Gaussian
transverse size around an average $\vev{R_l^2} \propto 1/Q^2$,
propagates through the nucleus and
fluctuates in size. According to colour transparency,
see Sect.~\ref{sec:Q2timeevol}, it interacts with the nucleus with a cross
section $\sigma_{\bar qq} = C(E_h) r^2$. All effects of fluctuations
and nuclear interactions are computed in a path-integral formalism for
dipole propagation in QCD matter
\cite{Kopeliovich:1998nw,Kopeliovich:1999am}. 
Finally, the effective in-medium
fragmentation function reads
\begin{align}
  D_A^h(z_h,Q^2,\nu) & = \int db^2 \int_{-\infty}^{\infty} dy
  \rho_A(\vec b,y) \int_0^\infty dt\, W(t,z_h,Q^2,\nu)\, Tr(b,y+t)
  \, 
\end{align}
where $Tr(b,y+t,\infty)$, called nuclear transparency in
\cite{Kopeliovich:2003py},  is the probability for a prehadron not to
be absorbed in the nucleus. The hadron attenuation ratio can then be
approximated as $R_M^h \approx D_A^h / D_D^h$.

Finally, medium-induced energy loss is included in the model as an increase in 
the parton energy loss proportional to the quark transverse
momentum broadening $\Delta p_T^2$,
according to the relation
$
   \Delta E_\text{ind} = \frac{3}{8} \alpha_s \Delta p_T^2 L 
$,
derived in Baier {\it et al.} (BDMPS) \cite{Baier:1996sk}.
(Note that this formula is valid for an 
asymptotically large medium, which is not the case in $e+A$ collisions; 
the authors of~\cite{Kopeliovich:2003py} argue that
finite medium size corrections are small at HERMES energy, but their
effect might in fact be non-negligible, see Fig.~\ref{fig:AAgeom}.)
The induced energy loss modifies the production time distribution via
\begin{align}\begin{split}
  & \Delta E \ra \Delta E + \Delta E_\text{ind} \\
  & Q^2 \ra Q^2 +  \Delta p_T^2 \ ,
\end{split}\end{align}
with $\Delta p_T^2$ computed in the colour dipole formalism
\cite{Baier:1998kq,Wiedemann:2000tf}: 
$
  \Delta p_T^2 = 2 C(E_q) \rho_A L
$.
This leads to an additional, but subleading, hadron suppression
compared to the effect of prehadron absorption, as shown by the two curves in
Fig.~\ref{fig:RMdipolemodel} left. 

The parameters of the model are fitted to other processes than hadron
production in $l+A$ collisions and, in this sense, the approach 
can be considered parameter-free. 
The comparison to HERMES experimental data for $\pi^\pm$
production is shown in Fig.~\ref{fig:RMdipolemodel}. The model can also
describe the smaller $K^+$ suppression compared to $\pi^\pm$ (not
shown here), as well as the EMC data at
$z_h > 0.5$. Use of the colour dipole formalism allows also the
computation of the Cronin effect, shown against data in
Fig.~\ref{fig:RMdipolemodel} right. The $Q^2$ dependence of $R_M$ is
discussed in Sect.~\ref{sec:Q2timeevol} (see Fig.~\ref{fig:Q2abs}).
An interesting consequence of this formalism is that the prehadron
production time is inversely proportional to $Q^2$:
\begin{align}
  \vev{t_{preh}} \propto (1-z_h) \frac{z_h \nu}{Q^2} \ .
 \label{eq:tstarQ2}
\end{align}
where $\nu \approx E_q$. Technically, this arises from the upper limit
of integration in Eq.~\eqref{eq:Wkopel}. Physically, it is interpreted saying
that a quark which is struck by a photon of large virtuality radiates
more intensely than for a lower virtuality: as a consequence, it will
be able to  travel only a shorter distance before hadronising at a given
$z_h$. In $p+A$ and $A+A$ collisions, $Q^2 \propto
p_{Th}^2$ and for midrapidity hadrons $E_q \approx p_{Th}$, hence 
$
  \vev{t_{preh}} \propto 1/p_T
$.
So, the prehadron production time actually shrinks with increasing
$p_T$, instead of increasing because of the large Lorentz boost. 
This would imply that prehadrons are formed well 
inside the hot medium produced in $A+A$ collisions at any $p_T$.  
Experimental confirmation of Eq.~\eqref{eq:tstarQ2} is thus very
important. It can be accomplished in a direct way by measuring the
hadron $p_T$-broadening, or indirectly with a scaling analysis of
$R_M$, see Sect.~\ref{sec:measuretp}.

\begin{figure}[tb]
  \centering
  \includegraphics[width=5.5cm,angle=270]{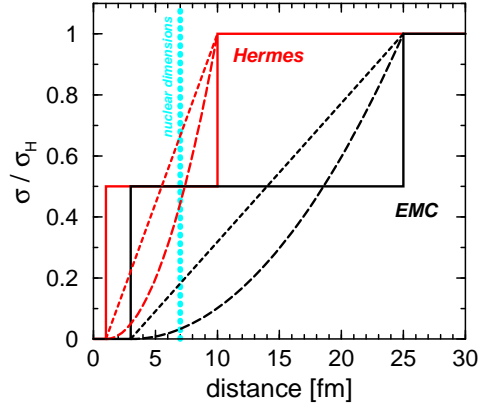}
  \caption{Sketch of the time evolution of the (pre)hadronic cross
    section for arbitrarily chosen formation times. Solid lines are
    for constant $\sigma_{preh}$, dotted and dashed lines are
    Eq.~\eqref{eq:linquad} for $n=1,2$. A typical nuclear distance of
    7 fm is indicated as a vertical dashed line. Figure taken from
    Ref.~\cite{Gallmeister:2007an}. } 
  \label{fig:sig*time}
\end{figure}

\subsection{The GiBUU transport Monte Carlo} \ \\
\label{sec:MCstringmodels}

In Refs.~\cite{Falter:2004uc,Gallmeister:2007an} a Monte Carlo
event generator has been used to explore hadron attenuation in
$\ell+A$ collisions in the framework of a transport model based on the
Boltzmann-Uehling-Uhlenbeck (BUU) equation.
In this approach, the lepton-nucleus interaction is split into
two parts: 
\begin{enumerate}
\item 
The  exchanged virtual photon produces a final state $X$ which
is determined using the Monte Carlo generators PYTHIA and
FRITIOF based on the Lund string model. 
Nuclear effects like binding energies, Fermi 
motion, Pauli blocking and coherence length effects that lead to
nuclear shadowing are also implemented.
\item 
The state $X$ is propagated through the nuclear target using a BUU
transport model \cite{Falter:2004uc,Effenberger:1999ay} which 
-- through  a probabilistic coupled-channel computation -- accounts for
particle creation, annihilation and elastic scattering in the final-state interactions. 
\end{enumerate}
The latest implementation of the code is known as GiBUU~\cite{Gallmeister:2007an}. 
The space-time evolution of hadronisation has been explored 
using two concepts of prehadron differing mainly in
the treatment of the production time $t_{preh}$: set to zero
in transport models~\cite{Effenberger:1999jc,Cassing:1999es,Bass:1998ca}, 
and depending on the energy and momentum of the string fragments
in the Lund model, see Sections~\ref{sec:formationtimes} and~\ref{sec:earlystrings}. 

In Ref.~\cite{Falter:2004uc}, the ``transport model'' view is adopted.
The string decay into colour neutral
prehadrons is assumed to happen instantaneously, hence
$t_{preh}=0$ fm. Hadrons are assumed to form after a fixed proper formation time
$\tau_f = 0.5$ fm in the hadron rest frame, 
which is then boosted to the laboratory frame,
\begin{align}
  t_h=\frac{z_h\nu}{m_h} \tau_f \ .  
 \label{eq:t_h_transport}
\end{align}
Between the prehadron and hadron formation times, 
only beam and target remnants, i.e.,
hadrons containing the valence quark of the struck nucleon or of the
resolved photon, are
allowed to interact with the rest of the nucleus. Specifically, the
prehadron cross section is assigned according to the constituent quark
model:
\begin{align}
  \sigma_{preh}^\text{baryon} = \frac{n_\text{org}}{3} \sigma_h^\text{baryon}
  \hspace*{0.8cm}
  \sigma_{preh}^\text{meson}  = \frac{n_\text{org}}{2} \sigma_h^\text{meson} 
 \label{eq:trmodxsec}
\end{align}
where $n_\text{org}$ denotes the number of quarks or antiquarks in the
hadron coming from the beam or target nucleon. This model can
describe fairly well HERMES data for $\pi^\pm$, $\pi^0$ and $K^\pm$ at
large $z_h \gtrsim 0.4$. For smaller $z_h$ some discrepancy for $K^+$
production on light nuclei emerges, related in part to the assignment
of the pre-kaon cross section \eqref{eq:trmodxsec}, in part to
feed-down due to the decay of diffractively produced $\rho$ mesons
into $K^+K^-$, or due to $\pi N$ interactions. The results for baryons
are less good, especially for $p$ production, which cannot explain the
strong rise of experimental $R_M^p$ at $z_h \lesssim 0.5$.
However, the biggest
challenge for this space-time scenario is in connection to EMC data,
whose suppression is vastly overestimated
due to the unlikely assumption of an instantaneous conversion
of string fragments into colour neutral prehadrons and of an
instantaneous jump from the prehadron to the hadron
cross-sections at time $t_h$. 

\begin{figure}[tb]
  \centering
  \includegraphics[width=10cm,angle=0]{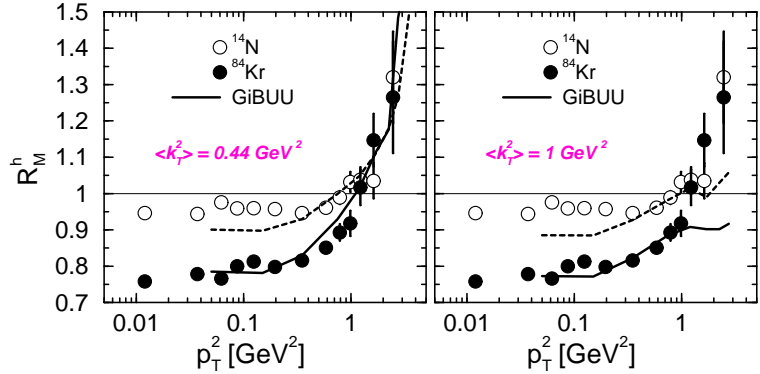}
  \caption{Multiplicity ratio in the GiBUU model as a function of
    $p_T$. Computations are done with parton intrinsic momentum 
    $\vev{k_T^2} = 0.44 \text{ GeV}^2$ ({\it left}) and 
    $\vev{k_T^2} = 1 \text{ GeV}^2$ ({\it right}). HERMES data are from
    \cite{Airapetian:2003mi}. Plot
    taken from \cite{Gallmeister:2007an}. 
  }
  \label{fig:GiBUUCronin}
\end{figure}

\begin{figure}[tb]
  \centering
  \includegraphics[width=\linewidth]{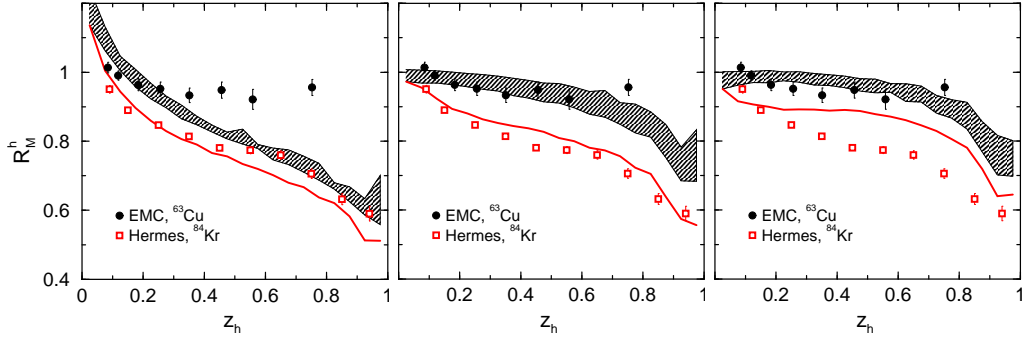}
  \caption{Multiplicity ratio in the GiBUU model, compared to
  HERMES at $E_{e}=27$ GeV \cite{Airapetian:2003mi} and EMC at
  $E_e = 100-280$ GeV \cite{Ashman:1991cx} data. The upper and lower
  bounds of the shaded band corresponds to higher and lower EMC beam
  energy, respectively. The prehadron cross section evolution is
  constant \eqref{eq:const}, linear and quadratic in time
  \eqref{eq:linquad}, from left to right. Plot
  taken from \cite{Gallmeister:2007an}.} 
  \label{fig:GiBUUtimeevol}
\end{figure}

\begin{figure}[t]
  \centering
  \parbox{4.4cm}{\hspace*{-.3cm}
    \includegraphics[width=4.4cm,angle=-90]{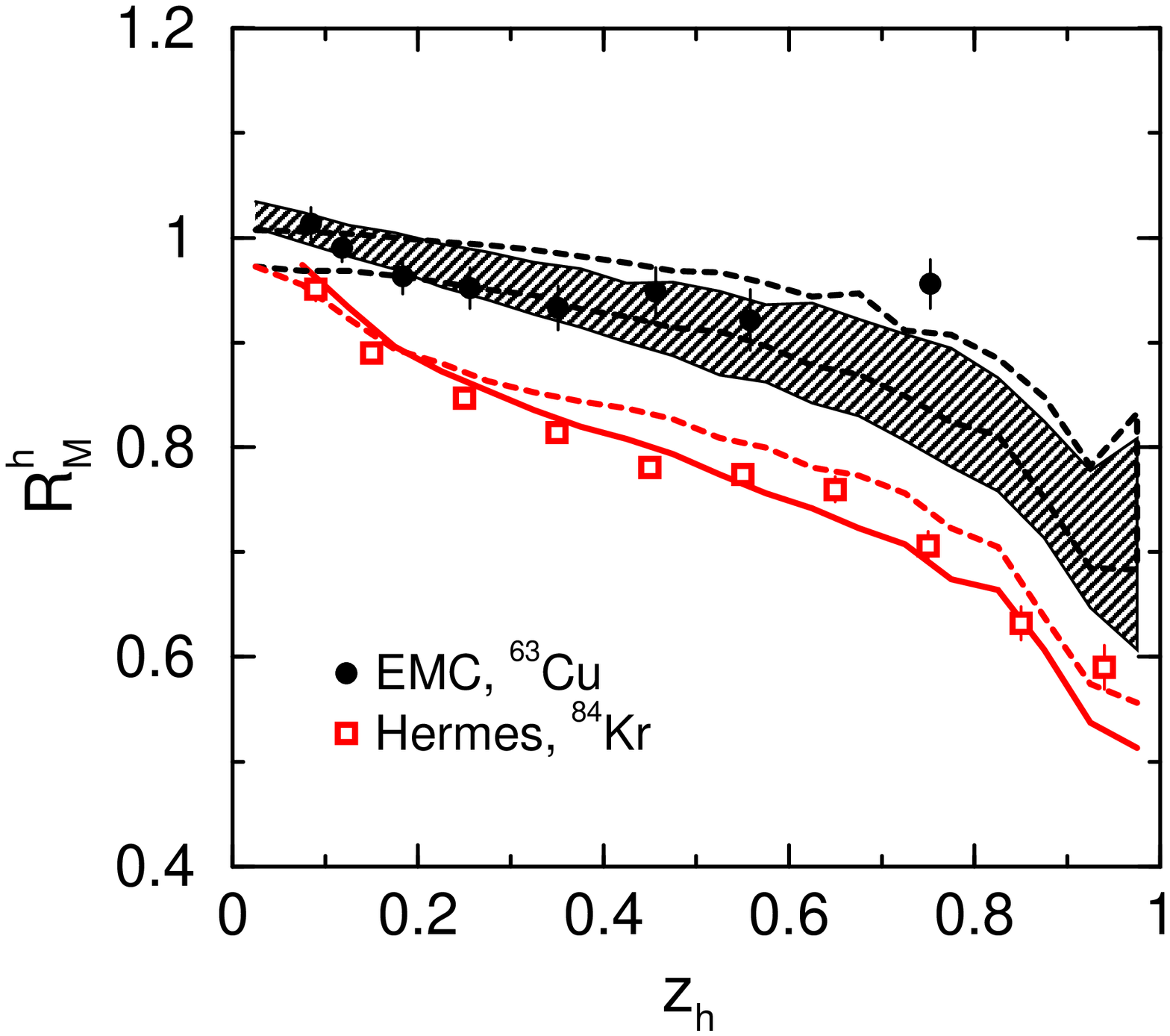}
  }
  \parbox{8.5cm}{\hspace*{.3cm}
    \includegraphics[width=8.5cm,height=4.5cm]{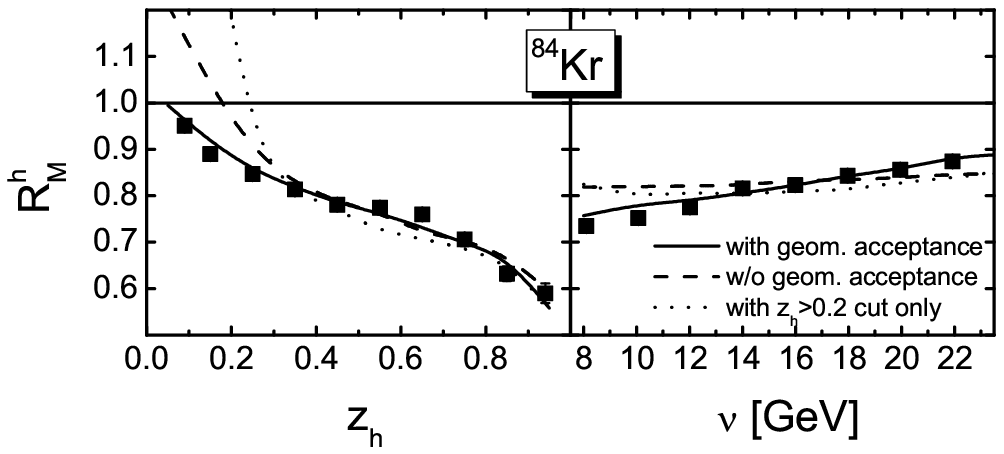}
  }
  \caption{{\it Left:} As in Fig.~\ref{fig:GiBUUtimeevol}, but for the quantum
    diffusion scenario \eqref{eq:quantumdiffusion} compared to the
    linear time-dependence scenario (dashed line). Plot taken from \cite{Gallmeister:2007an}.
   {\it Right:} Effect of the HERMES detector geometric acceptance on
    measurements of charged hadron attenuation. Solid line: full
    geometric and kinematic cuts. Dashed line: $4\pi$ detector with
    full kinematic cuts. Dotted line: $4\pi$ detector with $E_h \geq
    1.4$ GeV cut neglected. Figure taken from Ref.~\cite{Falter:2004uc}. 
  }
  \label{fig:GiBUUqd_detaccept}
\end{figure}

The solution proposed in
Ref.~\cite{Falter:2004uc} and fully worked out in
Ref.~\cite{Gallmeister:2007an} is to use the concept of prehadron
coming from the Lund model, and time-dependent prehadron cross
sections. The  PYTHIA generator 
is used as main source of information on the string fragmentation
process. In the Lund string model, we can actually associate three time
scales to each hadron: the times $t_{p1}$ and 
$t_{p2}$ at which the constituents of the hadron are created, and the
time $t_h$ at which the constituents meet and form the hadron. The
prehadron production time is then defined as 
\begin{align}
  t_{preh} = \min (t_{p1},t_{p2}) \ .
\end{align}
This definition is different from what is commonly used, see
Sections~\ref{sec:earlystrings} and
\ref{sec:modernstrings}. Most notably, the rank-1 hadron
contains the struck quark, which preexists the $\gamma^*N$ interaction:
hence, $t_{preh}^\text{rank-1} = 0$. The determination of the prehadron and
hadron formation time is carried out on an event-by-event basis by
extracting the corresponding information from the PYTHIA generator as
described in detail in \cite{Gallmeister:2005ad}. An interesting
result is that the prehadron and hadron formation times are quite
insensitive to the hadron mass, contrary to the common assumption that
they scale with a Lorentz boost factor $\propto 1/m_h$.
The prehadron cross section includes the main features of colour transparency, and
is allowed to vary in time according to linear and quadratic
evolution, or to a quantum diffusion picture, see
Eqs.~\eqref{eq:linquad}-\eqref{eq:quantumdiffusion}.
The requirement of reproducing both HERMES and EMC data rules out the
quadratic evolution, and slightly favours the quantum diffusion picture
with $\sigma_{preh} \propto 1/Q^2$ at $t=t_{preh}$, instead of $\sigma_{preh} =
0$, see Figs.~\ref{fig:GiBUUtimeevol} and \ref{fig:GiBUUqd_detaccept} left. The model
performs well in describing the flavour dependence of $R_M$ at HERMES,
with the exception of a slight underestimate of $K^-$ attenuation
over the whole $z_h$ range, and the lack of a strong rise in the
proton sector at $z_h \lesssim 0.4$.
It can also describe well the
Cronin effect on hadron $p_T$ distributions, without need for a
broadening of the intrinsic parton transverse momentum $\vev{k_T}$ as
commonly assumed, see Fig.~\ref{fig:GiBUUCronin}.

An important lesson coming from these simulations is that the
geometrical acceptance of the experiment does not fully cancel out in
the double ratio which defines the hadron attenuation ratio $R_M^h$
Eq.~\eqref{eq:att}, especially at low $z_h\lesssim 0.4$. This is shown in
Fig.~\ref{fig:GiBUUqd_detaccept} right, where unidentified charged 
hadron attenuation on $Kr$ at HERMES \cite{Airapetian:2003mi}
is computed with a $4\pi$ detector (dashed line) and 
the HERMES detector geometric acceptance (solid line), which is
important at $z_h \lesssim 0.4$ \cite{Falter:2004uc}. 
In both computations the full kinematic cuts have been included; the
dotted line shows the effect of neglecting the $E_h \geq 1.4$ GeV
cut. 

\subsection{$Q^2$ evolution and time-dependence of the
  prehadron cross section} \ \\
\label{sec:Q2timeevol}

In the context of string models, hadron attenuation is typically
constant in $Q^2$. However, 
a dependence of the prehadron cross section on $Q^2$ is to be expected on
general grounds due to 
colour transparency (see e.g. \cite{Jain:1995dd}), which is a direct consequence
of the composite nature of hadrons being made of quarks so that hadrons
fluctuate between quark configurations of different sizes. In
\cite{Low:1975sv,Nussinov:1975mw,Greenberg:1992xz,Greenberg:1994yg} 
it was found that in scattering of colour
neutral objects, configurations with small transverse size $b$ have
cross section $\propto b^2$. The size of a colour neutral prehadron
is likely to be determined by the hard scale $Q^2$ of the process, so
that we may expect, at least for the leading prehadron,
$\sigma_{preh} \propto 1/Q^2$. 

After its production, the
prehadron must evolve in size until hadron formation; correspondingly,
its cross section will evolve in time from $\sigma_{preh}$ to $\sigma_h$.
Let us picture the prehadron as a $\bar q q$ colour dipole for
simplicity. In Ref.~\cite{Dokshitzer:1991wu}, it was shown that in a
classical picture the size of a pair of colour charges with average
$\sqrt{\vev{k_T^2}}$ evolves linearly in time. Therefore, due to colour
transparency, 
its cross section evolves quadratically: $\sigma_{preh} \propto t^2$. 
If one includes the uncertainty principle in these considerations, the
assumption of fixed transverse momentum is not valid, since as $r\ra
0$ we have $k_T \ra \infty$. As a consequence, the
rise of the cross section is softened and becomes linear in time:
$\sigma_{preh} \propto t$. On the other hand, the struck quark is not a
bare particle, but is still accompanied by those parts of the
original nucleon field with Fourier components satisfying $k_T^2 >
Q^2$. Hence, the time dependence of $\sigma_{preh}$ may be expected to lie
between linear and quadratic.

In the context of semi-inclusive hadron production in 
$\ell+A$ collisions, this idea has been
phenomenologically explored in
Refs.~\cite{Gallmeister:2007an,Falter:2004uc} using the GiBUU Monte Carlo
generator (see Section~\ref{sec:MCstringmodels} for more details) and in
\cite{Akopov:2004ap} in the context of the two-scale string model, see 
Section~\ref{sec:modernstrings}. The time evolution of the prehadron
cross section is also implemented by means of the colour dipole model 
in the perturbative fragmentation model of   
\cite{Kopeliovich:2003py}, which also includes a $Q^2$-dependent 
production time $t_{preh} \propto (1-z_h) \nu / Q^2$,
see Section~\ref{sec:perthadrmodel}.

In Ref.~\cite{Gallmeister:2007an}, four scenarios have been explored. In
the first, the prehadron cross section is kept fixed,
\begin{align}
  \sigma_{preh} = \frac{1}{2} \sigma_h \ ,
  \label{eq:const}
\end{align}
where the proportionality factor has been chosen to reproduce HERMES
data. (Note that different proportionality factors arise in different 
models due to different assumptions on the underlying space-time
development of hadronisation, e.g., $\sigma_{preh}=0.67 \sigma_h$ in
\cite{Accardi:2005jd}, where prehadron interactions are treated
differently.) In the next two scenarios, 
\begin{align}
  \sigma_{preh}(t) = \left( \frac{t-t_{preh}}{t_h-t_{preh}} \right)^n \sigma_h 
    \qquad \quad n=1,2 
  \label{eq:linquad}
\end{align}
corresponding to the quantum mechanical linear time dependence and the
classical quadratic time dependence discussed in 
\cite{Dokshitzer:1991wu}. Note that at $t=t_{preh}$, the prehadron
is created with zero cross section, and that no dependence on $Q^2$ is
included. The last scenario implements the quantum
diffusion picture of Ref.~\cite{Farrar:1988me}, which combines the
linear rise in time with a non-zero and $Q^2$-dependent value for the
initial prehadron cross section:
\begin{align}
  \sigma_{preh}(t) = \sigma_0 + (1-\sigma_0) \left( \frac{t-t_{preh}}{t_h-t_{preh}}
    \right) \sigma_h  
  \qquad \quad \sigma_0 = r_{\text{\it lead}} \frac{1\text{GeV}^2}{Q^2} \ ,
  \label{eq:quantumdiffusion}
\end{align}
where $r_{\text{\it lead}}$ is the number of the struck nucleon
valence quarks contained in the produced hadron divided by its number
of valence quarks. E.g, a leading pion has $r_{\text{\it lead}} = 1/2$
and a subleading one has $r_{\text{\it lead}} = 0$, see
Section~\ref{sec:MCstringmodels}. 
In all scenarios, the prehadronic
cross section is 0 before $t=t_{preh}$ and reaches the hadronic $\sigma_h$
at $t=t_h$, and includes the basic features of colour
transparency. These cross sections are sketched in
Figure~\ref{fig:sig*time}. While HERMES data alone do not give enough
leverage to distinguish between these time evolution scenarios
\cite{Falter:2004uc,Akopov:2004ap}, the combined analysis of EMC and
HERMES data singles out the linear evolution scenario
\cite{Gallmeister:2007an}, see
Fig.~\ref{fig:GiBUUtimeevol}. However, it cannot distinguish between an
initial $\sigma_{preh}=0$ or $\sigma_{preh}\propto 1/Q^2$, despite a large
difference in the average $\vev{Q^2}$ in the two cases, even though the
latter is slightly preferred, see Fig.~\ref{fig:GiBUUqd_detaccept} left. 
This result is in line with the slow variation of the attenuation
ratio $R_M$ with $Q^2$ observed at HERMES \cite{Airapetian:2007vu}. 

\begin{figure}
  \centering
  \includegraphics[width=6.5cm]{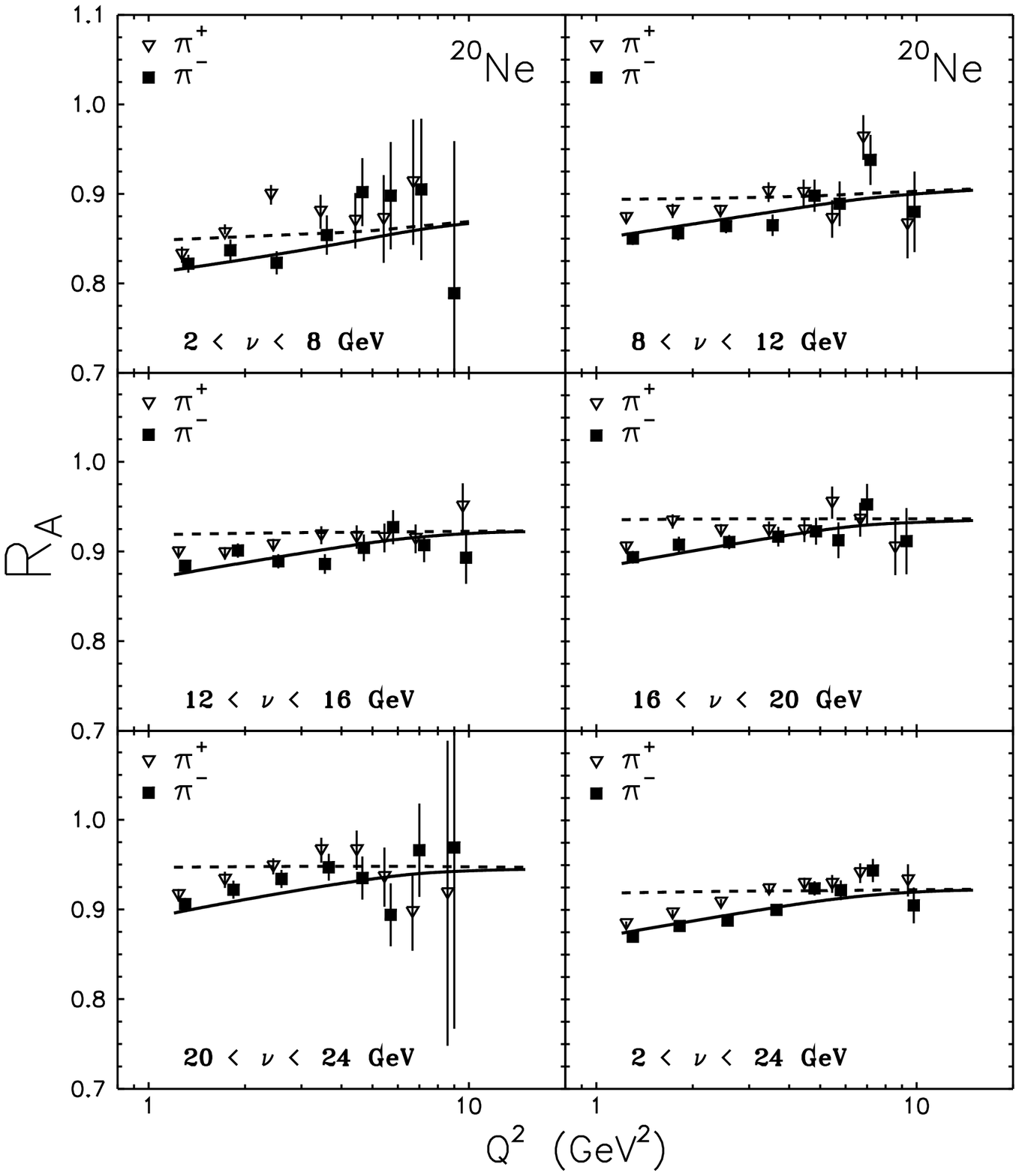}
  \parbox[b]{6.5cm}{
    \hspace*{.15cm}
    \includegraphics[width=6.5cm]{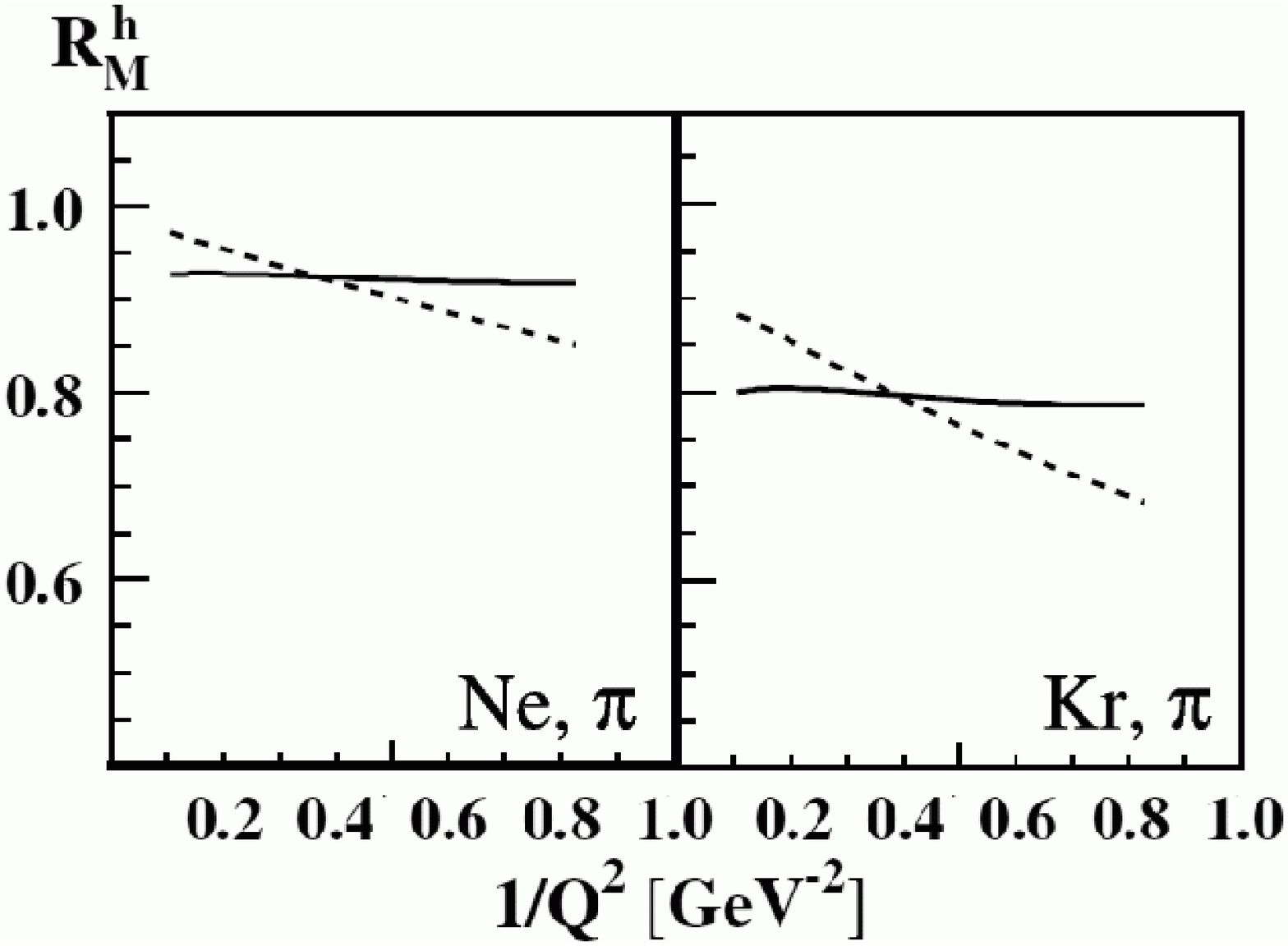} 
    \vskip.2cm
    \includegraphics[width=6.5cm]{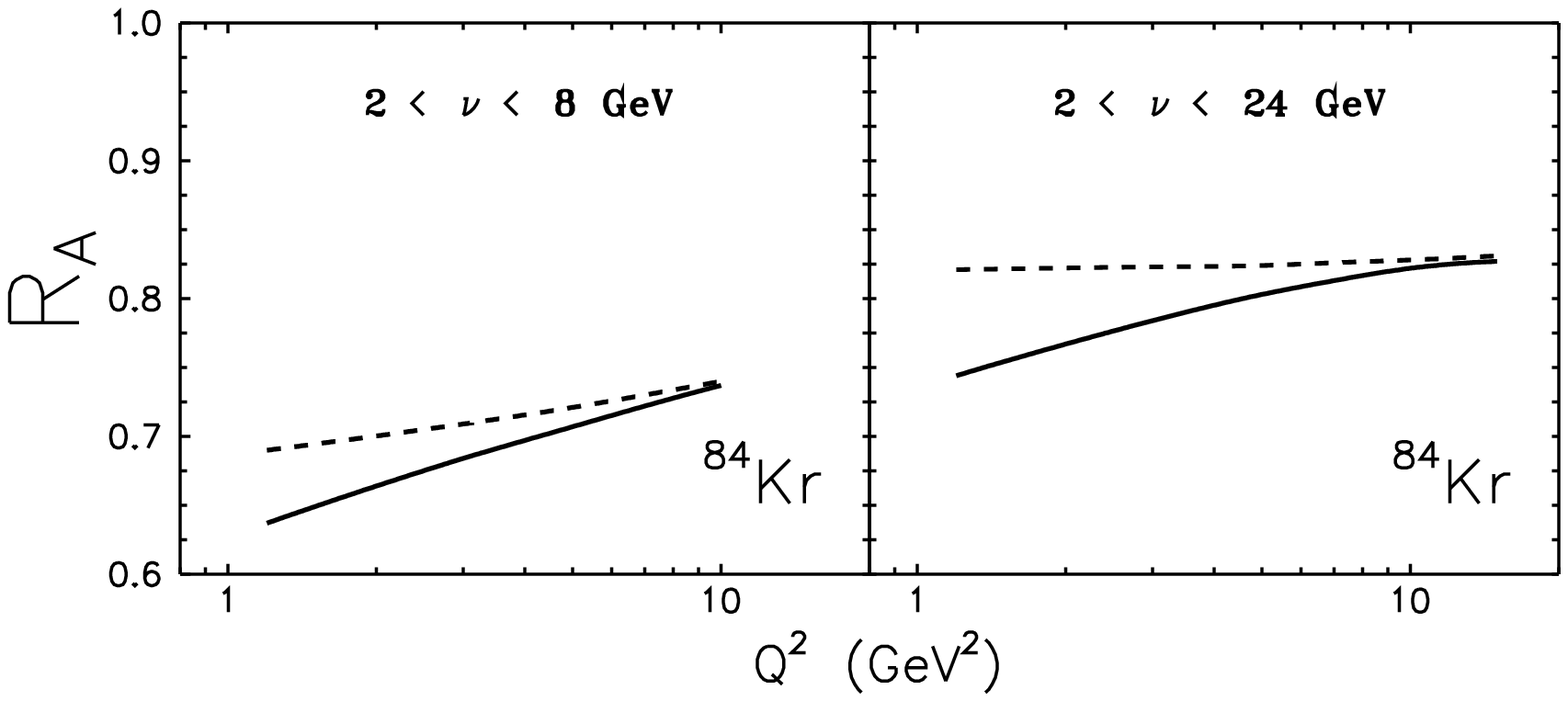}
  }
  \caption{
    {\it Left} and {\it bottom right}: $Q^2$ dependence of hadron
    attenuation in the colour dipole model of
    Ref.~\cite{Kopeliovich:2003py} for $Ne$ and $Kr$, respectively.
    Solid lines include medium-induced gluon bremsstrahlung, dashed
    lines do not. Points are preliminary HERMES data \cite{Airapetian:2007vu}.
    The two plots are taken from \cite{Kopeliovich:2003py}.
    {\it Top right:} $Q^2$ dependence of hadron attenuation in the
    absorption model of Ref.~\cite{Akopov:2004ap}. Solid lines
    computed in the improved two-scale model, with time evolution of
    the prehadron cross section. Dashed lines computed in the two-scale
    model without time evolution of the prehadron cross
    section. Plot adapted from \cite{Akopov:2004ap}.}
  \label{fig:Q2abs}
\end{figure}
 
An analysis of the $Q^2$ dependence of $R_M^h$ has been performed in 
\cite{Akopov:2004ap,Kopeliovich:2003py}. 
Based on colour transparency, one expects
$R_M = a + b/Q^2 + \mathcal{O}(1/Q^4)$ with negative $b$. However, if time
evolution of the prehadron cross section to reach the full hadronic one 
is sufficiently fast, its $Q^2$ dependence can become negligible.
This is shown in Fig.~\ref{fig:Q2abs}, top right, as difference
between the solid and dashed line, computed 
with a linear time dependence (Eq.~\eqref{eq:quantumdiffusion}) and no
time dependence respectively \cite{Akopov:2004ap}.  
In \cite{Kopeliovich:2003py}, where colour transparency and the time
evolution of the prehadron cross section are naturally implemented
in the light-cone dipole formalism, two further elements contribute to
the $Q^2$ dependence of $R_M^h$. The first is the $Q^2$ dependence of the
production time $t_{preh} \propto 1/Q^2$, which increases the
in-medium prehadron path length and consequently its attenuation with
increasing $Q^2$,
working in the opposite direction as colour transparency. The net
effect is an almost $Q^2$-independent attenuation ratio for light
nuclei, which gains a small slope on heavy targets or for low $\nu$
cuts, see Fig.~\ref{fig:Q2abs}. The second is
medium-induced gluon bremsstrahlung, which induces an additional 
attenuation at small $Q^2$, but whose effect disappears at large
$Q^2$, see Fig.~\ref{fig:Q2abs}.
A comparison of model computation with 
HERMES data on $Q^2$ distributions in light and medium-heavy nuclei
($Ne$, $Kr$, $Xe$) from Ref.~\cite{Airapetian:2007vu}, which show a
pretty slow variation of $R_M$ with $Q^2$, 
and the forthcoming JLab multi-differential
measurements of $z$ and $Q^2$ distributions will shed light on this problem.

In Ref.~\cite{Accardi:2002tv,Accardi:2005jd}, a different kind of
$Q^2$ dependence has been 
explored, namely, the possibility of a modification of
fragmentation functions induced by a partial deconfinement in the
nuclear wave function
\cite{DiasdeDeus:1985mg,Close:1984zn,Jaffe:1983zw}. It results in a
$Q^2$ rescaling of the FF: 
\begin{align}
  D\left(z_h,Q^2\right) \ra D\left(z_h,\xi_A(Q^2)Q^2\right) 
    \qquad \quad \xi_A(Q^2) = \left( \frac{\lambda_A}{\lambda_0} 
    \right)^{2\alpha_{s0}/\alpha_s(Q^2)} 
\end{align}
where $\alpha_{s0}$ is the strong coupling constant computed at the
scale at which DGLAP evolution starts, $\lambda_0$ is the nucleon
radius in the vacuum and $\lambda_A$ is the radius in the medium.
The rescaling factor $\xi_A$ depends on the amount of nucleon
overlap in the nuclear wave functions, and depends on $Q^2$ because
of DGLAP evolution \cite{Close:1984zn,Jaffe:1983zw}. 

The
phenomenological consequence is a 
multiplicity ratio which slightly decreases with $Q^2$
contrary to what is experimentally observed:
partial deconfinement does not seem to affect fragmentation functions
(even though it might modify the parton distribution functions, and
lead to a possible explanation of the EMC effect
\cite{Norton:2003cb}). This might have to do with the fact that
hadrons are typically formed outside the medium in the HERMES
kinematics, so that deconfinement is not affecting them.
At JLab, or for smaller values of $\nu$ at HERMES, 
hadrons will be in part formed inside the medium and partial
deconfinement may affect their fragmentation.

In summary, the $Q^2$ dependence of hadron attenuation is very
sensitive to the underlying dynamics of the hadronisation
process. However, since different physics inputs in theory models can
have similar effects on the computation of $R_M^h$, this observable
must be used in conjunction with others like the hadron 
transverse momentum broadening, and with the requirement to describe
attenuation in both low-energy experiments (CLAS,HERMES) and 
high-energy experiments (EMC), to sort out the physics 
underlying hadron attenuation.

\subsection{Prehadron absorption in hot QCD matter}\ \\
\label{sec:hotabsorption}

In Ref.~\cite{Cassing:2003sb,Gallmeister:2004iz}, prehadron absorption
in hot nuclear matter as a mechanism for jet quenching in $A+A$ collisions, 
has been considered in the context of a transport model similar
to the GiBUU model described in Sect.~\ref{sec:MCstringmodels}. 
The medium formation and evolution is described in the Hadron-String
Dynamics (HSD) approach \cite{Ehehalt:1996uq,Geiss:1998ki, Cassing:1999es}. 
The prehadron formation and
evolution is the same as in the GiBUU model, with the additional
assumption that hadrons are not allowed to be formed if the medium energy
density at their point in space and time is above the
critical energy density for a QGP formation, $\epsilon_c=1$ GeV/fm$^3$. 
This requirement mimics the effect of a QGP on high-energy hadrons,
even though the produced medium in the HSD simulations is always a hadron
gas. Note, however, that the energy-density cut is not assumed to
affect the prehadronic states. In addition, the model
phenomenologically includes the Cronin enhancement at intermediate
$p_{Th}$ by a nuclear enhancement of the intrinsic parton transverse
momentum. 

\begin{figure}
  \centering
  \includegraphics[width=7cm]{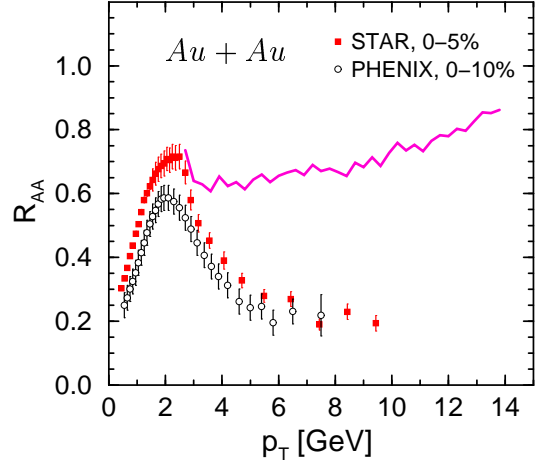}
  \caption{Suppression factor $R_{AA}$ as a
    function of $p_T\equiv p_{Th}$, for
    midrapidity charged hadrons in 0-5\% central $Au+Au$
    collisions at RHIC top energy, $\sqrtsnn=200$ GeV.
    The solid curve is the hadron quenching computed when 
    the leading prehadron 
    cross section rises linearly in time according to
    Eq.~\eqref{eq:linquad} with $t_{preh}=0$ and 
    $n=1$. 
    Experimental data are from Refs.~\cite{Adler:2003au,Adams:2003kv}.
    The figure is taken from Ref.~\cite{Cassing:2003sb}.
  }
  \label{fig:abshotmatter}
\end{figure}

Two different scenarios for the space-time
evolution of prehadrons have been used in the simulations.
In the first scenario, the leading prehadron cross section is constant
and the hadron formation proper time is kept fixed
at $\tau=0.5-0.8$ fm.
The resulting hadron suppression is mostly due to the interaction of
the leading prehadron with the medium, and is sufficient to explain
most of the hadron quenching at RHIC.
This is interpreted by the authors as evidence that prehadronic
final-state interactions can result in the observed jet quenching results.
However, when the leading prehadron cross section grows linearly with
time, as requested by a consistent description of HERMES and EMC data
for nuclear DIS \cite{Gallmeister:2007an}, the resulting hadron
quenching is clearly 
insufficient to explain data for RHIC central $Au+Au$ collisions,
see Fig.~\ref{fig:abshotmatter}. 
Therefore, prehadronic final state
interactions seem ruled out as main mechanism for hadron
quenching at RHIC, while they may in fact be responsible 
for hadron quenching in $e+A$ collisions. 

The reason may be that, contrary to the assumptions
in the HSD model, prehadrons cannot be formed in the QGP phase
($\epsilon > \epsilon_c$) because of the screening of the colour force
needed to bind them. If this is the case, the struck quark remains bare
and interacts with the medium until it escapes from it or until the medium enters the hadron gas
phase. Only at this point the hadronisation process can start 
but at this stage the medium is too dilute and the prehadronic
interactions are too weak to substantially contribute.
On the other hand, in the QGP phase, partonic energy loss can be important
because of the large colour charge density. 
If this explanation is correct, it would be interesting to confront the HSD
prehadron absorption model with $R_{PbPb}(p_T)$ data at SPS energies 
(note, however, that  we currently lack high-$p_T$ $p+p$ spectra at around 
$\sqrtsnn$~=~17~GeV~\cite{d'Enterria:2004ig})
where the QGP is formed in a more restricted range of centralities, if at all, and
is shorter lived than at RHIC. A good description of peripheral
collisions data may validate the model, so that a substantial overestimate
of central collision data might be interpreted as evidence for the
formation of a QGP.



\section{Challenges and outlook } 
\label{sec:future}
\label{sec:discussion}

In this last Section, we discuss observables that can help to better understand
parton propagation and fragmentation in QCD matter and, in particular, the relative
role of perturbative parton energy loss (Section~\ref{sec:parton}) and prehadron absorption 
(Section~\ref{sec:prehadron}) mechanisms. 
We conclude with a review of proposed future facilities and experiments.

\subsection{Partonic versus (pre)hadronic energy loss in nuclear DIS}\ \\ 
\label{sec:measuretp}

Despite very different assumptions on the
parton lifetime, the models presented in
Section~\ref{sec:parton}~(resp. \ref{sec:prehadron}) based on partonic (resp. prehadronic)
degrees of freedom, usually reproduce quite well the experimental
results on the multiplicity ratio $R_M^h$ in $e+A$ collisions.
It is therefore necessary to study observables which are more
directly sensitive to the space-time evolution of the hadronisation
process and can discriminate between partonic and pre-hadronic
contributions. 
Candidates suggested in the literature include the atomic
mass $A$ dependence of the nuclear attenuation, its production time
scaling, and the hadron transverse momentum broadening.

\subsubsection{\it Mass number dependence}-- 
\label{sec:Adep}
The mass number dependence of the nuclear attenuation ratio $1-R^h_M$
had been suggested as an observable sensitive to the different 
mechanisms involved in the hadronisation process. 
To leading order in $A^{1/3}$, one expects $1-R^h_M \propto
A^{2/3}$ in energy loss models because the average energy loss
$\Delta E_q \propto \vev{L_q^2} \propto A^{2/3}$ for partons with asymptotic 
energies, due to the LPM interference in QCD \cite{Baier:1996sk}.
On the other hand, in absorption models the survival probability is
proportional to the amount of traversed matter, so that $1-R^h_M
\propto \vev{L_A} \propto A^{1/3}$~\cite{Nikolaev:1979an,Bialas:1980at,Bialas:1983kn}. 
However, it has been recently
shown~\cite{Accardi:2005jd,Accardi:2005mm,Blok:2005vi,Gallmeister:2007an} 
that this simplistic argument does not really hold for absorption models, where
additional powers of $A^{1/3}$ need to be introduced to account for 
non-zero values of $\vev{t_{preh}}$, thus predicting
$R_M^h \propto A^{2/3} + \cO{A}$ as in energy loss models. 
Moreover, hadron attenuation is not likely to be describable by a simple
$A^\alpha$ power-law for nuclei with $A\gtrsim 80$ \cite{Accardi:2006ea}.

\begin{figure}[tb]
  \centering
  \parbox[c]{6.5cm}{
   \includegraphics[width=6.5cm,origin=t,clip=true]
                   {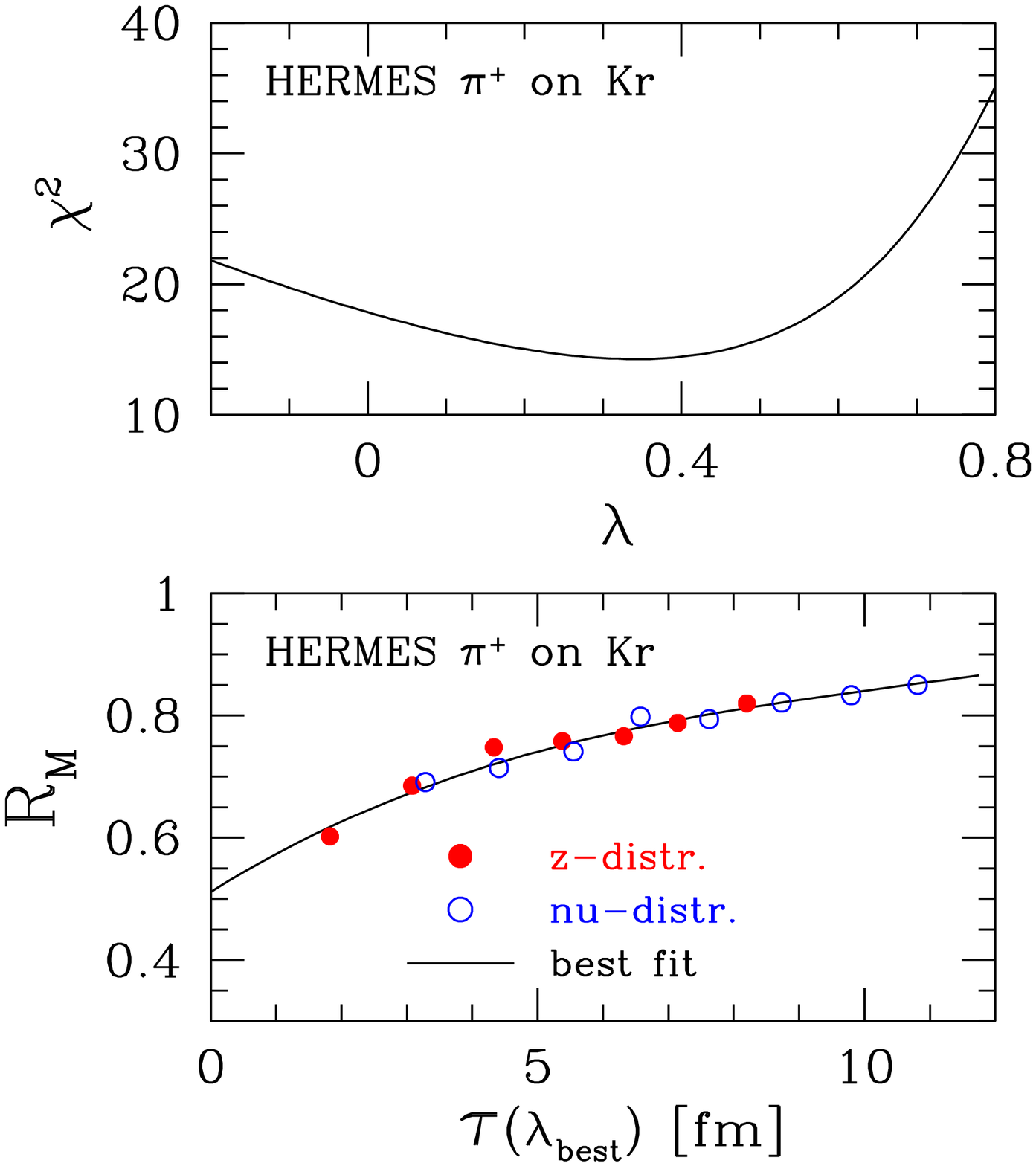}
  }
  \parbox[c]{6.5cm}{
    \includegraphics[width=7cm,origin=t,clip=true]
                    {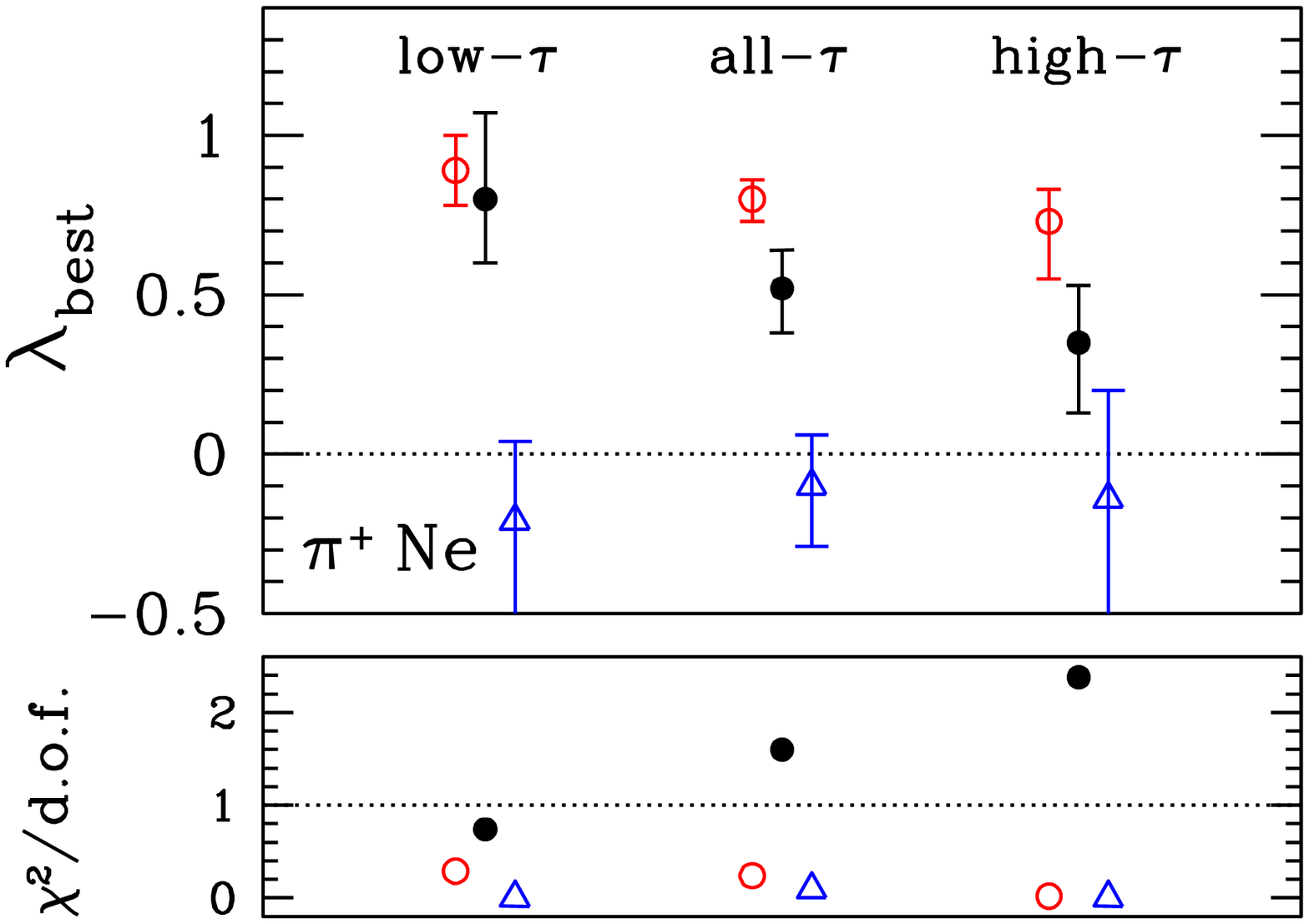}  
    \vskip.5cm  
  }
  \vspace*{0cm}
 \caption[]{
   Results of the production-time $\tau$ scaling analysis of HERMES
   data on $\pi^+$ production on $Kr$ targets \cite{Airapetian:2003mi}. 
   {\it Left:} $z$- and $\nu$- distributions as a function of $\tau$
   with corresponding $\chi^2$ as a function of the exponent
   $\lambda$.
   {\it Right:} Scaling exponent and $\chi^2$ per degree of freedom for   
   ``low-$\tau$'' and ``high-$\tau$'' data sets compared to the full
   data set for pion production on $Ne$ targets corresponding to
   ``medium-$\tau$''. Black disks: $\pi^+$ from
   preliminary HERMES data \cite{ElbakianDIS03}. Red circles: $\pi^+$ from the
   AGMP absorption model \cite{Accardi:2005jd,Accardi:2006ea}. Blue triangles:
   $\pi^\pm$ from the energy loss model of
   Refs.~\cite{Arleo:2003jz,Accardi:2006ea}. 
 }
 \label{fig:tauscaling}
\end{figure}

\subsubsection{\it Production time scaling in nDIS} --
\label{sec:scaling}
In Ref.~\cite{Accardi:2006qs} it is conjectured that $R_M$ should depend 
on a scaling variable $\tau$ defined as
\begin{align}
  \tau & = C\, z_h^\lambda (1-z_h)\mu \nu \ ,
 \label{eq:scalingvar}
\end{align}
rather than on $z_h$ and $\nu$ separately, i.e.
\begin{align}
  R_M = R_M\big[\tau(z_h,\nu)\big].
 \label{eq:RMscaling}
\end{align}
The scaling exponents $\lambda$ and $\mu$ can be obtained by a best fit
analysis of data or theoretical computations, whereas
the proportionality constant $C$ cannot be determined from the fit. 
Both absorption and energy loss models can be distinguished by 
the value of the scaling exponent: a positive  $\lambda$ value is
characteristic of absorption models, which assume a short
$\vev{t_{preh}}$ with a functional form similar to Eq.~\eqref{eq:scalingvar};
while a zero or negative $\lambda\lesssim0$ is predicted within energy loss
models due to energy conservation (a parton cannot radiate more energy
than $(1-z_h)\nu$).

The fits performed in Ref.~\cite{Accardi:2006qs} and similarly in
Ref.~\cite{Airapetian:2007vu} (with $\mu=1$ taken from analytical Lund
model computations \cite{Bialas:1986cf,Accardi:2002tv}) show that
$\tau$ is indeed a good scaling variable to analyze the HERMES
data,  see Fig.~\ref{fig:tauscaling} (left), 
and furthermore it lives up to the promise of separating
energy loss and absorption models.
The charged pion data exhibit globally a clear $\lambda = 0.5
\pm 0.15$, which, being positive and different from zero,
indicates that pre-hadronic effects significantly contribute to
$R_M^{\pi}$ and allows to interpret $\tau \approx \vev{t_{preh}}$. 
Moreover, it has been shown by applying different cuts on $z_h$ and $\nu$  
that the $\lambda$ exponent derived from experimental data decreases
with increasing $\tau$, thus suggesting an increasing contribution 
of partonic energy loss, see  Fig.~\ref{fig:tauscaling} (right). 
Finally, the slight pion-charge and nuclear-mass number dependence of
the  $\lambda$ exponent observed in Ref.~\cite{Accardi:2006qs} may
indicate additional dynamics beside production time and prehadron
absorption.  
Fits to other hadron flavors, such as kaons, protons and antiprotons, are less
conclusive due to the limited statistics. 

In conclusion, the $\tau$-scaling analysis of  $R_M$ seems
a promising tool for studying energy-loss and formation-time
contributions to hadron attenuation. 
However, new higher precision data sets in a wider range of
$\nu$, mass number and hadron flavors are necessary to determine a
possible scaling with $\mu\neq1$ and to fully exploit this observation.

\subsubsection{\it $p_T$-broadening and prehadron formation time} --
\label{sec:discussion-ptbroad}
The scaling analysis for $R_M$ just described 
gives only indirect evidence of the pre-hadron formation time effects, 
but does not allow to measure the absolute scale of $\vev{t_{preh}}$.
A more sensitive observable to the pre-hadron formation time
is the hadron transverse momentum broadening $\Delta\vev{p_T^2}$ in
nDIS compared to a proton or deuteron target
\cite{Kopeliovich:2003py,Kopeliovich:2006xy}. 
The $p_T$-broadening is commonly believed to arise
essentially during parton propagation because the prehadron is
the quark momentum
broadening $\Delta p_T^2$ is proportional to the quark path-length
in the nucleus \cite{Baier:1996sk}, and the prehadron is 
supposed to have a negligible elastic cross section~\cite{Kopeliovich:2003py}.
Under these assumptions, and remembering that hadron and parton
transverse momenta are related by $p_{T} = z k_T$
\cite{Collins:1981uw} with $z\approx z_h$ at LO, we obtain 
\begin{align}
  \vev{\Delta \vev{p_T^2}} \approx z_h^2 \qhat \vev{t_{preh}} 
\label{eq:ptbroadzz}
\end{align}
with $\qhat$ the partonic transport coefficient, and the prehadron
formation time obtained from the discussed scaling analysis:
\begin{align}
  \vev{t_{preh}} = C z_h^{0.5\pm0.15} (1-z_h) \nu \ ,
\label{eq:tpre}
\end{align}
with $C$ setting the overall scale of the production time
to be determined from broadening data.

\begin{figure}[tb]
  \begin{center}
    \includegraphics[width=6.5cm,clip=true,bb=20 330 600 730]
                   {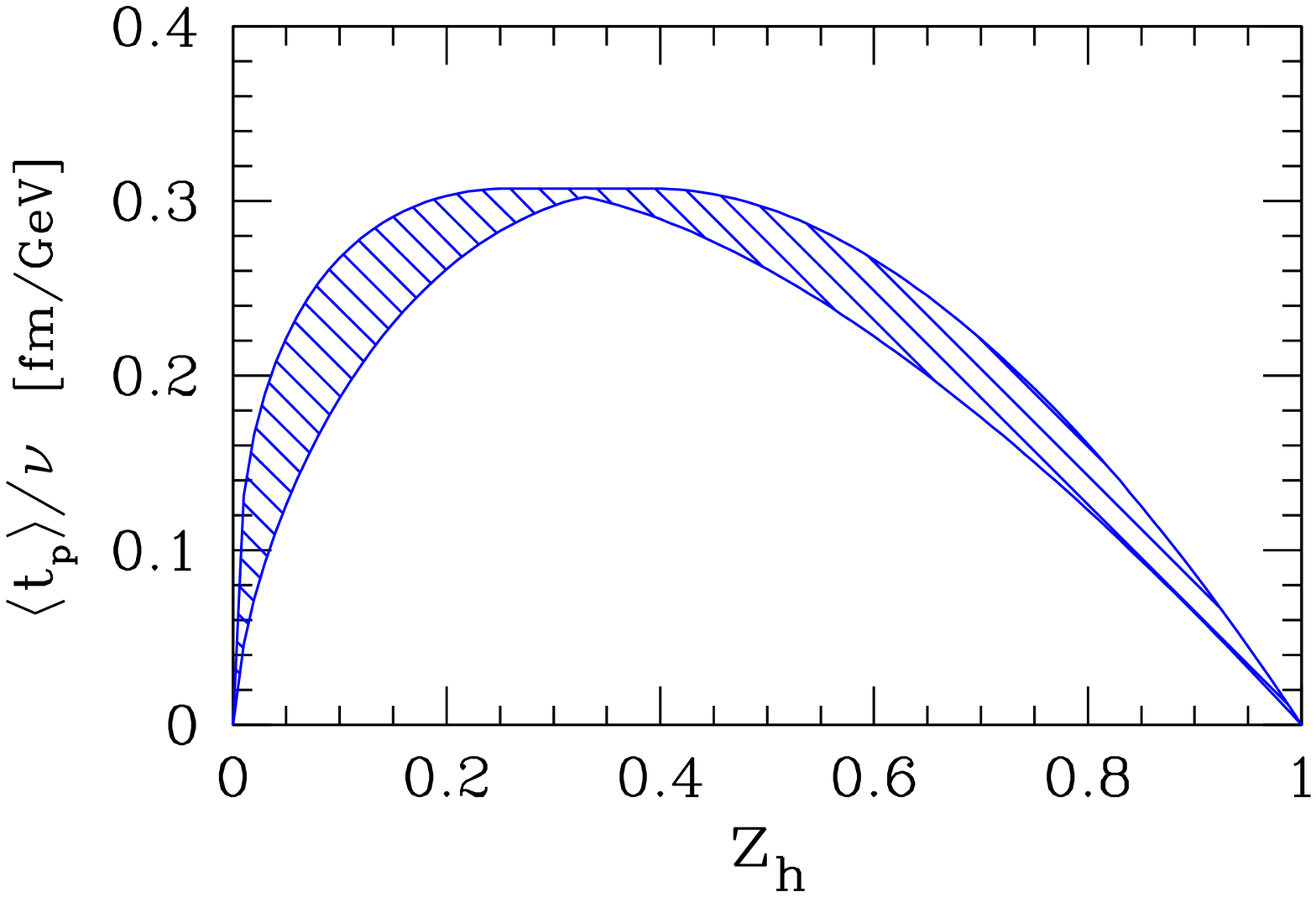}
    \includegraphics[width=6.5cm]{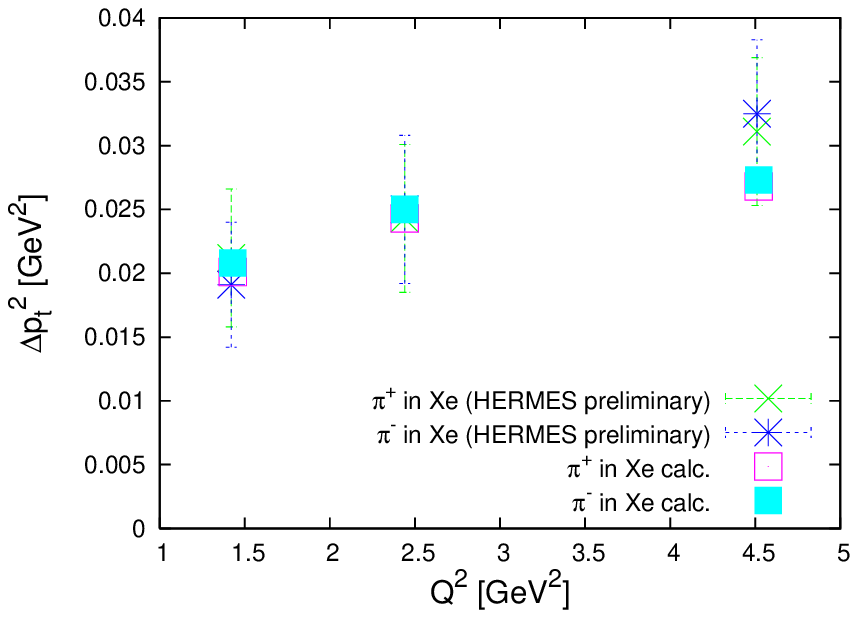}
   \caption{
     {\it Left:} Pre-pion formation time per unit $\nu$ estimated in
       Ref.~\cite{Accardi:2006qs}. 
     {\it Right:} hadron $p_T$-broadening at HERMES as a function of 
     $Q^2$ compared to the computations of Ref.~\cite{Domdey:2008aq}.
   }
  \label{fig:ptbroad-DGLAP}
  \end{center}
\end{figure}

The seemingly linear in $A^{1/3}$ dependence at HERMES
(Figure~\ref{fig:broad_A_hermes}) leads to the 
conclusion that the prehadron is formed close to the surface or
outside the heaviest nucleus, at least at the average kinematics
$\vev{z_h}=0.41$ and $\vev{\nu}=14.0$~GeV. Using an average in-medium
path length $\vev{L_{Xe}}\approx(3/4)R_{Xe}=4.3$ fm in
Eq.~\eqref{eq:tpre} one finds \begin{align} 
  C \gtrsim 0.8 \text{\ fm/GeV} \ .
\label{eq:Cest_naive}
\end{align}
The resulting prehadron
production time is plotted in Fig.~\ref{fig:ptbroad-DGLAP} left, and 
computed for each experimental bin in Table~\ref{tab:ptbroad}. 
However, the size of the $p_T$-broadening requires 
$\qhat \approx 0.03$~GeV$^2$/fm, which is much smaller than the 
$\qhat = 0.60$~GeV$^2$/fm obtained by fitting HERMES data on $R_M$ with a
pure energy loss model. Part of this discrepancy is due to the fact
that the energy loss model used in the fit of $R_M$ is known to
systematically overestimate the transport coefficient compared to other
implementations~\cite{Bass:2008rv}. However, this
can hardly explain the factor $\sim$20 discrepancy found here. One
possible explanation is that $\vev{t_{preh}}$ is in fact smaller than
just estimated.

\begin{table}[tb]
  \begin{center}
\begin{minipage}{8cm}
\begin{tabular}{cccccc}\hline
    & $\vev{Q^2}$ 
    & $\vev{\nu}$ 
    & $\vev{z}$
    & $\frac{\vev{Q^2}}{2m_N\vev{\nu}}$
    & $\vev{t_p}$ \\
    & \footnotesize [GeV$^2$]
    & \footnotesize [GeV]
    & 
    &
    & \footnotesize [fm]\\\hline
  $\vev{\Delta p_{Th}^2}$ vs $A$ & & & & & \\
  $Ne$ (2.3 fm) & 2.4 & 13.7 & 0.42 & 0.09 & 4.2 \\
  $Kr$ (3.7 fm) & 2.4 & 13.9 & 0.41 & 0.09 & 4.2 \\
  $Xe$ (4.3 fm) & 2.4 & 14.0 & 0.41 & 0.09 & 4.3 \\\hline
  $\vev{\Delta p_{Th}^2}$ vs $z$ 
     & 2.4 & 14.6 & 0.30 & 0.09 &  4.5 \\
     & 2.4 & 13.3 & 0.53 & 0.10 &  3.7 \\
     & 2.3 & 12.6 & 0.74 & 0.10 &  2.3 \\
     & 2.2 & 10.8 & 0.92 & 0.11 &  0.7 \\\hline
  $\vev{\Delta p_{Th}^2}$ vs $\nu$  
     & 2.1 &  8.1 & 0.48 & 0.14 &  2.4 \\ 
     & 2.5 & 12.0 & 0.42 & 0.11 &  3.7 \\
     & 2.6 & 15.0 & 0.40 & 0.10 &  4.6 \\
     & 2.4 & 18.6 & 0.36 & 0.07 &  5.8 \\\hline
  $\vev{\Delta p_{Th}^2}$ vs $Q^2$ 
     & 1.4 & 14.0 & 0.41 & 0.06 &  4.2 \\
     & 2.4 & 14.1 & 0.41 & 0.10 &  4.2 \\
     & 4.5 & 14.5 & 0.39 & 0.16 &  4.3 \\ \hline
\end{tabular}
\end{minipage}
\caption{Average HERMES kinematics for the $p_T$-broadening
  results \cite{vanHaarlem:2007zz,VanHaarlem:2007kj}. 
  In parentheses (next to the target nucleus symbol) is the
  average in-medium path length of the hadronising system 
  $\vev{L_A}\approx (3/4)R_A$, with $R_A=(1.12 \text{\ fm})A^{1/3}$. 
  The production time is computed according to
  Eqs.~\eqref{eq:ptbroadzz}-\eqref{eq:tpre}  with
  $C=0.8$~fm/GeV. The average $\vev{x_B}$ is very well
  approximated by $\vev{Q^2}/(m_N\vev{\nu})$.
}
  \label{tab:ptbroad}
  \end{center}
\end{table}

Turning to the $z_h$ dependence (Figure~\ref{fig:broad_hermes}), the observed
decrease as $z_h \ra 1$ is qualitatively compatible with the 
production times just derived (because $\vev{t_{pre}}\ra 0$ as
$z_h\ra1$),
but also with long formation times and a pure energy loss scenario
(because energy conservation requires $\Delta E_q \leq (1-z_h)\nu$). 
However, in either case it is very hard to reproduce the detailed
shape in $z_h$. Indeed, due to the $z_h^2$ factor in Eq.~\eqref{eq:ptbroadzz} 
in both cases $\vev{\Delta p_T^2}$ would have a peak at $z_h \gtrsim
0.7$, shifting to larger $z_h$ values as $A$ decreases. (Note that
this behaviour of the peak with $A$ is rather independent of the
detailed shape of $\vev{t_{preh}}$.)
This is at variance with HERMES data, which show a fixed peak around
$z_h\approx 0.5$, whose size increases with $A$.
One scenario not considered so far in literature involves short
formation times with large transport coefficients, to explain the
shape in $z_h$, and a non-negligible prehadron elastic cross section
to explain the $A$ dependence 

On the other hand, the basically flat in $\nu$ dependence of the
broadening would be compatible with the production time estimate of
Eq.~\eqref{eq:Cest_naive}. However, one should keep in mind that 
$\vev{x}$ decreases with $\nu$ so that NLO contributions increase;
since they tend to decrease the average energy of the fragmenting
quarks they might in fact contribute to tilt down the slope.

Finally, the linear increase of
$\Delta\vev{p_T^2}$ with $Q^2$ observed at HERMES (see e.g.
Fig.~\ref{fig:ptbroad-DGLAP}) remains a challenge to current
theoretical models.  
This dependence is in sharp contrast with the 
inverse power dependence from the colour dipole 
model of Ref.~\cite{Kopeliovich:2003py}, thus ruling out the
predicted $t_{preh} \propto E_{preh}/Q^2$ behaviour due to the stronger gluon
radiation. A flat $Q^2$ predicted
within most string-based models is also at variance with 
these data. Since a growing in $Q^2$ prehadron formation time seems
unlikely, dynamical effects at the parton level are needed in addition
to prehadron absorption to explain the $Q^2$-behaviour. 
Three mechanisms have been suggested in Ref.~\cite{Accardi:2008fz}:
\begin{enumerate}
\item 
{\it Medium-enhanced DGLAP evolution.}
The longer 
medium-enhanced DGLAP evolution
\cite{Ceccopieri:2007ek,Domdey:2008gp,Borghini:2005em,Armesto:2007dt}
at larger $Q^2$ would imply a stronger gluon radiation, hence a larger
$p_T$-broadening than at low $Q^2$.
\item 
{\it Next-to-leading order processes.}
At NLO, gluon fusion process,
$\gamma^*+g \ra q+\bar q$, imply $E_q < \nu$, hence a smaller
production time and $p_T$-broadening than at LO. The increasing
importance of LO vs. NLO processes as $x_B = Q^2/(2m_N\nu)$ increases 
could lead to the observed $p_T$-broadening dependence on $Q^2$. 
\item
{\it Prehadrons with very short production time,} and inelastic
cross section $\sigma_{preh}(t_p)\propto 1/Q^2$ slowly evolving in
time. Under these conditions, they will be the less absorbed the larger
$Q^2$, suffer a smaller surface bias  and contribute more to
$\Delta\vev{p_T^2}$.  
\end{enumerate}
Reference~\cite{Accardi:2008fz} also argues that studying the $p_T$-broadening
simultaneously binned in $x$ and $Q^2$ allows one to factor out the trivial 
$\nu=Q^2/(2m_Nx_B)$ kinematic correlations and to distinguish the
proposed scenarios.

The effect of medium modification of the DGLAP evolution
has been explored quantitatively in Ref.~\cite{Domdey:2008aq}, which
proposes  
\begin{align}
  \Delta\vev{p_T^2}(Q^2) = \Delta\vev{p_T^2}(\bar Q^2)
    + z_h^2 \,\nu \, \hat q \left( \frac{1}{\bar Q^2} - \frac{1}{Q^2}
    \right) \ .
\label{eq:mDGLAPbroad}
\end{align}
The first term is the $p_T$-broadening from quark multiple
scatterings, computed at $\bar Q^2 = 2.5$~GeV$^2$
using $\vev{t_p} = 1.19
z_h^{0.61}(1-z_h)^{1.09}\nu/\kappa$ from
\cite{Accardi:2002tv,Accardi:2005jd} and a transport coefficient 
$\hat q$~=~0.032~GeV$^2$/fm calculated in the dipole model for
quark-nucleon scattering. The second term models the effects
of the medium modified DGLAP evolution from Ref.~\cite{Domdey:2008gp}. 
Numerical results are presented in Fig.~\ref{fig:ptbroad-DGLAP} right. 
Note that Eq.~\eqref{eq:mDGLAPbroad} predicts a plateau in
$\Delta\vev{p_T^2}$ at $Q^2 \gtrsim 4$~GeV$^2$, which should be
experimentally verified. 

\subsubsection{\it A path forward} --
\label{sec:emergingpic}
As we have seen, HERMES data suggest relatively short
production times, such that the prehadron is formed in average close to the
nucleus surface, or even much more inside as the $z_h$ dependence of the
$p_T$-broadening seems to require. The details of the $\nu$ dependence of the
$p_T$-broadening and its linear increase with $Q^2$ demand for
parton-level dynamics typically not included in theoretical models. 

To clarify this complex interplay between parton dynamics and hadronisation, 
one needs detailed theoretical computations, combining energy
loss and prehadron absorption up to NLO, and accounting for surface
bias effects mostly neglected in these simplified considerations.
A production time scaling analysis of high precision $R_M$ data
coupled to a detailed study of the $p_T$ broadening dependence on all
kinematic variables is likely to reveal the size of the production
time and yield information on parton versus (pre)hadron dynamics.
Experimentally, progress will be
achieved by multi-dimensional binning, improved statistics for
different hadron flavours, and using an extended set of targets up to $Pb$. 

The CLAS experiment is ideally suited for such a programme, but  
has a reduced range in $\nu \lesssim 5$~GeV, which enhances the role of
prehadron absorption over parton energy loss, and in $x_B \gtrsim 0.1$. 
Its preliminary data
follow the same qualitative behaviour of HERMES data, but 
the measured $p_T$-broadening is larger than what would be
naively extrapolated from HERMES results based on the different
average $\vev{\nu}$. This might be due to a significant prehadron
elastic cross section, $\sigma_{preh}^{el} \propto \sigma_h^{el}$,
additionally contributing to the hadron $p_T$-broadening at CLAS, but
negligible at HERMES \cite{Domdey:2008aq}. The origin of the
CLAS/HERMES discrepancy needs clarification.

Finally, the  Electron-Ion Collider (EIC)
\cite{Aidala:2008ic,Deshpande:2005wd}, able to span from very small to
large $x_B$,  may clarify the role of LO and NLO processes. 
Furthermore, it will provide a large range in $\nu$,
which will allow for the study of purely partonic in-medium
propagation and the calibration of energy-loss models.

\subsection{$\pi^0$ vs. $\eta$ attenuation} \ \\
\label{sec:pi0vseta_att}

The measurement of $\pi^0$ and $\eta$ suppression,
which have a similar valence quark content 
but different masses and different hadronic cross sections, can provide an
interesting cross-check of the hadronisation picture and time-scales for 
(pre)hadron formation. 
In $Au+Au$ collisions at RHIC, the PHENIX experiment~\cite{Adler:2006bv} 
has shown a similar suppression $R_{AA}(p_T)$ ratio for the two species up to
$p_T \approx 10$~GeV/c, with the $\pi^0/\eta$ ratio independent of the
collision system or centrality of the $Au+Au$ collisions within uncertainties,
see Fig.~\ref{fig:RAA_rhic} (right). 
Such a result is naturally explained if the suppression takes place at the {\it partonic} 
level before the quenched parton fragments into final hadrons~\cite{Adler:2006bv,Adler:2006hu}, 
i.e., hadronisation occurs on timescales larger than the typical medium size. 
Yet, hadronisation in a deconfined QGP could be delayed compared to
cold QCD matter at least until the medium temperature falls below the critical
one. In fact results from 
$e+A$ collisions at HERMES, where the quark energy $\nu \approx
p_T|_\text{RHIC}$ but the virtuality $Q^2 = 2.5$~GeV$^2 \ll p_T^2 =
\mathcal{O}(100$~GeV$^2)$, seem to favour much shorter hadronisation timescales 
for hard partons (Sect.~\ref{sec:measuretp}).  

It would be thus important to measure and compare $\pi$ and $\eta$
attenuation in  nDIS experiments, where clean measurements are possible and no QGP is formed.
These measurements are planned for the CLAS experiment and will also be very
important in a hadronisation physics programme at the proposed EIC. 

\subsection{Baryon formation } \ \\
\label{sec:baryons}

Protons, and in general 3-quark baryon species, seem to follow a different
nuclear modification pattern than mesons in both cold and hot nuclear matter.
In nDIS, the difference observed in proton vs. mesons multiplicity
ratios resembles  the anomalous baryon enhancement reported at
intermediate $p_T\approx$~1.5 -- 8 GeV/c in proton-nucleus
and heavy-ion collisions. However, antiprotons follow the mesons
attenuation pattern, unlikely what happens in $p+A$ collisions where
both baryons and antibaryons are anomalously enhanced. 
Contrasting the baryon nuclear modifications in nDIS and $p+A$
collisions will clearly lead to a more profound understanding of the
hadronisation mechanism.

\subsection{High-$p_T$ hadrons at fixed-target energies}\ \\

The detailed interpretation of the nucleus-nucleus and proton-nucleus data
and the relative roles of high-$p_T$ hadron suppression and Cronin enhancement 
at around $\sqrtsnn$~=~20~GeV is obscured by the current absence of 
a direct measurement of the corresponding $p+p$ reference spectra at these energies. 
Use of various $p+p$ interpolations~\cite{d'Enterria:2004ig,Arleo:2008zd} compared 
with the $A+A$ data measured at top SPS~\cite{Aggarwal:2001gn,Aggarwal:2007gw}
or low RHIC~\cite{Adare:2008cx} energies, seems to indicate the onset of parton
energy loss in this energy range. Precise tests of the mechanisms of $p_T$ broadening 
in $p+A$ and of parton and/or (pre)hadron energy loss in $A+A$ require the measurement
of the baseline vacuum production of various hadron species in a perturbative regime 
($p_T\gtrsim$~2~GeV/c) in $p+p$ collisions at $\sqrtsnn$~=~20~GeV. Such 
large statistics measurements can hopefully be made in coming low energy scans at 
RHIC (and at higher luminosity in the planned RHIC-II upgrade).

\subsection{The heavy flavour puzzle } \ \\

The solution of the intriguing heavy-flavour results at RHIC, namely the
larger than expected suppression of heavy flavour production in central $A+A$ collisions, 
is likely to provide important insights in our theoretical
understanding of parton propagation and hadronisation, 
and on the nature of the Quark-Gluon Plasma.

On the theory side it has already prompted a revision of
established ideas and stimulated many new ones, see
Section~\ref{sec:heavyflavors}. However, no consensus has been
reached yet on the physical mechanism behind the large heavy-flavour 
suppression. 
Interestingly, heavy quark energy loss and in-medium propagation are
processes that can be 
addressed in AdS/CFT-based descriptions of the QGP,
and they are very promising to investigate the relevance of
string-theory methods for discussing QCD near the deconfinement phase
transition~\cite{Horowitz:2008zz}. 

On the experimental side, new clues will be offered in the near future
by experimental upgrades at RHIC,
aimed at directly identifying $D$ and $B$ mesons and at measuring their
individual suppression factors \cite{Frawley:2008kk}. 
At the LHC, heavy-flavour production will be more abundant, and its
nuclear modification one of the main areas of 
experimental interest because of their potential role in identifying and 
studying the QGP properties~\cite{Carminati:2004fp,Alessandro:2006yt,D'Enterria:2007xr}. 
At the Electron-Ion Collider, a
clean measurement of $D$ and $B$ suppression will also be able to
discriminate the proposed mechanisms, and remove the
uncertainties in the interpretation of the data due to the modelling
of the medium created in $A+A$ collisions, and to possible in-medium
regeneration processes.

\subsection{Dihadron and photon-hadron correlations} \ \\
\label{sec:hadcorr}

\begin{figure}[tb]
    \parbox[c]{7cm}{
      \includegraphics[width=7.cm]{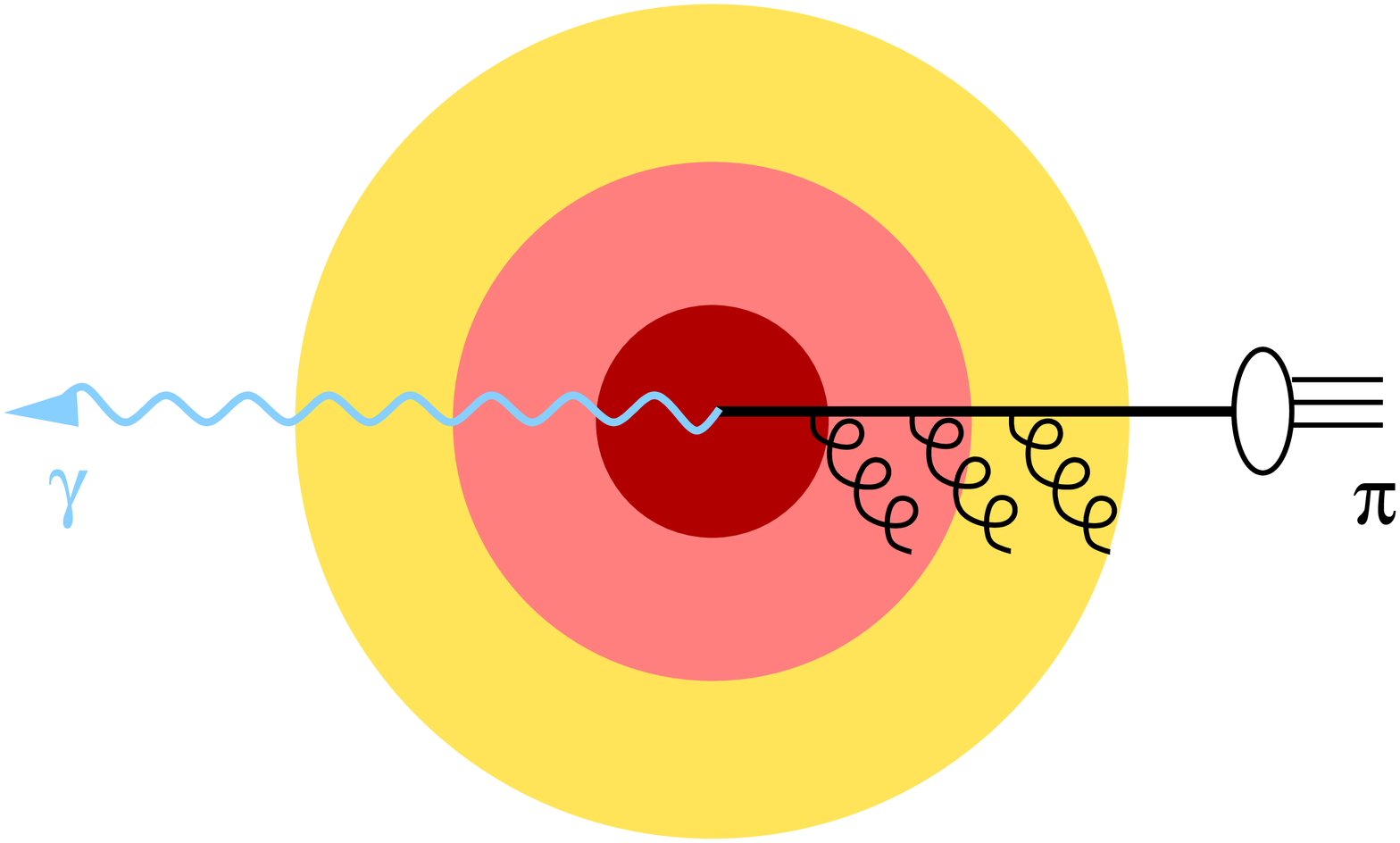}
    }\hfill
    \parbox[c]{6.2cm}{\centering
      \includegraphics[width=5.8cm]{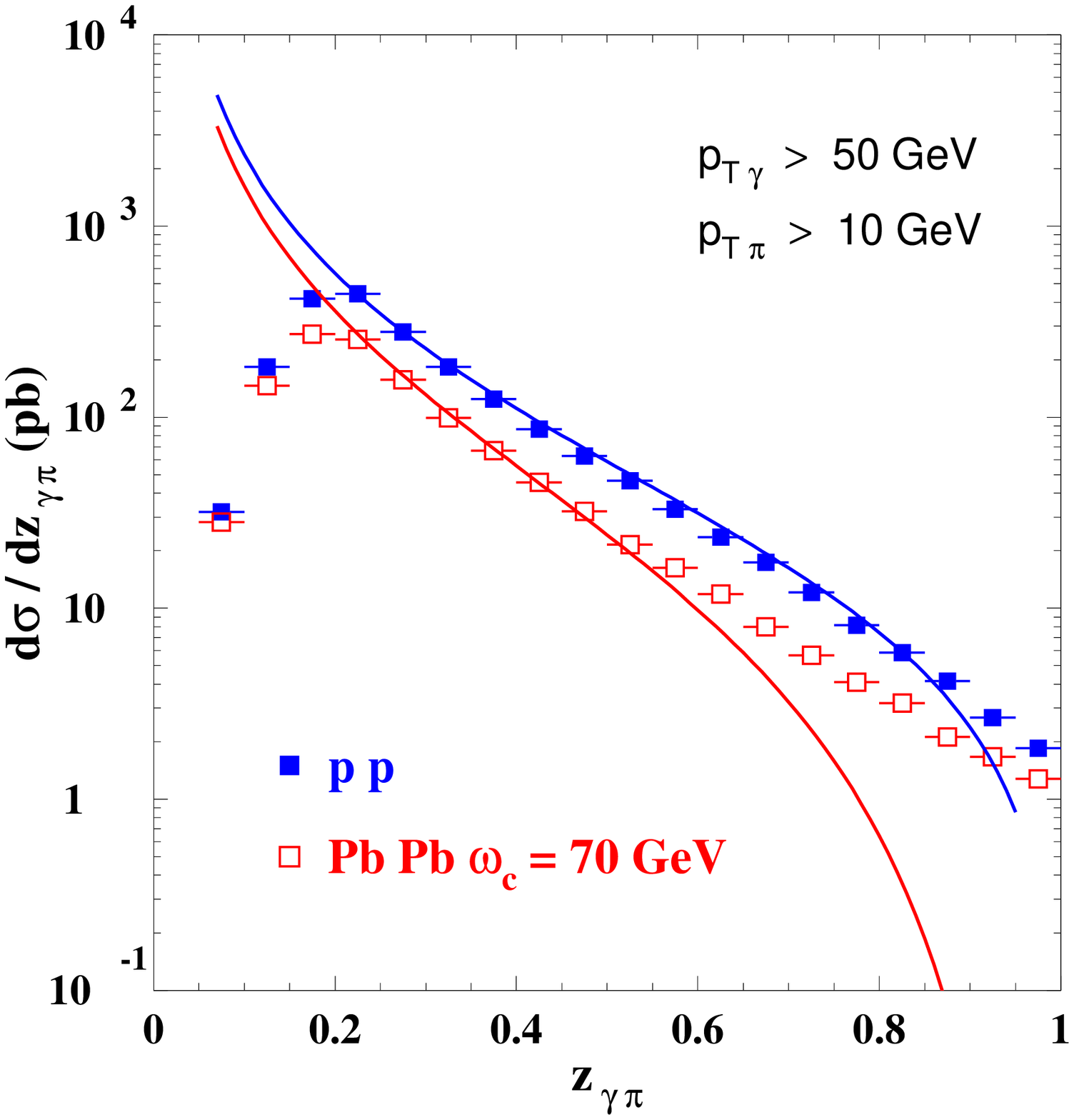}
    }
    \caption{{\it Left:} A leading-order channel for $\gampi$
      production. 
      {\it Right:} $\gampi$ imbalance distributions using
      $\picut=10$~GeV/c and $\gacut=50$~GeV. The solid lines show the
      (rescaled) quark fragmentation functions used in the
      calculation. 
    }
\label{fig:sketchgampi}
\end{figure}

High-$p_T$ two- and three-hadron correlations have proven a rich field 
to characterise medium modifications of parton fragmentation
in $A+A$ collisions, see Section~\ref{sec:dihadrons}. 
One can more directly access medium-modified FF in heavy-ion collisions by
tagging large-$\pt$ particle production with prompt photons, as
suggested in~\cite{Wang:1996yh} and later investigated in detail at
RHIC and LHC energy~\cite{Arleo:2004xj,Arleo:2006xb}. 
As shown in Fig.~\ref{fig:sketchgampi} left, at large $p_T$ and leading
order in $\alpha_s$, a photon is produced directly in the hard
subprocess, back-to-back to e.g. a quark which loses energy in the
medium before fragmenting into a pion. Because of momentum
conservation, the experimental $\gampi$ momentum imbalance variable, 
$
\z \equiv - {\bf \ptpi}\cdot{\bf \ptgamma}/{|{\bf \ptgamma}|^2},
$
reduces at LO to the theoretical hadron fractional momentum $z$ 
which can therefore be experimentally estimated. 
In Fig.~\ref{fig:sketchgampi} right, the $\gampi$
distributions is computed in QCD at LO in $p+p$ and $Pb+Pb$
collisions at the LHC. The observable tracks the quark FF into pions 
until the onset of the photon-fragmentation channel at large 
in a wide range of $\z$. This simple
connection between $\z$ and $z$, is however complicated 
by the effects of 
higher-order corrections or initial- and final-state soft
gluon radiation.
The most important background channel, to be reduced 
with appropriate kinematics and isolation cuts, is photon production 
by the collinear fragmentation of a hard parton.

In contrast to heavy-ions, only dihadron correlations in $z_h$ have
been measured in $e+A$ collisions, see Section~\ref{sec:2partcor_nDIS}.
They are the analog of the ``same-side'' (intrajet) 
dihadron correlation in $A+A$ integrated over the azimuthal angle, 
but are not affected by the complication of hot QCD matter effects.
In the HERMES kinematics, $\vev{x} \approx 0.1$ where the gluon density is
comparable to the quark density. Hence the processes 
$\gamma+q \ra q + g$ and $\gamma+g\ra q + \bar q$ may compete with the
LO $\gamma + q \ra q$ scattering. For NLO processes, the smaller parton
energy implies a larger nuclear suppression for either hadron, which
in turn biases the hard scattering towards the nuclear
surface. Moreover, the subleading hadron may be produced with 
larger $z$ than the leading hadron, 
which complicates a naive interpretation of the
observable. For these reasons, the relative importance of NLO compared
to LO hadron production may be amplified in dihadron correlation
compared to single-inclusive observables. 
An observable which is directly sensitive to $\mathcal{O}(\alpha_s)$ partonic cross
sections, and could cross-check the dihadron $z_h$ correlations,  
is the dihadron azimuthal correlation relative to the virtual photon
direction. 

Photon-hadron correlations  will be one of the key
measurements at CLAS, and very interesting for a hadronisation
programme at the EIC.
Moreover, the photon-hadron transverse momentum imbalance with respect
to the virtual photon axis will allow one to study the analog of the
same-side (intrajet) hadron correlations with the added advantage of
measuring the transverse momentum of the jet.

\subsection{Future perspectives } \ \\
\label{sec:openissues}
\label{sec:perspectives}

\begin{figure}[tb]
  \centering
  \includegraphics[width=12.cm]{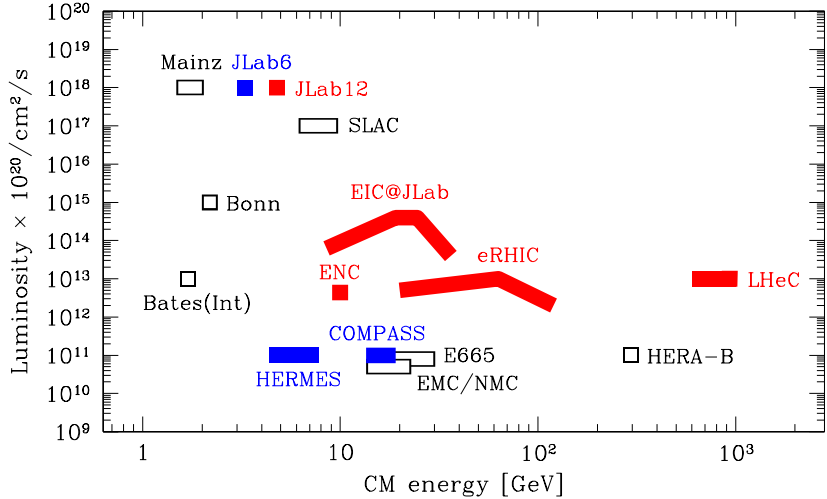}
\caption{
  Centre-of-mass energy vs. luminosity of past, present, approved and
  proposed lepton-nucleus facilities and experiments. 
}
\label{fig:Q2lumi}
\end{figure}

The challenges just discussed can
be addressed and likely solved at future facilities planned (or proposed) to become
operational within the next decade and beyond. The LHC heavy-ion
programme, the RHIC-II upgrade, the 12-GeV upgrade at Jefferson Lab, 
and the Electron Ion Collider are examples of planned new facilities
with capabilities that 
will significantly increase the existing experimental reach in terms
of energy and luminosity, see Fig.~\ref{fig:Q2lumi}. 

\begin{table}[tb]
\begin{center}
\begin{tabular}{ccccccc} \hline 
hadron & $c\tau$ & mass & flavour  & detection & Production rate \\
       &         &(GeV/c$^2$) & content &  channel &  per 1k DIS events \\
       &         &      &         &          &           \\ \hline 
$\pi^0$ & 25 nm & 0.13 & $u\bar{u}d\bar{d}$ & $\gamma\gamma$ & 1100  \\ 
$\pi^+$ & 7.8 m & 0.14 &   $u\bar{d}$ & direct & 1000  \\ 
$\pi^-$ & 7.8 m & 0.14 &   $d\bar{u}$  & direct & 1000  \\ 
$\eta$ & 0.17 nm & 0.55 & $u\bar{u}d\bar{d}s\bar{s}$&$\gamma\gamma$ & 120  \\ 
$\omega$ & 23 fm & 0.78 &  $u\bar{u}d\bar{d}s\bar{s}$ & $\pi^+\pi^-\pi^0$ & 170  \\ 
$\eta'$ & 0.98 pm & 0.96 &  $u\bar{u}d\bar{d}s\bar{s}$ & $\pi^+\pi^-\eta$ & 27  \\ 
$\phi$ & 44 fm & 1.0 &  $u\bar{u}d\bar{d}s\bar{s}$ & $K^+K^-$ & 0.8  \\ 
$f1$ & 8 fm & 1.3 &  $u\bar{u}d\bar{d}s\bar{s}$ & $\pi\pi\pi\pi$ & -  \\ 
$K^+$ & 3.7 m & 0.49 &  $u\bar{s}$ & direct & 75  \\ 
$K^-$ & 3.7 m & 0.49 &  $\bar{u}s$ & direct & 25  \\ 
$K^0$ & 27 mm & 0.50 &  $d\bar{s}$ & $\pi^+\pi^-$ & 42  \\ \hline
$p$ & stable & 0.94 &  $ud$ & direct & 530 \\ 
$\bar{p}$ & stable & 0.94 &  $\bar{u}\bar{d}$ & direct & 3 \\ 
$\Lambda$ & 79 mm & 1.1 &  $uds$ & $p\pi^-$ & 72 \\ 
$\Lambda(1520)$ & 13 fm & 1.5 &  $uds$ & $p\pi^-$ & - \\ 
$\Sigma^+$ & 24 mm & 1.2 &  $us$ & $p\pi^0$ & 6 \\ 
$\Sigma^0$ & 22 pm & 1.2 &  $uds$ & $\Lambda\gamma$  11 &\\ 
$\Xi^0$ & 87 mm & 1.3 &  $us$ & $\Lambda\pi^0$ & 0.6  \\ 
$\Xi^-$ & 49 mm & 1.3 &  $ds$ & $\Lambda\pi^-$ & 0.9  \\ \hline 
\end{tabular}
\end{center}
\caption{\small{Final-state hadrons potentially accessible for formation 
length and transverse momentum broadening studies in CLAS12. The 
rate estimates were obtained from the LEPTO event generator for an 11-GeV 
incident electron beam. (The criteria for selection of these particles 
was that $c\tau$ should be larger than the nuclear dimensions, and their 
decay channels should be measurable by CLAS12.)}}
\label{table:hadron_list}  
\end{table}

\subsubsection{\it The 12-GeV Jefferson lab upgrade}-- 
The 12-GeV accelerator upgrade of JLab will significantly extend existing 
experimental capabilities by doubling the accelerator electron-beam energy and by
increasing the luminosity by at least an order of magnitude for
large-acceptance fixed-target measurements. The CLAS12
spectrometer will operate with a continuous luminosity of $1-2 \times
10^{35}$ cm$^{-2}s^{-1}$ for 
nuclear target experiments with an 11-GeV electron beam. The planned
particle identification scheme, which will directly identify pions,
low-energy charged kaons, neutral kaons, protons, and electrons, can
be upgraded to include direct detection of higher-energy charged kaons
as well. With the increased luminosity comes access to rarer hadrons
whose attenuation in the nuclear medium has never before been
explored. 

\begin{figure}[tb]
  \centering
  \includegraphics[width=10.cm,height=6.8cm,clip]{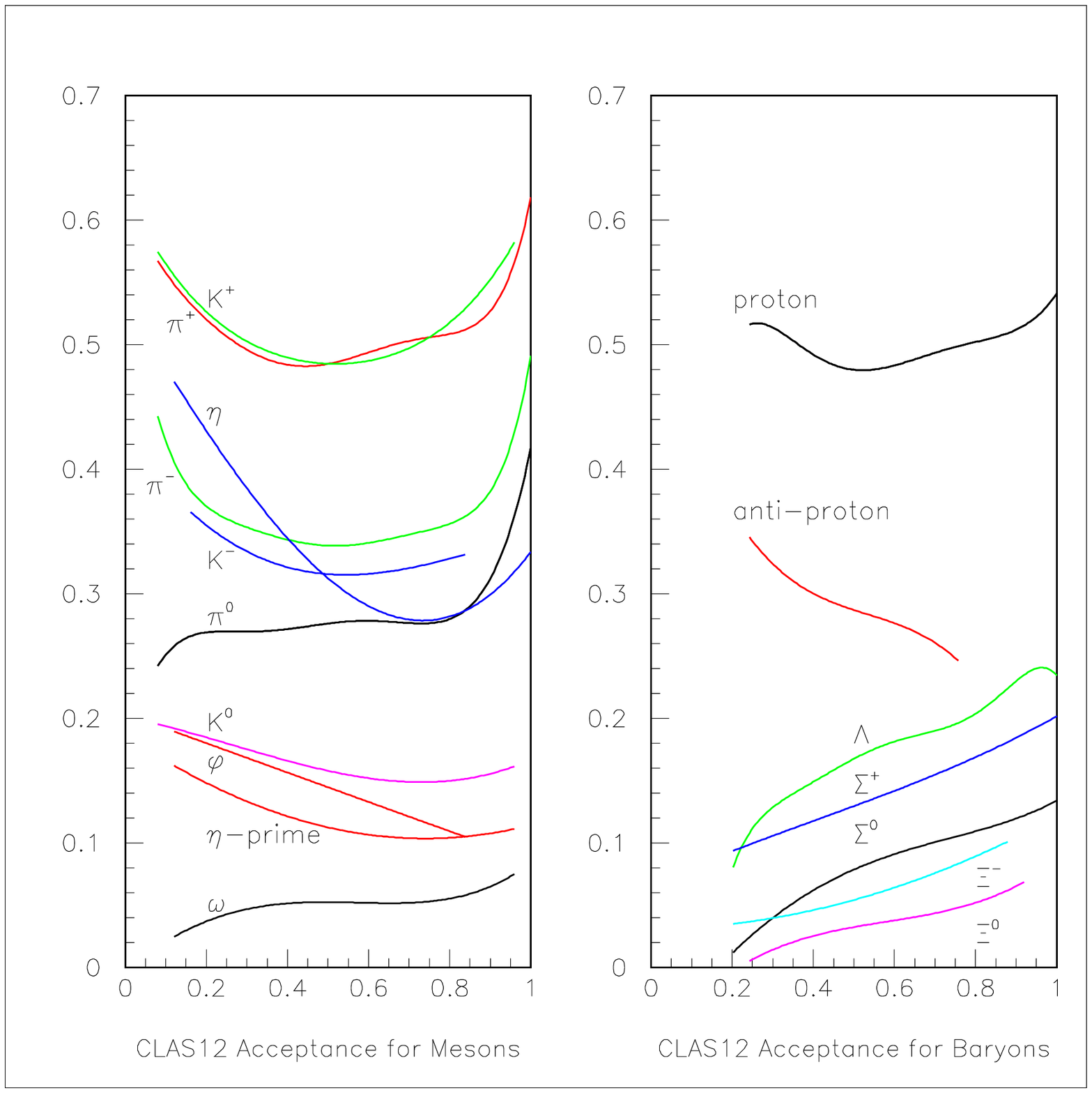}
\caption{Acceptances for mesons and baryons in the CLAS12 experiment at the proposed 12-GeV JLab facility.}
  \label{fig:CLAS12acceptances}
\end{figure}

Table~\ref{table:hadron_list} lists hadrons accessible with
CLAS12 that are stable over nuclear distance scales. Nuclear
attenuation measurements can be performed for all of these hadrons,
and transverse momentum broadening will be directly accessible for a
number of the listed hadrons. Estimations for the CLAS12 geometric
acceptances for these particles are plotted in
Fig.~\ref{fig:CLAS12acceptances}. As can be seen, a systematic study of
the mass dependence for two-quark systems is feasible for eight meson species
with a range of masses spanning nearly 1~GeV/c$^2$. Further, new insights
into the poorly-understood hadronisation mechanisms for baryons can be
gained by studying the series of eight baryon species indicated in this
table.  While experimentally challenging, the baryon measurements have the
potential to open up a completely new realm of studies of hadron
formation for three-quark systems.

\subsubsection{\it Other fixed target experiments: DIS and DY} --
The ongoing fixed-target COMPASS experiment~\cite{Abbon:2007pq} with
its up to 190 GeV muon beam, can cover a much larger $\nu$ and $Q^2$
range than HERMES or JLab.  
In particular, due to the higher reach in $\nu$, one  can
select partons hadronising outside the 
medium, and cleanly test parton energy loss models by studying the
hadron $p_T$-broadening. Therefore, a $\mu+A$ experimental 
programme at COMPASS would be very desirable to improve our
knowledge of the space-time evolution of hadronisation and
anticipate some of the results expected at the future electron-ion
colliders discussed in Sect.~\ref{sec:EIC}. At the same time the
availability of a high-energy proton or pion beam would allow
measuring nuclear modifications of DY processes and studying the
interplay of initial state and final state parton energy loss within the
same detector. 

Dedicated high-statistics medium energy DY experiments have been proposed in
\cite{Garvey:2002sn} to precisely measure initial-state 
parton energy loss in nuclear targets. The authors considered using the 
proton beams from the 120 GeV Fermilab Main Ring Injector and the
future 50 GeV Japan Hadron Facility (JPARC). The kinematic acceptance of a
dimuon spectrometer proposed for the Fermilab Experiment
E906~\cite{E906-proposal} and a similar spectrometer 
considered~\cite{Asakawa:2000ce,Peng:2000tz} at JPARC.
The nuclear partons are probed in the $0.1 < x < 0.5$ region
where the anti-shadowing and the EMC effect on nuclear PDFs interferes
with energy loss suppression of DY events. However, in this $x$-region
nuclear PDFs are well determined in global fits  
\cite{deFlorian:2003qf,Hirai:2007sx,Eskola:2009uj}. 

The interplay of hadron attenuation and $p_T$-broadening in nDIS and
DY nuclear modifications in $p+A$ collisions will be vital to
constrain energy-loss models and characterise the stopping power of
cold nuclear matter.

\subsubsection{\it Other fixed target experiments: high-$p_T$ hadron spectra} --
The study of high-$p_T$ hadron spectra in $p+p$ to $A+A$ collisions 
in the SPS energy range is fundamental for precisely detect and study 
QGP formation in heavy-ion collisions. Such a programme is one of the
aims of the NA61 experiments at CERN \cite{Posiadala:2009zf}, which upgrades
the NA49 spectrometer, and also by the NA60
\cite{Baldit:2000cq} and SELEX \cite{Engelfried:2002mh} experiments,
originally focused on quarkonium and 
charmed hadron production and nuclear modifications. 

The peculiarity of NA60 and SELEX is that they mount multiple nuclear
targets on the same beam: from Beryllium to Uranium in NA60; Copper
and Carbon in SELEX. As emphasised in \cite{Accardi:2005fu}
this target setup allows a high-precision measurement of the ratio of
{\it differential cross-sections} for inclusive hadron production 
on two different nuclei, 
$
  R^\sigma_{pA/pB} =  \frac{1}{A} \frac{d\sigma^{pA}}{dp_T^2 dy}
    \Big/ \frac{1}{B} \frac{d\sigma^{pB}}{dp_T^2 dy} 
$
where $B$ is the lightest available nucleus. 
Indeed, the beam luminosity as well as part of the systematic
errors cancel in the ratio.   
The $A$ dependence of the ratio is analogous to the centrality
dependence, but eliminates the
large experimental uncertainties due to the determination of 
the centrality and to the normalisation of the Cronin ratio.
Moreover, without
need of centrality cuts the statistics may be sufficient 
to probe the high-$p_T$ region and test the mechanism underlying the
Cronin effect.

\subsubsection{\it Electron-ion colliders}--
\label{sec:EIC}
An electron-ion collider would significantly complement and add new
dimensions to the experimental studies at fixed target facilities,
by extending the
range of accessible lepton, $E_e$, and ion, $E_A$, energies. 
Proposed designs for an electron-ion collider currently have a fairly
wide range of luminosities and centre-of-mass energies, see
Table~\ref{tab:EICparams}.   
The Electron-Ion Collider (EIC)  is a proposed
US-based facility, which would provide $E_e = 3-20$~GeV and
$E_A=15-100$~$A$GeV~\cite{Aidala:2008ic,Deshpande:2005wd}. 
There are currently two complementary conceptual
proposals\cite{Aidala:2008ic,ENC-EIC-workshop-2009}:  
(i) eRHIC would add an electron 
beam to the existing RHIC ion accelerator at BNL  -- the option of
staging it, and first realise a lower energy version with $E_e=2-4$~GeV, 
based on an energy recovery linear accelerator for the electron
beam is being actively discussed 
\cite{DeshpandePANIC08,Litvinenko-GSI2009}; 
(ii) EIC@JLab would add an ion accelerator to the CEBAF upgraded 12-GeV 
electron ring at Jefferson Lab \cite{Zhang-GSI2009,Ent-GSI2009}. 
The eRHIC concept emphasises the 
energy range; the EIC@JLab concept will reach up to 100 times more
luminosity, but with lower maximum energy.  In both cases polarised
proton and light-ion beams will be available.
The Large Hadron-electron Collider (LHeC) is a proposed upgrade of the
LHC at CERN \cite{Dainton:2006wd,LHeC-workshop-2008}, and will reach
much higher energies 
than the EIC still with a good luminosity, but only with unpolarised
hadron beams. 
Finally, the idea of building a low-energy electron-nucleon (ENC) collider at
FAIR has recently been advanced \cite{Jankowiak-GSI2009}: the goal is to
have a centre-of-mass energy between HERMES and COMPASS, but up to a
factor 100 higher luminosity and polarised hadron beams. 
Here, we will focus on the EIC
capabilities, but most of the discussed measurements will be in
principle possible also at the LHeC and, with the exception of jet 
measurements, at the ENC.

\begin{table}[t]
\center
\begin{tabular}{|l|c|ccc|ccc|c|}\hline
    & ENC\ \   
    & \multicolumn{3}{c|}{EIC@JLab} 
    & \multicolumn{3}{c|}{eRHIC} 
    & LHeC \ \ \\
  &   
  & low   & med    & high   \ \
  & med   & high   & full   \ \
  &  \\\hline
    $E_e$ [GeV] 
    & 3.3
    & 3     & 3-5    & 11   \ \      
    & 3     & 10     & 30   \ \
    & 70  \\
  $E_A$ [GeV/A] 
  &   7.5 
  &   6   & 30    &  30  \ \
  &  50   & 100   & 125  \ \
  & 2750  \\
    $\sqrt{s}$ [GeV] 
    & 10
    & 9     & 19-25  & 36    \ \
    & 24    & 98     & 122   \ \
    & 880 \\
  $\mathcal{L}$ [10$^{33}$ cm$^{-2}$s$^{-1}$]
  & 0.44  
  & 6     & 40    & 4     \ \
  & 0.5   &  1    & 0.2  \,
  & 1   \\\hline
\end{tabular}
\caption{Beam energies $E$, center of mass energies $\sqrt{s}$, and
  luminosity $\mathcal{L}$, for $e+A$ 
  collisions at the ENC (at FAIR), the EIC (BNL and JLab designs), 
  and the LHeC (ring-ring mode) proposed facilities.
}
\label{tab:EICparams}
\end{table}

The EIC will provide virtual photons with energies in
the $\nu=10-1600$~GeV range, large $Q^2$ up to 1000~GeV$^2$, and
low $x \gtrsim 10^{-5}$. Its high luminosity will allow
access to rare signals, multi-differential measurements, and dihadron
and $\gamma$-hadron correlations.  
It will be able to cross-check HERMES and CLAS measurements, while 
offering many more channels and an extended kinematic range 
to study hadronisation inside the nucleus at low $\nu$,
as well as testing basic QCD processes such as non-Abelian parton
energy loss and the space-time evolution of the DGLAP shower at high $\nu$.
EIC will be unique, compared to HERA (HERMES) and JLab (CLAS), in the
following areas:
\begin{itemize}
\item 
At the large $\nu$ accessible at the EIC, hadrons will clearly be
formed outside the nuclear medium, so that effects due to the
propagation of the struck quark can be experimentally isolated.
One will have new access to $p_T$-broadening studies, which can 
cleanly probe the parton radiative
energy loss as predicted by pQCD at asymptotic energy.
It will also be possible to study in detail the interplay of radiative
and collisional parton energy loss, medium modifications of the DGLAP
evolution, and test factorisation for the fragmentation functions.
\item 
Heavier mesons like $\eta$ and $\phi$ will be
more abundantly produced than at HERMES/JLab. At both
medium-low and large $\nu$, comparing their quenching to that of their
lighter $\pi$ and $K$ counterparts will provide important clues about  
parton propagation and hadronisation, see Sect.~\ref{sec:pi0vseta_att}.
\item
The EIC excellent low-$x$ coverage provides increased production of heavy
flavour and quarkonia. In particular, it will be possible to study the
heavy quarks energy loss through the $B$ and $D$ meson
suppression. Theoretical mechanisms proposed to explain 
charmonium suppression in $p+A$ and $A+A$ collisions can be put to the
test in a clean experimental environment by studying $J/\psi$, $\psi'$
and $\chi$ spectra. 
\item
The large $Q^2$ coverage will access a truly perturbative QCD regime,
whose prediction can be more confidently tested against the data, in
particular colour transparency effects and the $Q^2$-dependence of the
observables discussed in this review, like the hadron
$p_T$-broadening. 
\item
Baryon production through parton fragmentation will also be
accessible, because of the collider mode and the accessible 
final-state invariant masses. 
This will allow studying baryon transport in cold QCD matter 
and the baryon anomalies observed in fixed target $e+A$ collisions at
HERMES  and in heavy-ion collisions at RHIC. The ability to identify a good
variety of baryons, including the strange and charmed sector, will be a
key to this programme.
\item 
For the first time, jet physics in DIS with a nuclear target will be
experimentally accessible. 
In particular, medium modifications of the jet shape, and the
comparison of light-quark to heavy-quark and gluon initiated jets will
shed light on the mechanisms underlying parton energy loss.
These studies can also be extended to dijet or $\gamma$-jet correlations.
\end{itemize}
In summary, the collider kinematics and associated detectors with
excellent calorimetry, particle identification, vertex detection, and
rapidity coverage
will allow for a comprehensive programme to better understand how parton 
energy loss and fragmentation occur in cold QCD matter.

By contrast, the ENC will cover a smaller the $\nu=10-50$~GeV range,
which might or might not be sufficient to ensure hadronisation outside
the nuclear medium. The high luminosity and rather symmetric
kinematics would however ensure the possibility of a rich
hadronisation programme complementing and extending the HERMES results.

\subsubsection{\it RHIC II }--
In order to undertake extensive high-statistics studies of processes with low cross sections
in $A+A$ collisions, an upgrade of the RHIC luminosity will be
necessary as well as a comprehensive 
new detectors~\cite{Frawley:2006}. Key measurements will include two-particle correlations up 
to a $p_T$ range where fragmentation can be safely tested and extended to include heavy 
identified hadrons. In addition, a comprehensive set of measurements of photon-jet correlations, 
and heavy-quark tagged jets can provide valuable information on parton energy loss and 
hadronisation mechanisms.
The capabilities of the proposed new detector include: (i) excellent charged particle momentum 
resolution up to $p_T$~=~40~GeV/c in the central rapidity region, (ii) complete hadronic and 
electromagnetic calorimetry over a large phase space, (iii) particle identification up to large 
$p_T$ (20~--~30~GeV/c) including hadron and lepton separation in the central and forward regions, 
and (iv) high rate detectors, data acquisition, and trigger capabilities.

\subsubsection{\it LHC }--
\label{LHC}
The possibility of colliding heavy nuclei with high luminosity at the Large
Hadron Collider (LHC) offers a unique opportunity to investigate in detail the
behaviour of strongly interacting matter under extreme conditions of density 
and temperature. 
The factor 30 increase in $\sqrtsnn$ relative to RHIC corresponds to a huge 
increase in kinematic and statistical reach for hard probes, and new channels 
will become available~\cite{Abreu:2007kv} such as detailed jet studies 
(jet shapes, medium-modified fragmentation functions)~\cite{d'Enterria:2009am}.
\\

{\it ALICE:} --
The ALICE experiment was designed specifically to tackle as many measurements 
as possible in the high multiplicity environment of heavy-ion collisions. ALICE aims 
firstly at accumulating sufficient integrated luminosity in $Pb+Pb$ collisions at 
$\sqrtsnn =$~5.5~TeV, as well as to carry out studies of $p+p$ and $p+A$ collisions 
in order to establish the benchmark processes under the same experimental conditions.
ALICE can measure heavy-flavour production down to very low $p_T$ because its unique 
low transverse-momentum cutoff for particle detection. This can be achieved by using 
inclusive large impact-parameter lepton detection, and by reconstructing exclusive 
charm-meson decays at relatively low $p_T$. As a result, the measurement of the total 
heavy flavour cross section will require a smaller extrapolation and will thus show improved precision.
As discussed in \ref{sec:hadcorr} a very attractive methods to study hadron jets is to tag high 
energy hadrons or jets with prompt photons emitted opposite to the hadrons or jet directions.
The combined use of the PHOton Spectrometer (PHOS) -- with excellent energy resolution 
although in a limited acceptance -- and the ElectroMagnetic Calorimeter (EMCal) 
-- about seven times larger in acceptance but with slightly worse resolution and coarser granularity --
and the central tracking system for charged particles, will allow one to recover a large fraction of 
the jet energy, thus reducing sensitivity to the specific pattern of fragmentation, and to perform 
$\gamma$-hadron and $\gamma$-jet correlations measurements.\\

{\it ATLAS:} --
ATLAS is the largest particle detector ever constructed, and its design, like CMS, is aimed 
at addressing a broad variety of physics processes~\cite{Aad:2009wy}. 
The detector design has many features that make it ideal for studies of 
heavy-ion collisions~\cite{Wosiek:2007zz}. A central solenoid magnet and large outer 
toroidal field magnets provide momentum analysis for the tracking systems. 
The large acceptance electromagnetic and hadronic calorimeters have longitudinal segmentation and 
fine transverse segmentation, covering the range in pseudorapidity $|\eta| < 3.2$ (electromagnetic) 
and $|\eta| < 4.9$ (hadronic), providing excellent photon and jet physics coverage.
The high precision silicon tracking system covers the range $|\eta| < 2.5$ and the external muon 
spectrometers cover $|\eta| < 2.7$, for large-acceptance dimuon studies of heavy quark systems. A fast, 
three-level trigger system and a high-rate data acquisition system are designed for triggering on 
rare, high-$p_T$ particles and jets. 

This instrumentation provides excellent tools for study of several observables relevant to heavy-ion 
collisions. Study of hard interactions and jet quenching will be a strong focus, and are closely 
connected to the topics in this review. The acoplanarity of di-jet pairs should provide sensitivity 
to the angular diffusion of high-$p_T$ partons in the medium. Jet properties that will be measured 
include transverse momentum spectra, hadron fragmentation functions, 
and jet shapes; di-jet angle and energy correlations may be able to separately 
quantify collisional and radiative energy loss of hard-scattered partons. In addition, measurements 
of quarkonia in bottom and charm systems, as well as of $D$ and $B$ mesons, will open up 
detailed studies of heavy quark jet quenching at high $E_T$.\\

{\it CMS:} --
CMS is a general purpose experiment at the LHC designed to explore the physics at the TeV 
energy scale~\cite{Ball:2007zza} including comprehensive measurements in $Pb+Pb$ 
collisions~\cite{D'Enterria:2007xr}. The CMS detector features a 4 T solenoid surrounding 
central silicon pixel and microstrip tracking detectors and electromagnetic (ECAL, $|\eta|<$ 3) and 
hadronic (HCAL $|\eta|<$ 5) calorimeters, and muon detectors ($|\eta|<$ 2.4) embedded in the flux 
return iron yoke of the magnet. CMS is the largest acceptance detector at the LHC
with unique forward detection capabilities with the HF and CASTOR calorimeters
(5.1 $<|\eta|<$ 6.6). 
CMS is, by design, very well adapted to detect and reconstruct high-$p_T$ and high-mass particles.
The experiment can significantly extend the $p_T$ reach with respect to RHIC, thanks to the large 
hard cross sections at 5.5~TeV, the large acceptance of its tracking system ($|\eta|<2.5$), 
and its high-$p_T$ triggering capabilities. The silicon trackers have excellent reconstruction performances
starting above $p_{T}>$~0.3~GeV/c with a low fake track rates in central $Pb+Pb$.
Within $|\eta|<$1, the $p_T$ resolution is better than 2\% up to the highest $p_T$ values reachable.
The leading hadron suppression can thus be measured with low uncertainties all the way up to
300~GeV/c, allowing us to clearly discriminate between various model predictions 
(Fig.~\ref{fig:RAA_SPS_RHIC_LHC}, left).

Full jet reconstruction in $Pb+Pb$ collisions can be performed in CMS using the ECAL and HCAL calorimeters.
Jets start to be distinguishable above the background at $E_T\sim$30~GeV and can be fully reconstructed 
above 75~GeV (efficiency and purity close to 100\%) with a good energy resolution (better than 15\%). 
The expected jet $E_{T}$ spectrum after one month of $Pb+Pb$ running (0.5 nb$^{-1}$) --
taking into account High Level Trigger bandwidths and quenching effects as implemented in HYDJET~\cite{Lokhtin:2005px} --
reaches up to $E_T\approx$~0.5~TeV in central $Pb+Pb$, and will give access to detailed differential 
studies of jet quenching phenomena (jet shapes, energy-particle flows within the jet, ...).
The possibility of CMS not only to fully reconstruct jets,
but to tag them with prompt $\gamma$ and/or $Z$ bosons~\cite{Loizides:2008pb}
and to carry out detailed studies in the $c,\,b$ quark sector will be
very valuable to clarify the response of strongly interacting matter to fast heavy-quarks, 
and will provide accurate information on the transport properties of QCD matter.\\



\section{Conclusions}
\label{sec:Conclusions}

The physics of parton propagation, interaction and fragmentation in both cold and hot
strongly interacting matter has been reviewed. The most recent theoretical and 
experimental results on hadron production in deep inelastic lepton-nucleus scattering (nDIS), 
high-energy hadron-nucleus ($h+A$) and heavy-ion ($A+A$) collisions as well 
as Drell-Yan processes in hadron-nucleus interactions have been discussed. The main motivation 
of such studies is to provide new insights on the mechanisms of parton fragmentation and 
on the space-time evolution of quark and gluon hadronisation. Those are complex dynamical processes 
characterised at some stage by energy scales for which the QCD coupling becomes large 
and thus cannot be completely addressed with standard perturbative QCD tools.

The operational approach consists in comparing the modifications in hadron production 
observed in a ``cold QCD matter'' environment (as found in nDIS or $p+A$ collisions) as well 
as in ``hot QCD matter'' (created in $A+A$ collisions) with respect to the measurements in the 
``QCD vacuum'' (in DIS off protons or in proton-proton collisions). The quantitative comparisons 
are carried out using various phenomenological ratios (hadron multiplicity ratio $R_M^h$ in nuclear DIS, or 
nuclear modification factor $R_{pA,AA}$ in nuclear collisions) as a function of the relevant 
variables: hadron momentum, fraction of the photon 
energy, virtuality, hadron flavour, size of the medium, etc.
In the case of cold nuclear matter, the detailed study of the observed
deviations with respect to vacuum production 
allows one to constrain  various models describing the space-time
evolution of parton fragmentation and, in the case of nucleus-nucleus
collisions, also to characterise the (thermo)dynamical properties of
the hot and dense QCD matter produced in the reaction. 

A detailed comparison of the kinematics and phase-spaces in 
nDIS experiments at CERN, HERA and JLab and in proton-nucleus
(FNAL, SPS, RHIC) and heavy-ions (SPS, RHIC) collisions, indicates
that the range of relevant  hadron energies measured is
comparable, $E_h\approx$~2--20~GeV, as well as typical values of fractional momentum
of the parent parton energy carried away by the leading hadron, $z\approx$~0.4--0.9.
The same is not 
true, however, for the parton virtualities $Q$ involved which are at
least a factor of five larger for the hadrons produced at RHIC top
energies ($Q\lesssim$~20~GeV) than those typical 
at HERMES/JLab ($Q\lesssim$~3~GeV). 
Therefore a direct comparison of hadron suppression data from nDIS to $h+A$ and $A+A$
collisions is not possible, and theoretical models are needed to connect both colliding systems. 

The main theoretical approaches to account for
parton propagation and fragmentation in QCD matter can be roughly
divided in two groups: (i) models rooted in perturbative-QCD
studies of non-Abelian parton energy loss via gluon Bremsstrahlung or
in-medium modifications of the DGLAP equations for gluon
radiation, followed by hadronisation of the final state hadrons
in the vacuum, and (ii) approaches mostly based on modifications of
the non-perturbative Lund string fragmentation model that account for
the interaction in the nuclear medium of a colorless pre-hadronic stage. 

The range of application of these two classes of models depends on the value of the parton
lifetime, i.e., by the time $t_{preh}$ needed to produce a prehadron.
Simple perturbative estimates 
indicate that the typical parton lifetime at HERMES or RHIC strongly depends on the
hadron mass. For pions, one obtains lifetimes $t_{preh}\approx 20-30$~fm,
larger than the nuclear radius, although modified versions of the Lund 
and dipole fragmentation models indicate smaller lifetimes, $t_{preh} \lesssim 6$~fm.
For heavier particles such as $K,\eta,p$, one finds $t_{preh}\approx 6-9$~fm, 
and even smaller values for heavy-flavor mesons, $t_{preh}(B,D)\lesssim 1$~fm. 
Therefore it is in general necessary to incorporate prehadron-medium
interactions in the theoretical models, and to look for experimental
observables able to distinguish them from purely partonic interactions
and measure $t_{preh}$.

A simple way to test the different models and their underlying
assumptions is to study hadron production in nDIS and Drell-Yan processes 
in $h+A$ collisions, where the traversed medium is static, if not 
uniform, and its properties are well known.
This cold nuclear matter information serves then as baseline for $A+A$ collisions, 
where a number of new mechanisms modifying high-$p_T$ hadron production
can be used to ``tomographically'' study the thermodynamical and transport 
properties of the hot and dense matter created.\\

{\it Lepton-nucleus DIS} \\

The recent HERMES and older EMC data on hadron attenuation in nDIS,
encoded in the $R_M^h(z,\nu;A)$ ratio,  
support a picture where the space-time evolution of the fragmentation process
is modified by the surrounding matter, with  a sizeable inelastic
cross section between the colorless pre-hadronic state and the nuclear medium. 
However, when the detailed geometry and finite size of the target
nucleus are taken into account, 
nDIS data can also be described in terms of parton 
energy loss alone. 

Additional observables 
sensitive to the underlying dynamics of the hadronisation
process have been investigated.
In particular, the observed slow variation of  $R_M$ with $Q^2$ 
disfavours mechanisms like partial deconfinement and puts additional
constraints on other theoretical models.
The rising and seemingly linear behavior of the hadron $p_T$-broadening 
$\vev{\Delta p_T}$ vs. $A$ at HERMES supports the conclusion that the
prehadron is formed close to or just
outside the nucleus surface (at $\vev{\nu}\approx 14$ GeV and $Q^2\approx 2.5$
GeV$^2$) and allows to estimate the prehadron production time.
The details of the $\nu$ dependence of the
$p_T$-broadening, and especially its linear increase with $Q^2$ 
are better interpreted in terms of medium-modified DGLAP evolution, and demand 
for NLO computations so far neglected. 
However, the shape of $\vev{\Delta p_T}$ vs. $z_h$ cannot be
qualitatively understood by current prehadron absorption or energy
loss computations, and is challenging theory models. \\

{\it Hadron-nucleus collisions} \\

The observation of a Cronin enhancement ($R_{pA}>$~1) of the hadron yields
in the range $p_T = 2-8$~GeV/c is a characteristic feature of high-energy proton-nucleus 
collisions at fixed target energies (FNAL, SPS, HERA-B). At RHIC top energies, 
intermediate-$p_T$  proton, but barely meson, production is enhanced in $d+Au$
collisions and the typical $R_{dA}$ maximum around 2~GeV/c is progressively suppressed 
as the rapidity increases. The usual interpretation of the Cronin effect is based 
on multiple scatterings of the parton prior to its fragmentation.
Both initial- and final-state interactions must play a role in order to explain the 
various experimental observations. On the one hand, the disappearance of the enhancement at 
forward rapidities is easily explained by non-linear QCD evolution of the initial-state
parton distributions (though may also be due to effects occurring at the edge of phase space).
On the other hand, DY data suggest that parton rescatterings contribute only a small amount 
to the hadron enhancement and, in addition, final-state coalescence of the scattered 
parton with other soft partons in the nuclear environment naturally explains the larger 
enhancement observed for baryons compared to mesons. The Cronin effect helps thus to 
study the hadronisation mechanism in hadronic interactions at intermediate $p_T$, which 
may be dominated by parton recombination rather than by independent parton fragmentation.
Detailed control of the role of all these mechanisms is important to help 
identify the onset of high-$p_T$ hadron suppression in $A+A$ collisions 
in the $\sqrt s=10-100$ GeV range.\\

{\it  Nucleus-nucleus collisions} \\

The two most notable experimental results in $Au+Au$ collisions at
RHIC are (i) the observed factor of $\sim$5 suppression of high-$p_T$ hadrons,
and (ii) the strongly distorted azimuthal distributions of secondary hadrons
emitted in the away-side hemisphere of a high-$p_T$ trigger hadron.
Their properties such as magnitude and light flavour ``universality'',
and their dependence on $p_T$, centrality, path-length, and $\sqrtsnn$
are in quantitative agreement with the predictions of models based on
non-Abelian gluon radiation off hard scattered partons traversing the dense QCD medium.
The confrontation of these models to the data permits to derive the initial
gluon density $dN^g/dy\approx$ 1400 and transport coefficient
$\hat{q} =\mathscr{O}($10~GeV$^2$/fm$)$ of the produced medium at RHIC.
Yet, other observations such as the same suppression factor for light-
and heavy-quark mesons, and the less suppressed baryon compared to
meson production indicate that the standard factorisation assumption
of vacuum hadronisation after in-medium radiation, implicit in all
parton energy loss formalisms, may well not hold.
At lower collision energy, with shorter propagation times and a less dense
final state medium, hadron suppression in the cold nuclear targets may compete 
with that originating from the hot medium.\\

{\it Outlook } \\

We have reviewed a set of observables and new or improved
measurements that can answer several of the open questions mentioned here
and discussed in detail in this review. They include multi-dimensional
$p_T$-broadening measurements, hadron-hadron and hadron-photon
correlations, heavy flavor 
and jet shape modifications in nDIS, improved DY measurements and
large-$p_T$ hadron spectra in $h+A$ collisions, heavy quark tagging in
$A+A$ collisions. These observables can be addressed at new facilities
planned for the near or medium-term future, such as RHIC-II, JLab12,
and the Electron-Ion Collider (which in particular will open the study of
purely partonic in-medium processes in DIS). They can also be addressed 
by new experiments at existing
facilities, such NA61 at SPS, or by a creative use of data taken at
closed experiments such as NA60 or SELEX, among others.
The combination of new results, theoretical developments, and an open
mind to combine information from traditionally different fields such
as DIS and hadronic collisions will doubtlessly lead to a more
profound knowledge of parton propagation and fragmentation,
quark and gluon hadronisation, and the properties of strongly 
interacting QCD matter.

\acknowledgments

A.A. wishes to thank J.~Morf\'in for helpful discussions and
encouragement, and acknowledges support by DOE contract
No. DE-AC02-06CH11357, DOE contract
No. DE-AC05-06OR23177 (under which Jefferson Science Associates, LLC
operates Jefferson Lab), and NSF award No. 0653508. 
F.A. would like to thank the hospitality of CERN TH division, where
part of this work was carried out. 
D.d'E. acknowledges support by 6th EU Framework Programme under contract
MEIF-CT-2005-025073.
W.B. acknowledges support from DOE contract
No. DE-AC05-06OR23177 (under which Jefferson Science Associates, LLC
operates Jefferson Lab) and Conicyt/Fondecyt Grant 1080564.


\bibliographystyle{myJHEP-3} 
\bibliography{biblio}

\end{document}